\documentclass[a4papper,10pt,twoside,openright]{report}
\usepackage{suthesis-2e}
\usepackage{times}
\usepackage{amssymb}

\title{SPONTANEOUS SYMMETRY BREAKING ON NON-ABELIAN KALUZA-KLEIN AND
  RANDALL-SUNDRUM THEORIES}
\author{Rui Fernando Lima Matos}

\begin{document}

\newpage
\dept{Physics}
\copyrightfalse
\thesiscopyrightfalse
\figurespagefalse
\tablespagefalse

\beforepreface

\newpage

\begin{flushright}
 $ $
\vspace{2.in}

\textit{A Meus Pais,}

\vspace{.1in}
\textit{Fernando e Maria da Gl\'oria}
\end{flushright}
\newpage
\prefacesection{Acknowledgements}


I wish to acknowledge with much gratitude the guidance and inspiration of my
thesis advisor, Dr. Sergio Zerbini, and the warm hospitality of the Laboratorio di Fisica
Teorica at the Universit\`a degli Studi di Trento.

\vspace{.1in}
 
\noindent I would like to thank the precious contribution of Dra. Isabel Lopes without which this project wouldn't be accomplished.



\vspace{.1in}



\noindent I express my gratitude to Matilde, who has read and corrected the
early manuscript and that with her permanent encouragement has largely contributed to
the development of this work. 

\vspace{.1in}

\noindent Finally, I wish to express my most debt to my parents and brother, for their love and support.

\newpage
$ $

\newpage
\begin{center}
SPONTANEOUS SYMMETRY BREAKING ON NON-ABELIAN KALUZA-KLEIN AND
RANDALL-SUNDRUM THEORIES
\end{center}
\vspace{.1in}
\begin{center}
Rui Fernando Lima Matos

\textit{Departamento de F\'isica, Universidade de Coimbra, Portugal}
\end{center}
\vspace{.5in}
\begin{center}
ABSTRACT
\end{center}

\vspace{.3in}

The fibre bundle formalism inherent to the construction of non-abelian
Kaluza-Klein theories is presented and its associated dimensional reduction
process analysed: is performed the dimensional reduction of
$\mathcal{G}$-invariant matter and gauge fields over a multidimensional
universe; the harmonic decomposition of non-symmetric fields over its
internal space established and their dimensional reduction done. The spontaneous
compactification process is presented.

It is shown that during the dimensional reduction process two types of
symmetry breaking can occur: a geometric followed by a
spontaneous symmetry breaking. This last is connected to a scalar field
resulting from the dimensional reduction process itself. We determine
explicitly the scalar potential form leading to that symmetry breaking for the case
in which the internal space is symmetric and an analysis for the general case
is performed. 

The principal problems found in the construction of
realistic Kaluza-Klein models are discussed and some solutions reviewed. The
hierarchy problem is examined and its solution within Kaluza-Klein models with
a large volume internal space considered. 
As an alternative solution, the Randall-Sundrum model is presented and its
principal properties analysed. In special, the stability of the hierarchy is studied.

\vspace{.6in}

\newpage
\begin{center}
ROTTURA SPONTANEA DI SIMMETRIA IN TEORIE DI KALUZA-KLEIN NON-ABELIANE E DI RANDALL-SUNDRUM
\end{center}
\vspace{.1in}
\begin{center}
Rui Fernando Lima Matos

\textit{Departamento de F\'isica, Universidade de Coimbra, Portugal}
\end{center}
\vspace{.5in}
\begin{center}
SOMMARIO
\end{center}

\vspace{.3in}

Il formalismo dei spazi fibrati inerente alla construzione delle teorie
di Kaluza-Klein non-abeliane \`e presentato ed il suo processo di riduzione
dimensionale collegato analizzato: \`e effettuata la riduzione dimensionale dei
campi di materia e di gauge $\mathcal{G}$-invariabili in un'universo
multidimensionale; 
la decomposizione armonica dei campi non-simmetrici nello spazio interno di quello
\`e stabilita e la loro riduzione dimensionale fatta. Il processo di
compattificatione spontanea \`e presentato. 

\`E demonstrato che durante il processo di riduzione dimensionale due tipi di rotture di
simmetria possono accadere: una rottura di simmetria geometrica seguita da una
spontanea. Questa ultima \`e collegata ad un campo scalare derivato dal
processo di riduzione dimensionale stesso. Si determina esplicitamente la
forma del potenziale scalare che conduce a quella rottura simmetria nel caso
in cui lo spazio interno \`e simmetrico e un'analisi per il caso generale è
effettuata. 

I problemi principali trovati nella costruzione di modelli realistichi del
tipo Kaluza-Klein sono discussi ed alcune soluzioni sono riviste. 
Il problema della gerarchia \`e esaminato e la sua soluzione all'interno dei
modelli di Kaluza-Klein con grande volume di spazio interno 
\`e considerata. Come soluzione alternativa, il modello di
 Randall-Sundrum \`e presentato e le sue propriet\`a principali sonno analizzate. 
In speciale, la stabilit\`a della gerarchia \`e studiata.

\vspace{.6in}
PAROLE CHIAVE: Riduzione dimensionale; Compattificatione spontanea; 
Rottura di simmetria; Model Building; Gerarchia

\newpage
\begin{center}
QUEBRA ESPONT\^ANEA DE SIMETRIA EM TEORIAS DE KALUZA-KLEIN N\~AO-ABELIANAS
E DE RANDALL-SUNDRUM
\end{center}
\vspace{.1in}
\begin{center}
Rui Fernando Lima Matos

\textit{Departamento de F\'isica, Universidade de Coimbra, Portugal}
\end{center}
\vspace{.5in}
\begin{center}
RESUMO
\end{center}

\vspace{.3in}

O formalismo de espa\c cos fibrais inerente \`a constru\c c\~ao de teorias de
Kaluza-Klein n\~ao-abelianas \'e apresentado e o seu processo de redu\c c\~ao
dimensional associado anali-sado: \'e efectuada a redu\c c\~ao dimensional de
campos de mat\'eria e de gauge $\mathcal{G}$-invariantes sobre um universo
multidimensional; a decomposi\c c\~ao harm\'onica de campos assim\'etri-cos
sobre o espa\c co interno do \'ultimo \'e estabelecida e a sua redu\c c\~ao
dimensional feita. O processo de compactifica\c c\~ao espont\^anea \'e apresentado.

Mostra-se que durante o processo de redu\c c\~ao dimensional dois tipos de
quebra de simetria podem ocorrer: uma quebra de simetria geom\'etrica seguida
por uma espont\^a-nea. Esta \'ultima est\'a ligada a um campo escalar
resultante do pr\'oprio processo de redu\c c\~ao dimensional. Determina-se
explicitamente a forma do potencial escalar que d\'a origem \`a quebra de
simetria no caso em que o espa\c co interno \'e sim\'etrico e uma analise para
o caso geral \'e efectuada.

S\~ao discutidos os principais problemas encontrados na constru\c c\~ao de
modelos realistas do tipo Kaluza-Klein e algumas solu\c c\~oes s\~ao
revistas. O problema da hierarquia \'e estudado e a sua solu\c c\~ao no seio
de modelos \textit{\` a la} Kaluza-Klein com grande volume de espa\c co
interno consideradas. Como solu\c c\~ao alternativa, \'e apresentado o modelo
de Randall-Sundrum e as suas principais propriedades s\~ao analisadas. Em
especial, o problema da estabilidade da hierarquia \'e abordado.

\vspace{.6in}

PALAVRAS-CHAVE: Redu\c c\~ao dimensional; Compactifica\c c\~ao espont\^anea;
Quebra de simetria; Model Building; Hierarquia

\afterpreface


\chapter{INTRODUCTION}

\section{Early Kaluza-Klein Theories}

Space and time are usually considered to be fundamental concepts, to which
the roots of experience are solidly fixed. It is however through the
experience that their structure is perceived, and their definition
formed. Initially distinguished and relegated to the passive metaphysical domain, they
have regained their respect in Physics as a step to the unification was given:
the construction of the Electromagnetic Theory. The need to embrace the two
concepts into one was then foreseen and their properties reviewed. Not long
was needed to expect till that recent born hybrid concept called
space-time to have its ones
privileges in the Physical world: 
within General Relativity it
gains a freedom of movements, interacting with itself and with matter, curving
and distorcing, trebling with the common sense causality. Gravity was thus exchanged by a richer
space-time structure as the old rigidity property of space and time fell into oblition.
The time was then propitious for time and space together to achieve a new
level of sofistication. With the work of Kaluza \cite{kaluza}, space-time
arrives at a cult stage, as gravity and electromagnetism
are unified in the emptiness of pure geometry. Light becomes curvature,
gravity in the Einstein sense, of
an higher dimensional space-time left to darkness. 

Klein \cite{klein}
rediscovered the Kaluza theory, noticing the quantisation of the electric
charge as a result of the compactification of the circle and 
being the first to perform an harmonic analyse of a non-symmetric
field in a Kaluza theory context. His work was then continued by others
\cite{einstein,hoffmann1,jordan,pais,pauli,schmutzer1,schmutzer2}. 

\section{Non-Abelian Kaluza-Klein Theories}

After the development of the Yang-Mills theory in 1954 and its use in the
construction of the electroweak theory in the 60's and 70's it became clear that an
extension of the abelian, five-dimensional, Kaluza-Klein theory to the
non-abelian domain, unifying not only the electromagnetism to gravity but also
all the other forces in the world, should be done
\cite{deWitt1}. Kerner constructs such a theory \cite{kerner}, starting with
 a multidimensional universe $\mathcal{M}$ with a fibre bundle structure, such
 that its internal space is Lie group $\mathcal{G}$, defined over a
 four-dimensional manifold. 
By considering only $\mathcal{G}$ invariant metrics defined on
$\mathcal{M}$ and taking an Hilbert-Einstein action on the total space, he
arrives to an Hilbert-Einstein-Yang-Mills action in the base space, thus
unifying both theories. A more general construction was then made by Cho and
Freund \cite{cho}, where a final Hilbert-Einstein-Yang-Mills-Scalar theory
was derive. With \cite{luciani} the internal space of the multidimensional
universe was generalised to a homogeneous space. A formal dimensional
reduction process for non-symmetric fields in a non-abelian Kaluza-Klein
theory was firstly given in \cite{jadczyk2} and an extensive analyse of the
Kaluza-Klein scheme within the fibre bundle formalism was performed in
\cite{coquereaux1,coquereaux2,jadczyk1}. Modern applications can
be found in \cite{mourao,kapetanakis1}.

As such progresses to a unifying theory within the Kaluza-Klein scheme were
being made, a new type of theories were constructed: theories where mass could
be spontaneously generated by a spontaneously symmetry breaking of a gauge
group; theories where a new kind of symmetry - supersymmetry - could be
found, not making distinction between bosons and fermions; theories where
its objects were not punctual particles - local fields - but extensive domains - strings -,
as a solution to the non-renormalizable character of the most important
theories - such as General Relativity.
After the development of that theories, the pretending of the Kaluza-Klein theories as unifying theories could
not be let endure: supergravity and superstrings where better candidates to
that position. With the works of Witten \cite{witten,witten2} it was
shown that one could never construct a realistic model within the usual
Kaluza-Klein scheme since the left-right chiral asymmetry characteristic of
the Standard Model (SM) could not be achieved by dimensional
reduction. Several solutions where proposed
\cite{wetterich,witten2,randjbar}, but all them radical or
particular, decreasing in beauty in relation to the initial theory.

With the works of Cremmer and Scherk \cite{cremmer} and Luciani
 \cite{luciani} a new mechanism - that would be useful later for the 
construction of eleven-dimensional supergravity
 \cite{cremmer2,cremmer3,freund} - was developed: the
 spontaneous compactification. With this dynamical mechanism, by considering
 the action of Yang-Mills fields on a multidimensional universe, a manifold
 with no special structure could 
 achieve a fibre bundle structure, being then capable of being reduced through
 the Kaluza-Klein dimensional reduction process.

\section{Symmetry Breaking on non-Abelian Kaluza-Klein Theories}


The most important charateristic of Kaluza-Klein theories is not so much its
unifying character but the way as the unification is reached: all forces
in the four-dimensional observed world find their origin in the curvature of a
multidimensional universe freed from any force. But in the absence of any force,
the concept of mass becomes meanless since its measure becomes impossible. If
a matter field is defined over a free  multidimensional universe, it must be
massless. The mass must be then acquired by some dynamical process: the
spontaneous generation of mass by symmetry breaking becomes then a need for the
coerence of Kaluza-Klein theories where massive matter
fields are defined.


It is the purpose of this work to explain how can this symmetry breaking
occur, how can it be contained in the Kaluza-Klein theory itself, how can find
its origins on the dimensional reduction process.

%
%

\section[The Hierarchy
Problem and the Randall-Sundrum...]{The Hierarchy Problem and the Randall-Sundrum Solution}


The SM provides a remakably successful description of known phenomena,
describing strong and electroweak interactions in an elegant manner. However,
there are, from a theoretical point of view, some serious problems within this
model. The one that will be of interest to us is the hierarchy problem between the
electroweak scale and the Planck scale, that disables any attempt to
incorporate gravity into the model. Several solutions were presented:  
the introduction of supersymmetry (SUSY) at the electroweak scale was one
of them. By solving the hierarchy problem, the SUSY aproach would, however, inevitabily
introduce many extra fields to the SM, giving then rise to another problem: a possible
confict with the experimental results.

In recent years a new solution was given \cite{arkani}: by considering a
special Kaluza-Klein model, where only gravity could propagate into the
extra-dimensions, the hierarchy could be solved by taking for the only
fundamental scale in nature, the electroweak scale. The Planck scale should then be a
derivate scale, deriving precisely from the fact that gravity can propagate into
the extra-dimensions as  matter is retained to a four-dimensional world. 
This model presents however the inconvenient that the
internal space should be large when compared with the previous Kaluza-Klein
models, presenting new difficulties in the confrontation with the experience.

Emerging as an effective field theory \cite{lukas} from the M-theory
\cite{horava1,horava2,witten3,witten4,witten5}, the
Randall-Sundrum model \cite{randall1} as found a simplier description of the
world where the hierarchy problem is trivially solved. The four-dimensional
world perceptible to us is nothing but a brane embebed in a five-dimensional
universe with an orbifold structure. The hierarchy will, as in the previous
model, be connected to the fact that SM fields are located in the brane and
gravity propagates through the bulk of that multidimensional universe. The
stability of such construction as been discussed in \cite{goldberger2,grzadkowski} and the
need of the introduction of bulk fields presented. Quantum corrections due
to the Casimir effect of the extra, compact, dimensions can also be used to
stabilize the model, and so to stabilize the hierarchy \cite{garriga2,hoffmann1,gen}.
In \cite{goldberger}\cite{milton} the problem of renormalizating localized divergences in
brane models in a six-dimensional orbifold was studied.

In \cite{gomez} a new brane solution was given, having for background a bulk with a black
hole. This was extended in \cite{kraus} and \cite{csaki} to the case of an
AdS-Reissner-Nordstrom bulk's black hole and in \cite{brower} to the
six-dimensional bulk case.

\section{Dissertation Structure}

This dissertation is structured in the following way.
In the Chapter 2, a presentation of the basic formalism needed to the
development of any non-abelian Kaluza-Klein theory is made. Concepts as fibre bundle,
infinitesimal connection and absolute differential are defined.
Chapter 3 is dedicated to the dimensional reduction process inherent to
Kaluza-Klein theories. In a first stage, $\mathcal{G}$-invariant gravity,
gauge and matter fields are dimensionally reduced over a principal fibre
bundle and over a fibre bundle with an homogeneous internal space. Then an harmonic
abstract analyse of non-symmetric fields over a Lie group and a homogeneous
space is performed. Being written in terms of their $\mathcal{G}$-invariant
Fourier components, a non-symmetric field can then be subjected to dimensional
reduction.
In Chapter 4, the classification of the types of symmetry breakings that can
occur at the dimensional reduction process is performed and their analyse
made. 
The conditions for the construction of realistic models are discussed in
Chapter 5, and its main problems presented. One of them, the hierarchy problem,
naturally conduces to the introduction of the Randall-Sundrum model, theme of
the next chapter.
In the Chapter 6, the Randall-Sundrum model is reconstructed and its main properties
studied, in particular, the stability of the hierarchy within this model is discussed. 
Finally, a summary of the more important results is presented in Chapter 7.



\chapter{THE FORMALISM}

\textit{The fibre bundle formalism needed for the construction of non-Abelian Kaluza-Klein theories
is presented.}

\section{The Fibre Bundle Formulation}

\subsection{Basic Definitions}

Let $\mathcal{M}$ be a $C^\nu$-manifold $m$-dimensional with a 
differentiable fibre bundle structure $\mathcal{M}(V_n, M, \mathcal{G}, \pi, \Phi)$.
$V_n$ is the base space, quotient space $\mathcal{M}/R$, where R is
some equivalence relation on $\mathcal{M}$, being $V_n$ a $C^\nu$-manifold
$n$-dimensional. $M$ is also a $C^\nu$-manifold and $\mathcal{G}$ is a
Lie group that acts effectively on $M$, i.e., where there is defined a left action over $M$:
\begin{equation}
        \delta:\mathcal{G}\times M\longrightarrow M.
\end{equation}
Restrictions of $\delta$ to the first and to the second component of $\mathcal{G}\times M$ are denoted, respectively, by
\begin{eqnarray}
        \delta_g:M\longrightarrow M,\\
        \delta_y:\mathcal{G}\longrightarrow M,
\end{eqnarray}
where $y\in M$ and $g\in\mathcal{G}$, being the restrictions $\delta_y$ and $\delta_g$ defined by 
$\delta_y(g)=\delta(g,y),$ $g\in\mathcal{G}$ and $\delta_g(y)=\delta(g,y),$ $y\in M$. We will, in what follows,
simplify the notation of the action of $\mathcal{G}$ on $M$ by making $gy\equiv\delta(g,y)$, $g\in\mathcal{G},$
 $y\in M$.

Finally, $\pi$ is the projection of the 
bundle and $\Phi$ is a homomorphism family such that:

\begin{enumerate}
\item All $\phi\in\Phi$ are homomorphisms of $U\times M$ in $\pi^{-1}(U)$,
  where $U$ is an open set of $V_n$, called the domain of $\phi$, that we
  will represent by $\phi_U$;
\item If $x\in U$, $y\in M$, then $(\pi\circ\phi_U)(x,y)=x$;
\item If $\phi_U,\phi_V\in\Phi$ and $x\in U\cap V$, $\phi_U$ and $\phi_V$
  define homomorphisms of $\{x\}\times M$ and so of $M$ in $M_x=\pi^{-1}(x)$,
  homomorphisms that we will represent by $\phi_{Ux}$ and $\phi_{Vx}$; there is an
  element $g\in\mathcal{G}$ such that the automorphism $\phi_{Vx}^{-1}\circ \phi_{Ux}$ of
  $M$:
  $$\phi_{Vx}^{-1}\circ \phi_{Ux} = g;$$
\item The domains of the elements of $\Phi$ cover $V_n$.
\end{enumerate}

The fibres $M_x$, manifolds embedded in $\mathcal{M}$, are $C^\nu$-homomorphics
with $M$ by $\phi_{Ux}$, $\phi_{Vx}$ and so on.

Let $z$ be a point of $\mathcal{M}$ such that $\pi(z)=x\in V_n$. Then we can
take $z=\phi_{Ux}(y)$, with $y\in M$. But $\phi_{Vx}^{-1}\circ \phi_{Ux}=g$, for $g\in\mathcal{G}$, 
so $\phi_{Vx}^{-1}\circ \phi_{Ux}(y)=gy$ and $z=\phi_{Ux}(y)=\phi_{Vx}(gy)$.

A particular case of a fibre bundle is that in which $M$ and $\mathcal{G}$
are the same manifold, and where $\mathcal{G}$ acts in itself by 
left-translation. In this case the bundle is called a principal fibre bundle
and will be represented by $\mathcal{M}(V_n,\mathcal{G},\pi,\Phi)$. 
Most of the literature on non-abelian Kaluza-Klein theories confer this special structure 
to the $m$-dimensional space-time, being $V_n$, the base space, a 4-dimensional manifold ($n=4$) and
the structure group $\mathcal{G}$ a compact Lie group.
\cite{kerner,cho}.
In recent years a generalisation of this scheme has been proposed 
\cite{mourao,coquereaux2,coquereaux3,witten}. 
The spacetime
presents there a fibre bundle structure such that the internal space, $M$ in our notation, is an homogeneous
space. 
We will study the two models, starting by the first.

\subsection{The Principal Fibre Bundle $\mathcal{M}(V_n,\mathcal{G},\pi,\Phi)$}\label{principalfibrebundle}

Let us consider a principal fibre bundle $\mathcal{M}(V_n,\mathcal{G},\pi,\Phi)$ and let 
$\phi_{Ux}$ and $\phi_{Vx}$ be two homomorphisms of $\mathcal{G}$ in $G_x$, where
$x\in V_n$. A point $z\in\mathcal{M}$ over $x$, that is, a point in the
fibre $G_x$, can be represented by $z=\phi_{Ux}(\gamma)$, for some
$\gamma\in\mathcal{G}$. But we could also use $\phi_{Vx}^{-1}\circ \phi_{Ux} = g$, 
$g\in\mathcal{G}$ to obtain $z=\phi_{Vx}(g\gamma)$.
Let $j\in\mathcal{G}$ be another element of $\mathcal{G}$. Then we define 
a right-translation over $\mathcal{M}$
\begin{equation}
        zj=\phi_{Ux}(\gamma j).
\end{equation}
This definition is independent of the choice of the homomorphism, so we have a right action of the group $\mathcal{G}$
defined in the total space $\mathcal{M}$\footnote{We will use the same symbol $\Phi$ of the homomorphism family
of the bundle $\mathcal{M}(V_n,\mathcal{G})$ for the action of $\mathcal{G}$ on $\mathcal{M}$.}:
\begin{equation}
        ^{\leftarrow}{\Phi}:\mathcal{M}\times\mathcal{G}\longrightarrow\mathcal{M}.
\end{equation}
In a similar way, we can define a left-translation over $\mathcal{M}$:
\begin{equation}
        jz=\phi_{Ux}(j\gamma),
\end{equation}
and so we have a left action of $\mathcal{G}$ on $\mathcal{M}$:
\begin{equation}
        ^{\rightarrow}\Phi:\mathcal{G}\times\mathcal{M}\longrightarrow\mathcal{M}.
\end{equation}
Restrictions of the two components of these two actions will be denoted by $^{\leftarrow}\Phi_z$, $^{\leftarrow}\Phi_g$  
and $^{\rightarrow}\Phi_g$, $^{\rightarrow}\Phi_z$, respectively.
Since we have defined two actions, a left and a right, the symmetry group of the internal space will 
not be $\mathcal{G}$ but $\mathcal{G}\times\mathcal{G}$. We will use the symbol $\Phi$ to represent the action of 
$\mathcal{G}$ in $\mathcal{M}$ when the left or right character of the same is not important for the considerations
that will be done. $\Phi_g$ and $\Phi_z$ will represent, as usually, the restrictions of the two components of
this action.


\subsubsection{The Vertical Vector Space}

Let $T_z\mathcal{M}$ be the tangent vector space to $\mathcal{M}$ in $z$
and let $T_z G_z=\mathcal{V}_z$ the tangent vector subspace to the fibre
$G_z$ in the same point $z$. The vectors of $\mathcal{V}_z$ are said to
be vertical.

Since $\Phi_g$ preserves each fibre $G_x$, it transforms each tangent vector to $G_x$
in $z$ to a tangent vector in $zg$ to the same fibre. So if 
$\tau\in\mathcal{V}_z$ is vertical
so it is $\Phi_g^{*}\tau$, that is, $\Phi_g^{*}\tau\in\mathcal{V}_{zg}$.

\paragraph{Fundamental Vector Fields.}

Let $\tau\in\mathcal{V}_z$ be a vertical vector in $z=\phi_{Ux}(\gamma)$, for 
$\gamma\in\mathcal{G}$. Taking the differential of $\phi_{Ux}^{-1}$, we can 
establish a correspondence between $T_z\mathcal{M}$ and $T_\gamma\mathcal{G}$:

$$\phi_{Ux,z}^{-1*}:T_{z=\phi_{Ux}(\gamma)}\mathcal{M}\rightarrow T_\gamma\mathcal{G}.$$

And so, $\phi_{Ux,z}^{-1*}(\tau)\in T_\gamma\mathcal{G}$. Let us take 
$\theta_\gamma=\phi_{Ux,z}^{-1*}(\tau)$ for $z=\phi_{Ux}(\gamma)$. 

On the other hand, 

$$\phi_{Vx,z}^{-1*}:T_z\mathcal{M}\rightarrow T_{g\gamma}\mathcal{G},$$
and so $\phi_{Vx,z}^{-1*}(\tau)\in T_{g\gamma}\mathcal{G}$. It comes that 
$\theta_{g\gamma}=\phi_{Vx,z}^{-1*}(\tau)$.
But $\phi_{Vx}^{-1}\circ \phi_{Ux} = g$, and so $\phi_{Vx}^{-1}=g\phi_{Ux}^{-1}$. Taking the differential,
one obtains $\phi_{Vx,z}^{-1*}=g\phi_{Ux,z}^{-1*}$, and so,

$$\theta_{g\gamma}=g\theta_{\gamma},$$
that is, one obtains a left-invariant vector field in $\mathcal{G}$. In special, 
we can take $\gamma=e$, the unity of the group. Then

$$\theta_g=g\theta_e=g\lambda\equiv \frac{d}{dt}\left(g e^{t\lambda}\right),$$
where we have made $\theta_e=\lambda\in L_\mathcal{G}$, being $L_\mathcal{G}$ the Lie algebra defined
in $\mathcal{G}$.
In this way, the vector field left-invariant $\theta_g$ can be determined once
we know $\lambda\in L_\mathcal{G}$. To $\lambda$ one calls the element of the Lie algebra $L_\mathcal{G}$
generator of the vertical vector $\tau\in\mathcal{V}_{z=\phi_{Ux}(g)}$.
We can also follow the inverse path as we will see. Given a $\lambda\in L_\mathcal{G}$, we can deduce
a left-invariant vector field in $\mathcal{G}$ and from this a vertical vector in
$\mathcal{M}$.

In a similar way we can define a right-invariant vector field in $\mathcal{G}$, that is, a field $\theta_{\gamma}$
such that
$$\theta_{\gamma g}=\theta_{\gamma}g,$$
or, taking $\gamma=e$,
$$\theta_{g}=\lambda g \equiv \frac{d}{dt}\left(e^{t\lambda} g\right),$$
and from this field derive a vertical vector in $\mathcal{M}$.

We can also derive a left-invariant vertical vector field on $\mathcal{M}$. In fact, using the action $\Phi$ of
the group $\mathcal{G}$  on $\mathcal{M}$, one can associate to every element $\lambda\in L_\mathcal{G}$
a tangent vector to $\mathcal{M}$ at a point $z$:
$$\tau(z)=\Phi^*_z(\lambda)=\frac{d}{dt}(e^{t\lambda}z)|_{t=0}.$$

Let $\{\lambda_a\}_{a=n,\cdots,m-1}$ be a basis of the Lie algebra $L_\mathcal{G}$ and let $C_{ab}^c$ be
the structure constants of $\mathcal{G}$:
\begin{equation}\label{comut}
  [\lambda_a,\lambda_b]=C_{ab}^c\lambda_c, 
\end{equation}
where $[,]$ are the Lie brackets of the Lie Algebra $L_\mathcal{G}$. To each
element $\lambda_a$, $a=n,\cdots,m-1$,
we can associate a left-invariant vector field (or a left fundamental field):
$$\hat e^L_a=\frac{d}{dt}(e^{t\lambda_a}z)|_{t=0},\;\;\;(a=n,\cdots,m-1).$$
This left fundamental fields will then obey the following commutation rule
\begin{equation}
  [\hat e^L_a,\hat e^L_b]=-C_{ab}^c \hat e^L_c.
\end{equation}

An right-invariant vertical vector field on $\mathcal{M}$ could be derived in a similar way:
$$\hat e^R_a=\frac{d}{dt}(ze^{t\lambda_a})|_{t=0},\;\;\;(a=n,\cdots,m-1),$$
and
\begin{equation}
  [\hat e^R_a,\hat e^R_b]=C_{ab}^c \hat e^R_c.
\end{equation}

We have the important property that $\{\hat e^L_a\}_{a=n,\cdots,m-1}$ (and $\{\hat e^R_a\}_{a=n,\cdots,m-1}$) constitute
a basis of the vertical tangent space $\mathcal{V}_z$ at a point $z\in\mathcal{M}$.


\subsubsection{Infinitesimal Connection on a Principal Fibre Bundle}

To define an infinitesimal connection $\mathcal{H}_z$ at a point $z\in\mathcal{M}$
we must specify a field of vector subspaces $\mathcal{H}_z$ of $T_z\mathcal{M}$ that
satisfies the following conditions:

\begin{enumerate}
\item $\mathcal{H}_z$ is complementary to $\mathcal{V}_z$: and so all $\tau\in T_z\mathcal{M}$
  can be decomposed in a vertical part $\mathcal{V}\tau$ plus a horizontal part 
  $\mathcal{H}\tau\in\mathcal{H}_z$;
\item $\mathcal{H}_z$ depends differentially on $z$;
\item $\mathcal{H}_z$ it's $\mathcal{G}$-invariant in the sense that:
$$\mathcal{H}_{zg}=\Phi_g^*\mathcal{H}_z.$$
\end{enumerate}

To do this we define a $L_\mathcal{G}$-valued 1-form $\omega$ such that 

\begin{enumerate}
\item If $\tau$ is vertical, $\omega(\tau)$ is the element of $L_\mathcal{G}$ generated by $\tau$;
\item $\omega$ depends differentially on $z$;
\item $\omega(\Phi_g^*\tau)=Ad(g^{-1}) \omega(\tau)$.
\end{enumerate}

One can prove that for $\tau\in\mathcal{H}_z$, $\omega(\tau)=0$.

\paragraph{Local Sections.}

Let $\{U_i\}_{i=1,2,\cdots}$ be a covering of the base space $V_n$. For each set $U_i$ let
us suppose that there is defined a section $\sigma_i:U_i\subset V_n\longrightarrow\mathcal{M}$.
Since $\pi\circ\sigma_i=id_{V_n}$, we shall have, for $x\in U_i\cap U_j$, that $\sigma_i(x)$
and $\sigma_j(x)$ should be over the same fibre $G_x$. The two sections can then be connected
by an element $g_{ij}(x)$ of $\mathcal{G}$ (no sum over the indices is made):
\begin{equation}\label{sigma_i}
        \sigma_i(x)=g_{ij}(x)\sigma_j(x).
\end{equation}
Let $v\in T_x V_n$ be a tangent vector to $V_n$ at $x$. Then to this vector we can associate
the vectors $\tau_i=\sigma_{i,x}^*v\in T_{\sigma_i(x)}\mathcal{M}$ and 
$\tau_j=\sigma_{j,x}^*v\in T_{\sigma_i(x)}\mathcal{M}$. By taking the differential of (\ref{sigma_i}),
and noting that $g_{ij}(x)\sigma_j(x)=\phi_{U_jx}(g_{ij}\gamma)$ where $\sigma_j(x)=\phi_{U_jx}(\gamma)$,
$\phi_{U_jx}\in\Phi$ and $\gamma\in\mathcal{G}$,
we shall have
$$\sigma_{i,x}^*=g_{ij}(x)\sigma_{j,x}^*+dg_{ij}\sigma_j(x),$$
and using $\sigma_j(x)=g_{ij}^{-1}(x)\sigma_i(x)$,
\begin{equation}
        \sigma_{i,x}^*=g_{ij}(x)\sigma_{j,x}^*+g^{-1}_{ij}(x)dg_{ij}\sigma_i(x).
\end{equation}
By applying $\sigma_{i,x}^*$ to $v$ we get
\begin{equation}\label{tau_i}
        \tau_i=g_{ij}\tau_j+g^{-1}_{ij}dg_{ij}\sigma_i(x).
\end{equation}

Let us define a $L_\mathcal{G}$ valued local one-form $A_i$ over $U_i\subset V_n$ by
\begin{equation}
        A_i(v)=\omega(\tau_i),
\end{equation}
where $\omega$ is the infinitesimal connection form defined on the principal fibre bundle $\mathcal{M}$.
We shall call $A_i$ the local induced connection form on $U_i$.
By (\ref{tau_i}) we should have, for two local induced connection forms on two subsets $U_i$ and $U_j$ of
$V_n$, the following relation
\begin{equation}\label{omega_j}
        A_j=Ad(g^{-1}_{ij})A_i+g_{ij}^{-1}dg_{ij},
\end{equation}
where $A_j$ and $A_i$ are the local induced connection forms over $U_j$ and $U_i$, respectively,
and $g_{ij}$ is the element of $\mathcal{G}$ that connects the two sections defined over $U_i$ and 
$U_j$. These local induced connection forms will be used when we discuss the Yang-Mills potential.

The importance of this local induced connection forms is that not only an infinitesimal connection
form $\omega$ over the principal fibre bundle $\mathcal{M}$ can define them all, once given
a family $\{U_i,\sigma_i\}$, with $\{U_i\}$ a covering of $V_n$ and $\{\sigma_i\}$ a family
of local sections, $\sigma_i:U_i\subset V_n\longrightarrow\mathcal{M}$, but that the converse
it is also verified, that is, for a covering of $V_n$ provided with local
sections, the information consisting of a set of $L_\mathcal{G}$-valued local forms $A_i$
of $V_n$ and which satisfy (\ref{omega_j}) determines an infinitesimal connection on $\mathcal{M}$
\cite{lichnerowicz}.

In fact, let us consider a tangent vector $\tau\in T_z\mathcal{M}$ to $\mathcal{M}$ at a point $z$.
If $x=\pi(z)\in V_n$ is the point lying under $z$, then we can project $\tau$ in $V_n$ as $v=\pi_z^*\tau$.
If $x\in U_j$ then there is an element $g(x)\in\mathcal{G}$ such that we can write $z=g(x)A_j(x)$.
We shall define then, the infinitesimal connection form of a vector $\tau\in T_z\mathcal{M}$ in $z$  by
\begin{equation}\label{omega(tau)}
        \omega(\tau)=Ad(g^{-1})A_j(\pi_z^*\tau)+g^{-1}dg.
\end{equation}
One can then prove that this definition is independent of the local section chosen and that extending
(\ref{omega(tau)}) to all points $z$ of $\mathcal{M}$ we will get a $L_\mathcal{G}$-valued one-form
over $\mathcal{M}$ that satisfies all the properties of an infinitesimal connection form.

\paragraph{The Absolute Differential.}

We are now capable of defining the absolute differential of a $q$-form. Let $\alpha$ be a $M$-valued
$q$-form, where $M$ is a vectorial space. The absolute differential of $\alpha$, represented by
$D\alpha$, is a $(q+1)$-form defined as

$$D\alpha(\tau_0,\tau_1,\cdots, \tau_q)=d\alpha(\mathcal{H}\tau_0,\mathcal{H}\tau_1,\cdots,\mathcal{H}\tau_q),$$
where $d\alpha$ is the external differential of $\alpha$.
One of the properties of the absolute differential is that if one or more of the vectors 
$\tau_0,\tau_1,\cdots,\tau_q$ are vertical then $D\alpha(\tau_0,\tau_1,\cdots,\tau_q)=0$.

\paragraph{The Laplacian.}
Let us define an operator $\delta$ that acts on $p$-forms $(0\leq p\leq m)$ defined over $\mathcal{M}$ that
transforms every $p$-form $\alpha$ in a $(p-1)$-form $\delta\alpha$, called the codifferential of $\alpha$,
 and that it is defined by
$$\delta = (-1)^{p-1}\star D\star,$$
where $\star$ is the Hodge operator and $D$ the absolute differential in relation to some infinitesimal
connection. 

The Laplacian operator $\Delta$ is defined by means of the relation
$$\Delta=D\delta+\delta D=\{D,\delta\},$$
and it is an operator defined over $\mathcal{M}$ that brings into correspondence every $p$-form $\alpha$ to
a $p$-form $\Delta\alpha$. It reduces to the ordinary Laplacian of a function when applied to a $0$-form.

\paragraph{The Curvature.} 

The curvature of an infinitesimal connection is a $L_\mathcal{G}$-value $2$-form $\Omega$ defined as
\begin{equation}
        \label{Omega}
        \Omega=D\omega.
\end{equation}
One proves \cite{lichnerowicz,sharpe} that 
\begin{equation}
        \label{Omega2}
        \Omega=d\omega+\frac{1}{2}[\omega,\omega],
\end{equation}
where $[,]$ are the usual Lie brackets of the Lie algebra $L_\mathcal{G}$.

\subsection{The Fibre Bundle $\mathcal{M}(V_n, M, \mathcal{G}, \pi, \Phi)$ with $M$ a Homogeneous Space}

As it was noticed by Luciani \cite{luciani}, a possible generalisation of the previous scheme is the one in
which the spacetime possesses not a principal fibre bundle structure but a fibre bundle structure 
$\mathcal{M}(V_n, M, \mathcal{G}, \pi, \Phi)$ with fibres $M_x$
that are homomorphic to an homogeneous space $\mathcal{G}/\mathcal{H}$ in which a Lie group $\mathcal{G}$ acts 
from the left. The internal space
$M$ is identified with a homogeneous space for economical reasons since that is the space of lowest dimension that is possible to construct for a given symmetry group \cite{witten}. 
These types of spaces, in which a Lie group acts, are conveniently treated by the Klein geometry description, 
description that we will present next.

\subsubsection{Klein Geometries}

In the Klein approach to geometry \cite{sharpe}, one describes the geometry of a differentiable manifold $M$ on which a
Lie group $\mathcal{G}$ acts from the left in terms of the pair $(\mathcal{G},\mathcal{H})$, 
being $\mathcal{H}\subset\mathcal{G}$ a closed subgroup of $\mathcal{G}$, the
stabiliser of a fixed point $y_0$ of $M$
$$\mathcal{H}=\mathcal{H}_0=\{g\in\mathcal{G}:gy_0 = y_0\}.$$

To the pair $(\mathcal{G},\mathcal{H})$, where $\mathcal{G}$ is a Lie group and $\mathcal{H}\subset\mathcal{G}$ is
a closed subgroup such that $\mathcal{G}/\mathcal{H}$ is connected one calls a Klein geometry. 
The connected coset space $\mathcal{G}/\mathcal{H}$ is called the space of the Klein geometry, or simply the
homogeneous space. One can then study $(M,y_0;\mathcal{G})$ by studying its Klein geometry $(\mathcal{G},\mathcal{H})$\footnote{Here and in the following we suppose that all the 
stabilisers are conjugated to one only stabiliser $\mathcal{H}$.}.

Associated to a Klein geometry $(\mathcal{G},\mathcal{H})$ there is the principal $\mathcal{H}$ bundle
\newline $\mathcal{G}(\mathcal{G}/\mathcal{H},\mathcal{H},p,\Psi)$, where $p$ is the trivial projection that to an element
$g$ of $\mathcal{G}$ makes corresponds an element $[g]$ of $\mathcal{G}/\mathcal{H}$. $[g]$ is the class equivalence
such that two elements $g$ and $g'$ of $\mathcal{G}$ that differ from each other by the left product of an element of
$\mathcal{H}$ are considered equivalent, that is $[g]=g\equiv hg$, for $h\in\mathcal{H}$. In order to reduce
this principal bundle one must introduce the notion of the kernel of a Klein geometry $(\mathcal{G},\mathcal{H})$. 

The kernel of a
Klein geometry $(\mathcal{G},\mathcal{H})$ is the largest subgroup $\mathcal{K}$ of $\mathcal{H}$ that is normal in
$\mathcal{G}$, that is, such that $\mathcal{K}$ is an invariant subgroup of $\mathcal{G}$:
$$Ad(\mathcal{G})\mathcal{K}\subset\mathcal{K}.$$
One can prove that this subgroup is a 
closed Lie subgroup of $\mathcal{H}$ and that the left action of $\mathcal{G}$ on $\mathcal{G}/\mathcal{H}$ induces a
left action of $\mathcal{G}/\mathcal{K}$ on $\mathcal{G}/\mathcal{H}$, and that there is a diffeomorphism 
$(\mathcal{G}/\mathcal{K})/(\mathcal{H}/\mathcal{K})\rightarrow\mathcal{G}/\mathcal{H}$ commuting with the canonical 
left $\mathcal{G}/\mathcal{K}$ actions. So, for any closed subgroup $\mathcal{N}\subset\mathcal{K}$ that is normal in 
$\mathcal{G}$ one can always construct a Klein geometry $(\mathcal{G}/\mathcal{N},\mathcal{H}/\mathcal{N})$. The 
homogeneous space of this pair, that is, the space of this Klein geometry, 
$(\mathcal{G}/\mathcal{N})/(\mathcal{H}/\mathcal{N})$ is isomorphic with the homogeneous space of the pair 
$(\mathcal{G},\mathcal{H})$. The notion of effectiveness of a geometry comes then in to play.

A Klein geometry $(\mathcal{G},\mathcal{H})$ is effective if $\mathcal{K}=1$ and locally effective if $\mathcal{K}$
is discrete. It is now obvious that the geometry $(\mathcal{G}/\mathcal{N},\mathcal{H}/\mathcal{N})$ is ineffective
except when $\mathcal{N}=\mathcal{K}$. In this case one calls to the Klein geometry $(\mathcal{G}/\mathcal{K},\mathcal{H}/\mathcal{K})$ the associated effective Klein geometry of the pair
$(\mathcal{G},\mathcal{H})$. 

Since the two descriptions $(\mathcal{G},\mathcal{H})$ and $(\mathcal{G}/\mathcal{K},\mathcal{H}/\mathcal{K})$
are equivalent, we have reduce the principal $\mathcal{H}$ bundle to a principal $\mathcal{H}/\mathcal{K}$ bundle.
Since $(\mathcal{G}/\mathcal{K})/(\mathcal{H}/\mathcal{K})$ is isomorphic with $\mathcal{G}/\mathcal{K}$ the 
base space of this bundle is basicly the same as the base space of the previous principal bundle, changing only the
fibre, that is, the structure group (from $\mathcal{H}$ to $\mathcal{H}/\mathcal{K}$). 


The following notions will be useful later on.

A Klein geometry is called reductive if there is a decomposition $L_\mathcal{G}=L_\mathcal{H}\oplus L_\mathcal{P}$, where
$L_\mathcal{G}$ and $L_\mathcal{H}$ are the Lie algebras of $\mathcal{G}$ and $\mathcal{H}$, respectively.

An infinitesimal Klein geometry (or a Klein pair) is a pair of Lie algebras $(L_\mathcal{G},L_\mathcal{H})$ where
$L_\mathcal{H}$ is a subalgebra of $L_\mathcal{G}$. The kernel $L_\mathcal{K}$ of $(L_\mathcal{G},L_\mathcal{H})$
is the largest ideal of $L_\mathcal{G}$ contained in $L_\mathcal{H}$. If $L_\mathcal{H}=0$ the Klein pair is
effective. If there is a $L_\mathcal{H}$-module decomposition $L_\mathcal{G}=L_\mathcal{H}+L_\mathcal{P}$ then we
say that the Klein pair is reductive.


\subsubsection{The Principal Fibre Bundle Associated $\bar\mathcal{M}(V_n,\bar\mathcal{G},\bar\pi,\bar\Phi$)}

The action of $\mathcal{G}$ on $M$ can be extended to a left-action on the total space $\mathcal{M}$ by means
of the elements of the homomorphism's family $\Phi$. In fact, let $\phi_{Ux}$ and $\phi_{Vx}$ be two homomorphisms 
of $M$ in $M_x$ and let $z=\pi(x)\in M_x$ be a point over $x\in V_n$. Then there exists a point $y$ of $M$
such that $z=\phi_{Ux}(y)$. Using $\phi_Vx$ we can also make $z=\phi_{Vx}(gy)$ for some $g\in\mathcal{G}$.
We define the left-translation of $\mathcal{G}$ on $\mathcal{M}$ by means of
\begin{equation}
        jz=\phi_{Ux}(jy)=\phi_{Ux}\circ\delta(j,y),
\end{equation}
with $j\in\mathcal{G}$. One can prove that this definition is independent of the homomorphism chosen, and so
we have defined a left-action of $\mathcal{G}$ on $\mathcal{M}$.

Since there will be only a left-action of the Lie group $\mathcal{G}$ on $M$ and on $\mathcal{M}$ (and not
two actions as in the previous case), 
the symmetry group will be isomorphic with $\mathcal{G}$, and we shall
identify $\mathcal{G}$ with it.

To the fibre bundle $\mathcal{M}(V_n, M, \mathcal{G}, \pi, \Phi)$ we can always associate a principal fibre 
bundle $\bar\mathcal{M}(V_n,\bar\mathcal{G},\bar\pi,\bar\Phi$) that determines the first and that is determined by it
\cite{sharpe}. The necessary condition is that the fibre bundle $\mathcal{M}(V_n,M)$ be effective, that is, that
the action of $\mathcal{G}$ on $M$ be effective in the sense that for each automorphism of $M$ there is a correspondent
element $g$ of $\mathcal{G}$ that produces the same effect when acting on $M$ and that for each element $g$ of
$\mathcal{G}$ there corresponds an automorphism of $M$. Obviously this is not the case. There is a subgroup 
$\mathcal{H}$ of $\mathcal{G}$ that doesn't affect the internal space $M$ with its action. If $\mathcal{H}$ isn't
normal in $\mathcal{G}$ one can not perform a reduction of the structure group of $\mathcal{G}$ to 
$\mathcal{G}/\mathcal{H}$ and consider $\mathcal{M}(V_n, M)$ 
as a $\mathcal{G}/\mathcal{H}$ fibre bundle. There are two paths that we can follow: the first
is given by the Klein geometry description; the second has its beginning on
the introduction of the normaliser $\mathcal{N}_\mathcal{G}(\mathcal{H})$ of $\mathcal{H}$ on $\mathcal{G}$.

Following the Klein geometry description, the pairs 
$(\mathcal{G},\mathcal{H})$ and $(\mathcal{G}/\mathcal{K},\mathcal{H}/\mathcal{K})$, where $\mathcal{K}$
is the kernel of the first pair, are geometricly equivalent. The two homogeneous spaces being isomorphic are
basicly the same and so $M$ remain - by an isomorphism - unaltered. Since $\mathcal{G}/\mathcal{K}$ is normal in $\mathcal{G}$ 
(because both $\mathcal{G}$ and $\mathcal{K}$ are normal) we can perform the following group structure reduction 
$\mathcal{G}\longrightarrow\mathcal{G}/\mathcal{K}$. In this way the fibre bundle $\mathcal{M}(V_n, M)$
can be regarded as an effective $\mathcal{G}/\mathcal{K}$ fibre bundle. To this we can now associate a principal
fibre bundle $\bar\mathcal{M}(V_n,\bar\mathcal{G},\bar\pi,\bar\Phi)$, taking for the structure group 
$\bar\mathcal{G}=\mathcal{G}/\mathcal{K}$. This associated principal bundle can then be treated with the tools of 
the previous section. The Yang-Mills Fields will have their values not in the Lie algebra of $\mathcal{G}$ but
on $\mathcal{G}/\mathcal{K}$. Basicly $\mathcal{G}/\mathcal{K}$ is the smaller subgroup of $\mathcal{G}$ that
contains the homogeneous space $M=\mathcal{G}/\mathcal{H}$, and so what we have made is that we have ``enlarged''
a bit the homogeneous space in order to become a subgroup, being then possible to define a principal fibre bundle
in which we can define an infinitesimal connection.

The other path is not to ``enlarge'' the homogeneous space $M=\mathcal{G}/\mathcal{H}$ in order to obtain the smaller
subgroup of $\mathcal{G}$ on which $\mathcal{G}/\mathcal{H}$ is contained but to ``shrink'' the last to the maximal
subgroup contained on $\mathcal{G}/\mathcal{H}$. In order to do so we must define the normaliser of a subgroup.
The normaliser $\mathcal{N}_\mathcal{G}(\mathcal{H})$ of $\mathcal{H}$ on $\mathcal{G}$ is defined as the maximal subgroup of
$\mathcal{G}$ in which $\mathcal{H}$ is normal. It is explicitly defined as
\begin{equation}
  \mathcal{N}_\mathcal{G}(\mathcal{H})=\{n\in\mathcal{G}:n\mathcal{H}n\subset\mathcal{H}\}.
\end{equation}
Let it be defined the submanifold $\bar\mathcal{M}$ of $\mathcal{M}$ by
\begin{equation}
\bar\mathcal{M}=\{z\in\mathcal{M}:\mathcal{H}_z=\mathcal{H}\},
\end{equation}
with $H_z$ the stabiliser of the $\mathcal{G}$ action on the point
$z$. One proves that
$\bar\mathcal{M}(V_n,\bar\mathcal{G},\bar\pi,\bar\Phi)$ is a principal fibre bundle defined over $V_n$ and 
such that its structure group is 
$\bar\mathcal{G}=\mathcal{N}_\mathcal{G}(\mathcal{H})/\mathcal{H}$ (this is the maximal subgroup contained on 
$\mathcal{G}/\mathcal{H}$).  

The Yang-Mills Fields will then be valued on the Lie algebra of 
$\bar\mathcal{G}=\mathcal{N}_\mathcal{G}(\mathcal{H})/\mathcal{H}$. 
Since the dimension of $\mathcal{N}_\mathcal{G}(\mathcal{H})/\mathcal{H}$ is inferior
to that of $\mathcal{G}/\mathcal{K}$, one usually prefers, for economical reasons, the first as the structure group
of the associated bundle\footnote{we shall use always this version.}. 

A special case of the previous scheme is the one in which a global section of $\mathcal{M}$, 
$$\sigma:V_n\longrightarrow\mathcal{M},$$
that is $\mathcal{H}$-invariant is given, i.e., such that $\sigma(hx)=\sigma(x)$, for all $h\in\mathcal{H}$ and
$x\in\mathcal{M}$. In this case we can construct a submanifold of $\mathcal{M}$ that is $\mathcal{H}$-invariant.
By redefining the base space of $\mathcal{M}$ to be such space, we shall have a fibre bundle in which the normaliser
of $\mathcal{H}$ in $\mathcal{G}$ will be identical to $\mathcal{H}$ (by the supposed existence of $\sigma$).
Then $\bar\mathcal{M}$ will be a fibre bundle defined over the new base space of $\mathcal{M}$, having for
structure group $\mathcal{N}_\mathcal{G}({\mathcal{H}})/\mathcal{H}=\{e\}$. We can then identify this trivial bundle
with its own base space.

\paragraph{The Lie Algebra Decomposition.}
Let us suppose that the Klein geometry $(\mathcal{G},\mathcal{H})$ is reductive. Then we can make the following 
decomposition of the respective Lie algebra:
\begin{equation}
  L_\mathcal{G}=L_\mathcal{H}\oplus L_\mathcal{P},
\end{equation}
such that
\begin{equation}
  Ad(\mathcal{H})L_\mathcal{P}\subset L_\mathcal{P},
\end{equation}
or, since $\mathcal{H}$ is connected, at an infinitesimal level,
\begin{equation}\label{icaro}
  Ad(L_\mathcal{H})L_\mathcal{P}\subset L_\mathcal{P},
\end{equation}
that can also be written as $[L_\mathcal{H},L_\mathcal{P}]\subset L_\mathcal{P}$. If $[L_\mathcal{P},L_\mathcal{P}]\subset L_\mathcal{H}$
then the homogeneous space $\mathcal{G}/\mathcal{H}$ is called symmetric (see
next paragraph).

Another decomposition of the Lie algebra of $\mathcal{G}$ can be made if we consider the normaliser 
$\mathcal{N}_\mathcal{G}(\mathcal{H})$ of $\mathcal{H}$ on $\mathcal{G}$:

\begin{equation}
  L_\mathcal{G}=L_\mathcal{N}\oplus L_\mathcal{L},
\end{equation}
with
\begin{equation}
  Ad(\mathcal{N})L_\mathcal{L}\subset L_\mathcal{L},
\end{equation}
and where $L_\mathcal{N}$ denotes the Lie algebra of $\mathcal{N}_\mathcal{G}(\mathcal{H})$. If $\mathcal{N}$ is suposed connected we shall also have
\begin{equation}
  Ad(L_\mathcal{N})L_\mathcal{L}\subset L_\mathcal{L}.
\end{equation}

Since $\mathcal{H}$ is contained on $\mathcal{N}_\mathcal{G}(\mathcal{H})$, if the Klein pair $(L_\mathcal{N},L_\mathcal{H})$
is reductive we can also make
\begin{equation}
  L_\mathcal{N}=L_\mathcal{H}\oplus L_\mathcal{J},
\end{equation}
with 
\begin{equation}
  Ad(\mathcal{H})L_\mathcal{J}\subset L_\mathcal{J}.
\end{equation}
We can write the previous equation at an infinitesimal level,
\begin{equation}
  Ad(L_\mathcal{H})L_\mathcal{J}\subset L_\mathcal{J}.
\end{equation}

If $L_\mathcal{J}$ is orthogonal to the complement of $L_\mathcal{H}$ on $L_\mathcal{N}$ then we have that $L_\mathcal{J}$
will be also an invariant subspace in relation to $\mathcal{N}_\mathcal{G}(\mathcal{H})$, that is,
\begin{equation}
  Ad(\mathcal{N})L_\mathcal{J}\subset L_\mathcal{J},
\end{equation}
or
\begin{equation}\label{ioio}
  Ad(L_\mathcal{N})L_\mathcal{J}\subset L_\mathcal{J},
\end{equation}
The following the composition of the Lie algebra $L_\mathcal{G}$ is then possible
\begin{equation}
L_\mathcal{G}=L_\mathcal{H}\oplus L_\mathcal{J}\oplus L_\mathcal{L},
\end{equation}
and since we have $\mathcal{J}=\mathcal{N}/\mathcal{H}$, 
\begin{equation}
        L_\mathcal{G}=L_\mathcal{H}\oplus L_{\mathcal{N}/\mathcal{H}}\oplus L_\mathcal{L}.
\end{equation}
More, being 
\begin{equation}\label{irilo}
  L_\mathcal{P}=L_{\mathcal{N}/\mathcal{H}}\oplus L_\mathcal{L},
\end{equation}
 we can identify
$L_{\mathcal{N}/\mathcal{H}}\oplus L_\mathcal{L}$ with the tangent vector space to $M=\mathcal{G}/\mathcal{H}$
at the origin, that is, we have
$$T_e M=L_{\mathcal{N}/\mathcal{H}}\oplus L_\mathcal{L}.$$
This will be useful when we discuss the fibre bundle of adapted frames of $\mathcal{M}$.

A special case of Lie algebras that will be used through this paper are the
semisimple Lie algebras. 

Let $L_\mathcal{G}$ be a Lie algebra and $L_\mathcal{H}\subset L_\mathcal{G}$
be a linear subspace of $L_\mathcal{G}$. Then $L_\mathcal{H}$ is called an
ideal of $L_\mathcal{G}$ if $[L_\mathcal{G},L_\mathcal{H}]\subseteq
L_\mathcal{H}$. It is obvious that an ideal is also a subalgebra.

Let us take 
$$\mathcal{D}L_\mathcal{G}=[L_\mathcal{G},L_\mathcal{G}],$$
for the linear span of elements of the form $[u,v]$, with $u,v\in
L_\mathcal{G}$. This subalgebra is called the derived algebra of
$L_\mathcal{G}$. By defining inductively $\mathcal{D}^n L_\mathcal{G}$, $n\geq
0$,
$$\mathcal{D}^0=L_\mathcal{G},$$
$$\mathcal{D}^n=\mathcal{D}(\mathcal{D}^{n-1}L_\mathcal{G}),$$
we shall have the following sequence,
$$\mathcal{D}^0 L_\mathcal{G}\supseteq\mathcal{D}^1
L_\mathcal{G}\supseteq\cdots\supseteq\mathcal{D}^{n-1}
L_\mathcal{G}\supseteq\mathcal{D}^n L_\mathcal{G}.$$

A Lie algebra $L_\mathcal{G}$ is said solvable if $\mathcal{D}^n
L_\mathcal{G}=0$ for some $n\geq 1$. The radical $\mathcal{R}(L_\mathcal{G})$
of a Lie algebra $L_\mathcal{G}$ is defined as the largest solvable ideal of
$L_\mathcal{G}$ that contain all solvable ideals of $L_\mathcal{G}$. We shall
have that a Lie algebra $L_\mathcal{G}$ is solvable if and only if
$\mathcal{R}(L_\mathcal{G})=L_\mathcal{G}$.

A Lie algebra $L_\mathcal{G}$ is said semisimple if its radical is null,
$\mathcal{R}(L_\mathcal{G})=0$.
An equivalent way of defining a semisimple Lie algebra is by the introduction
of the its Killing-Cartan form 
\begin{equation}
  B(u,v)=Tr(Ad_u Ad_v).
\end{equation}
A Lie algebra is then semisimple if its Killing-Cartan form $B$ is
nondegenerate \cite{helgason}. A Lie algebra $L_\mathcal{G}\neq\{0\}$ is
simple if it is semisimple and has no ideals except $\{0\}$ and $L_\mathcal{G}$.

We now present some useful theorems on semisimple Lie algebras.

Let $L_\mathcal{G}$ be a semisimple Lie algebra, $L_\mathcal{H}$ an ideal in
$L_\mathcal{G}$ and $L_\mathcal{P}$ the set of elements $\lambda\in
L_\mathcal{G}$ which are orthogonal to $L_\mathcal{H}$ with respect to
$B$. Then $L_\mathcal{H}$ is semisimple, $L_\mathcal{P}$ is an ideal and
\begin{equation}
  L_\mathcal{G}=L_\mathcal{H}\oplus L_\mathcal{P}.
\end{equation}

A semisimple Lie algebra $L_\mathcal{H}$ as centre $\{0\}$ and can be decomposed as
\begin{equation}
  L_\mathcal{H}=\oplus_{\gamma=1}^{n} L_\mathcal{H}^\gamma,
\end{equation}
where $L_\mathcal{H}^\gamma$, $1\leq\gamma\leq n$, are simple ideals in
$L_\mathcal{G}$, and such that every ideal $L_H'$ of $L_\mathcal{H}$ can be
written as a direct sum of certain $L_\mathcal{H}^\gamma$.

Every compact Lie algebra $L_\mathcal{G}$ can be written as
\begin{equation}
  L_\mathcal{G}=\mathcal{C}(L_\mathcal{G})\oplus\mathcal{D}L_\mathcal{G},
\end{equation}
being $\mathcal{C}(L_\mathcal{G})$ the centre of $L_\mathcal{G}$ and
$\mathcal{D}L_\mathcal{G}$ semisimple and compact.

\paragraph{Symmetric Spaces.}
Let $\mathcal{G}$ be a connected Lie Group, $\mathcal{H}\subset\mathcal{G}$ a
closed subgroup of $\mathcal{G}$ and $\sigma$ an involutive automorphism of
$\mathcal{G}$. If $e\in\mathcal{H}\subset\mathcal{G}_\sigma$, with
$\mathcal{G}_\sigma\subset\mathcal{G}$
the closed subgroup of $\mathcal{G}$ consisting of all the elements left fixed
by $\sigma$, then the triple $(\mathcal{G},\mathcal{H},\sigma)$ is defined as a
symmetric space \cite{kobayashi3}.

A symmetric Lie algebra is a triple $(L_\mathcal{G},L_\mathcal{H},\sigma)$
with $L_\mathcal{G}$ a Lie algebra, $L_\mathcal{H}$ a subalgebra of
$L_\mathcal{G}$, and $\sigma$ an involutive automorphism of $L_\mathcal{G}$
such
that $L_\mathcal{H}$ consists of all elements of $L_\mathcal{G}$ that are left
fixed by $\sigma$.

Let $(\mathcal{G},\mathcal{H},\sigma)$ be a symmetric space. Then
$(L_\mathcal{G},L_\mathcal{H},\sigma^*_e)$ will be a symmetric Lie algebra.
Conversely, if $(L_\mathcal{G},L_\mathcal{H},\sigma^*)$ is a symmetric Lie
algebra and $\mathcal{G}$ is a simply connected Lie group with Lie algebra
$L_\mathcal{G}$, then the automorphism $\sigma^*$ of $L_\mathcal{G}$ will
induce an automorphism $\sigma$ of $\mathcal{G}$ and, for any $\mathcal{H}$
lying between $\mathcal{G}_\sigma$ and $e$, the triple
$(\mathcal{G},\mathcal{H},\sigma)$ will be a symmetric space.

Since $\sigma$ is an involutive, we can use it to perform a canonical
decomposition of $L_\mathcal{G}$: being $L_\mathcal{H}$ the eigenspace
associated to the eigenvalue $1$ of $\sigma$, if we take $L_\mathcal{P}$ to be
the eigenspace of $\sigma$ corresponding to $-1$ then we can take 
\begin{equation}
  L_\mathcal{G}=L_\mathcal{H}\oplus L_\mathcal{P}.
\end{equation}
Moreover, using $\sigma$ we can demonstrate that \cite{kobayashi3}
\begin{equation}
[L_\mathcal{H},L_\mathcal{H}]\subset L_\mathcal{H},
\end{equation}
\begin{equation}
[L_\mathcal{H},L_\mathcal{P}]\subset L_\mathcal{P}, 
\end{equation}
\begin{equation}
[L_\mathcal{P},L_\mathcal{P}]\subset L_\mathcal{H},
\end{equation}
and that we have
\begin{equation}
Ad(\mathcal{H})L_\mathcal{P}\subset L_\mathcal{P},
\end{equation}
or if $\mathcal{H}$ is connected, at an infinitesimal level,
\begin{equation}
  Ad(L_\mathcal{H})L_\mathcal{P}\subset L_\mathcal{P}.
\end{equation}

A symmetric Lie Algebra $(L_\mathcal{G},L_\mathcal{H},\sigma)$ is called orthogonal if the connected Lie group of
linear transformations of $L_\mathcal{G}$ generated by
$Ad_{L_\mathcal{G}}(L_\mathcal{H})$ is compact. If $L_\mathcal{H}$ is such
that has a trivial intersection with the centre $\mathcal{C}(L_\mathcal{G})$ of $L_\mathcal{G}$, then we
can write $(L_\mathcal{G},L_\mathcal{H},\sigma)$ as a direct sum of orthogonal
symmetric
Lie algebras $(L_\mathcal{G}^1,L_\mathcal{H}^1,\sigma^1)$ and 
$(L_\mathcal{G}^2,L_\mathcal{H}^2,\sigma^2)$ with the following properties
\cite{kobayashi}
\begin{enumerate}
\item If $L_\mathcal{G}^1=L_\mathcal{H}^1\oplus L_\mathcal{P}^1$ and 
$L_\mathcal{G}^2=L_\mathcal{H}^2\oplus L_\mathcal{P}^2$  are the
  canonical decompositions, then 
  \begin{equation}
    [L_\mathcal{P}^1,L_\mathcal{P}^1]=
    [L_\mathcal{P}^1,L_\mathcal{P}^2]=
    [L_\mathcal{P}^1,L_\mathcal{H}^2]=
    [L_\mathcal{H}^1,L_\mathcal{P}^2]=0.
  \end{equation}
\item $L_\mathcal{G}^2$ is semi-simple.
\end{enumerate}
If $L_\mathcal{H}$ does not have a trivial intersection with $\mathcal{C}(L_\mathcal{G})$ we can always write
\begin{equation}
  L_\mathcal{H}=L_\mathcal{H}^0\oplus L_\mathcal{H}',
\end{equation}
with $L_\mathcal{H}'$ having a trivial intersection with
$\mathcal{C}(L_\mathcal{G})$ and $L_\mathcal{H}^0=L_\mathcal{H}\cap \mathcal{C}(L_\mathcal{G})$.
We can then apply to $(L_\mathcal{G},L_\mathcal{H}',\sigma')$ - that will be
also a symmetric Lie algebra - the same theorem, obtaining then the following
decompositions of both $L_\mathcal{H}'$ and $L_\mathcal{P}'$,
\begin{equation}
  L_\mathcal{H}'=L_\mathcal{H}'^1\oplus L_\mathcal{H}'^2,
\end{equation}
\begin{equation}
  L_\mathcal{P}'=L_\mathcal{P}'^1\oplus L_\mathcal{P}'^2,
\end{equation}
with $L_\mathcal{P}'^1=L_\mathcal{P}'\cap\mathcal{R}(L_\mathcal{G})$, being
 $\mathcal{R}(L_\mathcal{G})$
the radical of $L_\mathcal{G}$, and
$L_\mathcal{P}'^2$ an $Ad({L_\mathcal{G}})L_\mathcal{H}'$-invariant subspace of
$L_\mathcal{P}'$.

We then have,
\begin{eqnarray}
  L_\mathcal{H}=L_\mathcal{H}^0\oplus L_\mathcal{H}'^1\oplus L_\mathcal{H}'^2,\\
  L_\mathcal{H}^0\oplus L_\mathcal{P}=L_\mathcal{P}'^1\oplus L_\mathcal{P}'^2.
\end{eqnarray}

A more general decomposition of an effective orthonormal symmetric Lie algebra
can then be performed \cite{helgason}: let
$(L_\mathcal{G},L_\mathcal{H},\sigma)$ be an effective orthonormal symmetric
Lie algebra. Then there exist ideals $L_\mathcal{G}^0,
L_\mathcal{G}^1,L_\mathcal{G}^2$ in $L_\mathcal{G}$ such that
\begin{enumerate}
\item $L_\mathcal{G}=L_\mathcal{G}^0\oplus L_\mathcal{G}^1\oplus
  L_\mathcal{G}^0$.
\item $L_\mathcal{G}^0,L_\mathcal{G}^1$ and $L_\mathcal{G}^2$ are invariant
  under $\sigma$ and orthogonal with respect to the Killing-Cartan form of
  $L_\mathcal{G}$.
\item By denoting the restrictions of $\sigma$ to
  $L_\mathcal{G}^0,L_\mathcal{G}^1$ and $L_\mathcal{G}^2$ as
  $\sigma^0,\sigma^1$ and $\sigma^2$, respectively, we shall have that
  $(L_\mathcal{G}^0,L_\mathcal{H}^0,\sigma^0),(L_\mathcal{G}^1,L_\mathcal{H}^1,\sigma^1)$
  and $(L_\mathcal{G}^2,L_\mathcal{H}^2,\sigma^2)$, with
  $L_\mathcal{H}^\gamma\subset L_\mathcal{G}^\gamma$, $\gamma=0,1,2$, ideals
  in $L_\mathcal{H}$, orthogonal with respect to $B$ and
  $L_\mathcal{H}=L_\mathcal{H}^0\oplus L_\mathcal{H}^1\oplus L_\mathcal{H}^2$,
  are effective orthogonal symmetric Lie algebras of the Euclidean type (that
  is locally euclidian),
  compact type and non-compact type, respectively.
\end{enumerate}
The decomposition of $L_\mathcal{H}$ into no more of tree ideals will be of
extreme importance in the scalar field analysis for determining spontaneously
symmetry breakings in Kaluza-Klein models with an internal symmetric space
(cf. chapter 4). The following relations will be very useful \cite{helgason},
\begin{enumerate}
\item $L_\mathcal{P}^{(0)}=\{g\in L_\mathcal{G}:B(g,g')=0\; for \; all \; g'\in L_\mathcal{G}\}.$
\item $[L_\mathcal{P}^{(0)},L_\mathcal{P}]=0$.
\item $L_\mathcal{H}^{(1)}=[L_\mathcal{P}^{(1)},L_\mathcal{P}^{(1)}].$
\item $L_\mathcal{H}^{(2)}=[L_\mathcal{P}^{(2)},L_\mathcal{P}^{(2)}].$
\item $L_\mathcal{H}^{(0)}$ is the orthogonal complement to
  $L_\mathcal{H}^{(1)}$ and $L_\mathcal{H}^{(2)}$ by $B$.
\item
  $[L_\mathcal{H}^{(0)},L_\mathcal{P}^{(1)}]=[L_\mathcal{H}^{(0)},L_\mathcal{P}^{(2)}]=0.$
\item
  $[L_\mathcal{H}^{(1)},L_\mathcal{P}^{(0)}]=[L_\mathcal{H}^{(1)},L_\mathcal{P}^{(2)}]=0$.
\item $[L_\mathcal{H}^{(2)},L_\mathcal{P}^{(0)}]=[L_\mathcal{H}^{(2)},L_\mathcal{P}^{(1)}]=0$.
\end{enumerate}

\subsection{The Fibre Bundle of Linear Frames}

In order to define a linear connection over $\mathcal{M}$, let us consider the tangent vector space $T_z\mathcal{M}$
in a point $z\in\mathcal{M}$. We will call a frame with origin $z$ to a basis 
$\{\hat e_{\hat\alpha}\}_{\hat\alpha=0}^{m-1}$ of $T_z\mathcal{M}$ and we will denote it by $\hat e^z$. To this frame in $z$
it corresponds a coframe in the same point, that is, a basis $\{\hat e^{\hat\beta}\}_{\hat\beta=0}^{m-1}$ of the cotangent vector
space $T^{*}_z\mathcal{M}$ such that $\hat e^{\hat\beta}(\hat e_{\hat\alpha})=\delta^{\hat\beta}_{\hat\alpha}$. We will 
represent this coframe with origin in $z\in\mathcal{M}$ by $\hat e_z$.

Let $\hat Q^z$ be the set of frames with origin in $z\in\mathcal{M}$ and $\hat Q_z$ the set of coframes with the
same origin. The fibre bundle of linear frames of $\mathcal{M}$
is the principal fibre bundle that has as its total space the differentiable manifold
\begin{equation}
        \label{hatE}
        \hat E(\mathcal{M})=\bigcup_{z\in\mathcal{M}}\hat Q^z,
\end{equation}
as its base space the differentiable manifold $\mathcal{M}$, as its the canonical projection $\hat p$ the mapping of $\hat E(\mathcal{M})$ into $\mathcal{M}$ such that
to a frame brings into correspondence its origin and that possesses as structural group the linear group, $GL(1,m-1;\mathcal{R})$. Taking in (\ref{hatE}) not $Q^z$ but the set of coframes with origin in $z\in\mathcal{M}$,
$Q_z$, and constructing in an analogous way a fibre bundle structure with the same structure group 
$GL(1,m-1;\mathcal{R})$ we will obtain the fibre bundle of linear coframes, bundle that we will represent by
$\hat E^*(\mathcal{M})$.

\paragraph{The Linear Connection.}
A linear connection on $\mathcal{M}$ is an infinitesimal connection on the principal bundle $\hat E(\mathcal{M})$.
Thus at the point $\hat e^z\in \hat E(\mathcal{M})$ the form $\hat\omega$ of the linear connection will take its values
on the Lie algebra of the structure group of $\hat E(\mathcal{M})$, that is, on the Lie algebra of $GL(1,m-1;\mathcal{R})$.
It can, in this way, be represented by a $m\times m$ matrix which we shall denote by 
$\hat\omega=(\hat\omega^{\hat\rho}_{\hat\sigma})$. Being $\hat\omega$ a one-form in $\mathcal{M}$, we can write it as $\hat\omega=\hat\omega_{\hat\gamma}\hat e^{\hat\gamma}$ or in terms of its components as

$$\hat\omega^{\hat\alpha}_{\hat\beta}=\hat\gamma^{\hat\alpha}_{\hat\beta\hat\gamma} \hat e^{\hat\gamma},$$
where $\hat\gamma^{\hat\alpha}_{\hat\beta\hat\gamma}$ are the coefficients of the linear connection at the point 
$z\in\mathcal{M}$ in the coframe $\hat e_z$. If the coframe is a natural coframe of local coordinates then
we represent the linear connection coefficients by $\hat\Gamma^{\hat\alpha}_{\hat\beta\hat\gamma}$.

\paragraph{The Covariant Derivative.}
Let $v$ be a contravariant vector field in $\mathcal{M}$ and consider two covering neighbourhoods $U$ and $V$ 
that contain both a point $z\in\mathcal{M}$, that is, $z\in U\cap V$. To each of these neighbourhoods one can
assign a different frame, $\hat e^z_U$ for $U$ and $\hat e^z_V$ for $V$. These two, appartening to the same vectorial 
space, $T_z\mathcal{M}$, can be connected by a regular $m\times m$ matrix,
\begin{equation}
        \hat e_{\hat\beta}^V = \hat\Lambda^{\hat\alpha}_{\hat\beta}\hat e^U_{\hat\alpha}\;\;\;\;\; 
        (\hat e_{\hat\beta}^V\in\hat e_V,\hat e_{\hat\alpha}\in\hat e_U),
\end{equation}
and so, if one writes the components of $v$ in these to frames as a row matrix $v^U$ in the first frame and
$v^V$ in the second and put $\hat\Lambda^V_U=(\hat\Lambda^{\hat\alpha}_{\hat\beta})$ then

$$v^V=\hat\Lambda^U_V v^U.$$
Differentiating the both sides

\begin{equation}
        \label{dv}
        dv^V=d\hat\Lambda^U_V v^U+\hat\Lambda^U_V dv^U,
\end{equation}
where $d\hat\Lambda^U_V$ is the matrix which components are the differentials of the elements of $\hat\Lambda^U_V$.

The linear connection form transforms as
\begin{equation}
        \label{omegaEV}
        \hat\omega_V=(\hat\Lambda^U_V)^{-1}\hat\omega_U \hat\Lambda^U_V+\hat\Lambda^V_U d\hat\Lambda^U_V,
\end{equation}
and so
$$\hat\omega_V v^V=\hat\Lambda^V_U\hat\omega_U \hat\Lambda^U_V v^v+\hat\Lambda^V_Ud\hat\Lambda^U_V v^V,$$
where $\hat\Lambda^V_U=(\hat\Lambda_V^U)^{-1}$. But $\hat\Lambda^U_V v^V=v^U$ and $0=d(\hat\Lambda^V_U \hat\Lambda^U_V)=
d\hat\Lambda^V_U \hat\Lambda^U_V+\hat\Lambda^V_U d\hat\Lambda^U_V$, so

\begin{equation}
        \label{omegav}
\hat\omega_V v^V = \hat\Lambda^V_U\hat\omega_U v^U-d\hat\Lambda^V_U v^U
\end{equation}

Adding (\ref{omegav}) to (\ref{dv}) we get

$$dv^V+\hat\omega_V v^V=\hat\Lambda^V_U(dv^V+\hat\omega_U v^U),$$
so, if we make
\begin{equation}
        \label{Dv}
        \hat Dv^U=dv^U+\hat\omega_U v^U,
\end{equation}
we will have
$$\hat Dv^V=\hat\Lambda^V_U \hat Dv^U.$$
We see that $\hat Dv$ transforms itself in the same manner as $v$, so it is a contravariant vector 1-form.
It is called the absolute differential of the vector field $v$. To the associated tensor one calls
the covariant derivative of $v$.

In terms of components, (\ref{Dv}) assumes the form
$$\hat Dv^{\hat\alpha}=dv^{\hat\alpha}+\hat\omega^{\hat\alpha}_{\hat\rho}v^{\hat\rho},$$
or, 
$$\hat Dv^{\hat\alpha}=(\partial_{\hat\beta}v^{\hat\alpha}+\hat\gamma^{\hat\alpha}_{\hat\rho\hat\beta}v^{\hat\rho})
\hat e^{\hat\beta}=\hat\nabla_{\hat\beta}v^{\hat\alpha}\hat e^{\hat\beta},$$
where we have made
$$\hat\nabla_{\hat\beta}v^{\hat\alpha}=\partial_{\hat\beta}v^{\hat\alpha}+
\hat\gamma^{\hat\alpha}_{\hat\rho\hat\beta}v^{\hat\rho}.$$

Apart from the covariant derivative of a vector field, or, by generalisation, of a tensor field, one can
construct two important local tensors: the torsion and the curvature tensors.

\paragraph{The Torsion Tensor.}
Let us take the differential of the transformation law for a coframe $\hat e^U$,
$$d\hat e^V=d(\hat\Lambda^V_U \hat e^U)=d\hat\Lambda^V_U\wedge \hat e^U+\hat\Lambda^V_U d\hat e^U,$$
and the exterior product between the linear connection form $\hat\omega_V$ and the coframe $\hat e^V$,
$$\hat\omega_V\wedge \hat e^V=\hat\Lambda^V_U \hat\omega_U \hat\Lambda^U_V\wedge \hat e^V+\hat\Lambda^V_Ud\hat\Lambda^U_V\wedge \hat e^V=\hat\Lambda^V_U\hat\omega_U\wedge \hat e^U-d\hat\Lambda^V_U\wedge\hat\Lambda^U_V \hat e^V$$
$$=\hat\Lambda^V_U\hat\omega_U\wedge \hat e^U-
\wedge d\hat\Lambda^V_U \hat e^U.$$
If we add the two we will get
$$d\hat e^V+\hat\omega_V\wedge \hat e^V=\hat\Lambda^V_U(d\hat e^U+\hat\omega_U\wedge \hat e^U),$$
that is $\hat\Sigma^U=d\hat e^U+\hat\omega_U\wedge \hat e^U$ transforms in the same way as the coframe, meaning, it
is a contravariant vector 2-form on $\hat E(\mathcal{M})$. It is called the torsion form of the linear 
connection.
In the frame $\hat e^U$, we have
\begin{equation}
        \label{SigmaE}
        \hat\Sigma=\hat\Sigma^{\hat\alpha}\hat e_{\hat\alpha}=(d\hat e^{\hat\alpha}+\omega^{\hat\alpha}_{\hat\rho}\wedge
\hat e^{\hat\rho})\hat e_{\hat\alpha},
\end{equation}
or, being $\hat\Sigma^{\hat\alpha}$ a 2-form,
\begin{equation}
        \hat\Sigma^{\hat\alpha}=\frac12\hat\Sigma^{\hat\alpha}_{\hat\beta\hat\gamma}\hat e^{\hat\beta}\wedge\hat 
        e^{\hat\gamma}=-\hat S^{\hat\alpha}_{\hat\beta\hat\gamma}\hat e^{\hat\beta}\wedge\hat e^{\hat\gamma},
\end{equation}
with $\hat S^{\hat\alpha}_{\hat\beta\hat\gamma}$ a tensor of rank $(1,2)$, the torsion tensor associated
to $\hat\Sigma$.

In a natural coframe of local coordinates we have, since $\hat e^{\hat\alpha}=dx^{\hat\alpha}$,
$$\hat\Sigma^{\hat\alpha}=\hat\omega^{\hat\alpha}_{\hat\beta}\wedge dx^{\hat\beta}=\hat\Gamma^{\hat\alpha}_{\hat\beta\hat\gamma}
dx^{\hat\gamma}\wedge dx^{\hat\beta}=-\frac12(\hat\Gamma^{\hat\alpha}_{\hat\beta\hat\gamma}-
\hat\Gamma^{\hat\alpha}_{\hat\gamma\hat\beta})dx^{\hat\gamma}\wedge dx^{\hat\beta}.$$
So the components of the torsion tensor will be
\begin{equation}
        \hat S^{\hat\alpha}_{\hat\beta\hat\gamma}=\frac12(\hat\Gamma^{\hat\alpha}_{\hat\beta\hat\gamma}-
        \hat\Gamma^{\hat\alpha}_{\hat\gamma\hat\beta}).
\end{equation}

\paragraph{The Curvature Tensor.}
Let us now take the differential of the linear connection form,
$$d\hat\omega_V=d(\hat\Lambda^V_U\hat\omega_U\hat\Lambda^U_V+\hat\Lambda^V_U d\hat\Lambda^U_V)=d\hat\Lambda^V_U\wedge\hat\omega_U
\hat\Lambda^U_V+d\hat\Lambda^V_U\wedge d\hat\Lambda^U_V+\hat\Lambda^V_Ud\hat\omega_U\hat\Lambda^U_V+\hat\Lambda^U_V\hat\omega_U\wedge
d\hat\Lambda^U_V,$$
and the external product of the linear connection form with it self,
$$\hat\omega_V\wedge\hat\omega_V=\hat\Lambda^V_U\hat\omega_U\wedge\hat\omega_U\hat\Lambda^U_V+\hat\Lambda^V_U\hat\omega_U\wedge
d\hat\Lambda^U_V-d\hat\Lambda^V_U\wedge\hat\omega_U\hat\Lambda^U_V-d\hat\Lambda^V_U\wedge d\hat\Lambda^U_V.$$
Adding the first equation to the second, we obtain
$$d\hat\omega_V+\hat\omega_V\wedge\hat\omega_V=\hat\Lambda^V_U(d\hat\omega_U+\hat\omega_U\wedge\hat\omega_U)\hat\Lambda^U_V,$$
or, if we write
\begin{equation}
        \label{OmegaE}
        \hat\Omega_U=d\hat\omega_U+\hat\omega_U\wedge\hat\omega_U,
\end{equation}
we have the following transformation law
$$\hat\Omega_V=(\hat\Lambda^U_V)^{-1}\hat\Omega_U\hat\Lambda^U_V,$$
that is, $\hat\Omega$ transforms itself as a tensor of rank $(1,1)$, that is, it is a tensor 2-form of 
type $Ad_{g^{-1}}$. It is called the curvature form of the linear connection. It could be directly derived
from the Maurer-Cartan equation (\ref{Omega2}).
Being the curvature a tensor of rank $(1,1)$ we can write
$$\hat\Omega=\hat\Omega^{\hat\alpha}_{\hat\beta}\hat e_{\hat\alpha}\otimes\hat e^{\hat\beta}=
(d\hat\omega^{\hat\alpha}_{\hat\beta}+\hat\omega^{\hat\alpha}_{\hat\rho}\wedge\hat\omega^{\hat\rho}_{\hat\beta})
\hat e_{\hat\alpha}\otimes\hat e^{\hat\beta}.$$
One defines the curvature tensor associated to $\hat\Omega$ by the relation
\begin{equation}
        \hat\Omega^{\hat\alpha}_{\hat\beta}=\frac12 \hat R^{\hat\alpha}_{\hat\beta\hat\gamma\hat\delta}\hat e^{\hat\gamma}
        \wedge\hat e^{\hat\gamma}.
\end{equation}  

\paragraph{The Biachi Identities.}
The Bianchi identities are obtained by taking the differential of both (\ref{SigmaE}) and (\ref{OmegaE}), and
writing the right side of both equations in terms of $\hat\Sigma$ and $\hat\Omega$. The final expressions are
\begin{eqnarray}
        d\hat\Sigma=\hat\Omega\wedge\hat e - \hat\omega\wedge\hat\Sigma, \\
        d\hat\Omega^E=\hat\Omega\wedge\hat\omega-\hat\omega\wedge\hat\Sigma.
\end{eqnarray}



\subsection{The Fibre Bundle of Affine Frames}

In each point $z$ of $\mathcal{M}$ the tangent vector space $T_z\mathcal{M}$ admits a natural affine
space structure. An affine frame $\hat S^z$ in $z\in\mathcal{M}$ will be defined by a vector $O\in T_z\mathcal{M}$, 
named the origin of the frame, and by a basis $\hat e^z$ of the tangent vector space $T_z\mathcal{M}$.

Let us consider two frames in the same point $z$, $\hat S^z = (O,\hat e^z)$ and $\hat S^{z}\,'=(O',\hat e^{z}\,')$. 
One wants to
know how to connect this to frames. If we denote by $\hat e_m$ and $\hat e_m'$ the origins of $S^z$ and
$\hat S^{z}$, respectively, where $m$ is the dimension of $T_z\mathcal{M}$, then the transformation rule is the
following
\begin{eqnarray}
        \label{ebeta}
        \hat e_{\hat\beta}'=\hat e_{\hat\alpha}\hat\Lambda^{\hat\alpha}_{\hat\beta'}\\
        \label{em}
        \hat e_m'= \hat e_m+v^{\hat\alpha}\hat e_{\hat\alpha},
\end{eqnarray}
where $v=O'-O=v^{\hat\alpha}\hat e_{\hat\alpha}$.

If we make the following definition
\begin{equation}
        \hat L=(\hat L_{\check\alpha\check\beta})=\left(
        \begin{array}{cc} 
                0 & \hat\Lambda^{\hat\alpha}_{\hat\beta'} \\
                1 & v^{\hat\alpha}
        \end{array}\right),
\end{equation}
where $\hat L$ is a $(m+1)\times(m+1)$ matrix, then we see that the equations (\ref{ebeta}, \ref{em}) can be written
as
\begin{equation}
        \hat e_{\check\beta}'=\hat e_{\check\alpha}\hat L^{\check\alpha}_{\check\beta'},
\end{equation}
or
$$\hat S'=\hat S\hat L.$$ 

The set of the matrix $\hat L$ make a group under matrix multiplication that is isomorphic to the affine group 
$GA(1,m;\mathcal{R})$, and we shall identify it with this group. 

Let $\hat \mathcal{Q}^z$ be the set of all affine frames in the point $z\in\mathcal{M}$ and let us construct the following
set
$$\hat\mathcal{E}(\mathcal{M})=\bigcup_{z\in\mathcal{M}}\hat \mathcal{Q}^z.$$
It can be shown that it is possible to define a natural differentiable manifold structure in $\hat \mathcal{E}$ and
from this a structure of fibre bundle, being the structure group the affine group $GA(1,m;\mathcal{R})$. To
this bundle $\hat \mathcal{E}(\mathcal{M})$ one calls the fibre bundle of affine frames. An affine connection 
$\hat\omega^\mathcal{E}$ in 
$\mathcal{M}$ will be defined as a infinitesimal connection on the fibre bundle of affine frames.

In general it is possible to associate to every linear connection on $\mathcal{M}$ an affine connection. One
has
\begin{equation}
        \hat\omega^\mathcal{E}=\left(
                \begin{array}{cc}
                        0 & \hat\omega \\
                        0 & \hat e_z + \hat DO
                \end{array},
        \right)
\end{equation}
where $\hat e_z$ is the coframe associated with $\hat e^z$ and $O$ is the origin of the affine frame.
\subsection{The Fibre Bundle of Orthonormal Frames}\label{fbof}

Once defined a structure of a Riemannian manifold over $\mathcal{M}$, it is possible to define in each
tangent vector space $T_z\mathcal{M}$, $z\in\mathcal{M}$, an orthonormal frame $\tilde e^z$ using the metric 
$\gamma$ of this manifold. It will be defined as a basis $\{\tilde e_{\hat\alpha}\}_{\hat\alpha=0}^{m-1}$
of $T_z\mathcal{M}$ such that the vectors $\tilde e_{\hat\alpha}$ satisfy the relations
$$\gamma(\tilde e_{\hat\alpha},\tilde e_{\hat\beta})=\delta_{\hat\alpha\hat\beta}.$$
If $\tilde e^z$ and $\tilde e^z\,'$ are two orthonormal frames at $z$, then exists an orthogonal $m\times m$
matrix $\tilde\Lambda$ such that
$$\tilde e^z\,'=\tilde e^z \tilde\Lambda,$$
or
$$\tilde e_{\hat\beta}'=\tilde e_{\hat\alpha}\tilde\Lambda^{\hat\alpha}_{\hat\beta'}.$$

As before, let us consider $\tilde Q^z$ the set of orthonormal frames with origin $z$. And let us consider
the following set
$$\tilde E(\mathcal{M})=\bigcup_{z\in\mathcal{M}}\tilde Q^z.$$
It is now possible to define on $\tilde E(\mathcal{M})$ a topology and a natural differentiable manifold structure
and, after defining a projection $p$ such that, to every frame $\tilde e^z$ of $\tilde E(\mathcal{M})$ brings
into correspondence its origin, that is, the point $z$, will be possible to construct a principal fibre bundle
over $\mathcal{M}$, being the total space $\tilde E(\mathcal{M})$ and the structure group the orthogonal group
$O(1,m-1;\mathcal{R})$. The bundle $\tilde E(\mathcal{M},O(1,m-1;\mathcal{R}),p,\Psi)$ is called the Fibre Bundle
of Orthonormal Frames. 

\paragraph{The Minkowski Connection.}
An Euclidean Connection, or a Minkowski Connection, on $\mathcal{M}$ will then be an infinitesimal connection on
$\tilde E(\mathcal{M})$.

It is possible to associate in a natural way a linear connection with every Minkowski connection. Representing
the last by $\tilde\omega$ and the first by $\hat\omega$ one sees that for a covering $U$ of $\mathcal{M}$ the
following definition
$$\hat\omega_U=\tilde\omega_U,$$
is independent of the covering $U$ chosen. In order to a linear connection be naturally associated with a
Minkowski connection, it is necessary and sufficient that the absolute differential of the metric tensor in
this connection be zero, that is, that the connection be metric compatible.
\subsection{The Fibre Bundle of Affine Orthonormal Frames}
The Fibre Bundle of Affine Orthonormal Frames of $\mathcal{M}$ or simply the Minkowski Affine Bundle, 
$\tilde\mathcal{E}(\mathcal{M})$, 
can be defined in the same way as the Fibre Bundle of Affine Frames taking for it's total space
the set of the affine frames that verify the orthonormality condition and for the transformation matrix
\begin{equation}
        \tilde L=(\tilde L_{\check\alpha\check\beta})=\left(
        \begin{array}{cc} 
                0 & \tilde\Lambda^{\hat\alpha}_{\hat\beta'} \\
                1 & v^{\hat\alpha}
        \end{array}\right).
\end{equation}
The structure group will be the affine orthogonal group $AO(1,m;\mathcal{R})$.
\subsection{The Fibre Bundle of Oriented Orthonormal Frames}
An orientation in a point $z\in\mathcal{M}$ is a pseudoscalar $\epsilon$ with square 1, that is, chosen
two frames $\hat e^z$ and $\hat e^z\,'$ at $z$ that are connected by the transformation rule
$$\hat e^z\,'=\hat e^z\hat\Lambda,$$
the value of $\epsilon$ is the same in the two or change signal if $\det(\hat\Lambda)>0$ or $\det(\hat\Lambda)<0$,
respectively $(\epsilon=\pm 1)$. We say that $\mathcal{M}$ is an orientable manifold if there is an orientation
in all the points $z\in\mathcal{M}$. 

If $\mathcal{M}$ is orientable then the fibre bundle of orthonormal frames $\tilde E(\mathcal{M})$ admits two
distinct components, that is, there exist at each point $z$ of $\mathcal{M}$ two arcwise connected families of
frames, the determinant $\det(\tilde\Lambda)$ associated with two frames of the same family are positive (the
 frames are in the "same side" of $\mathcal{M}$, that is, have the same orientation),
that associated with two frames from different families are negative. Each component admits a principal
fibre bundle structure over $\mathcal{M}$, with the component of the identity of the linear group as structural
group. We will denote the two components of the fibre bundle of the orthonormal frames of $\mathcal{M}$ by
$\tilde E_+(\mathcal{M})$ and $\tilde E_-(\mathcal{M})$. They are called the fibre bundles of oriented orthonormal
frames of $\mathcal{M}$.

In order to define a time-orientation, we must define over $\mathcal{M}$ a timelike vector field $\tau$ such that
$$\gamma(\tau,\tau)<0.$$
We are then capable of distinguishing between those frames that are transformed by a proper orthocronous Lorentz transformation, that preserves the nature of each vector of the frame, i.e., that transforms a timelike vector
into a timelike vector and a spacelike vector into a spacelike one, and those that don't. We will represent
the two subsets, respectively, by $\tilde E_+^\uparrow(\mathcal{M})$ and $\tilde E_+^\downarrow(\mathcal{M})$,
for the oriented component $\tilde E_+(\mathcal{M})$ of the bundle of orthonormal frames. Both components of
$\tilde E_+(\mathcal{M})$ will possess a structure of a differentiable principal fibre bundle, possessing for
the structure group $SO_+^\uparrow(1,m-1;\mathcal{R})$ or $SO_+^\downarrow(1,m-1;\mathcal{R})$. We will restrict ourselves
to the bundle $\tilde E_+^\uparrow(\mathcal{M})$ that possesses for structure group the Lorentz group of proper
orthocronous transformations $SO^\uparrow_+(1,m-1;\mathcal{R})$.


\subsection{The Fibre Bundle of Spinors}

Let us assume that $\mathcal{M}$ is an orientable riemannian manifold with metric $\gamma$. Then the
fibre bundle of orthonormal frames of $\mathcal{M}$ will be composed of two components $\tilde E_+(\mathcal{M})$ 
and $\tilde E_-(\mathcal{M})$. If one defines a time-orientation in $\mathcal{M}$ then these two will be
also composed by two components $\tilde E^\uparrow(\mathcal{M})$ and $\tilde E^\downarrow(\mathcal{M})$.
Let us consider the fibre bundle of
Lorentz proper orthocronous transformations $\tilde E_+^\uparrow(\mathcal{M})$.

Let $\tilde{\underline G}$ be a universal covering group of $SO^\uparrow_+(1,m-1;\mathcal{R})$
\footnote{It will be the spinor group Spin$^\uparrow_+(1,m-1;\mathcal{M})$. For the simple case $m=4$ one has 
$\tilde{\underline{\mathcal{G}}}=SL(2,\mathcal{C})$.} and 
$$\tilde\lambda:\tilde{\underline G}\longrightarrow SO^\uparrow_+(1,m-1;\mathcal{R})$$
a homomorphism covering.
A spin structure over $\mathcal{M}$ for $\tilde E_+^\uparrow(\mathcal{M})$ will be a differentiable principal fibre bundle
$\tilde{\underline{E}}(\mathcal{M},\tilde{\underline G},\tilde{\underline{p}},\tilde{\underline\Psi})$, with the structure group $\tilde{\underline G}$, the 
projection $\tilde{\underline{p}}$ together with a differentiable map
$$\tilde S:\tilde{\underline{E}}(\mathcal{M})\longrightarrow\tilde E^\uparrow_+(\mathcal{M}),$$
such that the following diagram commutes
$$
        \begin{array}{ccc}
                \tilde{\underline E}(\mathcal{M})\times\tilde{\underline G}
 & \stackrel{\tilde S\times\tilde\lambda}\longrightarrow & \tilde E^\uparrow_+(\mathcal{M})\times SO^\uparrow_+(1,m-1;\mathcal{R}) \\
                \downarrow & & \downarrow \\
                \tilde{\underline E}(\mathcal{M}) & \stackrel{\tilde S}{\longrightarrow} & \tilde E^\uparrow_+(\mathcal{M}) \\
                \tilde{\underline{p}}\searrow &  & \swarrow \tilde p\\
                & \mathcal{M} &
        \end{array}
$$
We have that $\tilde S$ commutes with the group operations, that is, for $\tilde{\underline \Lambda}\in\tilde{\underline G}$,
$$\tilde S\circ\tilde{\underline\Lambda}=\tilde\lambda(\tilde{\underline\Lambda})\circ\tilde S.$$

A spin connection over $\mathcal{M}$ is an infinitesimal connection on the spin fibre bundle 
$\tilde{\underline E}(\mathcal{M})$  and it will be represented by a $L_{\tilde{\underline G}}$-valued one-form $\tilde{\underline \omega}$.

Taking the pullback of $\tilde\lambda$ we can establish the following relation between the spin and the
Minkowski connection form:
$$\tilde{\omega}=\tilde\lambda_{*e}(\tilde{\underline\omega}),$$
where  the pullback is determinated in the unity $e$ of the Lie group $\tilde{\underline G}$. Since 
$\tilde\lambda:\tilde{\underline G}\longrightarrow SO^\uparrow_+(1,m-1;\mathcal{R})$ is a homomorphism of
Lie groups, $\tilde\lambda_{*e}:L_{\tilde{\underline G}}\longrightarrow L_{SO^\uparrow_+(1,m-1;\mathcal{R})}$ 
will be a homomorphism of the corresponding Lie algebras. And so we shall have
$$[\tilde\lambda_{*e}(\tilde{\underline \Lambda}),\tilde\lambda_{*e}(\tilde{\underline \Lambda'})]_{L_{SO^\uparrow_+(1,m-1;\mathcal{R})}}=
\tilde\lambda_{*e}([\tilde{\underline \Lambda},\tilde{\underline \Lambda'}]_{L_{\tilde{\underline G}}}),$$
where $[,]_{L_{SO^\uparrow_+(1,m-1;\mathcal{R})}}$ and $[,]_{L_{\tilde{\underline G}}}$ are the commutators of the
Lie algebras of $SO^\uparrow_+(1,m-1;\mathcal{R})$ and of $\tilde{\underline G}$, respectively, and
$\tilde{\underline \Lambda},\tilde{\underline \Lambda'}\in\tilde{\underline{\mathcal{G}}}$.

We shall now define $\tilde l=\tilde\lambda^{-1}_{*e}$, where $e$ is the unity of $SO^\uparrow_+(1,m-1;\mathcal{R})$.
Since $\tilde l$ establish a homomorphism between the Lie algebras of $\tilde{\underline G}$ and 
$SO^\uparrow_+(1,m-1;\mathcal{M})$ we can make, by $\tilde l$, a correspondence between $\tilde\omega$ and $\tilde{\underline\omega}$:
$$\tilde{\underline\omega}=\tilde l(\tilde\omega).$$

We will construct $\tilde l$ in the following way, being 
$\tilde{\underline G}$ a subgroup of a Clifford algebra $\underline C(1,m-1)$, let us consider the action of 
$\underline C(1,m-1)$ over each fibre of $\tilde{\underline E}(\mathcal{M})$. 
Let $\{\underline\gamma_{\hat\alpha}\}_{\hat\alpha=0}^{m-1}$ be
the linear operators on each fibre of the spin bundle, representing the generators of $\underline C(1,m-1)$:
\begin{equation}
        [\underline\gamma_{\hat\alpha},\underline\gamma_{\hat\beta}]=2\eta_{\hat\alpha\hat\beta},
\end{equation}
where $\eta_{\hat\alpha\hat\beta}=\textnormal{diag}(-1,1,1,...,1)$ is the Minkowski metric.
The generators of the local proper orthocronous Lorentz transformations, that is, a basis of
the Lie algebra of $\tilde{\underline G}$, will be related to the generators
of the Clifford algebra by
\begin{equation}
        \tilde{\underline\Sigma}_{\hat\alpha\hat\beta}=\frac14 
                [\underline\gamma_{\hat\alpha},\underline\gamma_{\hat\beta}],
\end{equation}
so if we make, for an element $\tilde\Lambda$ of the Lie algebra of $SO^\uparrow_+(1,m-1;\mathcal{R})$, the 
following definition 
\begin{equation}
        \tilde l(\tilde\Lambda) = \frac12 \tilde\Lambda^{\hat\alpha\hat\beta}
                \tilde{\underline\Sigma}_{\hat\alpha\hat\beta},
\end{equation}
we have establish a homomorphism between the two Lie algebras. In fact we can prove that 
$[\tilde l(\tilde{\Lambda}),\tilde l(\tilde{\Lambda'})]=
        \tilde l([\tilde{\Lambda},\tilde{\Lambda'}])$.

Furthermore, we will have that, for $\tilde\Lambda\in SO^\uparrow_+(1,m-1;\mathcal{R})$,
\begin{equation}
[l(\tilde\Lambda),\underline\gamma^{\hat\alpha}]=\underline\gamma^{\hat\beta}\Lambda^{\hat\alpha}_{\hat\beta}.
\end{equation}

In this way we have for a natural spin connection over $\mathcal{M}$,
$$\tilde{\underline\omega}=\tilde l(\tilde\omega)=\frac12\tilde\omega^{\hat\alpha\hat\beta}\tilde{\underline
\Sigma}_{\hat\alpha\hat\beta},$$
where $\tilde\omega^{\hat\alpha\hat\beta}$ are the components of the Minkowski connection for the basis
of the Lie algebra of $SO^\uparrow_+(1,m-1;\mathcal{R})$ related to $\tilde{\underline\Sigma}_{\hat\alpha\hat\beta}$
by $\tilde l$. Calculating the commutators and the anti-commutators we will get for the spin connection
\begin{equation}
        \tilde{\underline\omega}=
        \frac14\tilde\omega_{\hat\alpha\hat\beta}\underline\gamma^{\hat\alpha}\underline\gamma^{\hat\beta}.
\end{equation}

\paragraph{The Spinor Differential.}
Let $\phi$ be a spinor, i.e., $\phi\in\tilde{\bar E}(\mathcal{M})$. We define the spinor differential of $\phi$ as
the absolute differential of $\phi$ in the spinor connection $\tilde{\bar \omega}$:
\begin{equation}
  \tilde{\underline D}\phi=d\phi+\tilde{\underline\omega}\phi=d\phi+\frac14\tilde\omega_{\hat\alpha\hat\beta}
        \underline\gamma^{\hat\alpha}\underline\gamma^{\hat\beta}\phi,
\end{equation}
or in components $\tilde{\underline D}\phi=\tilde{\underline\nabla}_{\hat\beta}\phi \tilde e^{\hat\beta}$.

\paragraph{The Spinor Curvature.}
The curvature form of the spin connection is given by $\tilde{\underline\Omega}=\tilde{\underline D}\tilde{\underline\omega}$ and
it can be written in terms of $\tilde\omega$:
$$\tilde{\underline\Omega}=\tilde{\underline D}\tilde{\underline\omega}=\frac14(d+\frac14\tilde\omega_{\hat\alpha\hat\beta}\underline\gamma^{\hat\alpha}\underline\gamma^{\hat\beta})
\tilde\omega_{\hat\alpha\hat\beta}\underline\gamma^{\hat\alpha}\underline\gamma^{\hat\beta}.$$
The Riemann and Ricci tensors can be determined directly from the last expression. We have in a coordinate
basis
$$\tilde{\underline R}_{\hat\alpha\hat\beta}^{\hat\gamma}\,_{\hat\delta}=
\partial_{\hat\alpha}\underline\gamma_{\hat\beta}^{\hat\gamma}\,_{\hat\delta}-
\partial_{\hat\beta}\underline\gamma_{\hat\alpha}^{\hat\gamma}\,_{\hat\delta}+
\underline\gamma_{\hat\alpha}^{\hat\gamma}\,_{\hat\rho}\underline\gamma_{\hat\beta}^{\hat\rho}\,_{\hat\delta}.$$
From this we can calculate the Ricci tensor and the scalar of curvature in the usual way.

\paragraph{The Dirac Operator and the Spinor Laplacian.}
The Dirac operator is defined by\footnote{here the indices on the covariant derivative are of Lorentz type. In order to
write the Dirac operator in terms of coordinate indices we must use the vielbein (cf. section 2.1.11)},
\begin{equation}
        /\!\!\!\!\tilde D\phi=\underline\gamma^{\hat\beta}\tilde{\underline\nabla}_{\hat\beta}\phi.
\end{equation}
A spinor field that satisfies $/\!\!\!\!\tilde D\phi=0$ is called an harmonic spinor. The square of the Dirac
operator is called the spinor laplacian and is given by
\begin{equation}
        \tilde{\underline \Delta}=-\eta^{\hat\alpha\hat\beta}\tilde{\underline\nabla}_{\hat\alpha}
        \tilde{\underline\nabla}_{\hat\alpha}+\frac14\tilde{\underline R},
\end{equation}
where $\tilde{\underline R}$ is the scalar of curvature.

\subsection{The Fibre Bundle of Adapted Orthonormal Frames}\label{fbaof}

\subsubsection{The Principal Fibre Bundle $\mathcal{M}(V_n,\mathcal{G},\pi,\Phi)$}

Let us consider the fibre bundle of orthonormal frames $\tilde E(\mathcal{M})$ over 
$\mathcal{M}(V_n,\mathcal{G},\pi,\Phi)$ whose  structure group is the orthogonal group $O(1,m-1;\mathcal{R})$.
As we have seen in section \ref{principalfibrebundle} there is a natural lift of the action of $\mathcal{G}$
to $\mathcal{M}$ by $\sigma$. The same happens for the bundle $\tilde E(\mathcal{M})$:
since $\mathcal{G}$ acts on $\mathcal{M}$ by isometries, it is possible to define a natural lift of the action of $\mathcal{G}$ to $\tilde E(\mathcal{M})$,
\begin{eqnarray}
        ^{\rightarrow}\tilde\sigma:\mathcal{G}\times\tilde E(\mathcal{M})\longrightarrow\tilde E(\mathcal{M}),\\
        ^{\leftarrow}\tilde\sigma:\tilde E(\mathcal{M})\times\mathcal{G}\longrightarrow\tilde E(\mathcal{M}).
\end{eqnarray}
As $\tilde E(\mathcal{M})$ is now subject to the action of two Lie groups, $O(1,m-1;\mathcal{R})$ and $\mathcal{G}$,
we can reduce its structure group. In order to do so we must adapt the fibre bundle of orthonormal frames
to the reduction, choosing those frames of $\tilde E(\mathcal{M})$ that can be decomposed in two orthonormal frames
lying separately in the tangent vector space of the base space $V_n$ fibre $G_x$ and of the internal space 
$\mathcal{G}$. 
 

Let us consider a generic point $z$ of $\mathcal{M}$. 
The tangent vector space $T_z\mathcal{M}$ can be decomposed in two parts:
a vertical space $\mathcal{V}_z$, defined as the tangent space to the fibre $G_{x=\pi(z)}$ in the point $z$ (cf. 
section \ref{principalfibrebundle}), and a horizontal space $\mathcal{H}_z$ defined by an infinitesimal connection
on the principal fibre bundle $\mathcal{M}(V_n,\mathcal{G},\pi,\Phi)$. 
An orthogonal frame $\tilde e^z=\{\tilde e_{\hat\alpha}\}_{\hat\alpha=0}^{m-1}\in\tilde E(\mathcal{M})$ 
at $z\in\mathcal{M}$ 
is called adapted if 
$$\tilde e^z=\ddot e^z=\{\tilde e_{\alpha},\tilde e_{a}\}_{\alpha=0,\cdots,n-1; a=n,\cdots,m-1},$$ with
$\tilde e_{\alpha}\in\mathcal{V}_z$, $\tilde e_a\in\mathcal{H}_z$, where $n=\dim V_n$ and $m-n=\dim\mathcal{G}$.
The set $\ddot E(\mathcal{M})$ of all adapted orthogonal frames in all points of $\mathcal{M}$ is a subset 
of $\tilde E(\mathcal{M})$
and is obviously a principal fibre bundle over $\mathcal{M}$ with the structure group 
$O(1,n-1;\mathcal{R})\times O(m-n-1;\mathcal{R})$. It is called the Fibre Bundle of Adapted Orthonormal Frames 
of $\mathcal{M}$.
The induced infinitesimal connection $\ddot\omega$ will be the restriction of the connection
form $\tilde\omega$ of the fibre bundle of orthonormated frames $\tilde E(\mathcal{M})$ to $\ddot E(\mathcal{M})$.

\subsubsection{The Fibre Bundle $\mathcal{M}(V_n,\mathcal{G}/\mathcal{H},\mathcal{G},\pi,\Phi)$}
If we consider the case on which the fibre bundle of linear frames (and, obviously, the fibre bundle 
of orthonormal frames) is defined over a fibre bundle
$\mathcal{M}(V_n,\mathcal{G}/\mathcal{H},\mathcal{G},\pi,\Phi)$ in which the internal space is
homomorphic with a homogeneous space $\mathcal{G}/\mathcal{H}$ another definition of the 
fibre bundle of adapted orthonormated frames shall be given.
Let $z\in\mathcal{M}$ and let $\mathcal{V}_z$ and $\mathcal{H}_z$ be the vertical and the horizontal
vector spaces on $z$. We have then the decomposition of the tangent vector space of $\mathcal{M}$ at
z: 
$$T_z\mathcal{M}=\mathcal{V}_z\oplus\mathcal{H}_z.$$
 
Let $\tau\in\mathcal{V}_z$ be a vertical vector at $z$. To this we can associate a tangent vector $\theta_y\in T_y M$
to $M$ at $y\in M$ using an element $\phi_{Ux}$ of $\Phi$:
$$z=\phi_{Ux}(y)\in M_x\subset\mathcal{M},x=\pi(z)\in V_n: \theta_y=\phi^{-1*}_{Ux,y}(\tau)\in T_y M.$$
Using another element $\phi_{Vx}$ of $\Phi$ we can associate another tangent vector $\theta_{y'}\in T_{y'} M$
to $M$ at another point $y'\in M$:
$$z=\phi_{Vx}(y')\in M_x\subset\mathcal{M},x=\pi(z)\in V_n:\theta_{y'}=\phi^{-1*}_{Vx,y'}(\tau)\in T_{y'} M.$$
But since there must be an element $g$ of $\mathcal{G}$ such that $\phi_{Ux}(gy)=\phi_{Vx}(y')$, we have that
$\theta_y'=g\theta_y$.
Taking an element of $\Phi$ such that $z=\phi_{Wx}(e)$, with $e$ the unit of $M$, we have that there is an
element $g\in\mathcal{G}$ such
$\theta_y=g\theta_e,$
with $\theta_e$ lying on $T_e M$ and $\theta_y$ on $T_y M$. We can then make the following association
$\tau=\phi^*_{Wx,z}(\theta_e).$

Since we can decompose the tangent vector space of $M$ at $e$ as 
$$T_e M=L_{\mathcal{N}/\mathcal{H}}\oplus L_\mathcal{L}.$$
we can define an adapted frame of $T_e M$ as a frame $\lambda^e=\{\lambda_a\}_{a=n,\cdots,m-1}$
such that $\lambda_i\in L_{\mathcal{N}/\mathcal{H}}$, for $i=n,\cdots,(\textnormal{dim}\mathcal{N}/\mathcal{H}+n-1)$ and
$\lambda_o\in L_{\mathcal{L}}$, for $o=(\textnormal{dim}\mathcal{N}/\mathcal{H}+n),\cdots,m-n-1$.
To this adapted frame of $T_e M$ we can associate an adapted frame $\{\ddot e_a\}_{a=n,\cdots,m-1}$ of $\mathcal{V}_z$ 
by taking
$$\ddot e_i=\phi^*_{Wx,z}(\lambda_i),$$
$$\ddot e_o=\phi^*_{Wx,z}(\lambda_o),$$
where $i=n,\cdots,(\textnormal{dim}\mathcal{N}/\mathcal{H}+n-1)$ and 
$o=(\textnormal{dim}\mathcal{N}/\mathcal{H}+n),\cdots,m-n-1$. This is nothing but a decomposition of the vertical space
\begin{equation}
        \mathcal{V}_z=\mathcal{V}_z^{(\mathcal{N}/\mathcal{H})}\oplus\mathcal{V}_z^{(\mathcal{L})}
\end{equation}
An adapted orthonormal frame of $T_z\mathcal{M}$ is then defined as a frame 
$$\ddot e^z=\{\ddot e_\mu,\ddot e_i,\ddot e_o\}_{\mu=0,\cdots,n-1;i=n,\cdots,(\textnormal{dim}\mathcal{N}/\mathcal{H}+n-1);
o=(\textnormal{dim}\mathcal{N}/\mathcal{H}+n),\cdots,m-n-1},$$
where $\{\ddot e_\mu\}_{\mu=0,\cdots,m-1}$ is a orthonormal frame of the horizontal vector space $\mathcal{H}_z$
and where $\{\ddot e_i,\ddot e_o\}_{i=n,\cdots,(\textnormal{dim}\mathcal{N}/\mathcal{H}+n-1);
o=(\textnormal{dim}\mathcal{N}/\mathcal{H}+n),\cdots,m-n-1}$
is an adapted orthonormal frame of the vertical vector space $\mathcal{V}_z$.

As before we define the fibre bundle of adapted orthonormal frames as the fibre bundle that has as its
total space the set $\ddot E(\mathcal{M})$ of all adapted orthonormal frames of $\mathcal{M}$ in all of its points, as base
space $\mathcal{M}$ and as structure group the group 
$$O(1,n-1;\mathcal{R})\times O(\textnormal{dim}\mathcal{N}/\mathcal{H};\mathcal{R})\times
O(m-n-\textnormal{dim}\mathcal{N}/\mathcal{H}-1;\mathcal{R}).$$

As before, the induced infinitesimal connection $\ddot\omega$ will be the restriction of the connection
form $\tilde\omega$ of the fibre bundle of orthonormated frames $\tilde E(\mathcal{M})$ to $\ddot E(\mathcal{M})$.

\subsection{The Vielbein}

Let us consider a $m$-dimensional $C^\nu$-differentiable manifold $\mathcal{M}$ 
over which it is defined the principal fibre bundle of linear frames $\hat E(\mathcal{M})$.

There can be two types of tangent vectors to $\mathcal{M}$ at a point $z$:
those that transform as vectors under infinitesimal general coordinate
transformations and those that transform as Lorentz vectors under the
action of $SO(1,m-1)$.

Under infinitesimal general coordinate transformations
\begin{equation}
  z^{\hat\alpha}\rightarrow z^{\hat\alpha}+\xi^{\hat\alpha}(z),
\end{equation}
a covariant vector $v_{\hat\alpha}$ of the first type transforms according to
\begin{eqnarray}
v_{\hat\alpha}\rightarrow v_{\hat\alpha}+\delta_\xi v_{\hat\alpha},\\
  \delta_\xi v_{\hat\alpha}=-\partial_{\hat\alpha}\xi^{\hat\beta}v_{\hat\beta}-\xi^{\hat\beta}\partial_{\hat\beta}v_{\hat\alpha}.
\end{eqnarray}
Indices within this type of vectors are raised or lowered with the metric
$\gamma$.

Under a local Lorentz transformation,
\begin{equation}
  z^{\hat\alpha}\rightarrow \Lambda^{\hat\alpha}_{\hat\beta} z^{\hat\beta}
\end{equation}
a covariant vector $v_{\hat\alpha}$ of the second type transforms as
\begin{equation}
  v_{\hat\alpha}\rightarrow \Lambda^{-1\;\hat\beta}_{\hat\alpha}v_{\hat\beta}.
\end{equation}
Indices within this type of vectors are raised or lowered with the Minkowski
metric $\eta$.

Since they belong to the same tangent vector space, these two types of
vectors can be connected by one rotation, the vielbein 
$e^{\hat\alpha}_{\hat\beta}$.
If we use Latin indices to identify the Lorentz like components and
Greek indices to identify the covariant like components of a vector, the
two are related by
\begin{equation}
v_{\hat\alpha}=e_{\hat\alpha}^{\hat a} v_{\hat a}.
\end{equation}
If the vielbein has an inverse $e_{\hat a}^{\hat\alpha}$,
\begin{eqnarray}
e_{\hat a}^{\hat\alpha}e_{\hat\alpha}^{\hat b}=\delta_{\hat a}^{\hat b},\\
e_{\hat\alpha}^{\hat a} e_{\hat a}^{\hat\beta}=\delta_{\hat\alpha}^{\hat\beta},
\end{eqnarray}
it can be used to convert the covariant components of a vector into the
Lorentz ones
\begin{equation}
  v_{\hat a}=e_{\hat a}^{\hat\alpha}v_{\hat\alpha}.
\end{equation}

\section{Description of Matter Fields}

\subsection{The Configuration Space $\Gamma\mathcal{F}$}

Let $\mathcal{M}(V_n,\mathcal{G},\pi,\Phi)$ be the principal fibre bundle previously considered. Let be defined over
$\mathcal{M}$ a differentiable vector fibre bundle $\mathcal{F}(\mathcal{M},F,\mathcal{Y},p,\Psi)$. Let 
$\Gamma\mathcal{F}$ be the vector space of the sections $\phi$ of $\mathcal{F}$. This space has infinity dimension.
We will say that $\Gamma\mathcal{F}$ is the configuration space and that each $\phi\in\Gamma\mathcal{F}$,
$$\phi:\mathcal{M}\longrightarrow\mathcal{F}$$ represents a possible configuration of the matter field defined over $\mathcal{M}$. 

A field configuration 
$\phi\in\Gamma\mathcal{M}$ can be written as $\phi(z)=(z,\phi_z)$ in a neighborhood of $z$, 
with $z\in\mathcal{M}$ and $\phi_z\in F_z$,
being $F_z$ the fiber passing through $z$. We usualy refer ourselfs to $\phi_z$ as the field configuration on $z$.
Neglecting the point $z$ on which it is determinated, we will atribue to $\phi_\cdot$ also the designation of
matter field configuration. 

As presented above, a matter field configuration will be considered as a 0-form, that is, a scalar on $\mathcal{M}$ and
so $\phi\in\Lambda^0(\mathcal{M},\mathcal{F})$. A possible generalization of this scheme is to consider the matter
fields as $k$-forms on $\mathcal{M}$ with values on $\mathcal{F}$, that is, as elements of 
$\Lambda^k(\mathcal{M},\mathcal{F})$ (an example is the gravitational field $\gamma$). 
But since $\Lambda^k(\mathcal{M},\mathcal{F})$ is isomorphic with
$\Lambda^0(\mathcal{M},\Lambda^k\mathcal{M}\otimes F)$ our treatment can be applied also to this case (one changes
only the fibre of $\mathcal{F}$).

Let $\chi:\mathcal{Y}\times F\longrightarrow F$ be the action of $\mathcal{Y}$ on $F$ and let us represent a
field configuration $\phi\in\Gamma\mathcal{F}$ by $\phi(z)=(z,\phi_z)$, where $z\in\mathcal{M}$ and $\phi_z\in F_z$. 
Then, to every element $f_U$ of $F$ we can associate an element $\phi_z\in F_z$
by a homomorphism $\psi_U\in\Psi$, $U\subset\mathcal{M}$. 
If $\psi_{Uz}$ is the restriction of $\psi_U$ to $\{z\}\times F_z$ then we have 
$$\phi_z=\psi_{Uz}(f_U),$$
for some $f_U\in F$. Using another element $\psi_V$ of $\Psi$, $V\subset\mathcal{M}$ such that $z\in U\cap V$, we shall
have
$$\phi_z=\psi_{Vz}(f_V),$$
with $f_V\in F$ a different element of $F$. Since we must have $\psi^{-1}_{Uz}\circ\psi_{Vz}=y$ for some 
$y\in\mathcal{Y}$, we can establish the following relation
$$\phi_z=\psi_{Vz}(f_V)=\psi_{Uz}\circ\chi(y,f_U)=\psi_{Uz}(yf_U),$$
where we have made $yf_U=\chi(y,f_U)$ in order to simplify the notation. 

We can then define the following lift of the action of $\mathcal{Y}$ on $F$ to an action of $\mathcal{Y}$ on $\mathcal{F}$
by writing, for $\phi=(z,\phi_z)\in\mathcal{F}$ with $\phi_z=\psi_{Uz}(f_U)$, the map
\begin{equation}
        \begin{array}{c}
                \Psi:\mathcal{Y}\times\mathcal{F}\longrightarrow\mathcal{F}\\
                (y,\phi)\longmapsto y\phi=(z,\psi_{Uz}(yf)).
        \end{array}
\end{equation}

There will be lift of the action of $\mathcal{Y}$ to $\Gamma\mathcal{F}$ and it will be given by 
the following representation $T^\mathcal{Y}$
of the group $\mathcal{Y}$
\begin{equation}
        [T^{\mathcal{Y}}_y\phi](z)=\Psi(y,\phi)(y^{-1}z)=\Psi(y,(y^{-1}z,\phi_{y^{-1}z}))=(y^{-1}z,y\phi_{y^{-1}z}),
\end{equation}
for $y\in\mathcal{Y}$, $\phi=(z,\phi_z)\in\Gamma\mathcal{F}$ and where $y^{-1}z$ is the action of $\mathcal{Y}$ on
$\mathcal{M}$ (if there is one). We will then get $[T^{\mathcal{Y}}_y\phi]_z=y\phi_{y^{-1}z}$.

Besides this representation there can be another one\footnote{a lift of the action of $\mathcal{G}$ on $\mathcal{M}$
to $\mathcal{F}$ is supposed.}: a representation $T^\mathcal{G}$ of the action of the 
structure group of $\mathcal{M}$,
$\mathcal{G}$, on the bundle $\mathcal{F}$ (this action is supposed to be a lift of the action $\sigma$ of $\mathcal{G}$
on $\mathcal{M}$ to the bundle of automorphisms of $\mathcal{F}$). The induced representation will be
\begin{equation}\label{trans2}
        [T^\mathcal{G}_g\phi](z)=g\phi(g^{-1}z),
\end{equation}
with $g\in\mathcal{G}$ and $\phi\in\Gamma\mathcal{F}$.
Using $\phi(g^{-1}z)=(g^{-1}z,\phi_{g^{-1}z})$, we get
$$[T^\mathcal{G}_g\phi](z)=(z,g\phi_{g^{-1}z}),$$ and so we can make $[T_g^\mathcal{G}\phi]_z=g\phi_{g^{-1}z}$.

In general, $\mathcal{G}$ will represent some type of internal symmetry and so its action on a matter field
configuration should be reduced to the action of $\mathcal{G}$ on the field itself and not on the point
on where it is calculated. By other words, the action of an internal symmetry to a manifold representative of 
our universe should be reduced to the identity, that is, such manifold should be invariant to such action.
This is not the case, as one sees, since $\mathcal{M}$ is not invariant to the $\mathcal{G}$ action. This conduces
us directly to dimensional reduction, in order to find such symmetry invariant manifold. One of the main reasons for 
giving a $\mathcal{G}$-(principal) fibre bundle structure to $\mathcal{M}$ comes then easily from that fact: an external
symmetry of a larger universe, $\mathcal{M}$, can be seen as an internal symmetry of a reduced one, or
inversely, an internal symmetry in our measured universe can be regarded as an external symmetry of a larger one.

\subsection{The Absolute Differential of a Matter Field Configuration}

A matter field configuration $\phi\in\Lambda^k(\mathcal{M},\mathcal{F})$, with $\mathcal{M}(V_n,\mathcal{G},\pi,\Phi)$ 
a principal fibre bundle,
being a $k$-form over $\mathcal{M}$, has an absolute differential on $\mathcal{M}$ defined as
$$D\phi(z)(\tau_1,\ldots,\tau_k)=d\phi(z)(\mathcal{H}\tau_1,\ldots,\mathcal{H}\tau_k),$$
with $\tau_i\in T_z\mathcal{M}$, $(i=1,\ldots,k)$, and $\mathcal{H}\tau$ being the projection of $\tau$ in
the horizontal vector space $\mathcal{H}_z$.

Belonging to a representation of the structure group $\mathcal{Y}$ of $\mathcal{F}$ we can also define the 
absolute differential of a matter field configuration on $\mathcal{F}$. It will be given by
\begin{equation}
D\phi(z)(\tau_1,\tau_2\ldots)=D_\mathcal{M}\phi(z)(\mathcal{H}\tau_1,\mathcal{H}\tau_2,\ldots),
\end{equation}
with $\tau_i\in T_\phi\mathcal{F}$ and $\mathcal{H}\tau$ being the projection of $\tau$ in
the horizontal vector space $\mathcal{H}_z$ on the (principal) fibre bundle $\mathcal{F}$. $D_\mathcal{M}\phi(z)$
denotes the absolute differential defined on $\mathcal{M}$.

An observation must now be done. A matter field configuration will carry, in general, not only a representation
of the structure group $\mathcal{G}$ of $\mathcal{M}$ but also carry representations of other groups.
When the concept of absolute differential is used it is understooded that all the absolute differentials of
the matter field configuration in all the (principal) fibre bundles where it is defined were taken. In other words,
a principal fibre bundle has been constructed, being its structure group the group product of all
the groups that have a non trivial action over the matter field configuration (cf. section 3.1.4). The absolute differential
is then defined over this ``total'' principal fibre bundle. 

As an example, a common situation is that in which the fibre $F_z$ on which a scalar\footnote{scalar in relation to
a diffeomorphism.}
matter field configuration is valued
is isomorphic with the general linear group $\mathcal{Y}=GL(1,m-1;\mathcal{R})$ and over which the same group acts. 
We can then identify the two fibres and simply write $\mathcal{F}(\mathcal{M},F,\mathcal{Y})=\hat E(\mathcal{M})$. 
The matter field will then
be called a vector matter field. Possible reductions can be made by the introduction of a metric $\gamma$
on $\mathcal{M}$, reducing $\hat E(\mathcal{M})$ to $\tilde E(\mathcal{M})$. The absolute differential
of a $\mathcal{G}$-invariant matter field configuration will then be
\begin{equation}
        D\phi(z)=d\phi(z)+\tilde\omega\phi(z),
\end{equation}
where $\tilde\omega$ is the Minkowski connection form defined on the principal fibre bundle $\tilde E(\mathcal{M})$.
Chosen a frame $\tilde e^z$ of $\tilde E(\mathcal{M})$ at $z\in\mathcal{M}$, we can write
$$\phi(z)=\phi^{\hat\alpha}(z)\tilde e_{\hat\alpha},$$
where $\phi^{\hat\alpha}(z)$ are the components of $\phi(z)$ in that frame. In components,
\begin{equation}
        D\phi^{\hat\alpha}(z)=d\phi^{\hat\alpha}(z)+\tilde\omega^{\hat\alpha}_{\hat\beta}\phi^{\hat\beta},
\end{equation}
where $\tilde e_z$ is the coframe correspondent to $\tilde e^z$.

If we define $\mathcal{F}$ to be the fibre bundle of spinors, then the matter field will be of a
spinorial nature, transforming as a spinor. We have already given the absolute differential of a spinor:
\begin{equation}
        \tilde{\underline D}\phi=d\phi+\tilde{\underline\omega}\phi=d\phi+\frac14\tilde\omega_{\hat\alpha\hat\beta}
        \underline\gamma^{\hat\alpha}\underline\gamma^{\hat\beta}\phi
        =\tilde{\underline\nabla}_{\hat\beta}\phi \tilde e^{\hat\beta}.
\end{equation}
In a coordinate basis we shall have
\begin{equation}
\underline{\bar D}\phi = \tilde e^{\hat b}_{\hat\beta}\underline{\bar\nabla_{\hat b}}\phi \partial^{\hat\beta}.
\end{equation}

\subsection{The Action of a Form}\label{produtointerno}

\subsubsection{The Global Scalar Product over $\mathcal{M}$ and the Action of a Form}
Assuming that $\mathcal{M}$ is compact, we can define a global scalar product 
$<\cdot,\cdot>_\mathcal{M}:\Lambda^p\mathcal{M}\times\Lambda^p\mathcal{M}\longrightarrow\mathcal{R}$ 
over the $p$-forms of 
$\mathcal{M}$ by the relation
\begin{equation}
        <\alpha,\beta>_\mathcal{M}=\int_\mathcal{M}\alpha\wedge\star\beta=\int_\mathcal{M}\beta\wedge\star\alpha
\end{equation}

Let us define the global norm of a $p$-form over $\mathcal{M}$ by the relation $\left\|\alpha\right\|_\mathcal{M}=\sqrt{<\alpha,\beta>_\mathcal{M}}$. 

The classical action of two $p$-forms is defined by
\begin{equation}
        S_\mathcal{M}[\alpha,\beta]=<\alpha,\beta>_\mathcal{M}.
\end{equation}
In particular, the classical action of a gauge field will be the square of the curvature of the connection that 
defines it\footnote{in general one introduces a factor $\frac1{vol(\mathcal{Y})}$ in the following expression,
where $\mathcal{Y}$ is the gauge group and $vol(\mathcal{Y})$ its volume.},
\begin{equation}
        S_\mathcal{M}[\Omega]=\left\|\Omega\right\|^2_\mathcal{M}.
\end{equation}

\subsubsection{The Action of a $\mathcal{G}$-invariant form over a product space $\mathcal{G}\times_\mathcal{H}\bar\mathcal{M}$}

Let us consider a $\mathcal{G}$-invariant form $\alpha$ defined over a product space $\mathcal{G}\times_\mathcal{H}\bar\mathcal{M}$,
with both $\mathcal{G}$ and $\bar\mathcal{M}$ compact, and where $\mathcal{H}\subset\mathcal{G}$, and
let its action be given by
\begin{equation}
  S_{\mathcal{G}\times_\mathcal{H}\bar\mathcal{M}}[\alpha]=\parallel\alpha\parallel_{\mathcal{G}\times_\mathcal{H}\bar\mathcal{M}}.
\end{equation}
Then this action can be reduced to an action over $\bar\mathcal{M}$ by integrating over $\mathcal{G}/\mathcal{H}$,
\begin{equation}
  S_{\bar\mathcal{M}}[\alpha]=vol(\mathcal{G}/\mathcal{H})\parallel\alpha\parallel_{\bar\mathcal{M}},
\end{equation}
with $vol(\mathcal{G}/\mathcal{H})=\parallel 1 \parallel_{\mathcal{G}/\mathcal{H}}$ the volume of the coset 
space $\mathcal{G}/\mathcal{H}$.
A special case is the one in which $\mathcal{H}=\{e\}$:
$$S_{\bar\mathcal{M}}[\alpha]=vol(\mathcal{G})\parallel\alpha\parallel_{\bar\mathcal{M}}.$$

%
%



\chapter{DIMENSIONAL REDUCTION}

\textit{The dimensional reduction process of gravity, matter and gauge fields is presented. In the first section only $\mathcal{G}$-invariant
fields are considered. A $\mathcal{G}$-invariant metric defined over the total space $\mathcal{M}$ will give rise, after
performed the dimensional reduction of the Hilbert-Einstein action, to an induced metric on the base space together with
a Yang-Mills field and a finite number of scalar fields. In the second section, the behaviour of general fields defined
over $\mathcal{M}$ is studied with the help of abstract harmonic analysis. In general, every field defined on $\mathcal{M}$ will
manifest itself on the base space $V_n$ as an infinite tower of fields with increasing associated masses.}

\section{Dimensional Reduction of $\mathcal{G}$-invariant Fields}

\subsection{Dimensional Reduction of Gravity}

\subsubsection{The Principal Fibre Bundle $\mathcal{M}(V_n,\mathcal{G},\pi,\Phi)$}

Let $\mathcal{M}(V_n,\mathcal{G},\pi,\Phi)$ be an $m$-dimensional $C^\nu$-differentiable principal fibre bundle
defined over the base space $V_n$, an $n$-dimensional $C^\nu$-manifold, and that possesses as structure group
the Lie group $\mathcal{G}$.
As it was shown in the previous chapter, the action of $\mathcal{G}$ on itself will induce an action of the same
group over the manifold $\mathcal{M}$ (cf. section \ref{principalfibrebundle}). If we define over $\mathcal{M}$
the fibre bundle of frames $\hat E(\mathcal{M})$, then this action will be lift to an action of $\mathcal{G}$ on
$\hat E(\mathcal{M})$. It will be in this way subject to the action of two groups: the linear group 
$GL(1,m-1;\mathcal{R})$ and the group $\mathcal{G}$. The bundle $\hat E(\mathcal{M})$ will then be unnecessarily 
large and so it can be reduced.

Since we have already shown that $\hat E(\mathcal{M})$ can be reduced, by defining a metric $\gamma$ on $\mathcal{M}$
 to the fibre bundle of orthonormated frames $\tilde E(\mathcal{M})$ which possesses as structure group
 $O(1,m-1;\mathcal{R})$ (cf. section \ref{fbof}) and this last bundle
 can yet be reduced to the fibre bundle of adapted orthonormated frames $\ddot E(\mathcal{M})$ which has
 for the structure group $O(1,n-1)\times O(m-n-1)$ (cf. section \ref{fbaof}), we shall work directly with $\ddot E(\mathcal{M})$.

Let $(\mathcal{M},\gamma)$ be a $C^\nu$-riemannian manifold for which the metric $\gamma$ is bi-invariant to the
group $\mathcal{G}$, i.e., such that
\begin{equation}
        ^{\rightarrow}\Phi_g^*\gamma=\gamma=^{\leftarrow}\Phi_g^*\gamma,
\end{equation}
for all $g\in\mathcal{G}$, being $\Phi$ the action of $\mathcal{G}$ on $\mathcal{M}$ (cf. section \ref{principalfibrebundle}).

The bi-invariance property of $\gamma$ will imply that the metric will assume a constant value over each fibre $G_x$,
$x\in V_n$, of the principal fibre bundle $\mathcal{M}$. The metric can then depend only on the projection
of the point over which is defined, i.e., the point $x$ in $V_n$:
$$x\in V_n:z=\pi(x)\in G_x\subset\mathcal{M}\Rightarrow\gamma_z(z)=\gamma_z(x),$$
with $\gamma_z\in T_z^*\mathcal{M}\otimes T_z^*\mathcal{M}$.

It is possible then to define a metric $g$ on the base space $V_n$, obtaining in this way a riemannian structure 
$(V_n,g)$, and a family of $\mathcal{G}$-invariant metrics $\{\xi_x\}$, on each point $x\in V_n$, for the correspondent
fibre $G_x$. 

One proceeds as follows: let $\tau_1,\tau_2\in T_z\mathcal{M}$ be two tangent vectors to $\mathcal{M}$
and $\mathcal{H}\tau_1,\mathcal{H}\tau_2\in\mathcal{H}_z$ be their projection in the horizontal space $\mathcal{H}_z$
at $z$. If $x=\pi(z)\in V_n$, we will define the scalar product of two tangent vectors to $V_n$
in the point $x$, $v_1,v_2\in T_x V_n$, that are the projection of $\mathcal{H}\tau_1,\mathcal{H}\tau_2$, respectively,
as $g_x(v_1,v_2)=\gamma_z(\mathcal{H}\tau_1,\mathcal{H}\tau_2)$. $g$ will be a bilinear positive form and will be
$C^\nu$ in $V_n$ as it is obvious from the definition of $\gamma$. This definition is independent of the point
$z\in G_x$ chosen since the metric $\gamma$ is $\mathcal{G}$-bi-invariant.
For the same point $x\in V_n$, since $\gamma$ is $\mathcal{G}$-invariant, we can define the metric $\xi_x$ of $G_x$ as the restriction $\xi_x=\gamma_z|_{\mathcal{V}_z}$ with $z=\pi(x)\in G_x$ some point over $x$ and $\mathcal{V}_z$
the vertical vector space at $z$.

Since a $\mathcal{G}$-invariant metric on $\mathcal{M}$ defines an infinitesimal connection on the principal
fibre bundle $\mathcal{M}(V_n,\mathcal{G},\pi,\Phi)$ and since we have seen that the same metric induces a metric 
on the base space $V_n$ of this bundle and a family of $\mathcal{G}$-invariant metrics $\{\xi_x\}$ on each fibre 
$G_x$ of $\mathcal{M}$, we can perform the
following dimensional reduction
$$(\mathcal{M},\gamma)\longrightarrow(V_n,g,\omega,\xi).$$
The inverse of it is also possible, that is, given an infinitesimal connection $\omega$ on $\mathcal{M}$, a metric $g$ on $V_n$, and a family of $\mathcal{G}$-invariant metrics $\{\xi_x\}$ on the fibres of $\mathcal{M}$ an 
$\mathcal{G}$-invariant metric $\gamma$ of $\mathcal{M}$ can be constructed. In fact, let $\tau_1\tau_2\in T_z\mathcal{M}$ be
two tangent vectors to $\mathcal{M}$ at $z$ and let $\mathcal{H}\tau_1,\mathcal{H}\tau_2\in\mathcal{H}_z$ be their horizontal
part. Then we define the product between $\tau_1$ and $\tau_2$ in terms of $g$, $\xi_x$ and of $\mathcal{H}_z$ as
\begin{equation}\label{metrica}
        \gamma_z(\tau_1,\tau_2)=g_x(\pi^*_z\tau_1,\pi^*_z\tau_2)+\xi_x(\tau_1-\mathcal{H}\tau_1,\tau_2-\mathcal{H}\tau_2),
\end{equation}
where $\mathcal{V}\tau_i=\tau_i-\mathcal{H}\tau_i\in\mathcal{V}_z$, for $i=1,2$, are vertical vectors at $z$.

Since we have defined a metric $g$ over $V_n$, we can define the fibre bundle of orthonormated frames of $V_n$,
$\tilde E(1,n-1;\mathcal{R})$. The dimensional reduction process made can then be expressed by the sequence

$$\hat E(\mathcal{M})\stackrel{\gamma}{\rightarrow}\tilde E(\mathcal{M})\stackrel{\tilde\omega}{\rightarrow}
\ddot E(\mathcal{M})\stackrel{\ddot\omega,\pi}{\rightarrow}\tilde E(V_n),$$
or in terms of the structure groups of the previous principal fibre bundles,
$$GL(1,m-1;\mathcal{R})\stackrel{\gamma}{\rightarrow}O(1,m-1;\mathcal{R})
\stackrel{\tilde\omega}{\rightarrow} O(1,n-1;\mathcal{R})\times
O(m-n-1;\mathcal{R})\stackrel{\ddot\omega,\pi}{\rightarrow} O(1,n-1;\mathcal{R}).$$

\paragraph{Metric Decomposition.}

The differential of the bundle projection $\pi$ allows us to make a correspondence between the tangent vector space 
to the total space in a given point and the tangent vector space to base space in the point that is the projection
by $\pi$ of the previous:

$$\pi^*_z:T_z\mathcal{M}\rightarrow T_{x=\pi(z)}V_n.$$

Let $\ddot e^z=\{\ddot e_{\hat \alpha}\}_{\hat\alpha=0,1,\cdots,m-1}$ be an adapted orthonormated base 
of $T_z\mathcal{M}$
, and let $v\in T_z\mathcal{M}$. Then 
$v=v^{\hat\alpha}\ddot e_{\hat\alpha}$. Taking the differential application of this vector one gets

$$\pi^*_z(v)=\pi^*_z(v^{\hat\alpha}\ddot e_{\hat\alpha})=v^{\hat\alpha} \pi^*_z(\ddot e_{\hat\alpha}).$$

For each $\hat\mu$, $\pi^*_z(\ddot e_{\hat\alpha})$ will be a vector of $T_x V_n$, so if we take for the base of this
space $\{\ddot e_\alpha\}_{\hat\alpha=0,1,\cdots,n-1}$, we will get

$$\pi^*_z(\ddot e_{\hat\alpha})=\pi_z^{*\beta}(\ddot e_{\hat\alpha})\ddot e_\beta=
\pi_{z,\hat\alpha}^{*\beta}\ddot e_\beta,$$
where $\pi_z^{*\beta}$ will be the components of $\pi^*_z$ in the considered base. 

Let $\gamma_{\hat\alpha\hat\beta}$ and $g_{\mu\nu}$ be the covariant components of the metrics $\gamma$ and $g$ in the considered basis,
and $\gamma^{\hat\alpha\hat\beta}$ and $g^{\mu\nu}$ their contravariant components.
Then we can project the metric $\gamma$ of the total space in the base space through the differential of the bundle
projection. In components,
$$g^{\mu\nu}=\pi^{*\mu}_{z,\hat\alpha}\pi^{*\nu}_{z,\hat\beta}\gamma^{\hat\alpha\hat\beta}.$$

Let $\sigma$ be a global section of $\mathcal{M}$, defined as
$$\sigma:V_n\rightarrow\mathcal{M},$$
such that $\pi\circ\sigma=id_{V_n}$. Then $\sigma^*_{x=\pi(z)}:T_x V_n\rightarrow T_{z}\mathcal{M}$.

Let $v_1,v_2\in T_x V_n$ be two tangent vectors to $V_n$ at $x$ and let $\tau_1,\tau_2\in T_z\mathcal{M}$ be two tangent vectors to 
$\mathcal{M}$ at $z=\pi(x)$ such that
$\tau_i=\sigma^*_x v_i$ for $i=1,2$. We define the scalar product between $v_1$ and $v_2$ at $x\in V_n$ as
\begin{equation}
g(v_1,v_2)=\gamma(\tau_1,\tau_2)=[\gamma\circ(\sigma^*_x\otimes\sigma^*_x)](v_1,v_2).
\end{equation}
If $\sigma^{*\hat\alpha}_{x}$ are the components of $\sigma^*_x$ in the base $\{\ddot e^{\hat\alpha}\}_{\hat\alpha=0,\cdots,m-1}$, 
and if we make $\sigma^{*\hat\alpha}_{x}(\tilde e_\mu)=\sigma^{*\hat\beta}_{x,\nu}$, for $\{\tilde e_\mu\}\in\mathcal{H}_z$ then
we can write the components of $g$ in terms of the components of $\gamma$ on the coframes considered as
$$g_{\mu\nu}=\sigma^{*\hat\alpha}_{x,\mu} \sigma^{*\hat\beta}_{x,\nu} \gamma_{\hat\alpha\hat\beta}.$$

Let $z$ be a point in the fibre $G_x$. The metric $\xi_x$ of the fibre $G_x$ in that point will be given by the
restriction of the metric $\gamma$ of $\mathcal{M}$ to the vertical vector subspace $\mathcal{V}_z$ in the same point. Since $G_x$ is, through $\phi_{Ux}\in\Phi$, homomorphic to the structure group $\mathcal{G}$, a metric $\xi$ in 
$\mathcal{G}$ will define, through $\phi_{Ux}$, a metric $\xi_x$ on $G_x$. Since $\xi_x$ is $\mathcal{G}$-invariant, the metric $\xi$ shall also be $\mathcal{G}$-invariant, and so it can be completely determined by its values on the Lie algebra of the group, $L_\mathcal{G}$.

Let $z=\phi_{Ux}(\gamma)\in G_x\subset\mathcal{M}$ and let us take the differential of the homomorphism $\phi_{Ux}$ 
in a point $\gamma$ of $\mathcal{G}$: 
$$\phi^*_{Ux,\gamma}:T_\gamma\mathcal{G}\rightarrow T_{z=\phi_{Ux}(\gamma)}\mathcal{M}.$$
If $v_1,v_2\in T_\gamma\mathcal{G}$ are two tangent vectors to $\mathcal{G}$ on $\gamma\in\mathcal{G}$, then we can
associate the vertical vectors $\tau_1=\phi^*_{Ux,\gamma}(v_1),\tau_2=\phi^*_{Ux,\gamma}(v_2)\in\mathcal{V}_z$ on
$z\in G_x$.
We define the scalar product of $\tau_1$ with $\tau_2$ on $G_x$ by $\xi_x(\tau_1,\tau_2)=\xi_\gamma(v_1,v_2)$, being
$\xi_\gamma$ the metric of $\mathcal{G}$ in $\gamma\in\mathcal{G}$. If we choose for a basis of $T_\gamma\mathcal{G}$,
$\{\lambda_a\}_{a=n,\cdots,m-1}$ such that
$$\phi^*_{Ux,\gamma}(\lambda^a)=\ddot e^a,$$
with $a=n,\cdots,m-1$, then the relation between the two metrics $\xi_x$ and $\xi$ can be written in terms of its
components in the two basis $\{\ddot e^a\otimes\ddot e^b\}_{a,b=n,\cdots,m-1}$ and 
$\{\lambda^a\otimes\lambda^b\}_{a,b=n,\cdots,m-1}$
($\{\lambda^a\}_{a=n,\cdots,m-1}$ is a basis of 
$T^*_\gamma\mathcal{G}$ with $\lambda^a(\lambda_b)=\delta^a_b$):
\begin{equation}\label{metricaxi}
        (\xi_\gamma)_{ab}=\phi^{*c}_{Ux,\gamma\;a}\phi^{*d}_{Ux,\gamma\;b}(\xi_x)_{cd}.
\end{equation}

Since $(\xi_x)_{cd}=(\gamma_z)_{cd}$, we can rewrite the previous relation as
$$(\xi_\gamma)_{ab}=\phi^{*c}_{Ux,\gamma\;a}\phi^{*d}_{Ux,\gamma\;b}(\gamma_{z=\phi_{Ux}(\gamma)})_{cd}.$$

Due to the $\mathcal{G}$-invariance of $\gamma$ and of $\xi$, we can choose a point $z$ on the fibre $G_x$
such that $\gamma=e$, the unit of $\mathcal{G}$. We have then
\begin{equation}
        \xi_{ab}=\phi^{*c}_{Ux\;a}\phi^{*d}_{Ux\;b}\gamma_{cd},
\end{equation}
where we have neglected, in order to simplify the notation, the index $e$. $\xi$ in this last expression is
then considered as the metric of the Lie algebra $L_\mathcal{G}$.





Using the infinitesimal connection form $\omega$:
$$\omega:T_z\mathcal{M}\rightarrow L_\mathcal{G}=T_e\mathcal{G},$$
we can establish a relation between the contravariant components.  Let $\gamma^{\hat\alpha\hat\beta}$ be the contravariant
components of the metric of $\mathcal{M}$ in the chosen adapted orthonormal
frame. We have then, due to the adapted orthonormal nature
of the frame,
$$(\xi_x)^{ab}=\gamma^{ab}.$$
We can now use the infinitesimal connection form to relate the metric of the fiber $G_x$ to the metric of $\mathcal{G}$;
if $\omega^a_{\hat\alpha}$ are the components of the connection,
$$\xi^{ab}=\omega_c^a\omega_d^a(\xi_x)^{cd},$$
or, since $\omega_\mu^c=0$ (because if $\tau\in\mathcal{V}_z$ is vertical $\omega(\tau)=0$), 
$$\xi^{ab}=\omega^a_{\hat\alpha}\omega^b_{\hat\beta}\gamma^{\hat\alpha\hat\beta}.$$

In this way we obtain the following system of equations:
\begin{eqnarray}
        g^{\mu\nu}=\pi^{*\mu}_{\hat\alpha}\pi^{*\nu}_{\hat\beta}\gamma^{\hat\alpha\hat\beta},\\
  g_{\mu\nu}=\sigma^{*\hat\alpha}_{x,\mu} \sigma^{*\hat\beta}_{x,\nu} \gamma_{\hat\alpha\hat\beta},\\
  \xi^{ab}=\omega^a_{\hat\alpha}\omega^b_{\hat\beta}\gamma^{\hat\alpha\hat\beta},\\
  \xi_{ab}=\phi^{*c}_{Ux\;a}\phi^{*d}_{Ux\;b}\gamma_{cd}.
\end{eqnarray}
Together with the trivial relations on the fibre bundle:
\begin{eqnarray}
        \omega^a_{\hat\alpha}\phi_{Ux\;b}^{*\hat\alpha}=\delta^a_b,\\
        \omega^a_{\hat\alpha}\sigma_{x\mu}^{*\hat\alpha}=0,\\
        \pi^{*\mu}_{\hat\alpha}\sigma^{*\hat\alpha}_{x\nu}=\delta^\mu_\nu,\\
        \pi^{*\mu}_{\hat\alpha}\phi_{Ux\;a}^{*\hat\alpha}=0,
\end{eqnarray}
we determine the components of the metric of the total space, once known those of the base space and of the Lie group:
\begin{eqnarray}\label{a1}
        \gamma_{\hat\alpha\hat\beta}=\pi^{*\mu}_{\hat\alpha}\pi^{*\nu}_{\hat\beta}g_{\mu\nu}
        +\omega^a_{\hat\alpha}\omega^b_{\hat\beta}\xi_{ab},\\
        \label{a2}
        \gamma^{\hat\alpha\hat\beta}=\sigma^{*\hat\alpha}_{x,\mu} \sigma^{*\hat\beta}_{x,\nu} g^{\mu\nu}
        +\phi_{Ux\;a}^{*\hat\alpha}\phi_{Ux;b}^{*\hat\beta}\xi^{ab}.
\end{eqnarray}

Another way of deriving directly the equations (\ref{a1},\ref{a2}) is starting from the equation (\ref{metrica}).
Let $\ddot e^z\in\ddot E(\mathcal{M})$ be an adapted orthonormal frame at a point $z\in\mathcal{M}$ and let us 
consider two tangent vectors $\tau_1,\tau_2\in T_z\mathcal{M}$ at that point. Writing these two vectors in terms
of their components in the chosen frame,
$$\tau_i=\tau_i^{\hat\alpha}\ddot e_{\hat\alpha}\;\;(i=1,2),$$
we have that the vertical part of $\tau_1$ and of $\tau_2$ can be written in terms of the infinitesimal connection
form in the principal fibre bundle $\mathcal{M}$, $\omega$. Let us take 
$\omega(\tau_i)=\omega(\tau_i^{\hat\alpha}\ddot e_{\hat\alpha})=\tau_i^{\hat\alpha}\omega(\ddot e_{\hat\alpha})=\tau_i^{\hat\alpha}\omega_{\hat\alpha}$,
for $i=1,2$. We have that $\omega_{\hat\alpha}\in L_\mathcal{G}$ and so if $\{\lambda_a\}_{a=n,\cdots,m-1}$ is a base of the
Lie algebra $L_\mathcal{G}$, we have $\omega_{\hat\alpha}=\omega_{\hat\alpha}^a\lambda_a$, and so
$$\omega(\tau_i)=\tau_i^{\hat\alpha}\omega_{\hat\alpha}^a\lambda_a,\;\;(i=1,2).$$
By using an homomorphism $\phi_{Ux}\in\Phi$ we can write $\phi_{Ux,e}^*\omega(\tau_i)\in V_z$, that is, we get 
the vertical part of $\tau_1$ and of $\tau_2$. Writing in terms of components in the frame $\ddot e^z$ we will get
$$\phi_{Ux,e}^*(\omega(\tau_i))=\tau_i^{\hat\alpha}\omega_{\hat\alpha}^a\phi_{Ux,e}^*(\lambda_a)
=\tau_i^{\hat\alpha}\omega_{\hat\alpha}^a\phi_{Ux,e\;a}^{*c}\ddot e_c \;\;(i=1,2).$$
We can then write, using (\ref{metrica}),
$$\gamma_z(\tau_1,\tau_2)=g_x(\pi^*_x(\tau_1^{\hat\alpha}\ddot e_{\hat\alpha}),\pi^*_x(\tau_2^{\hat\beta}\ddot e_{\hat\beta}))
+\xi_x(\tau_1^{\hat\alpha}\omega_{\hat\alpha}^a\phi_{Ux,e\;a}^{*c}\ddot e_c,\tau_2^{\hat\beta}\omega_{\hat\beta}^a\phi_{Ux,e\;b}^{*d}\ddot e_d)\Leftrightarrow$$
$$\tau_1^{\hat\alpha}\tau_2^{\hat\beta}\gamma_z(\ddot e_{\hat\alpha},\ddot e_{\hat\beta})=
\pi^{*\mu}_{x\hat\alpha}\tau_1^{\hat\alpha}\pi^{*\mu}_{x\hat\beta}\tau_2^{\hat\beta}g_x(\ddot e_\mu,\ddot e_\nu)+
\tau_1^{\hat\alpha}\omega_{\hat\alpha}^a\phi_{Ux,e\;a}^{*c}\tau_2^{\hat\beta}\omega_{\hat\beta}^a\phi_{Ux,e\;b}^{*d}
\xi_x(\ddot e_c,\ddot e_d).$$
By writing $\gamma_{\hat\alpha\hat\beta}=\gamma_z(\ddot e_{\hat\alpha},\ddot e_{\hat\beta})$, $g_{\mu\nu}=g_x(\ddot e_{\mu},\ddot e_\nu)$
and $(\xi_x)_{cd}=\xi_x(\ddot e_c,\ddot e_d)$ we get
\begin{equation}
\gamma_{\hat\alpha\hat\beta}=
\pi^{*\mu}_{x\hat\alpha}\pi^{*\nu}_{x\hat\beta}g_{\mu\nu}+
\omega_{\hat\alpha}^a\omega_{\hat\beta}^b\phi_{Ux,e\;a}^{*c}\phi_{Ux,e\;b}^{*d}
(\xi_x)_{cd}.
\end{equation}
This expression can be further simplified by writing it not in terms of the metric $\xi_x$ of the fibre but of the metric of the 
Lie group $\mathcal{G}$. From (\ref{metricaxi}), by taking $\gamma=e$, we get
\begin{equation}
\gamma_{\hat\alpha\hat\beta}=
\pi^{*\mu}_{x\hat\alpha}\pi^{*\nu}_{x\hat\beta}g_{\mu\nu}+
\omega_{\hat\alpha}^c\omega_{\hat\beta}^d\xi_{cd}.
\end{equation}
Using the trivial relations on the bundle we could get the contravariant components of the metric also.

Let us now choose a trivial projection $\pi(x^0,x^1,\cdots,x^{m-1})=(x^0,x^1,\cdots,x^{n-1})$.
In a local coordinate basis, we have, for
$$\phi^{*a}_{Ux\;b}=\delta^a_b,$$
$$\pi^{*\mu}_{\nu}=\delta^\mu_\nu,$$
$$\phi^{*\mu}_{Ux\;\nu}=0,$$
$$\pi^{*\mu}_a=0,$$
that
\begin{equation}\label{gamma_}
\gamma_{\hat\alpha\hat\beta}=\left(
  \begin{array}{cc}
    g_{\mu\nu}+\xi_{ab}\omega^a_\mu\omega^b_\nu & \xi_{ab}\omega^a_\mu \\
    \xi_{ab}\omega^b_\nu & \xi_{ab}
  \end{array}
\right)
\end{equation}
\begin{equation}\label{gamma^}
\gamma^{\hat\alpha\hat\beta}=\left(
  \begin{array}{cc}
    g^{\mu\nu} & -g^{\mu\nu}\omega^b_\nu \\
    -g^{\mu\nu}\omega^a_\mu & \xi^{ab}+g^{\mu\nu}\omega^a_\mu\omega^b_\nu
  \end{array}
  \right)
\end{equation}
This form of the contravariant and covariant components of the metric is only
valid in the special coordinates that we have chosen.

\paragraph{Yang-Mills Potential and the Strength Tensor Field.}
Let $\{U_i\}_{i=1,2,\cdots}$ be a covering of $V_n$ and let be defined over each $U_i\subset V_n$ a local
section $\sigma_i:U_i\subset V_n\longrightarrow\mathcal{M}$. We define the Yang-Mills potential on $U_i$ as
the local induced infinitesimal connection form $A_i$ (cf. section \ref{principalfibrebundle}). As we
have seen, a family of Yang-Mills potentials $\{A_i\}_{i=1,2,\cdots}$ defined over the covering $\{U_i\}_{i=1,2,\cdots}$
define and is defined by an infinitesimal connection form $\omega$ defined over the principal fibre bundle
$\mathcal{M}$. When we have defined an global section on $\mathcal{M}$, $\sigma:V_n\longrightarrow\mathcal{M}$ then
only one Yang-Mills potential $A$ is needed. We can write the infinitesimal connection form $\omega$ at a
point $z$ of $\mathcal{M}$ in terms of $A$ and of an element $g$ of $\mathcal{G}$ dependent of the section $\sigma$
chosen ($z=g\sigma\circ\pi(z)$),
\begin{equation}
        \omega=Ad(g^{-1})(A\circ\pi^*)+g^{-1}dg.
\end{equation}
Let $\tau\in T_z\mathcal{M}$ be a tangent vector to $\mathcal{M}$ at $z$ and let $v=\pi^*_z\tau\in T_x V_n$, we define
$A(v)=\omega(\tau)$.

Due to the $\mathcal{G}$-invariance of $\gamma$, we can substitute (\ref{gamma_}, \ref{gamma^})
by
\begin{equation}\label{gamma_}
\gamma_{\hat\alpha\hat\beta}=\left(
  \begin{array}{cc}
    g_{\mu\nu}+\xi_{ab}A^a_\mu A^b_\nu & \xi_{ab}A^a_\mu \\
    \xi_{ab}A^b_\nu & \xi_{ab}
  \end{array}
\right)
\end{equation}
\begin{equation}\label{gamma^}
\gamma^{\hat\alpha\hat\beta}=\left(
  \begin{array}{cc}
    g^{\mu\nu} & -g^{\mu\nu}A^b_\nu \\
    -g^{\mu\nu}A^a_\mu & \xi^{ab}+g^{\mu\nu}A^a_\mu A^b_\nu
  \end{array}
  \right)
\end{equation}

We must now determine the components of the curvature of the induced connection, that is, the Yang-Mills Strength
Tensor Field $F=\sigma^*\Omega=DA$ in that base.





From the relation
$$F=dA+\frac12 [A,A],$$
one gets for the components of $F$, in the special base that we are using,

$$F^a_{\mu\nu}=dA^a_\mu\;_\nu+\frac12[A,A]^a_{\mu\nu}.$$
But we have $dA^a_\mu\;_\nu=\frac12(\partial_\mu A^a_\nu-\partial_\nu A^a_\mu)$, and from
the structure relations on the Lie group $\mathcal{G}$

$$[\lambda_b, \lambda_c]=C_{bc}^a \lambda_a,$$
one gets 
$$[A,A]^a_{\mu\nu}=[A^b \lambda_b,A^c\lambda_c]^a_{\mu\nu}=
A^b_\mu A^c_\nu[\lambda_b,\lambda_c]^a=A^b_\mu A^c_\nu C_{bc}^a.$$
And so,
$$F^a_{\mu\nu}=\frac12(\partial_\mu A^a_\nu-\partial_\nu A^a_\mu)+\frac12 C_{bc}^a A^b_\mu A^c_\nu.$$

\paragraph{Einstein-Yang-Mills equations.}
We will choose an affine connection for the riemannian manifold $\mathcal{M}$ that is metric compatible.
The Christoffel symbols are then given in the chosen local coordinates by

$$\Gamma_{\hat\beta\hat\gamma}^{\hat\alpha}=\frac12 \gamma^{\hat\alpha\hat\delta}(\partial_{\hat\beta}
\gamma_{\hat\gamma\hat\delta}+\partial_{\hat\gamma}\gamma_{\hat\delta\hat\beta}-\partial_{\hat\delta}
\gamma_{\hat\beta\hat\gamma}).$$

The Riemann's tensor components follow, in the usual way, from

$$R^{\hat\alpha}_{\hat\beta\hat\gamma\hat\delta}=\partial_{\hat\gamma}\Gamma^{\hat\alpha}_{\hat\beta\delta}-
\partial_{\hat\delta}\Gamma^{\hat\alpha}_{\hat\beta\hat\gamma}+\Gamma^{\hat\epsilon}_{\hat\beta\hat\delta}
\Gamma^{\hat\alpha}_{\hat\epsilon\hat\gamma}-\Gamma^{\hat\epsilon}_{\hat\beta\hat\gamma}\Gamma^{\hat\alpha}_{\hat\epsilon
\hat\delta}.$$

We then construct the Ricci tensor, $R_{\hat\alpha\hat\beta}=R^{\hat\lambda}_{\hat\alpha\hat\lambda\hat\beta}$, 
\begin{eqnarray*}
        R_{ab}=R_{ab}(\mathcal{G})+\frac14 F_{\mu\nu a} F^{\mu\nu}_b+\frac12\xi^{cd} D_\mu\xi_{ac}D^\mu\xi_{bd}-
        \frac14 \xi^{cd}D_\mu\xi_{ab}D^\mu\xi_{cd}-\frac12 D_\mu(D^\mu\xi_{ab}),\\
        R_{\mu\nu}=R_{\mu\nu}(V_n)-\frac12 F_{\mu\sigma a} F^{\mu a}_{\nu}-\frac14\xi^{ab}\xi^{cd}D_\mu\xi_{ac}D_\nu \xi_{bd}
        -\frac12 D_\mu(\xi^{ab}D_\nu\xi_{ab}),\\
        R_{\mu a}=\frac12 D^\sigma F_{\mu\sigma a}+\frac14F_{\mu\sigma a}\xi^{cd} D^\sigma\xi_{cd}-\frac12 C_{ab}^c
        \xi^{bd} D_\mu\xi_{cd},
\end{eqnarray*}
where $R(\mathcal{G})$ and $R(V_n)$ refers to the Ricci tensor of $\mathcal{G}$ and of $V_n$, respectively.
From the Ricci tensor we determine the scalar of curvature, $R=\gamma^{\hat\alpha\hat\beta}R_{\hat\alpha\hat\beta}$:
\begin{equation}
        R=R(\mathcal{G})+R(V_n)-\frac14F_{\mu\nu a} F^{\mu\nu a}-\frac14 D_\mu\xi_{ab}D^\mu\xi^{ab}-
        \frac14\xi^{ab}\xi^{cd}D_\mu\xi_{ab}D^\mu\xi_{cd}-D^\mu(\xi^{ab}D_\mu\xi_{ab}).
\end{equation}

The Hilbert-Einstein-Yang-Mills lagrangian density in $m$ dimensions with a cosmological term is defined as
\begin{equation}\label{L}
\begin{array}{c}
\mathcal{L}^{(m)}_{HEYM}=\sqrt{-\gamma}(R-2\Lambda_m)=
\sqrt{-g}\xi^\frac12[R(\mathcal{G})+R(V_n)-\frac14F_{\mu\nu a} F^{\mu\nu a}-\\\frac14 D_\mu\xi_{ab}D^\mu\xi^{ab}-
        \frac14\xi^{ab}\xi^{cd}D_\mu\xi_{ab}D^\mu\xi_{cd}-D^\mu(\xi^{ab}D_\mu\xi_{ab})-2\Lambda_m],
\end{array}
\end{equation}
where $\Lambda_m$ is the cosmological constant in $\mathcal{M}$ and $\gamma$, $g$ and $\xi$ are the determinants of the 
metrics of $\mathcal{M}$, $V_n$ and $\mathcal{G}$, respectively.
Since $\gamma$ depends only on $x\in V_n$, $R$, $g$, $\xi$ will only depend on $x\in V_n$. The same happens for the
Yang-Mills strength tensor field $F$. We can then reduce the previous lagrangian density to a $n$-dimensional one:
$$
\int_\mathcal{M}\mathcal{L}^{(m)}_{HEYM}d^m x=\textnormal{vol}(\mathcal{G})\int_{V_n}\mathcal{L}^{(n)}_{HEYM}d^n x.
$$

The Einstein-Yang-Mills equations, in the absence of matter fields and considering the cosmological term, 
are then the Euler-Lagrange equations of the
following action

$$S[\gamma]=\int_{\mathcal{M}}d^m x \sqrt{-\gamma} (R-2\Lambda_m)=\int_\mathcal{M}d^m\sqrt{-g}\xi^\frac12(R-2\Lambda_m),$$

Taking $\delta S=0$, one finds

$$(R_{\hat\alpha\hat\beta}-\frac12\gamma_{\hat\alpha\hat\beta}(R-2\Lambda))\delta\gamma^{\hat\alpha\hat\beta}=0.$$
The variations $\delta \gamma^{\hat\alpha\hat\beta}$ are not completely arbitrary: they should be in such way that
the special form that was given to $\gamma^{\hat\alpha\hat\beta}$ as a composition of the metrics of the
submanifolds of the bundle would be preserved in those variations.
If we varies $\delta S$ not in order to $\gamma$ but separatively with respect to
$g$, $A$ and $\xi$ we get:
\begin{eqnarray}
  R_{\mu\nu}-\frac12(R-2\Lambda_m)g_{\mu\nu}=0,\\
  R_{\mu a}=0, \\
  R_{ab}-\frac12(R-2\Lambda_m)\xi_{ab}=0.
\end{eqnarray}

\paragraph{Conformal Rescaling of the Hilbert-Einstein-Yang-Mills Lagrangian Density.} In order to identify, explicitly,
the kinetic terms of the scalar fields in the HEYM lagrangian density, we have to perform a conformal rescaling of
the last. To do this we change the metrics of the base space and of the internal space, multiplying both by a factor $\xi^j$, 
where $\xi$ is the determinant of the metric of the internal space $\mathcal{G}$ and $j$ is a real number to be chosen and
that can differ in the two cases, i.e., we have
\begin{equation}
\begin{array}{c}
        g'_{\mu\nu}=\xi^j g_{\mu\nu}\\
        \xi'_{ab}=\xi^k \xi_{ab},
\end{array}
\end{equation}
with $g'$ and $\xi'$ the new metrics expressed in terms of the old ones, $g$ and $\xi$, and with $j$ and $k$ real numbers.
These are chosen in order to remove the global factor $\xi^\frac12$ in (\ref{L}) as far as possible. We will choose them
by imposing the following requirements \cite{coquereaux1}:
\begin{enumerate}
\item $g^{\mu\nu}g^{\sigma\rho}\xi_{ab}\xi^\frac12 \sqrt{-g}=g'^{\mu\nu}g'^{\sigma\rho}\xi'_{ab}\sqrt{-g'},$
\item $g^{\mu\nu}\xi_{ab}\sqrt{-g}=g'^{\mu\nu}\sqrt{-g'}.$
\end{enumerate}

Using this procedure we get for the new HEYM lagrangian density conformally transformed
\begin{equation}
\begin{array}{c}
\mathcal{L}_{HEYM}=\sqrt{-g'}[R(\mathcal{G})+R(V_n)-\frac14 \xi'_{ab}g'^{\mu\sigma}g'^{\nu\rho}F_{\mu\nu}^aF_{\sigma\rho}^b
\\-\frac14 \xi'^{ab}\xi'^{cd}g'^{\mu\nu}(D_{\mu}\xi'_{ac}D_{\nu}\xi'_{bd}-\frac1{m-2}D_{\mu}\xi'_{ab}D_{\nu}\xi'_{cd})-
2\Lambda_m\xi'^{-\frac1{m-2}}],
\end{array}
\end{equation}
where $m$ is the dimension of the total space $\mathcal{M}$.
If we define the scalar potential as
\begin{equation}
        V(\xi)=2\Lambda_m\xi^{-\frac1{m-2}}-R(\mathcal{G}),
\end{equation}
then we have for the final HEYM lagrangian density,
\begin{equation}
\begin{array}{c}
\mathcal{L}_{HEYM}=\sqrt{-g}[R(V_n)-\frac14\xi_{ab}F^a_{\mu\nu}F^{b\;\mu\nu}
\\-\frac14 \xi^{ab}\xi^{cd}(D_{\mu}\xi_{ac}D^{\mu}\xi_{bd}-\frac1{m-2}D_{\mu}\xi_{ab}D^{\mu}\xi_{cd})-
V(\xi)].
\end{array}
\end{equation}

%
%


\subsubsection{The Fibre Bundle $\mathcal{M}(V_n,\mathcal{G}/\mathcal{H},\mathcal{G},\pi,\Phi)$}

Let us consider now the case in which the internal space is an homogeneous space $M=\mathcal{G}/\mathcal{H}$.
As we have seen in the previous chapter, the following reduction of the fibre bundle of linear frames $\hat E(\mathcal{M})$
defined over $\mathcal{M}(V_n,M,\mathcal{G},\pi,\Phi)$ is possible:
$$\hat E(\mathcal{M})\stackrel{\gamma}{\rightarrow}\tilde E(\mathcal{M})\stackrel{\tilde\omega}{\rightarrow}
\ddot E(\mathcal{M}),$$
where $\tilde E(\mathcal{M})$ is the fibre bundle of orthonormal frames, induced from $\hat E(\mathcal{M})$ once
a metric $\gamma$ of $\mathcal{M}$ is introduced and $\ddot E(\mathcal{M})$ is the fibre bundle of
adapted orthonormal frames. In terms of structure groups we have
$$GL(1,m-1;\mathcal{R})\stackrel{\gamma}{\rightarrow}O(1,m-1;\mathcal{R})
\stackrel{\tilde\omega}{\rightarrow} O(1,n-1;\mathcal{R})\times
O(k;\mathcal{R})\times O(m-k-n-1;\mathcal{R}),$$
where $k=\textnormal{dim}\mathcal{N}/\mathcal{H}$ and $\mathcal{N}$ is the normalizer of $\mathcal{H}$ in $\mathcal{G}$.

As before, in order to proceed with the gravitational dimensional reduction we must consider the bi-invariance of the metric
$\gamma$ of $\mathcal{M}$ to the structure group $\mathcal{G}$:
\begin{equation}
        ^{\rightarrow}\Phi_g^*\gamma=\gamma=^{\leftarrow}\Phi_g^*\gamma,
\end{equation}
for all $g\in\mathcal{G}$. We shall then have,
$$x\in V_n:z=\pi(x)\in G_x\subset\mathcal{M}\Rightarrow\gamma_z(z)=\gamma_z(x),\;\gamma_z\in T_z^*\mathcal{M}\otimes T_z^*\mathcal{M}$$
that is, the metric $\gamma$ assumes a constant value along the same fiber $G_x$, i.e., does not depend on the
point $z$ in which is determined but only on the projection $x=\pi(z)$ of that point.

A $\mathcal{G}$-invariant metric $\gamma$ on $\mathcal{M}$ will determine and be determined by 
a family of $\mathcal{G}$-invariant metrics $\{\xi_x\}$ on each fiber $M_x$ of $\mathcal{M}$, a $G$-invariant infinitesimal
connection $\mathcal{H}_{\bar z}$ on the associated principal fibre bundle $\bar\mathcal{M}(V_n,\bar\mathcal{G},\bar\pi,\bar\Phi)$
and a metric $g$ on the base space $V_n$.

Let us start by the family $\{\xi_x\}$ of metrics on the fibers $M_x$ of $\mathcal{M}$. Let $\tau_1,\tau_2\in\mathcal{V}_z$ be
two vertical vectors at a generic point $z$ of $\mathcal{M}$ that has as projection $x=\pi(z)\in V_n$. The metric $\xi_x$ of
the fiber $M_x$ at that point will be given by $\xi_x(\tau_1,\tau_2)=\gamma(\tau_1,\tau_2)$. Being $\gamma$ $\mathcal{G}$-invariant,
$\xi_x$ will be also $\mathcal{G}$-invariant over the fiber $M_x$.

Let us now consider the orthogonal complement $\mathcal{H}_z$ of the vertical space $\mathcal{V}_z$ at a point $z\in\mathcal{M}$, i.e.,
the subspace $\mathcal{H}_z\subset T_z\mathcal{M}$ such that $T_z\mathcal{M}=\mathcal{V}_z\oplus\mathcal{H}_z$, having 
for each pair of vectors $\tau\in\mathcal{V}_z,v\in\mathcal{H}_z$, $\gamma(\tau,v)=0$. Due to the $\mathcal{G}$-invariance
of the metric $\gamma$ we will have $\mathcal{H}_{gz}=\Phi^*_g\mathcal{H}_z$ for $g\in\mathcal{G}$. We must now prove that
the associated principal fibre bundle to $\mathcal{M}$, $\bar\mathcal{M}$, possesses $\mathcal{H}=\{\mathcal{H}_{\bar z}\}$
as infinitesimal connection.
As it was shown in the section \ref{fbaof}, the vertical space $\mathcal{V}_z$ at a point $z\in\mathcal{M}$ admits 
a decomposition of the form
\begin{equation}\label{V_z}
        \mathcal{V}_z=\mathcal{V}_z^{(\mathcal{N}/\mathcal{H})}\oplus\mathcal{V}_z^{(\mathcal{L})},
\end{equation}
resulting from the decomposition of the Lie algebra of $\mathcal{G}$,
\begin{equation}
        L_\mathcal{G}=L_\mathcal{H}\oplus L_{\mathcal{N}/\mathcal{H}}\oplus L_\mathcal{L}.
\end{equation}

Let us consider a point $\bar z$ of the associated principal fibre bundle $\bar\mathcal{M}$. $\bar\mathcal{M}$ being
a subset of $\mathcal{M}$ (we are considerating the case in which $\bar\mathcal{G}=\mathcal{N}/\mathcal{H}$) we have
that $\bar z\in\mathcal{M}$ also. We can then decompose the tangent vector space $T_{\bar z}\mathcal{M}$ at $\bar z$
as
$$
        T_{\bar z}\mathcal{M}=\mathcal{V}_{\bar z}\oplus\mathcal{H}_{\bar z},
$$
or, using the decomposition (\ref{V_z}),
\begin{equation}\label{dec}
T_{\bar z}\mathcal{M}=\mathcal{V}_{\bar z}^{(\mathcal{L})}\oplus\mathcal{V}_{\bar z}^{(\mathcal{N}/\mathcal{H})}\oplus\mathcal{H}_{\bar z}.
\end{equation}
Since de dimension of $\mathcal{V}_{\bar z}^{(\mathcal{N}/\mathcal{H})}$ is $k$ and that of $\mathcal{V}_{\bar z}^{(\mathcal{L})}$
is $m-n-k$, we have that $\mathcal{H}_{\bar z}$ has the same dimension as the base space $V_n$.

On the other side, the tangent vector space $T_{\bar z}\mathcal{M}$ contains the subspace $T_{\bar z}\bar\mathcal{M}$. 
Since $\mathcal{H}$ leaves the point $\bar z$ invariant, its action on $T_{\bar z}\bar\mathcal{M}$ is the trivial one.
But $\mathcal{V}_{\bar z}^{(\mathcal{L})}$ does not contain vectors invariant under $\mathcal{H}$ (because $\mathcal{L}$ 
does not have) and so $\mathcal{V}_{\bar z}^{(\mathcal{L})}$ and $T_{\bar z}\bar\mathcal{M}$ must be orthogonal.
We can then write
\begin{equation}
T_{\bar z}\mathcal{M}=\mathcal{V}_{\bar z}^{(\mathcal{L})}\oplus T_{\bar z}\bar\mathcal{M}.
\end{equation}
Confronting with (\ref{dec}) one gets the following decomposition of the tangent vector space to $\bar\mathcal{M}$
\begin{equation}
        T_{\bar z}\bar\mathcal{M}=
        \mathcal{V}_{\bar z}^{(\mathcal{N}/\mathcal{H})}\oplus\mathcal{H}_{\bar z}.
\end{equation}
Since $\mathcal{V}_{\bar z}^{(\mathcal{N}/\mathcal{H})}$ is the vertical space to $\bar\mathcal{M}$, we have that
$\mathcal{H}_{\bar z}$ is horizontal. We have then that $\mathcal{H}=\{\mathcal{H}_{\bar z}\}$ is an infinitesimal
connection over $\bar\mathcal{M}$. To this infinitesimal connection we can then associate an infinitesimal connection
form $\omega$. This will be an $L_{\mathcal{N}/\mathcal{H}}$-valued one form over $\bar\mathcal{M}$ as expected.
Since the Yang-Mills field is defined in terms of $\omega$ we have that the first will take its values on the Lie
algebra of $\mathcal{N}/\mathcal{H}$.

The metric $\gamma$ of $\mathcal{M}$ will induce a metric $g$ on $V_n$ as follows. Let us consider two vectors 
$\tau_1,\tau_2\in T_z\mathcal{M}$ and let $\mathcal{H}\tau_1,\mathcal{H}\tau_2\in\mathcal{H}_z$ be their projection
into $\mathcal{H}_z$. If $v_1=\pi(\mathcal{H}\tau_1)\in T_x V_n$ and $v_2=\pi(\mathcal{H}\tau_2)\in T_x V_n$ then
we can define the scalar product between $v_1$ and $v_2$ as $g(v_1,v_2)=\gamma(\mathcal{H}\tau_1,\mathcal{H}\tau_2)$.
Due to the $\mathcal{G}$-invariance of the metric $\gamma$ this definition is independent of the point $z\in M_x$ chosen
along the fiber $M_x$.

We can also proceed backwards, that is, given a metric $g$ on $V_n$, a family of $\mathcal{G}$-invariant metrics $\{\xi_x\}$
of all the fibers of $\mathcal{M}$ and an infinitesimal connection on the associated principal fibre bundle $\bar\mathcal{M}$ 
we can deduce a $\mathcal{G}$-invariant metric $\gamma$ for $\mathcal{M}$. In fact let us consider two tangent
vectors $\tau_1,\tau_2\in T_{\bar z}\bar\mathcal{M}$ to $\bar\mathcal{M}$ at a point $\bar z$ and let $z\in\mathcal{M}$ be
a point of $\mathcal{M}$ such that $z=\Phi(g,\bar z)=g\bar z$, for some element $g\in\mathcal{G}$ (we recall
that $\bar\mathcal{M}$ is a subset of $\mathcal{M}$). Then $v_i=\Phi^*_{\bar z}\tau_i\in T_z\mathcal{M}$, $i=1,2$, will be
tangent vectors to $\mathcal{M}$ at $z$. We can then define the scalar product between $v_1$ and $v_2$ at $z$ as
\begin{equation}\label{metrica2}
\gamma_z(v_1,v_2)=g_x(\pi^*_x v_1,\pi^*_x v_2)+\xi_x(v_1-\Phi^*_{\bar z}\mathcal{H}\tau_1,v_2-\Phi^*_{\bar z}\mathcal{H}\tau_2),
\end{equation}
where $\mathcal{H}\tau_i$, $i=1,2$, denotes the horizontal part of $\tau_i$ in the infinitesimal connection of the associated
principal fibre bundle $\bar\mathcal{M}$.

\paragraph{Metric Decomposition.}

Let us consider the tangent vector space $T_z\mathcal{M}$ at point $z\in\mathcal{M}$ with $\pi(z)=x\in V_n$ and let 
$\ddot e^z\in\ddot E(\mathcal{M})$ be an adapted orthonormal frame at that point. Using the same notations of the preceding
subsection, we have that the contravariant components of the metric $g^{-1}$ (in the base 
$\{\ddot e_\mu\otimes\ddot e_\nu\}_{\mu,\nu=0,\cdots,n-1}$) 
of the base space $V_n$ will be given in terms of those of the metric of the total space, $\gamma^{-1}$, and of the canonical 
projection differential, $\pi^*$: 
$$g^{\mu\nu}=\pi^{*\mu}_{z,\hat\alpha}\pi^{*\nu}_{z,\hat\beta}\gamma^{\hat\alpha\hat\beta}.$$
Introducing a global section $\sigma:V_n\longrightarrow\mathcal{M}$ we can also write for the covariant components of $g$:
$$g_{\mu\nu}=\sigma^{*\hat\alpha}_{x,\mu} \sigma^{*\hat\beta}_{x,\nu} \gamma_{\hat\alpha\hat\beta}.$$
We see that the treatment is the same as for the case of the principal fibre bundle $\mathcal{M}(V_n,\mathcal{G},\pi,\Phi)$,
as it should be.

Considerating the same frame, we can identify the covariant and the contravariant components of the metric of the fiber,
$\xi_x$ at that point with those corresponding to the metric of $\mathcal{M}$, that is, with the "vertical" components
of $\gamma$. We make $(\xi_x)_{ab}=\gamma_{ab}$ and $(\xi_x)^{ab}=\gamma^{ab}$ (we can do this because the frame is
adapted and orthonormal). Using an homomorphism $\phi_{Ux}\in\Phi$ we can write the covariant components of the
metric on the fiber $M_x$ in terms of that one of the homogeneous space $M=\mathcal{G}/\mathcal{H}$. We have, as in
previous section,
\begin{equation}\label{metricaxi2}
        (\xi_\gamma)_{ab}=\phi^{*c}_{Ux,\gamma\;a}\phi^{*d}_{Ux,\gamma\;b}(\xi_x)_{cd},
\end{equation}
and so, we get
\begin{equation}\label{metricaxi2}
        \xi_{ab}=\phi^{*c}_{Ux,e\;a}\phi^{*d}_{Ux,e\;b}\gamma_{cd}.
\end{equation}

In order to establish a relation between the contravariant components of the
metric on the fiber $M_x$ and of the metric on
$\mathcal{G}/\mathcal{H}$ we have to use the infinitesimal connection form $\bar\omega$ on the associated principal fibre
bundle $\bar\mathcal{M}$. It is here that the approach differs from that of the previous section.

Let us consider two tangent vectors $\tau_1,\tau_2\in T_z\mathcal{M}$ to $\mathcal{M}$ at $z$ and let $z=\Phi(g,\bar z)$ for
some point $\bar z\in\bar\mathcal{M}$ and some element $g\in\mathcal{G}$. Taking the differential of $\Phi$ and applying
it to the two previous vectors one gets
$$\Phi_g^*\tau_i\in T_{\bar z}\bar\mathcal{M}\;\;(i=1,2),$$
that is, we get two tangent vectors to the associated principal fibre bundle $\bar\mathcal{M}$ at $\bar z$.
On the other way, we have
$$\bar\omega(\Phi^*_g\tau_i)\in L_{\mathcal{N}/\mathcal{H}}\;\;(i=1,2),$$
and so, by applying $\delta_{g^{-1}}^*$ to the previous we get a tangent vector to the internal space 
$\mathcal{G}/\mathcal{H}$ at its origin:
$$\delta_{g^{-1}}^*\bar\omega(\Phi^*_g\tau_i)\in T_e(\mathcal{G}/\mathcal{H})\;\;(i=1,2).$$
We can now take the differential of an homomorphism $\phi_{Ux}\in\Phi$ between the internal space and the
fiber $M_x\subset\mathcal{M}$ at $x\in V_n$ and apply it to the previous vectors. The result will be
the two vertical parts of the vectors $\tau_i$, $i=1,2$, at $z\in\mathcal{M}$:
\begin{equation}\label{vertical}
  \mathcal{V}\tau_i=\phi^*_{Ux,e}\delta_{g^{-1}}^*\bar\omega(\Phi^*_g\tau_i)\in\mathcal{V}_z\;\;(i=1,2).
\end{equation}
We define then the scalar product of $\tau_1$ and $\tau_2$ at $z$ by
\begin{equation}
  \gamma_z(\tau_1,\tau_2)=g_x(\pi^*_x\tau_1,\pi^*_x\tau_2)+\xi_x(\mathcal{V}\tau_1,\mathcal{V}\tau_2),
\end{equation}
or, using (\ref{vertical}),
\begin{equation}\label{minhametrica}
  \gamma_z(\tau_1,\tau_2)=g_x(\pi^*_x\tau_1,\pi^*_x\tau_2)+\xi_x(\phi^*_{Ux,e}\delta_{g^{-1}}^*\bar\omega(\Phi^*_g\tau_1),
  \phi^*_{Ux,e}\delta_{g^{-1}}^*\omega(\Phi^*_g\tau_2)).
\end{equation}
We must now choose an adapted orthonormal frame at $z\in\mathcal{M}$. Let it be $\ddot e^z\in\ddot E(\mathcal{M})$
and let us write $\tau_i=\tau_i^{\hat\alpha}\ddot e_{\hat\alpha}$, $i=1,2$.

We can then do $$\Phi^*_g\tau_i=\tau_i^{\hat\alpha}\Phi^*_g\ddot e_{\hat\alpha}=\tau_i^{\hat\alpha}\Phi^*_{g\hat\alpha}
=\tau_i^{\hat\alpha}\Phi^{*\bar\alpha}_{g\hat\alpha}\ddot e_{\bar\alpha},$$ where we have chosen for a frame
of $T_{\bar z}\bar\mathcal{M}$, $\{\ddot e_{\bar\alpha}\}_{\bar\alpha=0,\cdots,n+k-1}$. By taking the infinitesimal connection,
$$\bar\omega(\Phi^*_g\tau_i)=\tau_i^{\hat\alpha}\Phi^{*\bar\alpha}_{\hat\alpha}\bar\omega(\ddot e_{\bar \alpha})=
\tau_i^{\hat\alpha}\Phi^{*\bar\alpha}_{\hat\alpha}\bar\omega_{\bar \alpha}=
\tau_i^{\hat\alpha}\Phi^{*\bar\alpha}_{\hat\alpha}\bar\omega_{\bar \alpha}^{\bar a}\lambda_{\bar a},$$ 
where $\{\lambda_{\bar a}\}_{\bar a=n,\cdots,n+k-1}$
is a basis of $L_{\mathcal{N}/\mathcal{H}}$. Applying $\delta^*_{g^{-1}}$ we get
$$\delta_{g^{-1}}^*\bar\omega(\Phi^*_g\tau_i)=\tau_i^{\hat\alpha}\Phi^{*\bar\alpha}_{\hat\alpha}\bar\omega_{\bar \alpha}^{\bar a}
\delta_{g^{-1}}^*(\lambda_{\bar a})=
\tau_i^{\hat\alpha}\Phi^{*\bar\alpha}_{\hat\alpha}\bar\omega_{\bar \alpha}^{\bar a}
\delta_{g^{-1}\bar a}^{*a}\lambda_a,$$ where $\{\lambda_a\}_{a=n,\cdots,m-1}$ is a basis of $T_e(\mathcal{G}/\mathcal{H})$.
We now obtain, by applying $\phi_{Ux}\in\Phi$,
$$\phi^*_{Ux,e}\delta_{g^{-1}}^*\bar\omega(\Phi^*_g\tau_i)=\tau_i^{\hat\alpha}\Phi^{*\bar\alpha}_{\hat\alpha}
\bar\omega_{\bar \alpha}^{\bar a}\delta_{g^{-1}\bar a}^{*a}\phi^*_{Ux,e}(\lambda_a)=\tau_i^{\hat\alpha}\Phi^{*\bar\alpha}_{\hat\alpha}
\bar\omega_{\bar \alpha}^{\bar a}\delta_{g^{-1}\bar a}^{*a}\phi^{*c}_{Ux,e\;a}\ddot e_c,$$
where $\{\ddot e_c\}\subset\ddot e^z$ is a frame of the vertical space at $z$.

By applying (\ref{minhametrica}) to $\tau_i$, $i=1,2$, we get
\begin{equation}
\begin{array}{c}
\tau_1^{\hat\alpha}\tau_2^{\hat\beta}\gamma_z(\ddot e_{\hat\alpha},\ddot e_{\hat\beta})=
\tau_1^{\hat\alpha}\tau_2^{\hat\beta}\pi^{*\mu}_{x\hat\alpha}\pi^{*\nu}_{x\hat\beta}g(\ddot e_\mu,\ddot e_\nu)+\\
\tau_1^{\hat\alpha}\tau_2^{\hat\beta}
\Phi^{*\bar\alpha}_{\hat\alpha}\bar\omega_{\bar \alpha}^{\bar a}\delta_{g^{-1}\bar a}^{*a}\phi^{*c}_{Ux,e\;a}
\Phi^{*\bar\beta}_{\hat\beta}\bar\omega_{\bar \beta}^{\bar b}\delta_{g^{-1}\bar b}^{*b}\phi^{*d}_{Ux,e\;a}
\xi_x(\ddot e_c,\ddot e_d),
\end{array}
\end{equation}
or, by writing $\gamma_{\hat\alpha\hat\beta}=\gamma_z(\ddot e_{\hat\alpha},\ddot e_{\hat\beta})$,
$g_{\mu\nu}=g_x(\ddot e_\mu,\ddot e_\nu)$ and $(\xi_x)_{cd}=\xi_x(\ddot e_c,\ddot e_d)$ we get for
the covariant components of the metric of the total space in the frame $\ddot e^z$
\begin{equation}
\gamma_{\hat\alpha\hat\beta}=
\pi^{*\mu}_{x\hat\alpha}\pi^{*\nu}_{x\hat\beta}g_{\mu\nu}+
\Phi^{*\bar\alpha}_{\hat\alpha}\bar\omega_{\bar \alpha}^{\bar a}\delta_{g^{-1}\bar a}^{*a}
\Phi^{*\bar\beta}_{\hat\beta}\bar\omega_{\bar \beta}^{\bar b}\delta_{g^{-1}\bar b}^{*b}\phi^{*c}_{Ux,e\;a}\phi^{*d}_{Ux,e\;b}
(\xi_x)_{cd}.
\end{equation}
We can now rewrite the previous relation by writing the covariant components of the metric $\xi_x$ of the fiber
$M_x$ in terms of those of the metric $\xi$ of the homogeneous space $\mathcal{G}/\mathcal{H}$ at its origin (we 
can do it since $\xi_x$ is $\mathcal{G}$-invariant). We have
$$\xi_{ab}=\phi^{*c}_{Ux,e\;a}\phi^{*d}_{Ux,e\;b}
(\xi_x)_{cd},$$
and so we get
\begin{equation}
\gamma_{\hat\alpha\hat\beta}=
\pi^{*\mu}_{x\hat\alpha}\pi^{*\nu}_{x\hat\beta}g_{\mu\nu}+
\Phi^{*\bar\alpha}_{\hat\alpha}\bar\omega_{\bar \alpha}^{\bar a}\delta_{g^{-1}\bar a}^{*a}
\Phi^{*\bar\beta}_{\hat\beta}\bar\omega_{\bar \beta}^{\bar b}\delta_{g^{-1}\bar b}^{*b}
\xi_{ab}.
\end{equation}
If we define $\omega_g=\delta_{g^{-1}}^*\bar\omega\Phi_g^*$ then, due to the $\mathcal{G}$ invariance of
$\bar\omega$, we shall have $$\omega_g=\bar\omega,$$ and so
\begin{equation}
\gamma_{\hat\alpha\hat\beta}=
\pi^{*\mu}_{x\hat\alpha}\pi^{*\nu}_{x\hat\beta}g_{\mu\nu}+
\bar\omega_{\hat\alpha}^{a}
\bar\omega_{\hat\beta}^{b}
\xi_{ab},
\end{equation}
a relation similar to the one obtained for the principal fibre bundle $\mathcal{M}(V_n,\mathcal{G},\phi,\Phi)$ - but where
the gauge field take is values on the Lie algebra of $\mathcal{N}(\mathcal{H})/\mathcal{H}$.
\paragraph{Einstein-Yang-Mills Equations.}
The results for the Ricci tensor are identical to those obtained for the principal fibre bundle $\mathcal{M}(V_n,\mathcal{G},\pi,\Phi)$.
We have,
\begin{eqnarray*}
        R_{ab}=R_{ab}(\mathcal{G}/\mathcal{H})+\frac14 F_{\mu\nu a} F^{\mu\nu}_b+\frac12\xi^{cd} D_\mu\xi_{ac}D^\mu\xi_{bd}-
        \frac14 \xi^{cd}D_\mu\xi_{ab}D^\mu\xi_{cd}-\frac12 D_\mu(D^\mu\xi_{ab}),\\
        R_{\mu\nu}=R_{\mu\nu}(V_n)-\frac12 F_{\mu\sigma a} F^{\sigma a}_{\nu}-\frac14\xi^{ab}\xi^{cd}D_\mu\xi_{ac}D_\nu \xi_{bd}
        -\frac12 D_\mu(\xi^{ab}D_\nu\xi_{ab}),\\
        R_{\mu a}=\frac12 D^\sigma F_{\mu\sigma a}+\frac14F_{\mu\sigma a}\xi^{cd} D^\sigma\xi_{cd}-\frac12 C_{ab}^c
        \xi^{bd} D_\mu\xi_{cd},
\end{eqnarray*}
where $R(\mathcal{G}/\mathcal{H})$ and $R(V_n)$ refers to the Ricci tensor of the internal space $\mathcal{G}/\mathcal{H}$ 
and of $V_n$, respectively.

As expected the scalar of curvature is
\begin{equation}
        R=R(\mathcal{G}/\mathcal{H})+R(V_n)-\frac14F_{\mu\nu a} F^{\mu\nu a}-\frac14 D_\mu\xi_{ab}D^\mu\xi^{ab}-
        \frac14\xi^{ab}\xi^{cd}D_\mu\xi_{ab}D^\mu\xi_{cd}-D^\mu(\xi^{ab}D_\mu\xi_{ab}).
\end{equation}

Taking for the Hilbert-Einstein-Yang-Mills lagrangian density, in the absence of matter,
\begin{equation}
\begin{array}{c}
\mathcal{L}^{(m)}_{HEYM}=\sqrt{-\gamma}(R-2\Lambda_m)=
\sqrt{-g}\xi^\frac12[R(\mathcal{G}/\mathcal{H})+R(V_n)-\frac14F_{\mu\nu a} F^{\mu\nu a}-\\\frac14 D_\mu\xi_{ab}D^\mu\xi^{ab}-
        \frac14\xi^{ab}\xi^{cd}D_\mu\xi_{ab}D^\mu\xi_{cd}-D^\mu(\xi^{ab}D_\mu\xi_{ab})-2\Lambda_m],
\end{array}
\end{equation}
where $\Lambda_m$ is the cosmological constant in $\mathcal{M}$ and $\gamma$, $g$ and $\xi$ are the determinants of the 
metrics of $\mathcal{M}$, $V_n$ and $\mathcal{G}/\mathcal{H}$, respectively, one gets from the variational principle the 
Einstein-Yang-Mills equations
\begin{eqnarray}
  R_{\mu\nu}-\frac12(R-2\Lambda_m)g_{\mu\nu}=0,\\
  R_{\mu a}=0, \\
  R_{ab}-\frac12(R-2\Lambda_m)\xi_{ab}=0.
\end{eqnarray}


\subsection{Classical Motion Equations}

\subsubsection{The Principal Fibre Bundle $\mathcal{M}(V_n,\mathcal{G},\pi,\Phi)$}


Let $\mathcal{C}(\mathcal{M})$ be the set of all $C^\nu$-curves on $\mathcal{M}$ and let $\alpha:\mathcal{R}_0^+\longrightarrow\mathcal{M}$ 
be a curve on $\mathcal{M}$. We can, using $\alpha$ and the action of $\mathcal{G}$ on $\mathcal{M}$, 
construct a family of curves $\{\alpha_g\}_{g\in\mathcal{G}}$ on $\mathcal{M}$ by defining
$\alpha_g(t)=g\alpha(t)$, for $t\in\mathcal{R}^+_0$, and by identifying $\alpha_e$ with $\alpha$. A class of equivalence can
then be made: let $\beta:\mathcal{R}^+_0\longrightarrow\mathcal{M}$ be a curve on $\mathcal{M}$; we say that $\beta$ and
$\alpha$ are equivalent, and write $\beta~\alpha$, if there is an element $g\in\mathcal{G}$ such that $\beta(t)=\alpha_g(t)$
for every $t\in\mathcal{R}^+_0$. We will, as usually, write the equivalence class of a curve $\alpha$ as $[\alpha]$.
All the elements of an equivalence class will possess the same projection ($\alpha$(t) and $\alpha_g$(t) lie in the same
fiber) and since the canonical projection is a
$C^\nu$-map that projection will be a $C^\nu$-curve of $V_n$. We then have $\pi(\alpha)=\pi([\alpha])\in\mathcal{C}(V_n)$.

Let 
$\mathcal{C}_0(\mathcal{M})\subset\mathcal{C}(\mathcal{M})$ be the set 
of all geodesics on $\mathcal{M}$, i.e., the set of all curves $\alpha\in\mathcal{C}(\mathcal{M})$ such that
\begin{equation}\label{geodesia}
\frac {D\dot\alpha(t)}{dt}=0,
\end{equation}
for all $t\in\mathcal{R}^+_0$, and where $\dot\alpha$ stands for $\frac{d\alpha}{dt}$. If a $\mathcal{G}$-invariant
metric $\gamma$ is chosen for the total space $\mathcal{M}$ then $\alpha\in\mathcal{C}_0(\mathcal{M})$ implies
that $[\alpha]\in\mathcal{C}_0(\mathcal{M})$, i.e., if $\alpha$ is a geodesic then all elements of the family 
$\{\alpha_g\}_{g\in\mathcal{G}}$ are equally geodesics on $\mathcal{M}$. In fact, since $\Phi^*_g\dot\alpha=\frac d{dt}\alpha_g$
and as the absolute differential commutes with the action of $\mathcal{G}$ on $\mathcal{M}$ - due to the $\mathcal{G}$-invariance
of $\gamma$ - we have
$$\frac D{dt}\dot\alpha_g=\frac D{dt}\Phi^*_g\dot\alpha=\Phi^*_g\frac D{dt}\dot\alpha=0,$$
and so $\alpha_g$ is a geodesic. 

One of the motivations for constructing Kaluza-Klein theories is that the dimensional reduction will break the geodesic condition
 (\ref{geodesia})
on the base space $V_n$, that is, particles will not only follow geodesics on
 this space but also geodesics in the total space
$\mathcal{M}$. In particular one is interested in the break of the geodesic condition on $V_n$ by the introduction of
an linear term on $\dot\alpha$ that could correspond to a Lorentz-like induced force on this space.

Let $\alpha\in\mathcal{C}_0(\mathcal{M})$ and $t\in\mathcal{R}^+_0$. Then $\dot\alpha(t)$ will be the tangent vector to $\alpha$ 
at the point $z=\alpha(t)\in\mathcal{M}$ and will in this way belong to the tangent space $T_{\alpha(t)}\mathcal{M}$.
Let $\ddot e^z=\{\ddot e_{\alpha},\ddot e_a\}_{\alpha=0,\cdots,n-1;a=n,\cdots,m-1}$ be an adapted orthonormal frame to $\mathcal{M}$ at $z$. 
Then we can write the velocity vector $\dot\alpha$ at $z$ as
$$\dot\alpha(t)=\dot\alpha^\alpha(t)\ddot e_{\alpha}^z+\dot\alpha^a(t)\ddot e_a^z.$$

Since the infinitesimal connection is, by definition, $\mathcal{G}$-invariant (cf. section \ref{principalfibrebundle}),
one element $\alpha_g\in [\alpha]$ will possess the same horizontal components for its velocity vector at the point
$gz=\alpha_g(t)=g\alpha(t)$ as the element $\alpha$, i.e.,
$$\mathcal{H}\dot\alpha_g(t)=\dot\alpha^\alpha(t)\ddot e_\alpha^{gz}.$$
Varying $g$ in $\mathcal{G}$ we conclude that all elements of the equivalence class $[\alpha]\in\mathcal{C}(\mathcal{M})$ have the 
same horizontal components in the adapted orthonormal frame family $\{\ddot e^{gz}\}_{g\in\mathcal{G}}$.
The vertical components of the vector $\dot\alpha(t)$ at $z\in\mathcal{M}$ can be obtained from a vector of the Lie algebra 
$L_\mathcal{G}$ through some element $\phi$ of $\Phi$. Let $q(t)\in L_\mathcal{G}$ be that vector so that
$\mathcal{V}\dot\alpha=\phi^*_{Ux,e}(q),$
with $\phi_U\in\Phi$, $x=\pi(z)\in U$ and $U\subset V_n$. In components, $\dot\alpha^a(t)=\phi^{*a}_{Ux}(q(t))$. 
Writing $q(t)$ in a basis $\{\lambda_a\}$ of $L_\mathcal{G}$, $q=q^b\lambda_b$, such that 
$\ddot e^z_a=\phi^*_{Ux,e}(\lambda_a)=\phi^*_{Ux,e\;a}$, we will have
$$\dot\alpha^a(t)=\phi^{*a}_{Ux,e\;b}q^b(t).$$

Fixing $q\in L_\mathcal{G}$ and changing $\phi\in\Phi$ we can associate to all the elements of $[\alpha]$ the same generator $q$.
Let $\phi_{V}\in\Phi$ be one of these elements such that $\phi_{Ux}^{-1}\circ\phi_{Vx}=g$ for $g\in\mathcal{G}$.
Then $g\phi_{Ux}(q)=\phi_{Vx}(q)$, and so $\Phi^*_g\dot\alpha(t)=\phi_{Vx}(q(t))$. Since through one point of $\mathcal{M}$ there passes
only one element of $[\alpha]$, we can make $\phi_{Vx}(q(t))=\dot\alpha_g(t)$,
being $\dot\alpha_g$ the tangent vector to $\alpha_g$ at $z$.

We will now define $q(x)=\phi_{U(g(t))x}(q(t))$, where $U(g)\subset V_n$ and where $g$ varies in $\mathcal{G}$. Then we
can write the velocity vector field of each element of $[\alpha]$ as
\begin{equation}
\dot\alpha_g(t)=\dot\alpha^\alpha(t)\ddot e^{gz}_\alpha+q^a(\pi(\alpha_g(t)))\ddot e^{gz}_a.
\end{equation}
By applying the geodesic condition (\ref{geodesia}) on $\mathcal{M}$ to each of this elements we will get the two
sets of equations
\begin{eqnarray}
        \frac {D\dot\alpha^\alpha}{dt}=q_a F^{a\;\alpha}_\beta\dot\alpha^\beta+\frac12 q^aq^b g^{\alpha\beta}D_\beta(\xi_{ab}),\\
        \frac {Dq^a}{dt}=C^a_{bc}q^bq^c.
\end{eqnarray}
By identifying the components $\dot x^\alpha$ of the velocity vector field $\pi^*_x\dot\alpha(t)$ of the projected geodesic
$\pi(\alpha)\in\mathcal{C}(V_n)$ with the horizontal components of $\dot\alpha$, and by writing explicitly the absolute
differential in terms of the coefficients of the linear connection on the bundle of adapted orthonormated frames,
the equations will be
\begin{eqnarray}
\frac{d^2x^\alpha}{dt}+\Gamma_{\beta\gamma}^\alpha\frac{dx^\beta}{dt}\frac{dx^\gamma}{dt}=
q_a F^{a\;\alpha}_\beta\frac{dx^\beta}{dt}+\frac12 q^aq^b g^{\alpha\beta}D_\beta(\xi_{ab}),\\
\frac{dq^a}{dt}=C^a_{bc}q^c\left(\frac{dx^\mu}{dt}A_\mu^b +q^b\right).
\end{eqnarray}

The first equation describes the motion of a classical particle in a curved space, subject to a Yang-Mills like force,
being the Yang-Mills vector charge given by the $q$ element of the Lie algebra of the internal symmetry group $\mathcal{G}$.
The second equation shows that such charge is not conserved along the tragectory of the particle.

\subsubsection{The Fibre Bundle $\mathcal{M}(V_n,\mathcal{G}/\mathcal{H},\mathcal{G},\pi,\Phi)$}

The treatment for the case in which the total space presents a fibre bundle structure with internal space a homogeneous
space is similar to the previous one. The only difference is that the tangent vector space to $\mathcal{G}/\mathcal{H}$ at
the origin can be decomposed into two Lie algebras:
\begin{equation}
T_e(\mathcal{G}/\mathcal{H})=L_{\mathcal{N}/\mathcal{H}}\otimes L_\mathcal{L},
\end{equation}
and so there will be not one charge vector but two: $q\in L_{\mathcal{N}/\mathcal{H}}$ and $h\in L_\mathcal{L}$.
They will be called, following \cite{coquereaux1}, the color charge vector and the Higgs charge vector, respectively.

The equations will be written in three groups, one for the base space $V_n$ and the others two for the internal
spaces (Lie algebras), $\mathcal{N}/\mathcal{H}$ and $\mathcal{L}$,
\begin{eqnarray}
\frac {D\dot\alpha^\alpha}{dt}=q_{\bar a}F_{\beta}^{\bar a\;\alpha}+\frac12 q^{\bar a}q^{\bar b}g^{\alpha\beta}D_\beta\xi_{\bar a\bar b}
+\frac12 h^a h^bg^{\alpha\beta}D_\beta\xi_{ab},\\
\frac {Dq^{\bar a}}{dt}=C^{\bar a}_{\bar b\bar c}q^{\bar a}q^{\bar b}+C^a_{bc}h^bh^c,\\
\frac {Dh^a}{dt}=C^a_{bc}h^bh^c,
\end{eqnarray}
where $C^{\bar a}_{\bar b\bar c}$ and $C^a_{bc}$ are the structure constants of $\mathcal{N}/\mathcal{H}$ and of 
$\mathcal{L}$, respectively.

\subsection{Dimensional Reduction of a Fibre Bundle}\label{redu}

\subsubsection{The Principal Fibre Bundle $\mathcal{M}(V_n,\mathcal{G},\pi,\Phi)$}

Let $\mathcal{M}(V_n,\mathcal{G},\pi,\Phi)$ be an $m$-dimensional $C^\nu$-differentiable principal
fibre bundle, representative of our multidimensional universe, on which $\mathcal{G}$ acts through
$\Phi$ and let it be defined over $\mathcal{M}$ the fibre bundle $\mathcal{F}(\mathcal{M},F,\mathcal{Y},p,\Psi)$
that possesses as structure group $\mathcal{Y}$ and as internal space $F$ (it could be one of the
fibre bundles of frames discussed in the previous chapter).

Let us suppose the existence of a lift of the action of $\mathcal{G}$ to $\mathcal{F}$:
\begin{equation}
\Phi_\mathcal{F}:\mathcal{G}\times\mathcal{F}\longrightarrow\mathcal{F},
\end{equation}
and of one $\mathcal{G}$-invariant global section of $\mathcal{M}$\footnote{This condition is very restrictive, in
fact it can be only applied to a small number of examples. This constrain is replaced by a more adequate resctriction
in the next subsection.},
\begin{equation}
\sigma:V_n\longrightarrow\mathcal{M},
\end{equation}
such that $
\Phi(g,\sigma(x))=\sigma(x),
$
for all $x\in V_n$ and $g\in\mathcal{G}$.

Using $\sigma$, we can define an $n$-dimensional submanifold of $\mathcal{M}$ by
$
\bar\mathcal{M}=\sigma(V_n),
$
and, through $p$ and $\bar\mathcal{M}$, the bundle
$
\bar\mathcal{F}=p^{-1}(\bar\mathcal{M}).
$
By defining the restrictions of the projection map $p$ and of the $\mathcal{Y}$-action to $\bar\mathcal{F}$ as
$\bar p=p|_{\bar\mathcal{F}}$ and $\bar\Psi=\Psi|_{\bar\mathcal{F}}$, we shall have that
$$\bar\mathcal{F}(\bar\mathcal{M},F,\mathcal{Y},\bar p,\bar\Psi)$$ is a fibre bundle that possesses as structure
group $\mathcal{Y}$.

Let it be notice that, since $\sigma$ is a $\mathcal{G}$-invariant section of $\mathcal{M}$, $\bar\mathcal{M}$
will be also $\mathcal{G}$-invariant, i.e., for each point $\bar z\in\bar\mathcal{M}$ we shall have
\begin{equation}
g\bar z=\bar\Phi(g,\bar z)=\bar z,
\end{equation}
where $\bar\Phi=\Phi|_{\bar\mathcal{M}}$ is the restriction of the $\mathcal{G}$-action on $\mathcal{M}$ to
$\bar\mathcal{M}$.

Turning upon $\bar\mathcal{F}$, we see that in general the dimension of $\bar\mathcal{F}$ will be inferior
to that of $\mathcal{F}$, and so, if $\bar\mathcal{F}$ defines completely - as will be shown - $\mathcal{F}$,
then a first dimensional reduction has taken place.

Let us take
\begin{equation}
\mathcal{F}=\mathcal{G}\times\bar\mathcal{F},
\end{equation}
and
\begin{equation}\label{odsio}
\begin{array}{c}
  i:\mathcal{G}\times\bar\mathcal{F}\longrightarrow\bar\mathcal{F}\\
  (g,\bar \phi)\longrightarrow\bar\Phi_\mathcal{F}(g,\bar\phi).
\end{array}
\end{equation}
Then over $\mathcal{F}$ will exist a $\mathcal{G}$-action given by $g'(g,\bar\phi)=(g'g,\bar\Phi(g',\bar\phi))$,
for $g'\in\mathcal{G}$ and $(g,\bar\phi)\in\mathcal{F}$. By extending the projections of $\mathcal{M}$ the fibre bundle 
$\mathcal{F}$ can be reconstructed.

Since the base space $\bar\mathcal{M}$ of $\bar\mathcal{F}$ is $\mathcal{G}$-invariant, the action of this
group on $\bar\mathcal{F}$ must be vertical, i.e., its action must be only over
the fibers $F_{\bar\phi}$ of $\bar\mathcal{F}$. Since over those fibers there is the action of another group,
$\mathcal{Y}$, an equivalence relation between the two actions could be made and in this way the fibre
bundle $\bar\mathcal{F}$ could be yet reduced.

To that purpose, let us consider the two actions
\begin{eqnarray}
\bar\Phi_{\bar\mathcal{F}}:\mathcal{G}\times\bar\mathcal{F}\longrightarrow\bar\mathcal{F},\\
\bar\Psi:\mathcal{Y}\times\bar\mathcal{F}\longrightarrow\bar\mathcal{F}.
\end{eqnarray}
Our goal is to establish a relation between these two. With this scope at sight, let us define the
following map
\begin{equation}\label{olara}
\begin{array}{c}\bar\psi:\mathcal{G}\times\bar\mathcal{F}\longrightarrow\mathcal{Y},\\
(g,\bar\phi)\longrightarrow\bar\psi(g,\phi)=\bar\Psi^{-1}_{\bar\phi}\circ\bar\Phi_{\bar\mathcal{F}}(g,\bar\phi).\\
\end{array}
\end{equation}
The restrictions of $\bar\psi$ to the first and the second components will be
denoted, respectively, by
\begin{eqnarray}
\bar\psi_g:\bar\mathcal{F}\longrightarrow\mathcal{Y},\\
\bar\psi_{\bar\phi}:\mathcal{G}\longrightarrow\mathcal{Y}.
\end{eqnarray}
These last can be used to establish a relation between the action of $\mathcal{G}$ on $\bar\mathcal{F}$ and
that of $\mathcal{Y}$ at fixed point $\bar\phi\in\bar\mathcal{F}$.
Then a problem takes place: we can establish an equivalence relation between the action of $\mathcal{G}$
and that of $\mathcal{Y}$ only at a local level! That is, the equivalence relation established through
$\bar\psi_{\bar\phi}$ is only valid at a single point $\bar\phi\in\bar\mathcal{F}$. Once we move away from this
point the equivalences made through $\bar\psi_{\bar\phi}$ will change because for two distinct points
$\bar\phi,\bar\phi'\in\bar\mathcal{F}$, $\bar\psi_{\bar\phi}$ and $\bar\psi_{\bar\phi'}$ will in general
differ.

In order to proceed, a restriction shall be made: fixing a point $\bar\phi_0\in\bar\mathcal{F}$ and
writing all the equivalence relations between the actions of the two groups through $\bar\psi_s=\bar\psi_{\bar\phi_0}$
the equivalence relations in all the other points $\bar\phi$ of $\bar\mathcal{F}$ can be deduced from
$\bar\psi_s$ through the following relation
\begin{equation}\label{relacao}
\bar\psi_{\bar\phi}=y(\bar\phi)\bar\psi_s y^{-1}(\bar\phi),
\end{equation}
for a convenient choice of an element $y$ of $\mathcal{Y}$ that will obviously depend on the point $\bar\phi$ on
which the equivalence relations are determined. The restriction is then the following: all the homomorphisms
$\bar\psi_{\bar\phi}$ should lie on the same orbit of the action of the inner automorphisms in Hom$(\mathcal{G},\mathcal{Y})$.

Through (\ref{relacao}) the equivalence relations between $\mathcal{G}$ and $\mathcal{Y}$ can be determined in all
points $\bar\phi\in\bar\mathcal{F}$. We will then be interested to know the points of $\bar\mathcal{F}$ that 
have the same equivalence relations as the fixed point $\bar\phi_0$, i.e., the points $\bar\phi\in\mathcal{F}$
such that, in all of them, to an element $g\in\mathcal{G}$ there is an element $y\in\mathcal{Y}$ - that is
the same for all the points - that their action on that points is the same. The set of these points is 
\begin{equation}
\mathcal{F}^r=\{\bar\phi\in\bar\mathcal{F}:\bar\psi_{\bar\phi}=\psi_s\}.
\end{equation}
A restriction of the action of $\mathcal{Y}$ (and, by equivalence, of $\mathcal{G}$) to $\mathcal{F}^r$ will
be such that a point on $\mathcal{F}^r$ is transformed, through that same action, in another point of
$\mathcal{F}^r$ (the action must be close). The subgroup of $\mathcal{Y}$ such that its elements satisfy
the previous request is the centralizer of $\bar\psi_s(\mathcal{G})$ in $\mathcal{Y}$, that is,
\begin{equation}
\mathcal{C}=\mathcal{C}_\mathcal{Y}(\bar\psi_s(\mathcal{G}))=\{y\in\mathcal{Y}:y\bar\psi_s(g) y^{-1}=\bar\psi_s(g),\;g\in
\mathcal{G}\}.
\end{equation}
Along with the restrictions $p^r=\bar p|_{\mathcal{F}^r}$, $\Psi^r=\bar\Psi|_{\mathcal{F}^r}$ we have that 
$$\mathcal{F}^r(\bar\mathcal{M},F,\mathcal{C},p^r,\Psi^r)$$
is a $\mathcal{C}$-fibre bundle. 
The inicial bundle can be reconstructed by taking $\bar\mathcal{F}=\mathcal{Y}\times_\mathcal{C}\mathcal{F}^r$.
A second and final reduction was then made.



\subsubsection{The Fibre Bundle $\mathcal{M}(V_n,\mathcal{G}/\mathcal{H},\mathcal{G},\pi,\Phi)$}

Let us now consider the case in which a fibre bundle $\mathcal{F}(\mathcal{M},F,\mathcal{Y},p,\Psi)$ is defined over
a fibre bundle $\mathcal{M}(V_n,\mathcal{G}/\mathcal{H},\pi,\Phi)$, whose internal space is homogeneous.

A lift of the action of $\mathcal{G}$ to $\mathcal{F}$,
\begin{equation}
\Phi_\mathcal{F}:\mathcal{G}\times\mathcal{F}\longrightarrow\mathcal{F},
\end{equation}
and of a $\mathcal{H}$-invariant global section of $\mathcal{M}$,
\begin{equation}
\sigma:V_n\longrightarrow\mathcal{M}
\end{equation}
are supposed to exist.

We make then the following definitions: $\bar\mathcal{M}=\sigma(V_n)\subset\mathcal{M}$, $\bar\mathcal{F}=p^{-1}
(\bar\mathcal{M})\subset\mathcal{F}$, $\bar p=p|_{\bar\mathcal{F}}$, $\bar\Psi=\Psi|_{\bar\mathcal{F}}$ and
\begin{equation}\label{op}
  \bar\Phi_{\bar\mathcal{F}}:\mathcal{G}\times\bar\mathcal{F}\longrightarrow\bar\mathcal{F}.
\end{equation}
A fibre bundle can then be constructed over $\bar\mathcal{M}$:
$$\bar\mathcal{F}(\bar\mathcal{M},F,\mathcal{Y},\bar p,\bar\Psi).$$
Since $\bar\mathcal{M}$ is $\mathcal{H}$-invariant, the action (\ref{op}) over $\bar\mathcal{F}$ will be vertical,
and so, if we define 
\begin{equation}
\begin{array}{c}
  i:\mathcal{G}\times_\mathcal{H}\bar\mathcal{F}\longrightarrow\bar\mathcal{F}\\
  ([g,\bar \phi])\longrightarrow\bar\Phi_\mathcal{F}(g,\bar\phi),
\end{array}
\end{equation}
we shall have that $i$ is a $\mathcal{G}$-invariant diffeomorphism, and so can be considered an embebing of 
$\bar\mathcal{F}$ in
\begin{equation}\label{ops}
\mathcal{F}=\mathcal{G}\times_\mathcal{H}\bar\mathcal{F}.
\end{equation}
With $\bar\Phi_{\bar\mathcal{F}}$ and (\ref{ops}) the inicial fibre bundle $\mathcal{F}(\mathcal{M},F,\mathcal{Y},
p,\Psi)$ can be reconstructed and so a first dimensional redution has been made.

Like in the previous section, in order to perform another reduction, let us define
\begin{equation}
\begin{array}{c}
\bar\psi:\mathcal{H}\times\bar\mathcal{F}\longrightarrow\mathcal{Y},\\
(h,\bar\phi)\longrightarrow\bar\psi(h,\bar\phi)=\bar\Psi_{\bar\phi}^{-1}\circ\bar\Phi_{\bar\mathcal{F}}(h,\bar\phi).
\end{array}
\end{equation}
The restrictions to the first and the second components will be, respectively,
\begin{eqnarray}
\bar\psi_h:\bar\mathcal{F}\longrightarrow\mathcal{Y},\\
\bar\psi_{\bar\phi}:\mathcal{H}\longrightarrow\mathcal{Y}.
\end{eqnarray}
By using the last restriction we can establish a correspondence between the vertical action of $\mathcal{H}$ on $\bar\mathcal{F}$
and that of $\mathcal{Y}$ for a fixed point $\bar\phi\in\bar\mathcal{F}$. As before, the equivalence relation that
can be made through $\bar\psi_{\bar\phi}$ will be local, and so we must introduce the restriction that all homomorphisms
$\bar\psi_{\bar\phi}$ should lie on the same orbit of the action of the inner automorphims in Hom$(\mathcal{H},\mathcal{Y})$,
i.e., once fixed a point $\bar\phi_0\in\bar\mathcal{F}$ and the restriction over it, $\bar\psi_{\bar\phi_0}=\bar\psi_s$, we shall have
\begin{equation}
\bar\psi_{\bar\phi}=y(\bar\phi)\bar\psi_s y^{-1}(\bar\phi),
\end{equation}
for all $\bar\phi\in\bar\mathcal{F}$ and for a conveniently choice of $y\in\mathcal{Y}$.

The previous relation allow us to construct all the equivalence relations over $\bar\mathcal{F}$, once known $\phi_s$ at
a point. As before we denote by
\begin{equation}
\mathcal{F}^r=\{\bar\phi\in\bar\mathcal{F}:\bar\psi_{\bar\phi}=\psi_s\},
\end{equation}
the subset of $\bar\mathcal{F}$ of those points $\bar\phi\in\bar\mathcal{F}$ such that their equivalence relations are the
same as those of the point $\bar\phi_0$ chosen.
Over this space it is defined a fibre bundle structure, 
$$\mathcal{F}^r(\bar\mathcal{M},F,\mathcal{C},p^r,\Psi^r),$$
 such that its structure group is given by
\begin{equation}
\mathcal{C}=\mathcal{C}_\mathcal{Y}(\bar\psi_s(\mathcal{H}))=
\{y\in\mathcal{Y}:y\bar\psi_s(h)y^{-1}=\bar\psi_s(h),\;h\in\mathcal{H}\},
\end{equation}
and $p^r=\bar p|_{\mathcal{F}^r}$, $\Psi^r=\bar\Psi|_{\mathcal{F}^r}$.
The inicial bundle can then be reconstructed by taking $\bar\mathcal{F}=\mathcal{Y}\times_\mathcal{C}\mathcal{F}^r$.


\subsection{Dimensional Reduction of Matter Fields}

\subsubsection{The Principal Fibre Bundle $\mathcal{M}(V_n,\mathcal{G},\pi,\Phi)$}

Let it be defined over $\mathcal{M}(V_n,\mathcal{G},\pi,\Phi)$ the fibre bundle of linear frames
$\hat E(\mathcal{M})$ and the fibre bundle $\mathcal{F}(\mathcal{M},F,\mathcal{Y},p,\Psi)$.
In general, a matter field configuration will carry a representation of the linear group $GL(1,m-1)$
and of the structure group $\mathcal{Y}$. It will be then defined as section of the following
fibre bundle $\hat\mathcal{Q}(\mathcal{M},F,\hat\mathcal{J},\hat\pi,\hat Q)$,
\begin{equation}
\hat\mathcal{Q} = \{(\hat e,\phi)\in \hat E(\mathcal{M})\times\mathcal{F}:\;\hat p(\hat e)=p(\phi)\},
\end{equation}
such that its structure group is $\hat\mathcal{J}=GL(1,m-1)\times\mathcal{Y}$ and has for canonical projection
and $\mathcal{J}$-action, $\hat\pi=\hat p\otimes p$ and $\hat Q$, respectively.

Once defined a metric $\gamma$ on $\mathcal{M}$, the fibre bundle of linear frames can be reduced
to $\tilde E(\mathcal{M})$, reducing $\hat\mathcal{Q}$ to the fibre bundle 
$\tilde\mathcal{Q}(\mathcal{M},\tilde\mathcal{J},\tilde\pi,\tilde Q)$, with 
$\tilde\mathcal{J}=O(1,m-1)\times\mathcal{Y}$. This one can yet be reduced through the introduction
of an adapted frame in every tangent space to $\mathcal{M}$. We will then get the bundle
$\ddot\mathcal{Q}(\mathcal{M},\ddot\mathcal{J},\ddot\pi,\ddot Q)$,
\begin{equation}
\ddot\mathcal{Q}=\{(\ddot e,\phi)\in \ddot E(\mathcal{M})\times\mathcal{F}:\;\ddot p(\ddot e)=p(\phi)\},
\end{equation}
with $\ddot\mathcal{J}=O(1,n-1)\times O(m-n)\times\mathcal{Y}$.

We will supose the existence of a lift of the $\mathcal{G}$ action to $\mathcal{F}$,
\begin{equation}
\Phi_\mathcal{F}:\mathcal{G}\times\mathcal{F}\longrightarrow\mathcal{F}.
\end{equation}
Since $\Phi$ naturally lifts to a $\mathcal{G}$-action on $\hat E(\mathcal{M})$, and so on $\ddot E(\mathcal{M})$,
we will have defined an action of $\mathcal{G}$ on $\ddot\mathcal{Q}$,
\begin{equation}\label{opoi}
\Phi_{\ddot\mathcal{Q}}:\mathcal{G}\times\ddot\mathcal{Q}\longrightarrow\ddot\mathcal{Q}
\end{equation}

Defining $\Gamma\ddot\mathcal{Q}$ as the matter field configuration space - the space of sections of $\ddot\mathcal{Q}$ -,
there will be in this space a representation of $\ddot\mathcal{J}$,
\begin{equation}
        [T^{\ddot\mathcal{J}}_{\ddot j}\ddot q](z)=\ddot Q(\ddot j,\ddot q)({\ddot j}^{-1}z),
\end{equation}
with $\ddot j\in\ddot\mathcal{J}$, $\ddot q\in\Gamma\ddot\mathcal{Q}$, and $z\in\mathcal{M}$.
By (\ref{opoi}) there will also be an induced representation of $\mathcal{G}$,
\begin{equation}
        [T^\mathcal{G}_g\ddot q](z)=\Phi_{\ddot\mathcal{Q}}(g,\ddot q)(g^{-1}z),
\end{equation}
for $g\in\mathcal{G}$ and $z\in\mathcal{M}$.

We will consider only $\mathcal{G}$-invariant matter fields configurations,
\begin{equation}
  [T^\mathcal{G}_g\ddot q](z)=\ddot q(z).
\end{equation}

Let it be defined a $\mathcal{G}$-invariant global section on $\mathcal{M}$,
\begin{equation}
\sigma:V_n\longrightarrow\mathcal{M},
\end{equation}
such that $\Phi(g,\sigma(x))=\sigma(x)$, for all $x\in V_n$  and $g\in\mathcal{G}$, and
let us define the $\mathcal{G}$-invariant submanifold $\bar\mathcal{M}=\sigma(\mathcal{M})\subset\mathcal{M}$.
As in the previous section, we will take $\bar{\ddot\mathcal{Q}}={\ddot\pi}^{-1}(\bar\mathcal{M})$ and
construct the fibre bundle $\bar{\ddot\mathcal{Q}}(\bar\mathcal{M},\ddot\mathcal{J},\bar{\ddot\pi},\bar{\ddot Q})$
over $\bar\mathcal{M}$. This bundle, by the results of the previous section, completely defines the initial
fibre bundle $\ddot\mathcal{Q}$.

In order to proceed, we will assume that the metric $\gamma$ defined over $\mathcal{M}$ is $\mathcal{G}$-invariant.
Then we can define a metric over $\bar\mathcal{M}$ by projecting $\gamma$ through $\ddot p$.
An action of $O(1,n-1)$ over $\bar\mathcal{M}$ will then be given and so we can define over $\bar\mathcal{M}$
the fibre bundle of orthonormated frames, $\tilde E(\bar\mathcal{M})$. We can then take
\begin{equation}
\bar{\ddot\mathcal{Q}}=\{(\tilde e,\ddot e,\bar\phi)\in\tilde E(\bar\mathcal{M})\times\mathcal{V}\ddot E(\bar\mathcal{M})\times\bar\mathcal{F}:\;
\mathcal{V}p(\ddot e)=\bar p(\bar\phi)\},
\end{equation}
where $\mathcal{V}\ddot E(\bar\mathcal{M})\subset\ddot E(\bar\mathcal{M})$ is the principal fibre bundle constructed from
$\ddot E(\bar\mathcal{M})$ by taking only the vertical vectors over $\bar\mathcal{M}$, that possesses for structure 
group $O(m-n)$ and where $\bar\mathcal{F}=p^{-1}(\bar\mathcal{M})$ is the total space of the fibre bundle
$\bar\mathcal{F}(\bar\mathcal{M},F,\mathcal{Y},\bar p,\bar \Psi)$.

We can now perform the dimensional reduction of $\bar{\ddot\mathcal{Q}}$ by reducing the fibre bundles by which it is defined.
The bundle $\mathcal{V}\ddot E(\bar\mathcal{M})$ will lead to a fibre bundle with a structure group
$\mathcal{C}_{O(m-n)}(\mathcal{G})$ (cf. previous section). The reduction of $\bar\mathcal{F}$ will conduce to the bundle
$\mathcal{F}^r$ that possesses as structure group $\mathcal{C}_{\mathcal{Y}}(\mathcal{G})$. 
We shall then have for the reduced form of $\bar{\ddot\mathcal{Q}}$, $\mathcal{Q}'^r$, a structure group
\begin{equation}
\mathcal{J}'^r=O(1,n-1)\times\mathcal{C}_{O(m-n)}(\bar\lambda_s(\mathcal{G}))\times\mathcal{C}_{\mathcal{Y}}(\bar\psi_s(\mathcal{G})),
\end{equation}
where
\begin{equation}
\begin{array}{c}
\bar\lambda:\mathcal{G}\times\mathcal{V}\ddot E(\bar\mathcal{M})\longrightarrow O(m-n)
\end{array}
\end{equation}
is the analogous of (\ref{olara}) for the bundle $\mathcal{V}\ddot E(\bar\mathcal{M})$ and $\bar\lambda_s = \bar\lambda(\ddot e_0)$,
for some fixed point $\ddot e_0\in\mathcal{V}\ddot E(\bar\mathcal{M})$.
$\mathcal{Q}^r$ will be given explicity by
\begin{equation}
\mathcal{Q}'^r=\{\bar{\ddot q}\in\bar{\ddot\mathcal{Q}}:\;\bar\beta(g,\bar{\ddot q})=\bar\beta_s(g),\;g\in\mathcal{G}\},
\end{equation}
with 
\begin{equation}
\bar\beta=\bar\lambda\otimes\bar\psi:\mathcal{G}\times\bar{\ddot\mathcal{Q}}\longrightarrow O(m-n)\times\mathcal{Y},
\end{equation}
and $\bar\beta_s=\bar\lambda_s\otimes\bar\psi_s$.

We will then have a matter field configuration space $\Gamma\mathcal{Q}'^r$ on which three representations are defined:
a representation of the Lorentz group, a representation of $\mathcal{C}_{O(m-n)}(\bar\lambda_s(\mathcal{G}))$ and of 
the gauge group $\mathcal{C}_{\mathcal{Y}}(\bar\psi_s(\mathcal{G}))$. Since we are considering only $\mathcal{G}$-invariant
matter fields, the representation of $\mathcal{C}_{O(m-n)}(\bar\lambda_s(\mathcal{G}))$ must be trivial, and so,
over this space of $\mathcal{G}$-invariant matter field configurations we can neglect the action of this group. Then we can
reduce further the matter field fibre bundle $\mathcal{Q}'^r$ to a bundle $\mathcal{Q}^r$ with a structure group given
by
\begin{equation}
\mathcal{J}^r=O(1,n-1)\times\mathcal{C}_{\mathcal{Y}}(\bar\psi_s(\mathcal{G})).
\end{equation}

We have then concluded that a $\mathcal{G}$-invariant matter field configuration, in a given representation of
a gauge group $\mathcal{Y}$, defined over a multidimensional universe
with a principal fibre bundle structure $\mathcal{M}(V_n,\mathcal{G})$ is equivalent to a matter field configuration,
in a representation of the gauge group $\mathcal{C}_{\mathcal{Y}}(\bar\psi_s(\mathcal{G}))$, defined over the base space 
$\bar\mathcal{M}\subset\mathcal{M}$.


\subsubsection{The Fibre Bundle $\mathcal{M}(V_n,\mathcal{G}/\mathcal{H},\mathcal{G},\pi,\Phi)$}

The treatment for a matter field configuration defined over $\mathcal{M}(V_n,\mathcal{G}/\mathcal{H},\mathcal{G},\pi,\Phi)$
is very similar to the treatment given in the previous section, and so we will present only the results.

A matter field configuration over $\mathcal{M}$ will be a section of a fibre bundle 
$\mathcal{Q}(\mathcal{M},\mathcal{J})$, with $\mathcal{J}=GL(1,m-1)\times\mathcal{Y}$, where $\mathcal{Y}$ is the
gauge group of the bundle $\mathcal{F}(\mathcal{M},\mathcal{Y})$. After defining a $\mathcal{H}$-invariant
global section $\sigma:V_n\longrightarrow\mathcal{M}$ and considering the $\mathcal{G}$-invariance of the matter field
configurations and of the metric $\gamma$ defined on $\mathcal{M}$, several dimensional reductions of $\mathcal{Q}$ will
take place. The final result will be a bundle $\mathcal{Q}^r$ with a structure group
\begin{equation}
\mathcal{J}^r=O(1,n-1)\times\mathcal{C}_{\mathcal{Y}}(\bar\psi_s(\mathcal{H})).
\end{equation}


\subsection{Dimensional Reduction of Gauge Fields}

\subsubsection{The Principal Fibre Bundle $\mathcal{M}(V_n,\mathcal{G},\pi,\Phi)$}

Let us consider a principal fibre bundle $\mathcal{F}(\mathcal{M},\mathcal{Y},p,\Psi)$ defined
over $\mathcal{M}(V_n,\mathcal{G},\pi,\Phi)$ and let us supose the existence of a lift of
the action of $\mathcal{G}$ to $\mathcal{F}$:
\begin{equation}
\Phi_\mathcal{F}:\mathcal{G}\times\mathcal{F}\longrightarrow\mathcal{F}.
\end{equation}
Restrictions to the first and second components will be denoted, as usually, by $\Phi_{\mathcal{F}\,g}$
and $\Phi_{\mathcal{F}\,\phi}$, respectively.

Let it be defined an infinitesimal connection over $\mathcal{F}$, described by an infinitesimal
connection form $\omega$ that we will supose to be $\mathcal{G}$-invariant:
\begin{equation}\label{opii}
\Phi_{\mathcal{F}\,g}^*\omega =\omega.
\end{equation}

Let it be defined a $\mathcal{G}$-invariant global section of $\mathcal{M}$,
\begin{equation}
\sigma:V_n\longrightarrow\mathcal{M},
\end{equation}
and let it be defined the principal fibre bundle $\bar\mathcal{F}(\bar\mathcal{M},\mathcal{Y})$ 
over $\bar\mathcal{M}=p^{-1}(\mathcal{M})$ (cf. section 3.1.3). $\bar\mathcal{F}$ will
completely define $\mathcal{F}$ through the dimensional reduction process, so it is expected
that the infinitesimal connection form $\omega$ can be subject to reduction over $\bar\mathcal{F}$.

To that purpose, let us consider a generic point $\phi\in\mathcal{F}$ over $z=p(\phi)\in\mathcal{G}_x\subset\mathcal{M}$,
with $\pi(z)=x\in V_n$, and let $T_\phi\mathcal{F}$ be the tangent vector space to $\mathcal{F}$ in
that point. Let $\bar\phi\in\bar\mathcal{F}$ be over $\bar z\in\bar\mathcal{M}$ such that $\phi=\Phi_{\mathcal{F}\,g}(\bar\phi)$
for a convenientely choice of $g\in\mathcal{G}$. Then $\bar z$ will lie in the same fiber of $z$, i.e., $\bar z\in\mathcal{G}_x$.
Let $\tilde e^{\bar\phi}\in\tilde E(\bar\mathcal{F})$ be a orthonormated frame in the tangent vector space $T_{\bar\phi}\bar\mathcal{F}$.
We can induce a decomposition of the cotangent space $T_\phi\mathcal{F}$ by choosing an adapted frame 
$\ddot e^\phi\in\ddot E(\mathcal{F})$ such that 
\begin{equation}\label{decompo}
\ddot e^\phi = \Phi_{\mathcal{F}\,g}^*(\tilde e^{\bar\phi})\oplus\Psi_y^*(\tilde e^z),
\end{equation}
where $\tilde e^z\subset\tilde E(\mathcal{G}_x)$ is an orthonormated frame of the tangent space $T_z\mathcal{G}_x$ and
$y\in\mathcal{Y}$ such that $\bar\phi=\Psi(y,z)$. 

Associated with the decomposition (\ref{decompo}) there will
be an identical decomposition of the cotangent space $T^*_\phi\mathcal{F}$. Let us write the infinitesimal connection
form $\omega$ in such adapted orthonormal frame $\ddot e_z\in\ddot E^*(\mathcal{F})$
\begin{equation}
\omega=\omega_1+\omega_2,
\end{equation}
with $\omega_1$ and $\omega_2$ lying, respectively, in the subspace spawned by $\tilde e_{\bar\phi}$
and $\tilde e_z$.

Let us define 
\begin{eqnarray}
\bar\omega=\omega_1|_{\bar\mathcal{F}},\\
\bar\xi(\bar\phi)=\Phi^*_{\mathcal{F}\,\bar\phi}\omega_2|_{\bar\mathcal{F}}.
\end{eqnarray}
We will then have that $\bar\omega$ will obey all the conditions presented in \ref{principalfibrebundle}, being 
an infinitesimal connection form defined on $\bar\mathcal{F}$. $\bar\xi$ will be a $\mathcal{Y}$-invariant 
map from $L_\mathcal{G}$ to the Lie algebra of the gauge group, $L_\mathcal{Y}$. By $\mathcal{Y}$-invariance
we will have
\begin{equation}
\bar\xi\circ\Psi_y=Ad_{y^{-1}}\circ\bar\xi,
\end{equation}
and by $\mathcal{G}$-invariance of $\omega$, (\ref{opii}),
\begin{equation}
\bar\xi(\bar\phi)\circ Ad_g = Ad_{\bar\psi_{\bar\phi}(g)}\circ\bar\xi(\bar\phi),
\end{equation}
where $\bar\psi$ is given by (\ref{olara}). By (\ref{opii}), $\bar\omega$ will also be $\mathcal{G}$-invariant,
\begin{equation}
\Phi_{\mathcal{F}\,g}^*\bar\omega=\bar\omega.
\end{equation}

We must be able to reconstruct the initial connection $\omega$ from the knowledge of $(\bar\omega,\bar\xi)$ and
of the $\mathcal{G}$ and $\mathcal{Y}$ actions in order to the dimensional reduction process be complete.
From \ref{redu} we can construct $\mathcal{F}$ by taking $\mathcal{F}=\mathcal{G}\times\bar\mathcal{F}$ and
using $\bar\Phi_\mathcal{F}$.
A tangent vector $v$ to $\mathcal{F}$ at a point $\phi=(g,\bar\phi)$ can then be written as
$$v=\bar v+\theta$$
with $\bar v\in T_{\bar\phi}\bar\mathcal{F}$ and $\theta\in T_g\mathcal{G}$. In special, we can write $\theta = g\lambda$,
with $\lambda\in L_\mathcal{G}$ the vector of the Lie algebra of $\mathcal{G}$ generated by $\theta$
(cf. section \ref{principalfibrebundle}).
In this way, we shall have
\begin{equation}
v=\bar v+g\lambda.
\end{equation}
We will define the action of a infinitesimal connection form $\omega$ over $v$ as
\begin{equation}
\omega(v)=\bar\omega(\bar v)+\bar\xi(\bar\phi)(\lambda).
\end{equation} 
It can be proved that this connection obeys all the properties given in \ref{principalfibrebundle}.

From a principal fibre bundle $\mathcal{F}(\mathcal{M},\mathcal{Y})$ over which it is defined a $\mathcal{G}$-invariant infinitesimal
connection form $\omega$ we get then, by dimensional reduction, a principal fibre bundle $\mathcal{F}(\bar\mathcal{M},\mathcal{Y})$
with a $\mathcal{G}$-invariant infinitesimal connection form $\bar\omega$ and an $\mathcal{Y}$ and $\mathcal{G}$-invariant
map $\bar\xi$. A dimensional reduction of an Yang-Mills-type action defined over $\mathcal{F}$,
\begin{equation}
  S_\mathcal{F}[\Omega]=\frac1{vol(\mathcal{Y})}\parallel\Omega\parallel^2_{\mathcal{F}},
\end{equation}
will then occur. By taking a decomposition of $\Omega=D\omega$ according to (\ref{decompo}) and using the homomorphism 
(\ref{odsio}) we can write
\begin{eqnarray}
i^*\left(D\omega\right)^{(1,1)}=i^*\left(d\omega_1+\frac12[\omega_1,\omega_1]\right)
=d\bar\omega+\frac12[\bar\omega,\bar\omega]=\bar\Omega,\\
i^*\left(D\omega\right)^{(1,2)}=i^*\left(\frac12d\omega_2+\frac12[\omega_1,\omega_2]\right)=\frac12p^*\left(d\bar\xi+[\bar\omega,
\bar\xi]\right)=\frac12p^*\left(\bar D\bar\xi\right),\\
i^*\left(D\omega\right)^{(2,2)}=i^*\left(\frac12[\omega_2,\omega_2]\right)=\frac12p^*\left([\bar\xi,\bar\xi]\right)=
p^*\left(\alpha(\bar\xi)\right).
\end{eqnarray}
The reduced action can then be determined:
\begin{equation}
S_{\mathcal{G}\times\bar\mathcal{F}}[\bar\Omega,\bar\xi]=
\frac1{vol(\mathcal{Y})}\left(\parallel\bar\Omega\parallel_{\mathcal{G}\times\bar\mathcal{F}}+
\frac12 \parallel p^*\left(\bar D\bar\xi\right) \parallel_{\mathcal{G}\times\bar\mathcal{F}}+
\parallel p^*\left(\alpha(\bar\xi)\right)\parallel_{\mathcal{G}\times\bar\mathcal{F}}\right),
\end{equation}
and since $\bar\omega$ and $\bar\xi$ are $\mathcal{G}$-invariant, we can integrate over $\mathcal{G}$
(cf. section 2.2.3),
\begin{equation}
S_{\bar\mathcal{F}}[\bar\Omega,\bar\xi]=
\frac{vol(\mathcal{G})}{vol(\mathcal{Y})}\left(\parallel\bar\Omega\parallel_{\bar\mathcal{F}}+
\frac12 \parallel \bar p^*\left(\bar D\bar\xi\right) \parallel_{\bar\mathcal{F}}+
\parallel \bar p^*\left(\alpha(\bar\xi)\right)\parallel_{\bar\mathcal{F}}\right).
\end{equation}
This action will trivially be reduced to the following action over
$\mathcal{F}^r$,
\begin{equation}
S_{\mathcal{F}^r}[\Omega^r,\xi^r]=
\frac{vol(\mathcal{G})vol(\mathcal{Y}/\mathcal{C})}{vol(\mathcal{Y})}\left(\parallel\Omega^r\parallel_{\mathcal{F}^r}+
\frac12 \parallel p^{r*}\left(D\xi^r\right) \parallel_{\mathcal{F}^r}+
\parallel p^{r*}\left(\alpha(\xi^r)\right)\parallel_{\mathcal{F}^r}\right),
\end{equation}
with $\xi^r=\bar\xi|_{\mathcal{F}^r}$, $\omega^r =\bar\omega|_{\mathcal{F}^r}$ and $\Omega^r=D\omega^r$.
Moreover, the previous action can yet be written (integrating over the fibre of $\mathcal{F}^r$ and by
defining a local section $s$ on $\mathcal{F}^r$) as
\begin{equation}
S_{\bar\mathcal{M}}[A,\phi]=\int_{\bar\mathcal{M}}d^n x \sqrt{-g}\left(\frac14 tr(F_{\mu\nu}F^{\mu\nu})+
\frac12 tr(\bar D_\mu\phi\bar D^\mu\phi)-V(\phi)\right),
\end{equation}
where $F=s^*\Omega$, $A=s^*\omega$, $\phi=p^{r*}\xi$, $$V(\phi)d^n
x=-a(\phi)\wedge\star a(\phi),$$ with $a=s^*\alpha$, and $g$ is the determinant of the
metric of $\bar\mathcal{M}$ and where the internal volume factor was rescalled to one.


\subsubsection{The Fibre Bundle $\mathcal{M}(V_n,\mathcal{G}/\mathcal{H},\mathcal{G},\pi,\Phi)$}

Let $\mathcal{F}(\mathcal{M},\mathcal{Y},p,\Psi)$ be a principal fibre bundle defined over 
$\mathcal{M}(V_n,\mathcal{G}/\mathcal{H},\mathcal{G},\pi,\Phi)$ in the conditions of section 2.1.3.
A lift of the $\mathcal{G}$-action of $\mathcal{M}$ to $\mathcal{F}$ is suposed,
\begin{equation}
\Phi_\mathcal{F}:\mathcal{G}\times\mathcal{F}\longrightarrow\mathcal{F},
\end{equation}
and a $\mathcal{H}$-invariant global section $\sigma$ of $\mathcal{M}$ and an infinitesimal connection
form $\omega$ over $\mathcal{F}$ that is also $\mathcal{G}$-invariant, are given. 
The base space $\bar\mathcal{M}$ of/and the principal
fibre bundle $\bar\mathcal{F}(\bar\mathcal{M},\mathcal{Y},\bar p,\bar\Psi)$ are constructed in
the same way as in the previous subsection. Following, again, the previous subsection we define
in the same way a decomposition of the tangent vector space $T_\phi\mathcal{F}$ of $\mathcal{F}$
in a point $\phi\in\mathcal{F}$:
\begin{equation}\label{decompo}
\ddot e^\phi = \Phi_{\mathcal{F}\,g}^*(\tilde e^{\bar\phi})\oplus\Psi_y^*(\tilde e^z),
\end{equation}
where $\tilde e^z\subset\tilde E(\mathcal{G}/\mathcal{H}_x)$ is an orthonormated frame of the tangent space $T_z\mathcal{G}/
\mathcal{H}_x$ and
$y\in\mathcal{Y}$ such that $\bar\phi=\Psi(y,z)$. 
Using the metric an identical decomposition of the cotangent vector space at the same point $\phi\in\mathcal{F}$ will
occur. Let us then write the infinitesimal connection form $\omega$, following such decomposition:
\begin{equation}
\omega=\omega_1+\omega_2.
\end{equation}
We define, as before,
\begin{eqnarray}
\bar\omega=\omega_1|_{\bar\mathcal{F}},\\
\bar\xi(\bar\phi)=\Phi^*_{\mathcal{F}\,\bar\phi}\omega_2|_{\bar\mathcal{F}},
\end{eqnarray}
and with the same lines of thought as exposed in the previous subsection, $\bar\omega$ will be a $\mathcal{G}$-invariant
infinitesimal connection form over $\bar\mathcal{F}$ and $\bar\xi$ will be, for each $\bar\phi\in\bar\mathcal{F}$,
 an $\mathcal{Y}$-invariant map from $L_\mathcal{P}$ to $L_\mathcal{Y}$. It will further obey
\begin{equation}\label{osio}
  \bar\xi(\bar\phi)\circ Ad_h = Ad\bar\psi_{\bar\phi}(h)\circ\bar\xi(\bar\phi),
\end{equation}
for all $h\in\mathcal{H}$.

With a reasoning similar to that of the previous subsection we shall have that, for a tangent vector to $\mathcal{F}$
at $\phi$ that can be written as
\begin{equation}
  v = \bar v +\lambda_\mathcal{H}\theta_\mathcal{H}+\lambda_\mathcal{P}\theta_\mathcal{P},
\end{equation}
with $\lambda_\mathcal{H}\in L_\mathcal{H}$ and $\lambda_\mathcal{P}\in L_\mathcal{P}$ generators,
once given $\bar\omega$ and $\bar\xi$ in the associated point on $\bar\mathcal{F}$ and with the action of $\mathcal{G}$
on $\mathcal{F}$, we can define a $\mathcal{G}$-invariant infinitesimal connection form $\omega$ of $\mathcal{F}$ at that point as
\begin{equation}
  \omega(v)=\bar\omega(\bar v)+\bar\psi_{\bar\phi}^*(\lambda_\mathcal{H})+\bar\xi(\bar\phi)(\lambda_\mathcal{P}),
\end{equation}
and following this procedure to define an infinitesimal connection on $\mathcal{F}$.

We now proceed to the dimensional reduction of an Yang-Mills-type action defined over $\mathcal{F}$:
\begin{equation}
  S_\mathcal{F}[\Omega]=\frac1{vol(\mathcal{Y})}\parallel\Omega\parallel^2_{\mathcal{F}}.
\end{equation}
By decomposing $\Omega$ according to the same decomposition of $\omega$, we shall have
\begin{eqnarray}
i^*\left(D\omega\right)^{(1,1)}=i^*\left(d\omega_1+\frac12[\omega_1,\omega_1]\right)=d\bar\omega+\frac12[\bar\omega,\bar\omega]=
\bar\Omega,\\
i^*\left(D\omega\right)^{(1,2)}=i^*\left(\frac12d\omega_2+\frac12[\omega_1,\omega_2]\right)=\frac12p^*\left(d\bar\xi+[\bar\omega,
\bar\xi]\right)=\frac12p^*\left(\bar D\bar\xi\right),\\
\begin{array}{r}
i^*\left(D\omega\right)^{(2,2)}=i^*\left(\frac12[\omega_2,\omega_2]\right)=\frac12p^*\left([\bar\psi^*+\bar\xi,\bar\psi^*+\bar\xi]\right)\\
=\frac12p^*\left([\bar\xi,\bar\xi]-\bar\xi\circ[\cdot,\cdot]|_{L_\mathcal{P}}-\bar\psi^*\circ[\cdot,\cdot]|_{L_\mathcal{H}}\right)=
p^*\left(\alpha(\bar\xi)\right),
\end{array}
\end{eqnarray}
where in the last expression we have used (\ref{osio}), i.e., the relation (written in another but equivalent form)
\begin{equation}
[\bar\psi^*(\lambda_\mathcal{H}),\bar\xi(\lambda_\mathcal{P})]=\bar\xi\circ[\lambda_\mathcal{H},\lambda_\mathcal{P}],
\end{equation}
for $\lambda_\mathcal{H}\in L_\mathcal{H}$ and $\lambda_\mathcal{P}\in L_\mathcal{P}$.
The action can then be writen as
\begin{equation}
S_{\mathcal{G}\times_\mathcal{H}\bar\mathcal{F}}[\bar\Omega,\bar\xi]=
\frac1{vol(\mathcal{Y})}\left(\parallel\bar\Omega\parallel_{\mathcal{G}\times_\mathcal{H}\bar\mathcal{F}}+
\frac12 \parallel p^*\left(\bar D\bar\xi\right) \parallel_{\mathcal{G}\times_\mathcal{H}\bar\mathcal{F}}+
\parallel p^*\left(\alpha(\bar\xi)\right)\parallel_{\mathcal{G}\times_\mathcal{H}\bar\mathcal{F}}\right),
\end{equation}
which can be reduced to
\begin{equation}
S_{\bar\mathcal{F}}[\bar\Omega,\bar\xi]=
\frac{vol(\mathcal{G}/\mathcal{H})}{vol(\mathcal{Y})}\left(\parallel\bar\Omega\parallel_{\bar\mathcal{F}}+
\frac12 \parallel \bar p^*\left(\bar D\bar\xi\right) \parallel_{\bar\mathcal{F}}+
\parallel \bar p^*\left(\alpha(\bar\xi)\right)\parallel_{\bar\mathcal{F}}\right),
\end{equation}
and this to
\begin{equation}
S_{\mathcal{F}^r}[\Omega^r,\xi^r]=
\frac{vol(\mathcal{G}/\mathcal{H})vol(\mathcal{Y}/\mathcal{C})}{vol(\mathcal{Y})}\left(\parallel\Omega^r\parallel_{\mathcal{F}^r}+
\frac12 \parallel p^{r*}\left(D\xi^r\right) \parallel_{\mathcal{F}^r}+
\parallel p^{r*}\left(\alpha(\xi^r)\right)\parallel_{\mathcal{F}^r}\right).
\end{equation}

Finally, by defining a local section $s$ of $\mathcal{F}^r$, we can give $S$ its definitely form
\begin{equation}
S_{\bar\mathcal{M}}[A,\phi]=\int_{\bar\mathcal{M}}d^n x \sqrt{-g}\left(\frac14 tr(F^2)+
\frac12 tr(D\phi^2)-V(\phi)\right),
\end{equation}
where $F=s^*\Omega^r$, $A=s^*\omega^r$, $\phi=p^{r*}\xi^r$,
$$V(\phi)d^n x=- a(\phi)\wedge\star a(\phi),$$ with $a=s^*\alpha$ and $g$ is the determinant of the
metric of $\bar\mathcal{M}$ and where the internal volume factor was, as before, rescalled to one.


\section{Dimensional Reduction of General Fields}

\subsection{Harmonic Analysis}

\subsubsection{Unitary Representation of the Internal Symmetry Group $\mathcal{G}$.}
Let $\mathcal{U}(\mathcal{G})$ be the set of all continuous irreducible unitary representations $U$ of 
$\mathcal{G}$\footnote{The internal symmetry group $\mathcal{G}$ is compact.} \cite{hewitt1,kirillov}
and let us suppose that there is an intertwining transformation $T$ \cite{hewitt2} such that to two 
irreducible unitary representations, $U$ and $U'$ that act, respectively, on the Hilbert spaces
$\mathcal{H}$ and $\mathcal{H}'$, we can establish the relation
$U_gT=TU_g',$
for all $g\in\mathcal{G}$. If $T$ is an isometry then we say that $U$ and $U'$ are equivalent and
we write $U\sim U'$. The dual object of $\mathcal{G}$, $\hat\mathcal{G}$, is defined as the set of all equivalence classes of
irreducible unitary representations $U$ of $\mathcal{G}$. To each $\sigma\in\hat\mathcal{G}$ we can then associate
various representations $U\in\mathcal{U}(\mathcal{M})$. We will represent a generic element of the class $\sigma$ by
$U^{(\sigma)}$ and denote its (Hilbert) representation space by $\mathcal{H}_{(\sigma)}$\footnote{All representations
$U^{(\sigma)}\in\sigma$ operate on Hilbert spaces of the same finite dimension and so we can consider that all
this representations operate on a single Hilbert space $\mathcal{H}_{(\sigma)}$.}. 
Due to the compactness of $\mathcal{G}$ this space has a finite dimension $d_\sigma$.

Let $\sigma\in\hat\mathcal{G}$, $U^{(\sigma)}$ be the representative of $\sigma$ and let $\{\zeta_i\}_{i=1,\cdots,d_\sigma}$
be a fixed but arbitrary orthonormal basis in the representation space $\mathcal{H}_{(\sigma)}$. We
define the coordinate functions of $U^{(\sigma)}$ in the previous base as
\begin{equation}
u^{(\sigma)}_{jk}(g)=<U^{(\sigma)}_g\zeta_k,\zeta_j>,
\end{equation}
for $j,k\in\{1,\cdots,d_\sigma\}$, $g\in\mathcal{G}$ and where $<,>$ is the internal product defined on the Hilbert space
$\mathcal{H}_{(\sigma)}$.

We can now present the Weyl-Peter theorem \cite{hewitt2}:
\paragraph{Theorem.} For all $\sigma\in\hat\mathcal{G}$ and $j,k\in\{1,2,\cdots,d_\sigma\}$, let it be defined the coordinate
functions $u^{(\sigma)}_{jk}$ of the reprentatives $U^{(\sigma)}$ in a basis $\{\zeta_1,\cdots,\zeta_{d_\sigma}\}$
of the respective representation spaces $\mathcal{H}_{(\sigma)}$. The set of functions $d_\sigma^{\frac12}u_{jk}^{(\sigma)}$
is an orthonormal basis for $\mathcal{L}_2(\mathcal{G})$. Thus for $\phi\in\mathcal{L}_2(\mathcal{G})$, we have
\begin{equation}\label{weylpeter}
        \phi=\sum_{\sigma\in\hat\mathcal{G}}\sum_{j,k=1}^{d_\sigma}d_\sigma <\phi,u_{jk}^{(\sigma)}>u_{jk}^{(\sigma)},
\end{equation}
where $<\phi,u_{jk}^{(\sigma)}>=\int_\mathcal{G}\phi u^{(\sigma)*}_{jk}d\lambda$, with $d\lambda$ the de Haar measure on
$\mathcal{G}$, and the series in (\ref{weylpeter})
converges in the metric of $\mathcal{L}_2(\mathcal{G})$.
\paragraph{} The character $\chi_U$ of a finite-dimensional representation $U$ of $\mathcal{G}$ is defined as
the function $\chi_U(g)=tr(U_g)$, where the trace is defined as $tr(U_g)=\sum_{k=1}^d u_{kk}(g)$,
being $d$ the dimension of the representation space and $u_{kk}$, $k=1,\cdots,d$, the coordinate functions associated to
$U\in\mathcal{U}(\mathcal{G})$ in some basis of the representation space. One can prove that for two irreducible unitary
representations appartening to the same class $\sigma$ their character is the same. We can then associate to each $\sigma
\in\hat\mathcal{G}$ the corresponding character of the class $\chi_\sigma$.
We will now present a corolarium of the Weyl-Peter theorem,
\paragraph{Corolarium.} Let $\phi\in\mathcal{L}_2(\mathcal{G})$. Then the equalities
\begin{equation}\label{corolarium}
\phi=\sum_{\sigma\in\hat\mathcal{G}} d_\sigma\chi_\sigma * \phi = \sum_{\sigma\in\hat\mathcal{G}}d_\sigma \phi * \chi_\sigma,
\end{equation}
hold, the series converging in the metric of $\mathcal{L}_2(\mathcal{G})$.
\paragraph{}The convolution of $\chi_\sigma$ with $\phi$ is explicitly given by \cite{hewitt1},
\begin{equation}
\chi_\sigma * \phi(g)=\int_\mathcal{G}\chi_\sigma(\lambda)\phi(\lambda^{-1}g)d\lambda,
\end{equation}
where $d\lambda$ is the de Haar measure on $\mathcal{G}$. We can then write (\ref{corolarium}) as
\begin{equation}
\phi(g)=\sum_{\sigma\in\hat\mathcal{G}}d_\sigma\int_\mathcal{G}\chi_\sigma(\lambda)\phi(\lambda^{-1}g)d\lambda.
\end{equation}
\subsubsection{Harmonic Expansion}
Let $\Gamma\mathcal{F}$ be the configuration space, i.e., the space of the sections of the fibre bundle $\mathcal{F}(\mathcal{M},F,\mathcal{Y},\Psi)$. Locally we can write the total space $\mathcal{M}$ as a product space $\mathcal{M}=V_n\times\mathcal{G}$ of
the base space $V_n$ and the internal space $\mathcal{G}$. We can then write, at a local level, for an arbitrary
point $z\in\mathcal{M}$, in a local coordinate system,
$$z=(x,g)\equiv (x,\phi_{Ux}(g)),$$
where $x=\pi(z)\in U\subset V_n$ and $z=\phi_{Ux}(g)$ for $g\in\mathcal{G}$ and $\phi_U\in\Phi$.
Then, at a local level, we have
$$\phi(z)=\phi(x,g).$$
Let us now make the restriction
$$\phi_x(g)=\phi(x,g),$$
for fixed $x\in V_n$. Then we will have that $\phi_x(g)$ is an application that has as support $\mathcal{G}$.
Let us now write $\phi(z)=(z,\phi_z)$, such that, $\phi_x(g)=(z,\phi^x_g)$. Then $\phi^x_g\in\mathcal{L}_2(\mathcal{G})$, 
and so we can then apply the corolarius of the Weyl­Peter theorem to $\phi^x_g$. We have
$$\phi_x(g)=\sum_{\sigma\in\hat\mathcal{G}}d_\sigma\int_\mathcal{G}\chi_\sigma(\lambda)\phi_{\lambda^{-1}g}d\lambda,$$
or, since we have by (\ref{trans2}), $$[T_g\phi]_z=[T_\lambda\phi]_{(x,g)}=\phi^x_{\lambda^{-1}g},$$ we get
$$\phi^x_g=\sum_{\sigma\in\hat\mathcal{G}}d_\sigma\int_\mathcal{G}\chi_\sigma(\lambda)[T_\lambda\phi]^x_gd\lambda.$$
Appearing $x$ only as a free parameter we can write directly
\begin{equation}
\phi_z=\sum_{\sigma\in\hat\mathcal{G}}d_\sigma\int_\mathcal{G}\chi_\sigma(\lambda)[T_\lambda\phi]_zd\lambda.
\end{equation}
Writing $[T_g\phi](z)=(z,\phi_{g^{-1}z})=(z,[T_g\phi]_z)$, we finally get
\begin{equation}\label{phiz}
\phi(z)=\sum_{\sigma\in\hat\mathcal{G}}d_\sigma\int_\mathcal{G}\chi_\sigma(\lambda)[T_\lambda\phi](z)d\lambda.
\end{equation}

We can then define a projection operator over the configuration space $\Gamma\mathcal{F}$:
\begin{equation}
\begin{array}{ccc}
\pi_\sigma:\Gamma\mathcal{F} & \longrightarrow & \Gamma_\sigma\mathcal{F}\\
\phi & \rightarrow & \pi_\sigma\phi(z)
\end{array},
\end{equation}
with
$$\pi_\sigma\phi(z)=d_\sigma\int_\mathcal{G}\chi_\sigma(\lambda)[T_\lambda\phi](z)d\lambda.$$
The subspace $\Gamma_\sigma\mathcal{F}$ consists of the field configurations that transforms with $\mathcal{G}$
in the same manner as the irreducible representation $U^{(\sigma)}$ of $\sigma$.
We can then rewrite (\ref{phiz}) as
\begin{equation}
\phi(z)=\sum_{\sigma\in\hat\mathcal{G}}\pi_\sigma\phi(z).
\end{equation}
This equation expresses the fact that $\oplus_{\sigma\in\hat\mathcal{G}}\Gamma_{(\sigma)}\mathcal{F}$ is isomorphic
with $\Gamma\mathcal{F}$.
Following \cite{coquereaux2} we will call to an element $\phi\in\Gamma_{(\sigma)}\mathcal{F}$ an harmonic of type
$\sigma$.

We will now choose an arbitrary but fixed basis $\{\zeta_i\}_{i=1,\cdots,d_\sigma}$ of the representation
space $\mathcal{H}_{(\sigma)}$ of the irreducible unitary representation $U^{(\sigma)}$ for $\sigma\in\hat\mathcal{G}$.
The character of this representation can be written, as we have seen, in terms of the associated coordinate functions
of $U^{(\sigma)}$, $u_{jk}^{(\sigma)}$. In fact, it depends only on the diagonal coordinate functions $u_{kk}^{(\sigma)}$.
We can then write, going backwards,
\begin{equation}
\chi_\sigma(\lambda)=\sum_{k=1}^{d_\sigma}u_{kk}^{(\sigma)}(\lambda)=\sum_{k=1}^{d_\sigma}<U^{(\sigma)}\zeta_k,\zeta_k)>.
\end{equation}
Substituting this in the expression of $\pi_\sigma\phi(z)$, we get
\begin{equation}
\pi_\sigma\phi(z)=\sum_{k=1}^{d_\sigma}d_\sigma\int_\mathcal{G}<U^{(\sigma)}_\lambda\zeta_k,\zeta_k>[T_\lambda\phi](z)d\lambda.
\end{equation}
By defining, following \cite{coquereaux2}, the Fourier component of $\phi$ along a vector $\zeta\in\mathcal{H}_{(\sigma)}$
in the representation class $\sigma$ as
\begin{equation}
\phi^{(\sigma)}_\zeta(z)=d_\sigma\int_\mathcal{G}U^{(\sigma)}_\lambda\zeta[T_g\phi](z)d\lambda,
\end{equation}
we have
\begin{equation}\label{fourier}
\pi_\sigma\phi(z)=\sum_{k=1}^{d_\sigma}<\phi^{(\sigma)}_{\zeta_k},\zeta_k>.
\end{equation}

Let us define the fibre bundle $\mathcal{F}_{(\sigma)}(\mathcal{M},\mathcal{H}_{(\sigma)}\otimes F,\mathcal{Y},\Phi_\sigma)$ 
defined over $\mathcal{M}$ and that possesses for fibers the tensor product space $\mathcal{H}_{(\sigma)}\otimes F_z$, where $F_z$ is 
the typical fiber of $\mathcal{F}$. We have that the Fourier components within one representation of the type $\sigma$
belong to a very special subspace of the space $\Gamma\mathcal{F}_{(\sigma)}$ of all sections of $\mathcal{F}_{(\sigma)}$: belong to a
subspace $\Gamma_{inv}\mathcal{F}_{(\sigma)}\subset\Gamma\mathcal{F}_{(\sigma)}$ that is $\mathcal{G}$-invariant. This is the main
reason for introducing the Fourier components of some configuration field: being $\mathcal{G}$-invariant, the
Fourier components can be dimensionally reduced (the prescription was given in the previous section).
The equation (\ref{fourier}) then traduces the fact that $\Gamma_{(\sigma)}\mathcal{F}$ is isomorphic with
the space $\Gamma_{inv}\mathcal{F}_{(\sigma)}\otimes\mathcal{H}_{(\sigma)}$:
\begin{equation}
\Gamma_{(\sigma)}\mathcal{F}=\Gamma_{inv}\mathcal{F}_{(\sigma)}\otimes\mathcal{H}_{(\sigma)}
\end{equation}
We have then established the following isomorphism:
\begin{equation}
\Gamma\mathcal{F}=\oplus_{\sigma\in\hat\mathcal{G}}\Gamma_{inv}\mathcal{F}_{(\sigma)}\otimes\mathcal{H}_{(\sigma)}.
\end{equation}

\subsubsection{Harmonic Analysis over an Homogeneous Space}

In order to replace in the previous analysis the internal space by an homogeneous space $\mathcal{G}/\mathcal{H}$
we must have to apply, for all the representations
$U^{(\sigma)}$, $\sigma\in\hat\mathcal{G}$, over the Hilbert space $\mathcal{H}_{(\sigma)}$, the following condition
\begin{equation}
U^{(\sigma)}_{gh} = U^{(\sigma)}_{g} U^{(\sigma)}_{h},
\end{equation}
valid to all $h\in\mathcal{H}\subset\mathcal{G}$. In terms of coordinate functions we shall have
\begin{equation}
u_{jk}^{(\sigma)}(gh)=\sum_{mn}u_{jm}^{(\sigma)}(g)u_{nk}^{(\sigma)}(h).
\end{equation}

In order to proceed we supose that
\begin{equation}
|\det Ad_\mathcal{G}(h)|=|\det Ad_\mathcal{H}(h)|,
\end{equation}
for $h\in\mathcal{H}$. Then there will be a positive $\mathcal{G}$-invariant
measure $d\lambda_\mathcal{H}$ on $\mathcal{G}/\mathcal{H}$ \cite{helgason}
and such measure will be unique up to a constant factor. Moreover, we shall
have for $f\in\mathcal{L}_2(\mathcal{G})$,
\begin{equation}
\int_\mathcal{G} f(\lambda)d\lambda
=\int_{\mathcal{G}/\mathcal{H}}\left(\int_\mathcal{H} f(\lambda h)dh\right)d\lambda_\mathcal{H}.
\end{equation}
Let us define
\begin{equation}
\hat f([g])=\int_\mathcal{H}f(gh)dh.
\end{equation}
Then the mapping $f\rightarrow\hat f$ will be a linear mapping from
$\mathcal{L}_2(\mathcal{G})$ onto $\mathcal{L}_2(\mathcal{G}/\mathcal{H})$ and
so can be used to write the decomposition (\ref{phiz}) over the homogeneous
space $\mathcal{G}/\mathcal{H}$. 

We write the Fourier components of $\phi$ along a vector
$\zeta\in\mathcal{H}_{(\sigma)}$ as
\begin{equation}
\phi^{(\sigma)}_\zeta(z)=d_\sigma\int_{\mathcal{G}/\mathcal{H}}U_{\lambda_\mathcal{H}}\hat\phi_{\lambda_\mathcal{H}}(z)d\lambda_\mathcal{H},
\end{equation}
with 
\begin{equation}
\hat\phi_{\lambda_\mathcal{H}}(z)=\int_\mathcal{H}U_h^{(\sigma)}\zeta
\left[T_{\lambda_\mathcal{H}h}\phi\right](z)dh
\end{equation}

The analysis will then be identical, only changing the internal space for an
homogeneous space.



\subsection{Dimensional Reduction of Gravity and Matter Fields}

Let it be defined over $\mathcal{M}(V_n,\mathcal{G}/\mathcal{H})$ the fibre
bundle $\mathcal{F}(\mathcal{M},F,\mathcal{Y})$ and the correspondent space of
local sections $\Gamma\mathcal{F}$. A field, in general, can be written as an
element $\phi\in\Gamma\mathcal{F}$, and so it can be decomposed as
\begin{equation}
\phi(z)=\sum_{\sigma\in\hat\mathcal{G}}\pi_\sigma\phi(z).
\end{equation}
We then write $\pi_\sigma\phi(z)$ in terms of the field configuration's Fourier components. 

The Fourier components of a field configuration for all the representation classes $\sigma\in\hat\mathcal{G}$
determine completely the second and are determined by it. Being $\mathcal{G}$-invariant, each Fourier 
component can be subjected to the dimensional reduction process analysed in the previous
sections. 
Since the dimensional reducted Fourier components
determine completely the unreduced Fourier components, we have performed the dimensional reduction of
the field configuration itself.

\chapter{SYMMETRY BREAKING}

\textit{A general approach to the study of symmetry breaking processes in Kaluza-Klein type theories is presented.
It is shown that in general the dimensional reduction process inherent to such type of theories will conduce,
first, to a geometrical symmetry breaking, being the gauge group of the Yang-Mills fields obtained a subgroup
of the structure group of the fibre bundle structure of the initial manifold; and second, to a spontaneous 
symmetry breaking of that reduced gauge group led by a scalar field also obtained from the reduction process.
It is explicitly present the form of the breaking scalar potential when the internal space is symmetric.
Finally, the spontaneous compactification process, by which a fibre bundle structure can be dynamicly achieved
by a general manifold - and so without imposing it} \`a priori \textit{- is presented.}

\section{Quantisation}

\subsection{Quantisation of Matter Fields}
\subsubsection{The Matter Field Configuration Manifold}
\paragraph{Definitions.}
Let $\Gamma\mathcal{F}$ be the configuration space defined over $\mathcal{M}$. Since $\Gamma\mathcal{F}$ is
a vectorial space it possesses the structure of an infinitely dimensional manifold. 
A point over $\Gamma\mathcal{F}$ will
be represented by $\phi$. 
The coordinates of a point $\phi$ will be given by the correspondent components of $\phi$ in some basis $\{\phi_i\}_{i=0}^\infty$
of $\Gamma\mathcal{F}$ (usually the set of eigenvectors of some operator):
$$\phi=\sum_{i=0}^{\infty} c_i \phi_i,$$
$$\phi=(c_1,c_2,c_3,\cdots).$$


\paragraph{The Tangent and Cotangent Bundles.}
From each $c_i$ one constructs the differential $dc_i$, creating in this way a basis $\{dc_i\}_{i=0}^\infty$ of the cotangent 
vector space $T^{*}_\phi\Gamma\mathcal{F}$. In a similar way, $\{\partial_i\}_{i=0}^\infty$ will constitute a basis of the
tangent vector space $T_\phi\Gamma\mathcal{F}$ in the same point $\phi$. We then construct the tangent and cotangent
vector bundles, $T\Gamma\mathcal{F}$ and $T^{*}\Gamma\mathcal{F}$.


\paragraph{The Classical Action and the Dynamical Space.}
The action $S$ of the field will be a real-value scalar function (functional) on the configuration space, that is, 
$S\in\Lambda^0(\Gamma\mathcal{F},\mathcal{R})$. The variational derivative $\delta S/\delta \phi$ will then be a one-form on $\Gamma\mathcal{F}$, 
 being the dynamical field configurations defined as the field configurations $\phi\in\Gamma\mathcal{F}$ that obey
 
 $$\frac{\delta S}{\delta \phi}=0,$$
for some boundary conditions. The set of all (classical) dynamical field configurations is entitled the dynamical (sub)space of 
the configuration space and will be represented by $\Gamma\mathcal{F}_S$.

\subsubsection{The Quantisation Process}\label{quantizationproc}

The quantisation of the matter field configurations can be settled in the following three steps:
\begin{enumerate}
\item Let $\Gamma\mathcal{F}^*$ be the dual space to $\Gamma\mathcal{F}$ and let us fix an action $S$ and 
define the following functional:


\begin{equation}
\mathcal{Z}_S(j)=\int_{\Gamma\mathcal{F}}\mathcal{D}\phi\; e^{i\left(
S(\phi)+<j,\phi>_\mathcal{M}\right)},
\end{equation}
for $j\in\Gamma\mathcal{F}^*$ and where $$<j,\phi>_\mathcal{M}=\int_\mathcal{M}\phi\wedge *j$$ is the internal product
defined over $\mathcal{M}$ (cf. section \ref{produtointerno}). 

\item From $\mathcal{Z}_S\in\Lambda^0(\Gamma\mathcal{F}^*,\mathcal{R})$ we define the connected generating functional through
\begin{equation}
        \exp(i\mathcal{W}_S(j))=\mathcal{Z}_S(j),
\end{equation}
with $j\in\Gamma\mathcal{F}^*$.

By expanding $\mathcal{W}_S$ in a functional Taylor development,
\begin{equation}
  \begin{array}{c}
  \mathcal{W}_S(j)=\sum_{n=1}^\infty\frac1{n!}\left<\cdots\left<G^{(n)},j\right>_\mathcal{M}\cdots,j\right>_\mathcal{M} \\
  =\sum_{n=1}^\infty\frac1{n!}\int_\mathcal{M}d^mz_1\sqrt{-\gamma}\cdots d^mz_n\sqrt{-\gamma}
   G^{(n)}_{i_1\cdots i_n}(z_1,\cdots,z_n)j^{i_1}(z_1)\cdots j^{i_n}(z_n),
  \end{array}
\end{equation}
we get the connected Green's functions $G^{(n)}$.

The background matter field configuration is defined through the $\mathcal{W}_S\in\Lambda^0(\Gamma\mathcal{F}^*,\mathcal{R})$ as

\begin{equation}
        \phi_c=\frac{\delta\mathcal{W}_S(j)}{\delta j}.
\end{equation}
$\phi_c$ can be defined as a section of a fibre bundle $\mathcal{F}_c(\mathcal{M},F_c,\mathcal{Y}_c,p_c,\Psi_c)$,
 $$\phi_c\in\Gamma\mathcal{F}_c.$$

\item Finally, the effective action $\Gamma\in\Lambda^0(\Gamma\mathcal{F}_c,\mathcal{R})$ is defined as the functional 
Legendre transformation of the connected generating functional in terms of the background field,
\begin{equation}
        \Gamma(\phi_c)=\mathcal{W}[j(\phi_c)]-<j(\phi_c),\phi_c>.
\end{equation}

The (quantum) dynamical (sub)space of the background matter field configuration space $\Gamma\mathcal{F}_{\Gamma}$ is defined
as the subspace of the background fields $\phi_c\in\Gamma\mathcal{F}_c$ that satisfies
\begin{equation}
\frac{\delta\Gamma(\phi_c)}{\delta\phi_{c}}+j(\phi_c)=0.
\end{equation}
Or, in the absence of currents, as the solutions of 
$$\frac{\delta\Gamma(\phi_c)}{\delta\phi_c}=0,$$
that is the quantum formulation of the identical expression for the classical action $S$.

By expanding $\Gamma$ in a functional Taylor development,
\begin{equation}
  \begin{array}{c}
  \Gamma(\phi_c)=\sum_{n=1}^\infty\frac1{n!}\left<\cdots\left<\Gamma^{(n)},\phi_c\right>_\mathcal{M}\cdots,\phi_c\right>_\mathcal{M} \\
  =\sum_{n=1}^\infty\frac1{n!}\int_\mathcal{M}d^mz_1\sqrt{-\gamma}\cdots d^mz_n\sqrt{-\gamma}
   \Gamma^{(n)}_{i_1\cdots i_n}(z_1,\cdots,z_n)\phi_c^{i_1}(z_1)\cdots 
\phi_c^{i_n}(z_n),
  \end{array}
\end{equation}
we get the 1PI Green's functions or the proper vertices $\Gamma^{(n)}$.

\end{enumerate}

Since the quantum fibre bundle $\mathcal{F}_c$ usually coincides with the classical one, $\mathcal{F}$ - this
neglecting that $\mathcal{F}$ is in fact an operator space over an Hilbert space and that $\mathcal{F}_c$ isn't - 
we can make $\Gamma\mathcal{F}_c=\Gamma\mathcal{F}$,
and thus the quantum process is resumed to an automorphism in the space of functions defined over the matter configuration
field manifold, that is, we write the previous quantisation process as an automorphism 
$$\mathcal{T}:\Lambda^0(\Gamma\mathcal{F},\mathcal{R})\longrightarrow\Lambda^0(\Gamma\mathcal{F},\mathcal{R}),$$
such that the effective action $\Gamma$ is written in terms of the classical action $S$ through $\Gamma=\mathcal{T}S$.

The quantisation process, and so the quantisation operator, is dependent on the intermediate connected generating functional
$\mathcal{W}_S$. It is this functional that induces the quantum transformation $\phi\rightarrow\phi_c$ in $\Gamma\mathcal{F}$.
This being non-local (referred to the configuration manifold $\Gamma\mathcal{F}$), it differs from the usually (active)
"coordinate" transformations on a generic manifold, and it is the non-locality of the quantum operator $\mathcal{T}$ that
set its nature as a statistical one.

An expression to the quantisation operator $\mathcal{T}$ is desirable if one wishes to study the breaking of a symmetry of
the action due to the quantum process (anomalous or spontaneously symmetry breaking).

\subsubsection{Perturbative Expansion}

Let us suppose that the background matter field configuration is close to the matter field configuration, i.e,
that
$$\phi=\phi_c+\tilde\phi,$$
where $\tilde\phi$ it's a small quantum perturbation to $\phi_c$. Then we can write for the classical action
\begin{equation}
S(\phi)=S(\phi_c)+\left<\frac{\delta S}{\delta\phi}(\phi_c),\tilde\phi\right>_\mathcal{M}+\frac12\left<\tilde\phi,\left<\frac{\delta^2 S}
{\delta\phi^2}(\phi_c),\tilde\phi\right>_\mathcal{M}\right>_\mathcal{M}+\cdots
\end{equation}

If the quantum effects are small, as they are supposed to, the effective action shall be given, approximatively by the classical one:
\begin{equation}
\Gamma(\phi_c)=S(\phi_c)+\sum_{k=1}^\infty\Gamma^{(k)}(\phi_c),
\end{equation}
where $\Gamma^{(k)}$ is the $k$-loop correction to the classical action due to quantum effects. We will be interested only
in the one-loop correction term $\Gamma^{(1)}$.

From the Legendre transformation
$$\mathcal{W}_S[j(\phi_c)]=\Gamma(\phi_c)+<j,\phi_c>_\mathcal{M},$$
the background equations
$$j(\phi_c)=-\frac{\delta\Gamma}{\delta\phi_c}(\phi_c),$$
from the measure transformation
$$\mathcal{D}\phi=\mathcal{D}\tilde\phi,$$
and from the functional development of both the classical and effective actions we obtain
\begin{equation}
\begin{array}{c}
\exp\left[i\left(S(\phi_c)+
\sum_{k=0}^\infty\Gamma^{(k)}(\phi_c)+
\left<\frac{\delta S}{\delta\phi_c}(\phi_c),\phi_c\right>_\mathcal{M}+
\sum_{k=0}^\infty\left<\frac{\delta\Gamma^{(k)}}{\delta\phi_c}(\phi_c),\phi_c\right>_\mathcal{M}\right)\right]=\\
\int_{\Gamma\mathcal{F}}\mathcal{D}\tilde\phi\;\exp\left[i\left(S(\phi_c)+
\frac12\left<\tilde\phi,\left<\frac{\delta^2 S}{\delta\phi^2}(\phi_c),\tilde\phi\right>_\mathcal{M}\right>_\mathcal{M}+\cdots
+\left<\frac{\delta S}{\delta\phi_c}(\phi_c),\phi_c\right>_\mathcal{M}+\sum_{k=1}^\infty\left<\frac{\delta\Gamma^{(k)}}{\delta\phi_c}
(\phi_c),\phi_c\right>_\mathcal{M}\right)\right]
\end{array}
\end{equation}

Cancelling the common terms we will get
$$\prod_{k=1}^\infty e^{i\Gamma^{(k)}(\phi_c)}=\int_{\Gamma\mathcal{F}}\mathcal{D}\tilde\phi\;e^{-i\left<\tilde\phi,\left<\mathcal{A},\tilde\phi
\right>_\mathcal{M}\right>_\mathcal{M}},$$
where we made
\begin{equation}
        \mathcal{A}=-\frac{\delta^2 S}{\delta\phi_c^2}(\phi_c).
\end{equation}
We get for the one-loop term
\begin{equation}
e^{i\Gamma^{(1)}(\phi_c)}=\int_{\Gamma\mathcal{F}}\mathcal{D}\tilde\phi\;e^{-i\left<\tilde\phi,\left<\mathcal{A},\tilde\phi
\right>_\mathcal{M}\right>_\mathcal{M}}.
\end{equation}

We now write the quantum perturbation in terms of its coordinates:
$$\tilde\phi=\sum_{k=1}^\infty c_n\tilde\phi_n,$$
where $\{\phi_n\}$ is an orthonormal base of $\Gamma\mathcal{F}$ and its elements are eigenvectors from the operator $\mathcal{A}$:
$$\mathcal{A}\tilde\phi_n=A_n\tilde\phi_n.$$
Writing for the measure
$$\mathcal{D}\tilde\phi=\prod_{k=1}^\infty \frac\mu{\sqrt{\pi}}dc_k,$$
with $\mu$ a normalisation constant,
we will have
$$e^{i\Gamma^{(1)}(\phi_c)}=\prod_{l=1}^\infty\int dc_l\;\frac\mu{\sqrt{\pi}}
e^{-i\sum_{m,n}c_mc_nA_n<\tilde\phi_m,\tilde\phi_n>_\mathcal{M}}.$$
And since $<\tilde\phi_m,\tilde\phi_n>_\mathcal{M}=\delta_{mn}$ we will have
$$e^{i\Gamma^{(1)}(\phi_c)}=\prod_{k=1}^\infty\int dc_k\;\frac\mu{\sqrt\pi} e^{-iA_kc_k^2}.$$
We can now perform a Wick rotation, changing the eigenvalues $A_k$ into $-iA_k$, and so we get
$$e^{i\Gamma^{(1)}(\phi_c)}=\prod_{k=1}^\infty\int dc_k\;\frac\mu{\sqrt\pi} e^{-A_kc_k^2}.$$
Each integral in this product is a Gaussian integral with value $\sqrt{\pi\mu^2/A_n}$ so we finally get
\begin{equation}
\Gamma^{(1)}(\phi_c)=\frac{1}2\sum_{n=1}^\infty \ln\left(\frac{A_n}{\mu^2}\right).
\end{equation}

Introducing the zeta-function
\begin{equation}
\zeta(\frac{\mathcal{A}}{\mu^2}|s)=\sum_{n=1}^\infty \left(\frac{A_n}{\mu^2}\right)^{-s},
\end{equation}
and taking its derivate,
\begin{equation}
\zeta'(\frac{\mathcal{A}}{\mu^2}|s)=-\sum_{n=1}^\infty \ln\left(\frac{A_n}{\mu^2}\right)\left(\frac{A_n}{\mu^2}\right)^{-s},
\end{equation}
we can write the one-loop correction as
\begin{equation}\label{zeta}
        \Gamma^{(1)}(\phi_c)=-\frac{1}2\zeta'(\frac{\mathcal{A}}{\mu^2}|0).
\end{equation}


\section{Symmetries}

\subsection{Basic Definitions}
Let $\mathcal{M}(V_n,\mathcal{G}/\mathcal{H},\mathcal{G},\pi,\Phi)$ and $\mathcal{F}(\mathcal{M},F,\mathcal{Y},p,\Psi)$
be differentiable fibre bundles, being the first the base space of the second, and let us consider the space of the sections
of $\mathcal{F}$, i.e., the matter field configuration space $\Gamma\mathcal{F}$. Let 
$S,\Gamma\in\Lambda^0(\Gamma\mathcal{F},\mathcal{R})$ be, respectively, the classical and the effective actions of the
matter and background fields.

We say that the classical (or effective) action of a matter (or background) field configuration possesses a given symmetry if
there is an action of a Lie group $\mathcal{B}$ over $\Gamma\mathcal{F}$ that lifts to an action of the same group over
$\Lambda^0(\Gamma\mathcal{F},\mathcal{R})$ such that the induced representation $T^\mathcal{B}$ of the symmetry group
transforms $S$ (or $\Gamma$) into a functional that is linear dependent on $S$ (or $\Gamma$), i.e., we have, for
a $\mathcal{B}$-type symmetry the following condition
\begin{equation}\label{simetria}
[T^\mathcal{B}_b S](\phi)=c(b)S(\phi),
\end{equation}
for all $b\in\mathcal{B}$ and for $c\in\Lambda^0(\mathcal{B},\mathcal{R})$ different from the null function.

We will be interested not in the symmetries of the action $\mathcal{S}$ over the multidimensional universe $\mathcal{M}$
but in the symmetries of the reduced and effective action $S^r$ over the base space $\bar\mathcal{M}$. In general,
if the action over $\mathcal{M}$ obeys (\ref{simetria}), then the reduced one will obey an identical relation
\begin{equation}
[T^{\mathcal{B}^r}_b S^r](\phi)=c^r(b)S^r(\phi),
\end{equation}
where $\mathcal{B}^r\subset\mathcal{B}$ is the symmetry group of $S^r$ - that will be, in general, of lower dimension
than that of the unreduced action, $S$.

\section{Symmetry Breaking Process}
\subsection{Characterisation}\label{caracterization}

\subsubsection{Symmetry Breaking Process.}
A symmetry breaking process is defined as an automorphism in the space of functionals over $\Gamma\mathcal{F}$ that
does not preserve a given symmetry, i.e., it can be written as an operator $\mathcal{T}\in Aut(\Lambda^0(\Gamma\mathcal{F},\mathcal{R}))$
such that the transformed functional $\mathcal{T}S$ does not obey the condition (\ref{simetria}) even if the original functional
does so. In general we can write
\begin{equation}\label{odio}
[T^{\mathcal{B}'}_{b'} \mathcal{T}S](\phi)=c(b')\mathcal{T}S(\phi),
\end{equation}
for $b'\in\mathcal{B}'$ and where $\mathcal{B}'\subset\mathcal{B}$ is a subgroup of $\mathcal{B}$. 

\subsubsection{Geometrical Symmetry Breaking Process.}
A geometrical symmetry breaking process is defined by an homomorphism $\mathcal{T}$ 
between the space of functionals over $\Gamma\mathcal{F}$
and $\Gamma\mathcal{F}^r$, with $\mathcal{F}^r\subset\mathcal{F}$, such that $\mathcal{T}$ 
is not an eigenvector of a representation $T^\mathcal{B}$
of $\mathcal{B}$ but it is an eigenvector of a representation of a subgroup $\mathcal{B}^r$ of $\mathcal{B}$, i.e.,
\begin{equation}
T^{\mathcal{B}^r}\mathcal{T}=c^r\mathcal{T}.
\end{equation}
This expression is identical to (\ref{odio}) when $\mathcal{F}$ and $\mathcal{F}^r$ are the same fibre bundle,
and it presents, in fact, a generalisation of the previous case. 

In general, there will be always a geometric symmetry
breaking when the initial fibre bundle $\mathcal{F}$ can be reduced to a fibre bundle $\mathcal{F}^r$ that
completely determines it. We shall have in this case $\mathcal{B}=\mathcal{Y}$ and $\mathcal{B}^r=\mathcal{C}$
(cf. chap. 3). In this chapter we will only study  symmetry breaking processes of the reduced action, that
will be, in general, preceded by a geometrical symmetry breaking from the unreduced one.

\subsubsection{Quantum Symmetry Breaking Process.}

\paragraph{Anomalous Symmetry Breaking.}
A very common type of symmetry breaking occurs when quantising a given field theory and the final action - the
effective one, $\Gamma$ - does not
retain an internal symmetry of the classical one - $S$. In this case, when the symmetry breaking is due to the
quantisation process, we are in the presence of an anomalous symmetry breaking.

\paragraph{Spontaneously Symmetry Breaking.}
A particular case of anomalous symmetry breaking is the one in which the quantisation process does not
break directly an internal symmetry but reveals itself incomplete. When quantising
a theory there is a possibility that in the end a degeneracy in the lower state of the theory be found.
In order to theory be defined it is necessary that such state be unique: one of the states in the degeneracy
must be chosen as the vacuum of the theory. By doing so, that is, by completing the quantum transformation
with a direct translation of the background fields configurations on $\Gamma\mathcal{F}$,
there will be, in general, a symmetry breaking: a spontaneously symmetry breaking.

An spontaneously symmetry breaking will occur if the quantum motion equation
\begin{equation}\label{accao}
\frac{\delta\Gamma}{\delta\phi_c}=0,
\end{equation}
holds for some $v$ nonzero value of $\phi_c$, being necessary a redefinition of the dynamical subspace $\Gamma\mathcal{F}_\Gamma$ 
on $\Gamma\mathcal{F}$,
$$\phi_c\longrightarrow \phi_c'=T(v)\phi_c,\;\;(\phi_c\in\Gamma\mathcal{F}_\Gamma,\phi_c'\in\Gamma\mathcal{F}'_\Gamma),$$
with $T(v):\Gamma\mathcal{F}_\Gamma\rightarrow\Gamma\mathcal{F}_\Gamma'$,
in order to the theory to have a vanishing vacuum expectation value. 
The action on the functional space $\Lambda^0(\Gamma\mathcal{F},\mathcal{R})$ will be
\begin{equation}
\Gamma'(\phi_c)=\Gamma(T(v)\phi_c).
\end{equation}

The complete quantum transformation can then be written as
\begin{equation}
\Gamma=(\mathcal{T}S)\circ T(\upsilon).
\end{equation}
It must be noted that the value of $v$ is obtained from (\ref{accao}), being necessary to know the expression of
the effective action. In general this is not possible and only approximations can be given to $\Gamma$ (cf. section
\ref{quantizationproc}). 


\subsection{Spontaneous Symmetry Breaking}

\subsubsection{Spontaneous Symmetry Breaking at the Tree-Level}

At a tree-level analysis, we say that a spontaneous symmetry breaking takes place if the classical action possesses
roots in points away from the origin, i.e., that the equation
\begin{equation}\label{classicalaction}
\frac{\delta S}{\delta \phi_c}(\phi_c)=0,
\end{equation}
has for solution a non-null value of $\phi_c$.

In general the classical action can be written in terms of a Lagrangian density $\mathcal{L}$ that by its part
can be written as the sum of two terms: a kinetic and a potential term.
We can then write
$$S(\phi_c)=\int_\mathcal{M}d^mz\sqrt{-\gamma}\mathcal{L}=\int_\mathcal{M}d^mz\sqrt{-\gamma}\left[K(D\phi_c,\phi_c)-V_0(\phi_c)\right],$$
where $K(D\phi_c,\phi_c)$ and $V_0(\phi_c)$ are, respectively, the kinetic and the potential term.
Since we are mainly interested in cases where the vacuum expectation value is translational invariant the condition
of tree-level spontaneously symmetry breaking can be written as
\begin{equation}
\frac{dV_0}{d\phi_c}(\phi_c)=0,
\end{equation}
for a non-null solution $\phi_c$. It expresses the fact that $V_0(\phi_c)$ has a minimum for a non null value
of the classical field $\phi_c$.

It is then the form of the classical potential to dictate if it will occur or not
a tree-level spontaneously symmetry breaking.

\subsubsection{Induced Spontaneous Symmetry Breaking from One-Loop Radiative Corrections}
Let us consider the trunked quantisation process to the one-loop correction of the classical action, that is,
let us consider the following effective action approximation,
\begin{equation}\label{oneloop}
\Gamma(\phi_c)=S(\phi_c)+\Gamma^{(1)}(\phi_c).
\end{equation}

Induced spontaneous symmetry breaking due to one-loop radiative corrections occurs if (\ref{accao}) is verified
for a non-null $\phi_c$ after the addition of the one-loop correction term $\Gamma^{(1)}$ to the classical action $S$.
That is, it is supposed that (\ref{classicalaction})
has for its only solution $\phi_c=0$ and that is the correction one-loop term that comports the symmetry breaking.
One must say that this type of analysis is in some sense unjustified since one can not say that a symmetry breaking
occurs without calculating all radioactive correction terms. It can be that from the addition of all terms the
symmetry breaking is lost, that is, that the quantisation process is complete and the symmetry breaking occurs
from neglecting the high order terms in the perturbative expansion of that one. An example of this is given
in \cite{frampton}.

Turning to (\ref{oneloop}) we have that the one-loop correction can in general be written as a correction
to the classical potential since it will only depend on $\phi_c$ and not on in its derivatives. By writing
$$\Gamma^{(1)}(\phi_c)=-\int_\mathcal{M}d^mz\sqrt{-\gamma}V^{(1)}(\phi_c),$$
the spontaneously symmetry breaking condition can be rewritten as
\begin{equation}
\frac{d}{d\phi_c}\left[V_0+V^{(1)}\right](\phi_c)=0,
\end{equation}
for a non-null value $\phi_c$.

\subsection{The Goldstone Problem and the Higgs Mechanism}

Let $\mathcal{B}$ be an $d_\mathcal{B}$-dimensional symmetry group of the classical action and 
let $\mathcal{B}'\subset\mathcal{B}$ be the $d_{\mathcal{B}'}$-dimensional symmetry group
of the effective action. And let us consider the two point connected and 1PI Green functions $G^{(2)}$ and $\Gamma^{(2)}$
\footnote{the notation could be misleading: it is not the two-loop correction to the effective action that we are considering.}.
The two are related by
\begin{equation}
\int_\mathcal{M}d^mz\sqrt{-\gamma}G^{(2)}_{ij}(x-z)\Gamma^{(2)jk}(z-y)=\delta^k_i\delta^{(m)}(x-y),
\end{equation}
for $x,y\in\mathcal{M}$.

In order to proceed we must suppose that the Lie algebra $L_\mathcal{B}$ is reductive,
\begin{equation}\label{decomp}
  L_\mathcal{B}=L_{\mathcal{B}'}\oplus L_{\mathcal{B}''},
\end{equation}
where $L_{\mathcal{B}'}$ is the Lie algebra of $\mathcal{B}'$ and $L_{\mathcal{B}''}$ is the complement to the last in 
$L_\mathcal{B}$.

Let us choose an adapted orthonormal base of $L_\mathcal{B}$, 
$$\{\lambda_i\}_{i=1,\cdots,d_\mathcal{B}}=\{\lambda_{i'},\lambda_{i''}\}_{
i'=1,\cdots,d_{\mathcal{B}'};i''=d_{\mathcal{B}'}+1,\cdots,d_{\mathcal{B}}}.$$

Let 
\begin{eqnarray}
\beta:\mathcal{B}\times\Gamma\mathcal{F}\longrightarrow\Gamma\mathcal{F},\\
\beta':\mathcal{B}'\times\Gamma\mathcal{F}\longrightarrow\Gamma\mathcal{F},
\end{eqnarray}
be the actions of $\mathcal{B}$ and $\mathcal{B}'$ on the configuration space $\Gamma\mathcal{F}$.
Then we shall have $\beta(b,\phi)\in\Gamma\mathcal{F}$ for $b\in\mathcal{B}$ and $\phi\in\Gamma\mathcal{F}$.
In special, for $b$ in a neighbourhood of the unit $e$ of $\mathcal{B}$, we can take, using the exponential map,
$$b=e^{-i\lambda_jb^j},$$
with $b^j$, $j=1,\cdots,d_\mathcal{B}$, in $\mathcal{R}$. For $b^j$ infinitesimal we can make
\begin{equation}
b=e-i\lambda_jb^j.
\end{equation}
Then the action of this element in $\Gamma\mathcal{F}$ will be given by
$$\beta(b,\phi)=\beta(e-i\lambda_jb^j,\phi)=\phi-ib^j\beta(\lambda_j,\phi).$$
The variation in $\phi$ induced by the infinitesimal transformation represented by $b$ will be
$$\delta_b\phi=\gamma(b,\phi)-\phi=-ib^j\beta(\lambda_j,\phi).$$
By using (\ref{decomp}) we can write
\begin{equation}
\delta_b\phi=-ib^{j'}\beta(\lambda_{j'},\phi)-ib^{j''}\beta(\lambda_{j''},\phi)=\delta_{b'}\phi+\delta_{b''}\phi.
\end{equation}

The associated variation on $\Gamma\mathcal{F}^*$ will be given by
\begin{equation}
\begin{array}{c}
\delta_bj(x)=\int_\mathcal{M}d^m\sqrt{-\gamma}z\frac{\delta j(x)}{\delta\phi(z)}\delta\phi(z)=\int_\mathcal{M}
d^mz\sqrt{-\gamma}\frac{\delta^2\Gamma}{\delta\phi(z)\delta\phi(x)}\delta\phi(z)\\=\int_\mathcal{M}d^mz\sqrt{-\gamma}
\Gamma^{(2)}(x-z)(\delta_{b'}\phi(z)+\delta_{b''}\phi(z))\end{array}
\end{equation}
Since $\phi$ will remain invariant for $\mathcal{B}'$-type transformations, we will have $\delta_{b'}\phi=0$.
The same will not occur for $\mathcal{B}''$, and so, if the symmetry breaking is not to be followed by the 
generation of currents, we should have
\begin{equation}
\delta_bj=0\Rightarrow \Gamma^{(2)}(x-z)\delta_{b''}\phi(z)=0.
\end{equation}
But this gives,
\begin{equation}
\Gamma^{(2)}(x-z)=0=G^{(2)}(x-z),
\end{equation}
and so the symmetry breaking will be followed by the production of $d_{\mathcal{B}''}=d_\mathcal{B}-d_{\mathcal{B}'}$
massless Goldstone bosons.
In order to solve
this problem we have to consider that there will be the production of currents by the symmetry breaking process:
\begin{equation}
\delta_bj\neq 0,
\end{equation}
but that such currents will be absorbed into a gauge field of the $\mathcal{B}$-symmetry, generating a mass vector field. 
This is the Higgs mechanism and it consists on the promotion of the global $\mathcal{B}$-symmetry into a local one.




\section{Symmetry Breaking from Dimensional Reduction}

\subsection{The Scalar Field's Potential}

Let us consider a principal fibre bundle $\mathcal{F}(\mathcal{M},\mathcal{Y},p,\Psi)$ defined over a multidimensional
universe $\mathcal{M}(V_n,M,\mathcal{G},\pi,\Phi)$ that possesses as internal space a symmetric homogeneous space
$\mathcal{G}/\mathcal{H}$, with $\mathcal{H}\subset\mathcal{G}$, and let it be defined an $\mathcal{G}$-invariant
infinitesimal connection form $\omega$ on $\mathcal{F}$. 

From previous results (cf. section 3.1.5), this bundle can 
be subjected to dimensional reduction, leading to a bundle $\mathcal{F}^r(\bar\mathcal{M},\mathcal{C},p^r,\Psi^r)$,
with $\mathcal{C}$ the centraliser of $\bar\psi(\mathcal{H})$ in $\mathcal{Y}$. The infinitesimal connection form
$\omega$ will define over $\mathcal{F}^r$ an $\mathcal{H}$-invariant infinitesimal connection $\omega^r$ and an
$\mathcal{Y}$-invariant mapping $\xi^r$ from $\mathcal{F}^r$ to
$L_\mathcal{P}^*\otimes L_\mathcal{Y}$. The reduced action will be
\begin{equation}
S_{\mathcal{F}^r}[\Omega^r,\xi^r]=
\frac{vol(\mathcal{G}/\mathcal{H})vol(\mathcal{Y}/\mathcal{C})}{vol(\mathcal{Y})}\left(\parallel\Omega^r\parallel_{\mathcal{F}^r}+
\frac12 \parallel p^{r*}\left(D\xi^r\right) \parallel_{\mathcal{F}^r}+
\parallel p^{r*}\left(\alpha(\xi^r)\right)\parallel_{\mathcal{F}^r}\right),
\end{equation}
with
\begin{equation}
\alpha(\xi^r)=\frac12[\xi^r,\xi^r]-\frac12\xi^r\circ[\cdot,\cdot]|_{L_\mathcal{P}}-\frac12\bar\psi^*\circ[\cdot,\cdot]|_{L_\mathcal{H}}.
\end{equation}
By defining a section $s$ of $\mathcal{F}^r$, we shall get an action defined
over the base space of $\mathcal{F}^r$, i.e., over $\bar\mathcal{M}$,
\begin{equation}
S_{\bar\mathcal{M}}[A,\phi]=\int_{\bar\mathcal{M}}d^n x \sqrt{-g}\left(\frac14 tr(F^2)+
\frac12 tr(D\phi^2)-V(\phi)\right),
\end{equation}
where the scalar potential is given by
\begin{equation}
  V(\phi)d^n x = - a(\phi)\wedge\star a(\phi),
\end{equation}
with $a=s^*\alpha$ and $\phi=p^{r*}\xi^r$. 
By choosing a local coordinate basis $\{dx^\mu\}_{\mu=0,\cdots,n-1}$ on
$\bar\mathcal{M}$ and a basis $\{\lambda_a\}_{a=1,\cdots,dim(\mathcal{P})}$ on
  $L_\mathcal{P}$ we can write for the action
\begin{equation}
S_{\bar\mathcal{M}}[A,\phi]=\int_{\bar\mathcal{M}}d^n x \sqrt{-g}\left(\frac14 tr(F_{\mu\nu}F^{\mu\nu})+
\frac12 tr(D_\mu\phi D^\nu\phi)-V(\phi)\right),
\end{equation}
with $V$ expressed in terms of $a$ in those basis,
\begin{equation}\label{potential1}
  V(\phi)=<a_{ab},a^{ab}>_{L_\mathcal{Y}},
\end{equation}
having
\begin{equation}
a_{ab}=\frac12[\phi(\lambda_a),\phi(\lambda_b)]-\frac12\phi\circ[\lambda_a,\lambda_b]|_{L_\mathcal{P}}-\frac12\bar\psi^*\circ[\lambda_a,\lambda_b]|_{L_\mathcal{H}},
\end{equation}
and being $<\cdot,\cdot>_{L_\mathcal{Y}}$ an inner product defined on $L_\mathcal{Y}$.

%
%
%

We can then write for the scalar potential,
\begin{equation}\label{sis}
V(\phi)=\sum_{k=0}^4 V^{(k)}(\phi),
\end{equation}
with
\begin{eqnarray}
V^{(0)}=\frac14<\bar\psi^*([\lambda_a,\lambda_b]_{L_\mathcal{H}}),\bar\psi^*([\lambda^a,\lambda^b]_{L_\mathcal{H}})>_{L_\mathcal{Y}},\\
V^{(1)}(\phi)=\frac12<\phi([\lambda_a,\lambda_b]_{L_\mathcal{P}}),\bar\psi^*([\lambda^a,\lambda^b]_{L_\mathcal{H}})>_{L_\mathcal{Y}},\\
\begin{array}{c}
V^{(2)}(\phi)=\frac14<\phi([\lambda_a,\lambda_b]_{L_\mathcal{P}}),\phi([\lambda^a,\lambda^b]_{L_\mathcal{P}})>_{L_\mathcal{Y}}\\-
\frac12<[\phi(\lambda_a),\phi(\lambda_b)]_{L_\mathcal{Y}},\bar\psi^*([\lambda^a,\lambda^b]_{L_\mathcal{H}})>_{L_\mathcal{Y}},
\end{array}\\
V^{(3)}(\phi)=-\frac12<[\phi(\lambda_a),\phi(\lambda_b)]_{L_\mathcal{Y}},\phi([\lambda^a,\lambda^b]_{L_\mathcal{P}})>_{L_\mathcal{Y}},\\
V^{(4)}(\phi)=\frac14<[\phi(\lambda_a),\phi(\lambda_b)]_{L_\mathcal{Y}},[\phi(\lambda^a),\phi(\lambda^b)]_{L_\mathcal{Y}}>_{L_\mathcal{Y}}.
\end{eqnarray}


\subsection{Scalar Fields Analysis}

\subsubsection{The Lie Algebra Decomposition}

The form (\ref{potential1}) given for the scalar fields potential is still unappropriated since it does not take into account
that $\phi$ must be an interwining operator \cite{hewitt2,kirillov},
\begin{equation}\label{beubeu2}
  \phi\circ Ad(\mathcal{H})(L_\mathcal{P}) = Ad(\bar\psi(\mathcal{H}))(L_\mathcal{Y})\circ\phi.
\end{equation}
From now on we will consider that both $\mathcal{H}$ and $\mathcal{Y}$ are connected in a way that we can write
the infinitesimal version of (\ref{beubeu2}) \cite{coquereaux1}
\begin{equation}
  \phi\circ Ad(L_\mathcal{H})(L_\mathcal{P})=Ad(\bar\psi(L_\mathcal{H}))\circ\phi(L_\mathcal{P}).
\end{equation}
We will also consider the Lie algebras $L_\mathcal{G}$ and $L_\mathcal{Y}$ to be simple.

In order to identify the representation of $\mathcal{C}_{\mathcal{Y}}(\bar\psi(\mathcal{H})$ in which the scalar fields $\phi$ belong 
we must decompose both the representations $Ad(L_\mathcal{H})$ and $Ad(\bar\psi(L_\mathcal{H}))$ that act, respectively, in the 
Lie algebras $L_\mathcal{P}$ and $L_\mathcal{Y}$, into irreducible representations in order to apply the Schur theorem to the 
common representations \cite{hewitt1,hewitt2,kirillov}.

To that purpose, we must be able to write both $L_\mathcal{P}$ and $L_\mathcal{Y}$ in subspaces that will be invariant under
$Ad(L_\mathcal{H})$ and $Ad(\psi(L_\mathcal{H}))$. We begin with $L_\mathcal{P}$.

By (\ref{icaro}) we shall have that $L_\mathcal{P}$ is an invariant space under $Ad(L_\mathcal{H})$ and by (\ref{irilo})
we can write the following decomposition of $L_\mathcal{P}$,
\begin{equation}
  L_\mathcal{P}=L_{\mathcal{N}/\mathcal{H}}\oplus L_\mathcal{L}.
\end{equation}
By (\ref{ioio}) and considering that $\mathcal{H}\subset\mathcal{N}$ we shall have that
\begin{equation}
  Ad(L_{\mathcal{H}})L_{\mathcal{N}/\mathcal{H}}\subset L_{\mathcal{N}/\mathcal{H}},
\end{equation}
and so will be an invariant subspace of $L_\mathcal{P}$. Moreover, it will be a subspace over which only a trivial representation
of $Ad(L_\mathcal{H})$ acts.

In order to find a decomposition of $L_\mathcal{L}$ we must first decompose the Lie algebra of $\mathcal{H}$ into invariant subspaces under
$Ad(L_\mathcal{H})$.
We shall have
\begin{equation}
  L_\mathcal{H}=\oplus_{\gamma=0}^N L_\mathcal{H}^{(\gamma)},
\end{equation}
getting,
\begin{equation}
  Ad(L_\mathcal{H})L_\mathcal{P}=[L_\mathcal{H},L_\mathcal{P}]=\oplus_{\gamma=0}^N [L_\mathcal{H}^\gamma,L_\mathcal{P}],
\end{equation}
and so we find the following decomposition of the representation
$U_\mathcal{H}$ of $Ad(L_\mathcal{H})$ over $L_\mathcal{P}$ - we suppose that
a basis as been chosen -,
\begin{equation}
U_\mathcal{H}=U_\mathcal{H}^0\oplus\left(\oplus^N_{\gamma=1} U_\mathcal{H}^{\gamma}\right),
\end{equation}
with $U_\mathcal{H}^{\gamma}$ acting over $[L_\mathcal{H}^\gamma,L_\mathcal{P}]$ and $U_\mathcal{H}^0$ being the trivial representation. 
There can be that
some of the representations $U_\mathcal{H}^{\gamma}$ will be equivalent to a fixed representation $W_\mathcal{H}$ of $Ad(L_\mathcal{H})$
\cite{hewitt2}.
We shall then have that $W_\mathcal{H}$ is contained in $U_\mathcal{H}$. If $\{W^{\delta}\}_{\delta=1,\cdots,n}$ is a family 
of mutually inequivalent representations $U_\mathcal{H}^\gamma$ contained in $U_\mathcal{H}$ and if $m_\delta$ is
the multiplicity of an arbitrary $W^\delta_\mathcal{H}$ in $U_\mathcal{H}$, then we can write
\begin{equation}
  U_\mathcal{H}=W_\mathcal{H}^0\oplus\sum_{\delta=1}^n m_\delta W_\mathcal{H}^\delta,
\end{equation}
with $W_\mathcal{H}^0=U_\mathcal{H}^0$.
This will then induce the following decomposition of $L_\mathcal{H}$ (that is only an arrangement of the terms of
the previous decomposition):
\begin{equation}
L_\mathcal{H}=L_\mathcal{H}^0\oplus\oplus_{\delta=1}^n L_\mathcal{H}^\delta,
\end{equation}
with
\begin{equation}
L_\mathcal{H}^\delta = \oplus_{i=1}^{m_\delta} L_{\mathcal{H}i}^\delta,
\end{equation}
where $L_\mathcal{H}^{(\delta)}$ is the collection of subspaces of $\mathcal{H}$ that are invariant under $W_\mathcal{H}^\gamma$.

Under this decomposition of the Lie algebra of $\mathcal{H}$ a similar one can be presented to $L_\mathcal{L}$:
\begin{equation}
L_\mathcal{L}=\oplus_{\delta=1}^n L_\mathcal{L}^\delta,
\end{equation}
with
\begin{equation}
L_\mathcal{L}^\delta = \oplus_{i=1}^{m_i} L_{\mathcal{L}i}^\delta.
\end{equation}
The decomposition of $L_\mathcal{P}$ will then be
\begin{equation}
L_\mathcal{P}=L_{\mathcal{N}/\mathcal{H}}\oplus_{\delta=1}^n\left[\oplus_{i=1}^{m_i} \left(L_{\mathcal{L}i}^\delta\right)\right],
\end{equation}
corresponding a decomposition of the $Ad(L_\mathcal{H})$ over it,
\begin{equation}\label{oidii}
Ad(L_\mathcal{H})=W_\mathcal{H}^0\oplus \sum_{\delta=1}^n m_\delta W_\mathcal{H}^\delta,
\end{equation}
with $W_\mathcal{H}^0$ the trivial representation.

We turn now to the decomposition of
$Ad(\bar\psi^*(L_\mathcal{H}))L_\mathcal{Y}$. 
\begin{equation}
Ad(\bar\psi^*(L_\mathcal{H}))L_\mathcal{Y}=[\bar\psi^*(L_\mathcal{H}),L_\mathcal{Y}]
\end{equation}
We make the first
decomposition of $L_\mathcal{Y}$ as
\begin{equation}
L_\mathcal{Y}=\bar\psi^*(L_\mathcal{H})\oplus L_\mathcal{Y}'.
\end{equation}
We will then get
\begin{equation}
Ad(\bar\psi^*(L_\mathcal{H}))L_\mathcal{Y}=[\bar\psi^*(L_\mathcal{H}),\bar\psi^*(L_\mathcal{H})]\oplus[\bar\psi^*(L_\mathcal{H}),L_\mathcal{Y}'].
\end{equation}
Since $\bar\psi^*$ is a homomorphism of Lie algebras we have
\begin{equation}
[\bar\psi^*(L_\mathcal{H}),\bar\psi^*(L_\mathcal{H})]=\bar\psi^*\circ
[L_\mathcal{H},L_\mathcal{H}]=\bar\psi^*\circ Ad(L_\mathcal{H}).
\end{equation}
By further decompose $L_\mathcal{Y}'$,
\begin{equation}
L_\mathcal{Y}'=\phi(L_\mathcal{P})\oplus L_\mathcal{Y}'',
\end{equation}
and by noticing that
\begin{equation}
[\psi^*(L_\mathcal{H}),\phi(L_\mathcal{P})]=\phi\circ[L_\mathcal{H},L_\mathcal{P}]=\phi\circ Ad(L_\mathcal{H})L_\mathcal{P}
\end{equation}
we obtain
\begin{equation}
Ad(\bar\psi^*(L_\mathcal{H}))L_\mathcal{Y}=\bar\psi^*\circ
Ad(L_\mathcal{H})\oplus\phi\circ Ad(L_\mathcal{H})L_\mathcal{P}\oplus [\bar\psi^*(L_\mathcal{H}),L_\mathcal{Y}''].
\end{equation}

By writing 
\begin{equation}
  L_\mathcal{Y}''=L_\mathcal{Y}^{(0)}\oplus L_\mathcal{Y}^{(3)},
\end{equation}
where $L_\mathcal{Y}^{(0)}$ is the subspace of $L_\mathcal{Y}$ where a trivial
action of $Ad(L_\mathcal{H})$ occurs, and decomposing
\begin{equation}
  L_\mathcal{Y}^{(2)}=\phi\circ Ad(L_\mathcal{H})L_\mathcal{P},
\end{equation}
in the same manner as in (\ref{oidii}) but without the trivial representation
$W^0_\mathcal{H}$,
\begin{equation}
L_\mathcal{Y}^{(2)}=\sum_{\delta=1}^n m'_\delta W_\mathcal{H}^\delta,
\end{equation}
we will get
\begin{equation}
Ad(\bar\psi^*(L_\mathcal{H}))L_\mathcal{Y}=L_\mathcal{Y}^{(0)}\oplus
\bar\psi^*(Ad(L_\mathcal{H}))\oplus \sum_{\delta=1}^n m'_\delta
W_\mathcal{H}^\delta\oplus [\bar\psi^*(L_\mathcal{H}),L^{(3)}].
\end{equation}

From the Schur theorem we can write $\phi$ as a sum over the different representations of $\mathcal{H}$,
\begin{equation}
\phi=\phi^0\oplus\left(\oplus_{\delta=1}^n \phi^\delta\right),
\end{equation}
with
\begin{eqnarray}
\phi^0 : L_{\mathcal{N}/\mathcal{H}}\longrightarrow L_\mathcal{Y}^{(0)},\\
\phi^\delta : L_\mathcal{L}^\delta\longrightarrow L_\mathcal{Y}^{(2)\,\delta}.
\end{eqnarray}

Let us now define a family of basic interwining operators:
\begin{equation}
\iota^\delta_{i,k}: L_{\mathcal{L}i}^\delta\longrightarrow L_{\mathcal{Y}k}^{(2)\,\delta}.
\end{equation}
We can then write
\begin{equation}
\phi^\delta = \sum_{i=1}^{m_i}\sum_{k=1}^{m_k'} \iota^\delta_{i,k} \phi^{\delta\,i,k}.
\end{equation}
Since over $L_{\mathcal{Y}k}^{(1)\,\delta}$, for $k=1,\cdots,m_\delta'$, it acts the same
representation $W^\delta$ of $\mathcal{H}$,
\begin{equation}
\phi^{\delta i} = \sum_{k=1}^{m_k'} \iota^\delta_{i,k} \phi^{\delta\,i,k},
\end{equation}
will be transformed by $\mathcal{H}$ as a $m_k'$-multiplet within the representation $W^\delta$.
For each $\delta$, there will be $m_\delta(m_\delta+1)/2$ multiplets,
and so there will be a total of $N_{mult}$ multiplets of scalar fields, with $N_{mult}$ given by
\begin{equation}
N_{mult} = \frac12\sum_{\delta=1}^n m_\delta(m_\delta+1).
\end{equation}
There will be also a singlet if $L_{\mathcal{N}/\mathcal{H}}$ is non-trivial.

\subsubsection{Case when $\mathcal{G}/\mathcal{H}$ is Symmetric}

When $\mathcal{G}/\mathcal{H}$ is symmetric, $L_\mathcal{H}$ can be decomposed into
\begin{equation}\label{iops}
L_\mathcal{H}=L_\mathcal{H}^{0}+L_\mathcal{H}^{1}+L_\mathcal{H}^{2},
\end{equation}
with $L_\mathcal{H}^{\delta}$, $\delta=0,1,2$, simple subalgebras. We shall then have
three distinct representations of $\mathcal{H}$, $W^0$, $W^1$ and $W^2$. Those will induce 
the following decomposition of $L_\mathcal{P}$:
\begin{equation}
L_\mathcal{P}=L_{\mathcal{N}/\mathcal{H}}\oplus L_\mathcal{L}^{1}\oplus L_\mathcal{L}^{2}.
\end{equation}
Without losing generality, we can take $m_\delta=1$, $\delta=1,2$, having then for the
multiplets
\begin{equation}
\phi^{\delta }= \sum_{k=1}^{m_\delta'}\iota^\delta_k\phi^{\delta,k},
\end{equation}
with $\iota^\delta_k:L_\mathcal{L}^{\delta}\rightarrow L_{\mathcal{Y}k}^{(2)\,\delta}.$
We shall then get $N_{mult}=2$ multiplets of scalar fields and possibly one singlet.


\subsection{Symmetry Breaking Analysis}

\subsubsection{Case when $\mathcal{G}\subset\mathcal{Y}$}

Let it be defined the mapping
\begin{equation}
\begin{array}{c}
  \Lambda:\bar\mathcal{M}\times L_\mathcal{G}\longrightarrow L_\mathcal{Y}\\
  (\bar z,\lambda)\longrightarrow \{\begin{array}{c}
        \bar\psi^*_e(\lambda),\; if\;\; \lambda\in L_\mathcal{H},\\
        \phi(\bar z)(\lambda),\; if\;\; \lambda\in L_\mathcal{P}.
    \end{array}
\end{array}
\end{equation}
We can then express the potential through $\Lambda$, by taking
\begin{equation}
a_{ab}(\bar z)=\frac12[\Lambda(\bar z,\lambda_a),\Lambda(\bar z,\lambda_b)]_\mathcal{Y}-\frac12\Lambda(\bar z,[\lambda_a,\lambda_b]_\mathcal{G}).
\end{equation}

Since the potential is the sum of non-negative internal products in
$\mathcal{Y}$, it must always be non-negative and so if we take
$\Lambda$ to be an homeomorphism of Lie algebras by defining an appropriated $\phi$, we shall have 
$$V(\phi)=0,$$
with $\phi$ as a minimum of the potential. The group that leaves this minimum invariant will be the effective gauge group of the
theory. Since it is the homomorphism $\Lambda$ that defines this minimum, it shall be the group $\mathcal{R}$ that leaves invariant 
$\Lambda$. 

By taking $L_\mathcal{R}$ to be the Lie algebra of that group $\mathcal{R}$, we shall have
\begin{equation}
  L_\mathcal{R}=\mathcal{C}_{L_\mathcal{Y}}(\Lambda(L_\mathcal{G})),
\end{equation}
with $\mathcal{C}_{L_\mathcal{Y}}(\Lambda(L_\mathcal{G}))$ the centraliser of the Lie algebra of $\mathcal{G}$ in $L_\mathcal{Y}$.

Moreover we shall have $L_\mathcal{R}\subset L_\mathcal{C}$, with $\mathcal{C}$ the centraliser of $\bar\psi(\mathcal{H})$ 
in $\mathcal{Y}$. That is, there will be a spontaneous symmetry breaking of the reduced gauge group $\mathcal{C}$ to
$\mathcal{R}$.

\subsubsection{Case when $\mathcal{G}/\mathcal{H}$ is symmetric}

When $\mathcal{G}/\mathcal{H}$ is symmetric we have $[L_\mathcal{P},L_\mathcal{P}]\subset L_\mathcal{H}$ and since $L_\mathcal{G}$
is simple, $[L_\mathcal{P},L_\mathcal{P}]=L_\mathcal{H}$. We shall further
have that $[\lambda_a,\lambda_b]_{L_\mathcal{P}}=0$ and
$\phi([\lambda_a,\lambda_b]_{L_\mathcal{P}})=0$. Thus the only surviving terms in (\ref{sis}) shall be
\begin{eqnarray}
V^{(0)}=\frac14<\bar\psi^*([\lambda_a,\lambda_b]_{L_\mathcal{H}}),\bar\psi^*([\lambda^a,\lambda^b]_{L_\mathcal{H}})>_{L_\mathcal{Y}},\\
V^{(2)}(\phi)=-\frac12<[\phi(\lambda_a),\phi(\lambda_b)]_{L_\mathcal{Y}},\bar\psi^*([\lambda^a,\lambda^b]_{L_\mathcal{H}})>_{L_\mathcal{Y}},\\
V^{(4)}(\phi)=\frac14<[\phi(\lambda_a),\phi(\lambda_b)]_{L_\mathcal{Y}},[\phi(\lambda^a),\phi(\lambda^b)]_{L_\mathcal{Y}}>_{L_\mathcal{Y}}.
\end{eqnarray}
Moreover we will suppose that the basis $\{\lambda_a\}$ of $L_\mathcal{P}$ will be adapted to the decomposition of $L_\mathcal{P}$.
We can then proceed to the analysis of each of these terms.

\paragraph{$V^{(0)}$.} Since $\bar\phi^*:L_\mathcal{H}\rightarrow L_\mathcal{Y}$ is an homomorphism of Lie algebras and the
internal product of $L_\mathcal{Y}$ is, by force, $Ad(L_\mathcal{Y})$-invariant, we can define an $Ad(L_\mathcal{H})$-invariant
internal product on $L_\mathcal{H}$ by simply taking
\begin{equation}
<\lambda_\mathcal{H},\lambda_\mathcal{H}'>_{L_\mathcal{H}}=<\bar\psi^*(\lambda_\mathcal{H}),\bar\psi^*(\lambda_\mathcal{H}')>_{L_\mathcal{Y}}.
\end{equation}
We shall then have for $V^{(0)}$,
\begin{equation}
V^{(0)}=\frac14<[\lambda_a,\lambda_b]_{L_\mathcal{H}},[\lambda^a,\lambda^b]_{L_\mathcal{H}}>_{L_\mathcal{H}}.
\end{equation}
Being a basis of $L_\mathcal{P}$, $\{\lambda_a\}$ shall obey $[\lambda_a,\lambda_b]=\lambda_{ab}$, for
some $\lambda_{ab}\in L_\mathcal{H}$. Then
\begin{equation}
V^{(0)}=\frac14<\lambda_{ab},\lambda^{ab}>_{L_\mathcal{H}}.
\end{equation}
Since the norm on $L_\mathcal{H}$ is always positive, we shall have
\begin{equation}
V^{(0)}>0.
\end{equation}

Let us now proceed to the analysis of $\lambda_{ab}$ considering the symmetric
decomposition (\ref{iops}). Since $[\lambda_a,\lambda_b]$ is non null only if
$\lambda_a$ and $\lambda_b$ are in the same representation and belong to the
same space $L_\mathcal{P}^{(1)}$ or $L_\mathcal{P}^{(2)}$, we shall have that
$\lambda_{ab}\in L_\mathcal{P}^{(1)}$ or $\lambda_{ab}\in
L_\mathcal{P}^{(2)}$, respectively. Since $\{\lambda_a\}$ is adapted, we shall have
\begin{equation}
V^{(0)}=\frac14 \sum_{a,b\in
  L_\mathcal{H}^{(1)}}<\lambda_{ab},\lambda_{ab}>_{L_\mathcal{H}}+\frac14\sum_{a,b\in L_\mathcal{H}^{(2)}}<\lambda_{ab},\lambda_{ab}>_{L_\mathcal{H}},
\end{equation}
where we have used the fact that $L_\mathcal{H}^{(1)}$ and
$L_\mathcal{H}^{(2)}$ are orthogonal.


\paragraph{$V^{(2)}$.} Let us write $\phi$ in terms of the basic interwining operators $\iota^\delta_k$:
\begin{equation}
  \phi=\sum_{\delta=0}^{2}\sum_{k=0}^{m_k'}\phi^{\delta k}\iota^\delta_k = \sum_{\delta k}\phi^{\delta k} \iota^\delta_k
  =\sum_\Delta \phi^\Delta\iota_\Delta,
\end{equation} 
where we have taken $\Delta=(\delta,k)$ and $m_0'=1$.

The second term in the potential will then be given by
\begin{equation}
V^{(2)}=-\frac12 \sum_{\Delta\Delta'}\phi^\Delta\phi^{\Delta'}<[\iota_{\Delta}(\lambda_a),\iota_{\Delta'}(\lambda_b)]_{L_\mathcal{Y}},
\bar\psi^*(\lambda^{ab})>_{L_\mathcal{Y}},
\end{equation}
with $\lambda^{ab}=[\lambda^a,\lambda^b]\in L_\mathcal{H}$. Since we have chosen the decomposition of $L_\mathcal{Y}$ to be 
orthogonal with respect to $<\cdot,\cdot>_{L_\mathcal{Y}}$, the above internal product will be null unless 
$$[\iota_{\Delta}(\lambda_a),\iota_{\Delta'}(\lambda_b)]_{L_\mathcal{Y}}\in\bar\psi^*(L_\mathcal{H}).$$
There must then be an element $\lambda_{ab}'\in L_\mathcal{H}$ such that
\begin{equation}\label{cvb}
\bar\psi^*(\lambda_{ab}')=[\iota_{\Delta}(\lambda_a),\iota_{\Delta'}(\lambda_b)]_{L_\mathcal{Y}}.
\end{equation}
We will then have
\begin{equation}
<[\iota_{\Delta}(\lambda_a),\iota_{\Delta'}(\lambda_b)]_{L_\mathcal{Y}},
\bar\psi^*(\lambda^{ab})>_{L_\mathcal{Y}}=<\bar\psi^*(\lambda_{ab}'),\bar\psi^*(\lambda^{ab})>_{L_\mathcal{Y}}=
<\lambda_{ab}',\lambda^{ab}>_{L_\mathcal{H}}.
\end{equation}
We can now take, since $[L_\mathcal{P},L_\mathcal{P}]=L_\mathcal{H}$,
\begin{equation}
\lambda_{ab}'=[\lambda_a',\lambda_b'],
\end{equation}
with the $\lambda_a'$ connected to the $\lambda_a$ through a rotation:
\begin{equation}
\lambda_a'=R_a^c\lambda_c.
\end{equation}
So we get $\lambda_{ab}'=R_a^c R_b^d \lambda_{cd}$, having
\begin{equation}
<\lambda_{ab}',\lambda^{ab}>_{L_\mathcal{H}}=R_a^c R_b^d <\lambda_{cd},\lambda^{ab}>_{L_\mathcal{H}}
\end{equation}
This internal product will only be non-null if both $\lambda_{cd}$ and
$\lambda^{ab}$ belong to the same representation, which shall be
$L_\mathcal{H}^{(1)}$ or $L_\mathcal{H}^{(2)}$. From this we conclude that
(\ref{cvb}) is non null only when $\Delta=\Delta'$.
The sum will then be restricted to a sum over the identical representations, which implies that (since $\{\lambda_a\}$ is adapted):
\begin{equation}
<\lambda_{ab}',\lambda^{ab}>_{L_\mathcal{H}}=\delta_{\Delta\Delta'}\sum_{\delta=1}^2\sum_{a,b,c,d\in
  L_\mathcal{H}^{(\delta)}}R_a^cR_b^d<\lambda_{cd},\lambda_{ab}>_{L_\mathcal{H}},
\end{equation}
that is,
\begin{equation}
<\lambda_{ab}',\lambda^{ab}>_{L_\mathcal{H}}=c(R)\delta_{\Delta\Delta'},
\end{equation}
with 
\begin{equation}\label{ocar}
  c(R)=\sum_{\delta=1}^2\sum_{a,b,c,d\in
  L_\mathcal{H}^{(\delta)}}
  R_a^cR_b^d<\lambda_{cd},\lambda_{ab}>_{L_\mathcal{H}}.
\end{equation}

The second term of the potential has then the form
\begin{equation}
V^{(2)}(\phi)=-\frac12 c\parallel\phi\parallel^2,
\end{equation}
with
\begin{equation}
\parallel\phi\parallel^2=\sum_{\Delta}(\phi^{\Delta})^2.
\end{equation}

\paragraph{$V^{(4)}$.} As before, let us write
\begin{equation}
  \phi=\sum_{\delta=0}^{2}\sum_{k=0}^{m_k'}\phi^{\delta k}\iota^\delta_k = \sum_{\delta k}\phi^{\delta k} \iota^\delta_k
  =\sum_\Delta \phi^\Delta\iota_\Delta.
\end{equation} 
We get for the fourth term in the scalar potential
\begin{equation}
V^{(4)}(\phi)=\frac14 \sum_{\Delta\Delta'\Delta''\Delta'''}\phi^{\Delta}\phi^{\Delta'}\phi^{\Delta''}\phi^{\Delta'''}
<[\iota_\Delta(\lambda_a),\iota_{\Delta'}(\lambda_b)]_{L_\mathcal{Y}},
[\iota_{\Delta''}(\lambda^a),\iota_{\Delta'''}(\lambda^b)]_{L_\mathcal{Y}}>_{L_\mathcal{Y}}.
\end{equation}

As before, the commutators in the internal product are non null only when they
act over the same representation:
\begin{equation}
V^{(4)}(\phi)=\frac14 \sum_{\Delta\Delta'}(\phi^{\Delta})^2(\phi^{\Delta'})^2
<[\iota_\Delta(\lambda_a),\iota_{\Delta}(\lambda_b)]_{L_\mathcal{Y}},
[\iota_{\Delta'}(\lambda^a),\iota_{\Delta'}(\lambda^b)]_{L_\mathcal{Y}}>_{L_\mathcal{Y}}.
\end{equation}
There will be then two elements, $\lambda_{ab},\lambda_{ab}'\in L_\mathcal{H}$
such that
\begin{equation}
\bar\psi^*(\lambda_{ab})=[\iota_\Delta(\lambda_a),\iota_{\Delta}(\lambda_b)]_{L_\mathcal{Y}},
\end{equation}
and
\begin{equation}
\bar\psi^*(\lambda_{ab}')=[\iota_{\Delta'}(\lambda_a),\iota_{\Delta'}(\lambda_b)]_{L_\mathcal{Y}}.
\end{equation}
We shall then have, with identical definitions as in the previous paragraph,
\begin{equation}
V^{(4)}(\phi)=\frac14 \sum_{\Delta\Delta'}(\phi^{\Delta})^2(\phi^{\Delta'})^2
R_a'^c R_b'^d<\lambda_{cd},\lambda^{ab}>_{L_\mathcal{H}}.
\end{equation}
By the same line of thought, we get
\begin{equation}
V^{(4)}(\phi)=\frac14 c(R') \parallel\phi\parallel^4,
\end{equation}
with $c$ given by (\ref{ocar}) and
\begin{equation}
\parallel\phi\parallel^4 =\sum_{\Delta} (\phi^\Delta)^4.
\end{equation}

\paragraph{V.}
The potential is then given by
\begin{equation}
V(\phi)=\frac14 c(R')\parallel\phi\parallel^4-\frac12
c(R)\parallel\phi\parallel^2+\frac14 c(\delta).
\end{equation}
By taking $\upsilon=c(R)/c(R')$ and
$$\frac{c(\delta)}{c(R')}=\upsilon+\frac{4\beta}{c(R')},$$
we can write the scalar potential in its final form,
\begin{equation}
V(\phi)=\frac14 c(R')\left(\parallel\phi\parallel^2-\upsilon\right)^2+\beta.
\end{equation}

By applying the methods of the first sections of this chapter to this potential,
one imediatly finds out that this potential conduces to a spontaneous symmetry
breaking of the gauge group $\mathcal{C}(\mathcal{Y})$.

Some examples for the explicit determination of the scalar potential can be found in \cite{kapetanakis1} and in \cite{mourao}.


\section{Spontaneous Compactification}

In the third chapter was presented a dimensional reduction
scheme of a multidimensional universe that naturally follows from 
the fibre bundle structure that the last possesses. One of the
questions that may be asked is whether such structure should be
given \textit{a priori}, or rather should result from some dynamical
mechanism.

As presented in the literature \cite{cremmer,luciani,mourao},
such dynamical mechanism can be achieved only thought the introduction
of matter fields in an Einstein multidimensional universe since Einstein 
equations in the absence of matter fields terms can not lead
to a manifold whose structure would be a fibre bundle with a non-Abelian
symmetry group. 

The standard procedure is to add Yang-Mills fields in the Hilbert-Einstein
action \cite{cremmer,luciani}. This represents an apparently withdraw
in the Kaluza-Klein unification idea but with the advent of supergravity
theories in multidimensional universes \cite{cremmer2,cremmer3},
such procedure becomes inherent to the need that the action be supersymmetric.

\subsection{Compactification of an Homogeneous Space}

Let us consider a m-dimensional Riemannian manifold $(\mathcal{M},\gamma)$ over which
is defined a principal fibre bundle $\mathcal{F}(\mathcal{M},\mathcal{Y},p,\Psi)$.
A local infinitesimal connection form defined on $\mathcal{F}(\mathcal{M},\mathcal{Y})$ will
be considered, as usually, as an Yang-Mills field $A$ over $\mathcal{M}$ whose gauge symmetry
group will be $\mathcal{Y}$.
The local curvature of the infinitesimal connection form will be, aside a factor of 2, the
Yang-Mills field strength, $F$.

The action will be given as the sum of two terms: the Hilbert-Einstein
action with a cosmological term, $S_{HE}$, and the Yang-Mills action, $S_{YM}$,
\begin{equation}
\label{accaocomp}
S=S_{HE}+S_{YM}.
\end{equation}

The two will be given by
\begin{eqnarray}
S_{EH}=\int_\mathcal{M}d^mz\sqrt{-\gamma}\frac1{2\kappa^2}\left(R-2\Lambda_m\right),\\
S_{YM}=\frac1{4g^2}<F,F>_\mathcal{M},
\end{eqnarray}
where $R$ is the scalar of curvature of $\mathcal{M}$, $\Lambda_m$ the
cosmological constant in $m$-dimensions and
\begin{equation}
F_{\hat\alpha\hat\beta}^{(\hat a)}=\partial_{\hat\alpha}A_{\hat\beta}^{(\hat a)}-
\partial_{\hat\beta}A^{(\hat a)}_{\hat\beta}+C^{(\hat a)}_{(\hat b)(\hat c)}A^{(\hat b)}_{\hat\alpha}A^{(\hat c)}_{\hat\beta}
\end{equation}
is the Yang-Mills field strength of the Yang-Mills field $A^{(\hat a)}_{\hat\alpha}$, and where $C^{\hat a}_{\hat b\hat c}$ are
the structure constants of the $k$-dimensional gauge symmetry group $\mathcal{Y}$\footnote{In this section, in order 
to distinguish between indices of the gauge group $\mathcal{Y}$ and those of the internal space we write the firsts
between brackets.}.

From (\ref{accaocomp}) the following equations can be deduced:
\begin{eqnarray}
R_{\hat\alpha\hat\beta}-\frac12\gamma_{\hat\alpha\hat\beta}R+\gamma_{\hat\alpha\hat\beta}\Lambda_m=\kappa^2 
T_{\hat\alpha\hat\beta},\label{ad}\\
\nabla_{\hat\alpha}F^{\hat\alpha\hat\beta (\hat a)}+C^{(\hat a)}_{(\hat b)(\hat c)}A^{(\hat b)}_{\hat\alpha}F^{\hat\alpha\hat\beta (\hat c)}=0,
\label{da}
\end{eqnarray}
where
\begin{equation}
T_{\hat\alpha\hat\beta}=-\frac1{g^2}F_{\hat\alpha\hat\gamma}^{(\hat a)}F_{\hat\beta\ (\hat a)}^{\hat\gamma}+\gamma_{\hat\alpha\hat\beta}
F_{\hat\gamma\hat\delta}^{(\hat a)} F_{(\hat a)}^{\hat\gamma\hat\delta},
\end{equation}
is the Yang-Mills stress-energy tensor.

We seek a solution of the previous equations in which $\mathcal{M}$ adopts a fibre bundle structure $\mathcal{M}(V_n,
\mathcal{G}/\mathcal{H},\mathcal{G},\pi,\Phi)$, with $V_n$ as base space and a $d$-dimensional homogeneous space $\mathcal{G}/\mathcal{H}$ that is $\mathcal{G}$-symmetric as the internal space, such that $d=m-n\leq k$, the structure group $\mathcal{G}$ being a subgroup of the gauge field group $\mathcal{Y}$:
$$\mathcal{G}\subset\mathcal{Y}.$$
We suppose that the Lie algebra of the gauge group $\mathcal{Y}$ is reductive,
\begin{equation}
L_{\mathcal{Y}}=L_\mathcal{G}\oplus L_\mathcal{I},
\end{equation}
being $\mathcal{I}$ the orthogonal complement of $\mathcal{G}$ on $\mathcal{Y}$.

Let us choose an adapted orthonormal base of $L_\mathcal{Y}$, $\{\tilde\lambda_{(\hat a)}\} = \{\tilde\lambda_{(\tilde a)},
\tilde\lambda_{(\bar a)}\}$, such that $\{\tilde\lambda_{(\tilde a)}\}$ is a base of the Lie algebra of
$\mathcal{G}$ and $\{\tilde\lambda_{(\bar a)}\}$ a base of $L_{\mathcal{I}}$, and let us write the Yang-Mills field
in that base.

In order to solve the equations (\ref{ad},\ref{da}) we give the following ansatz:
\begin{enumerate}
\item The metric on $\mathcal{M}$ is reductive
  \begin{equation}
    \gamma=g\oplus\xi,
  \end{equation}
  where $g$ is the metric on the base space $V_n$ such that the base space presents a null curvature $R_{\alpha\beta}$ 
  and $\xi$ is an $\mathcal{G}$-invariant metric of
  the internal space $\mathcal{G}/\mathcal{H}$;
\item The gauge potential is orthogonal to $\mathcal{I}$,
  \begin{equation}
    A_{\hat\alpha}^{(\hat a)}\tilde\lambda_{\hat a}=A_{\hat\alpha}^{(\tilde a)}\tilde\lambda_{(\tilde a)}\in
    L_\mathcal{G},   
  \end{equation}
  that is, $A_{\hat\alpha}^{(\bar a)}=0$.
\item The gauge potential $A^{(\hat a)}_{\hat\alpha}$ vanishes on $V_n$,
  \begin{equation}
    A^{(\hat a)}_{\alpha}=0,\;\;\; (\alpha=0,\cdots,n-1);
    \end{equation}
\item The gauge potential does not depend on the coordinate system over $V_n$\footnote{here we use latin indices to indicate
    internal space components; $\mathcal{Y}$ gauge group components are denoted by latin indices with an hat.},
  \begin{equation}
    A^{(\hat a)}_{\hat\alpha}=A^{(\hat a)}_{\hat\alpha}(x^b),\;\;\; (b=n,\cdots,m-1);
  \end{equation}
\item The gauge potential is $\mathcal{G}$-symmetric on $\mathcal{G}/\mathcal{H}$.
\end{enumerate}

With such ansatz the motion equations become
\begin{eqnarray}
  -\frac12 g_{\alpha\beta}R+g_{\alpha\beta}\Lambda_m=\kappa^2 T_{\alpha\beta},\label{cd}\\
  R_{ab}-\frac12 \xi_{ab} R+\xi_{ab}\Lambda_m = \kappa^2 T_{ab},\label{dfs}\\
  \nabla_{a}F^{ab (\hat a)}+C^{(\hat a)}_{(\hat b)(\hat c)}A^{(\hat b)}_{d}F^{ab (\hat c)}=0,\label{dfsd}
\end{eqnarray}
where 
\begin{eqnarray}
T_{\alpha\beta}=\frac1{4g^2}g_{\alpha\beta}F_{ab}^{(\hat a)}F^{ab}_{(\hat a)},\\
T_{ab}=\frac1{g^2}F_{ac}^{(\hat a)}F^{c}_{b(\hat a)}-\frac1{4g^2}\xi_{ab}F_{cd}^{(\hat a)}F^{cd}_{(\hat a)},
\end{eqnarray}
and $R=R_a^a$. From (\ref{cd}) we get
$$R=-\frac{\kappa^2}{g^2}F_{ab}^{(\hat a)}F^{ab}_{(\hat a)}-2\kappa^2\Lambda_m,$$
and by inserting this result in (\ref{dfs}) we have
\begin{eqnarray}
R_{ab}=-\frac{\kappa^2}{g^2}F_{ac}^{(\hat a)}F_{b (\hat a)}^c,\label{fds}\\
\Lambda_m=-\frac1{4\kappa^2}R.\label{fsd}
\end{eqnarray}
Finally we can rewrite (\ref{fds}) as
\begin{equation}\label{cons}
R_{ab}=-\frac{\kappa^2}{g^2}F_{ac}^{(\tilde a)}F_{b(\tilde a)}^c,
\end{equation}
since the components of $F$ lying on $\mathcal{I}$ are all null.

The fifth condition of our ansatz, in which the Yang-Mills is $\mathcal{G}$-symmetric, wasn't yet used. To do it let us consider the
following Lie algebra reduction
\begin{equation}
  L_\mathcal{G}=L_\mathcal{H}\oplus L_\mathcal{P}
\end{equation}
an let it be defined an adapted orthonormal base $\{\ddot\lambda_{\tilde a}\}=\{\ddot\lambda_{\dot a},\ddot\lambda_a\}$ on $L_\mathcal{G}$,
with $\{\ddot\lambda_{\dot a}\}$ and $\{\ddot\lambda_a\}$ orthonormated bases of $L_\mathcal{H}$ and $L_\mathcal{P}$, respectively.
The elements of this base will generate a set of (left or right) fundamental vector fields over $\mathcal{G}$:
\begin{eqnarray}
\tilde\theta_{g\;\tilde a}^R=\frac{d}{dt}(e^{t\ddot\lambda_{\tilde a}}g)|_{t=0},\\
\tilde\theta_{g\;\tilde a}^L=\frac{d}{dt}(ge^{t\ddot\lambda_{\tilde a}})|_{t=0}.
\end{eqnarray}
These fields can be projected on $\mathcal{G}/\mathcal{H}$, conducing to fundamental fields on this space,
\begin{eqnarray}
\theta_a^R=p_{*\;\tilde a}^a\tilde\theta^{\tilde a\;R},\\
\theta_a^L=p_{*\;\tilde a}^a\tilde\theta^{\tilde a\;L}.
\end{eqnarray}
As fundamental fields they will be $\mathcal{G}$-symmetric vectors on $\mathcal{G}/\mathcal{H}$ and will constitute a base in the
tangent vector space at any point of that space, and so every vector field on
$\mathcal{G}/\mathcal{H}$ can be written as a linear
combination of these fundamental fields. In special a $\mathcal{G}$-symmetric vector field will be written as a linear combination
on these fields with coefficients that are $Ad(\mathcal{G})$-invariants, i.e., for a $\mathcal{G}$-symmetric vector field $v$ we
have
\begin{eqnarray}
  v=v^a\theta_a,\\
  Ad(g)^a_b v^b=v^a,
\end{eqnarray}
for every $g\in\mathcal{G}$ and we have neglected the left-right nature of the field. We can then write, for the Yang-Mills field,
\begin{equation}
A=A^{(\tilde a)}\ddot\lambda_{(\tilde a)}=A^{(\tilde a)a}\ddot\lambda_{(\tilde a)}\theta_a,
\end{equation}
and since it is $\mathcal{G}$-symmetric,
\begin{equation}
Ad(g)^a_b A^{(\tilde a)b}=A^{(\tilde a)a}.
\end{equation}

We now proceed to the construction of an $\mathcal{G}$-invariant metric $\xi$ on $\mathcal{G}/\mathcal{H}$ \cite{coquereaux1}. 
By taking its inverse
we can write it in the base $\{\theta_a\otimes\theta_b\}$:
\begin{equation}
\xi^{-1}=\xi^{ab}\theta_a\otimes\theta_b.
\end{equation}
The condition that the metric be $\mathcal{G}$-symmetric is then
\begin{equation}
Ad(g)_c^a Ad(g)_d^b \xi^{cd}=\xi^{ab}.
\end{equation}

We will consider that the group $\mathcal{G}$ is simple. Then once an $\mathcal{G}$-invariant metric is given
 all the other bi-invariant metrics will be proportional to that, i.e., if $\xi_0$ is such a metric we can write 
all the others as $\xi=\rho^2\xi_0$ of real $\rho$. A special choice is to make $\xi_0^{ab}=\delta^{ab}$, that is,
\begin{equation}
  \xi_0^{-1}=\sum_{a}\theta_a\otimes\theta_a.
\end{equation}
In a local coordinate system we should then have
\begin{equation}
  \xi_0^{-1}=\sum_{a}\theta_a^b\theta_a^c\partial_b\otimes\partial_c,
\end{equation}
where $\theta_a=\theta_a^b\partial_b$.

We can now add the following conditions to our ansatz:
\begin{enumerate}
\item[6.] The gauge field is orthogonal to $\mathcal{H}$,
\begin{equation}
  A^{(\tilde a)}_b\ddot\lambda_{(\tilde a)}=A^{(a)}_b\ddot\lambda_{(a)}\in L_\mathcal{P},
\end{equation}
that is, $A^{(\dot a)}_b=0$.
\end{enumerate}
\begin{enumerate}
\item[7.] We finally take
\begin{equation}
  A^{(a)}=\mu\theta_{(a)},
\end{equation}
i.e., $A=\mu\sum_a\ddot\lambda_{a}\otimes\theta_a$.
\end{enumerate}
Then we can write every $\mathcal{G}$-invariant metric of $\mathcal{G}/\mathcal{H}$ as
\begin{equation}\label{abd}
  \xi^{-1}=\frac{\rho^2}{\mu^2}\sum_{(a)} A^{(a)}\otimes A^{(a)},
\end{equation}
for $\rho,\mu\in\mathcal{R}$.

Writing $A$ in a local coordinate system,
\begin{equation}
A=\mu\sum_{(a)}\theta_{(a)}^b\ddot\lambda_{(a)}\otimes\partial_b,
\end{equation}
we can rewrite (\ref{abd}) in terms of components in such local coordinate system as
\begin{equation}\label{boi}
\xi^{bc}=\frac{\rho^2}{\mu^2}\sum_{(a)} A^{b(a)}A^{c(a)}.
\end{equation}

We have written our solution in terms of three constants: $\rho$, $\mu$ and $\Lambda_m$. We must
now introduce this results in the equations obtained in order to find specific values
for the three. If we achieve to do so, spontaneous compactification has taken place.

We start by calculating the Ricci tensor of (\ref{cons}) by using its proper definition.
Since $A^{(a)}$ are, for $(a)=1,\cdots,d$, covariant vectors on $\mathcal{G}/\mathcal{H}$, the Riemann tensor
is defined by
\begin{equation}
[\nabla_b,\nabla_c]A_d^{(a)}=-R_{bcd}^eA_e^{(a)}-S_{bc}^e\nabla_e A_d^{(a)}.
\end{equation}
Being the geometry riemannian, the torsion tensor will be null, $S_{bc}^e=0$, and so, 
\begin{equation}
\nabla_b\nabla_c A_d^{(a)}-\nabla_c\nabla_b A_d^{(a)}=-R_{bcd}^eA_e^{(a)}.
\end{equation}
By summing over $b=d$, we get
\begin{equation}
\nabla_c\nabla^b A_b^{(a)}-\nabla^b\nabla_c A_b^{(a)}=R^e_c A_e^{(a)}.
\end{equation}

We have then to determine the covariant derivative of the Yang-Mills vector field in the ansatz that we are using.
Since we have written $A$ in terms of the fundamental fields $\theta$, the first must obey the following commutation
rules (cf. eq. (\ref{comut})),
\begin{equation}
\frac{1}{\mu^2}[A^{(a)},A^{(b)}]=\frac{1}{\mu}C^{(a)(b)}_{(c)}A^{(c)},
\end{equation}
and so, by writing $A^{(a)}$ in a local coordinate system, we have
\begin{equation}
A^{d(a)}\partial_d A^{e(b)}-A^{d(b)}\partial_d A^{e(a)}=\mu C^{(a)(b)}_{(c)} A^{e(c)}.
\end{equation}
Since the first term is antisymmetric we can replace $\partial_d$ by its covariant partner, $\nabla_d$,
\begin{equation}
  A^{d(a)}\nabla_d A^{e(b)}-A^{d(b)}\nabla_d A^{e(a)}=\mu C^{(a)(b)}_{(c)} A^{e(c)}.
\end{equation}

We can now lower the space-like indices of $A^{e(b)}$ and $A^{e(a)}$ by using metric $\xi^{ef}$, and by considering
that its covariant derivative is null,
$$\xi^{ef}(A^{d(a)}\nabla_d A_f^{b}-A^{d(b)}\nabla_d A^{(a)}_f)=\mu C^{(a)(b)}_{(c)}A^{e(c)}.$$
Multiplying and summing by $\xi_{ek}$ we get
$$A^{d(a)}\nabla_d A_f^{(b)}-A^{d(b)}\nabla_dA_f^{(a)}=\mu C^{(a)(b)}_{(c)}A^{(c)}_f.$$


By noticing that the fundamental vector fields $\theta_a$ are Killing vector fields on $\mathcal{G}/\mathcal{H}$,  
obeying the Killing equation,
\begin{equation}
  \nabla_b \theta_{c\;a}+\nabla_c\theta_{b\;a}=0,
\end{equation}
we shall have that the Yang-Mills vector field components on $L_\mathcal{P}$ will also be Killing vectors:
\begin{equation}\label{killing}
  \nabla_b A_c^{(a)}+\nabla_c A_b^{(a)}=0,
\end{equation}
for $(a)=1,\cdots,d$, and so we can take
$$-A^{d(a)}\nabla_f A_d^{(b)}+A^{d(b)}\nabla_f A_d^{(a)}=\mu C^{(a)(b)}_{(c)}A_f^{(c)}.$$
Multiplying and summing both members by $A_{e(a)}$,
$$A^{d(b)}A_{e(a)}\nabla_f A_d^{(a)}-A_{e(a)}A^{d(a)}\nabla_f A_d^{(b)}=\mu C^{(a)(b)}_{(c)}A_f^{(c)}A_{e(a)},$$
and using (\ref{boi}) in the second term of the first member, we find
$$A^{d(b)}(\nabla_f(A_{e(a)}A^{(a)}_d)-A^{(a)}_d\nabla_f A_{e(a)})-\frac{\mu^2}{\rho^2}\xi^d_e\nabla_f A^{(b)}_d=\mu C^{(a)(b)}_{(c)}A_f^{(c)}.$$
The first term inside the brackets is, up to a factor, the covariant derivative of the metric $\xi_{ed}$, and so it is null.
By defining
\begin{equation}
\chi^{(b)(a)}=\frac{\rho^2}{\mu^2}A^{d(b)}A_{d}^{(a)},
\end{equation}
we have
$$-\chi^{(b)(a)}\nabla_f A_{e(a)}-\nabla_f A_e^{b}=\frac{\rho^2}{\mu} C^{(a)(b)}_{(c)}A^{c}_f A_{e(a)},$$
i.e.,
\begin{equation}
(\delta^{(b)}_{(a)}+\chi^{(b)}_{(a)})\nabla_f A_e^{(a)}=\frac{\rho^2}{\mu}C^{(b)(d)}_{(c)} A^{(c)}_fA_{e(d)},
\end{equation}
where we have used $\chi^{(a)(b)}=\chi^{(b)(a)}$. Since $\chi^{(a)}_{(b)}\chi^{(b)}_{(c)}=\chi^{(a)}_{(c)}$, we
shall have
\begin{equation}\label{ola}
\nabla_f A_e^{(a)}=\frac{\rho^2}{\mu}C^{(b)(d)}_{(c)} A^{(c)}_fA_{e(d)}\left(\delta^{(a)}_{(b)}-\frac12\chi^{(a)}_{(b)}\right),
\end{equation}

We now make use of the following relations,
\begin{eqnarray}
\nabla_e A_e^{(a)}=0,\\
\nabla^e\nabla_f A_e^{(a)}=-\nabla^e\nabla_e A_f^{(a)}=-\nabla^2 A_f^{(a)},
\end{eqnarray}
derived from (\ref{killing}). The Ricci tensor is given by
\begin{equation}
R_f^dA_d^{(a)}=\nabla^e\nabla_fA_e^{(a)}-\nabla_f\nabla^e A_e^{(a)},
\end{equation}
and so
\begin{equation}
R_f^dA_d^{(a)}=-\nabla^2A_f^{(a)}.
\end{equation}

By taking the divergence of (\ref{ola}) we have
$$\nabla^2 A_f^{(a)}=\frac{\rho^2}{\mu}C^{(b)(d)}_{(c)}A_e^{(c)}\nabla^eA_{f(d)}\left(\delta^{(a)}_{(b)}-\frac12\chi^{(a)}_{(b)}\right),$$
and using again (\ref{ola}),
$$\nabla^2 A_f^{(a)}=\frac{\rho^4}{\mu^2}C^{(b)(d)}_{(c)}C_{(e)(f)(g)}A_e^{(c)}A^{e(g)}A_f^{(f)}
\left(\delta^{(d)}_{(e)}-\frac12\chi^{(d)}_{(e)}\right)
\left(\delta^{(a)}_{(b)}-\frac12\chi^{(a)}_{(b)}\right),$$
i.e.,
$$\nabla^2 A_f^{(a)}=\rho^2 C^{(b)(d)}_{(c)}C_{(e)(f)(g)}\chi^{(g)}_{(c)}A_f^{(f)}
\left(\delta^{(d)}_{(e)}-\frac12\chi^{(d)}_{(e)}\right)
\left(\delta^{(a)}_{(b)}-\frac12\chi^{(a)}_{(b)}\right).$$

Performing all the calculations, we get
\begin{equation}
R_{ab}=-\frac34\rho^2\xi_{ab}+\frac{\rho^4}{2\mu^2}A_a^{(a)}A_b^{(b)}C_{(a)(c)(d)} C_{(b)}^{(c)(f)} \chi^{(d)}_{(f)},
\end{equation}
or by defining, following \cite{luciani},
\begin{equation}
K_{(a)(b)}=C_{(a)(c)(d)} C_{(b)}^{(c)(f)} \chi^{(d)}_{(f)},
\end{equation}
the Ricci tensor will be given by
\begin{equation}
R_{ab}=-\frac34\rho^2\xi_{ab}+\frac{\rho^4}{2\mu^2}A_a^{(a)}A_b^{(b)}K_{(a)(b)}.
\end{equation}

The Yang-Mills strength tensor non null components can be written as
$$F_{ab}^{(a)}=\nabla_a A_b^{(a)}-\nabla_b A_a^{(a)}+C^{(a)}_{(b)(c)}A_a^{(b)}A_b^{(c)},$$
and using (\ref{killing}),
\begin{equation}
F_{ab}^{(a)}=2\nabla_a A_b^{(a)}+C^{(a)}_{(b)(c)}A_a^{(b)}A_b^{(c)}.
\end{equation}

From (\ref{ola}), we have
\begin{equation}
F_{ab}^{(a)}=\left(1+\frac{2\rho^2}{\mu}\right)C^{(a)}_{(b)(c)}A^{(b)}_a A^{(c)}_b-\frac{\rho^2}{\mu}C^{(b)}_{(c)(d)}A^{(c)}_a
A_b^{(d)}\chi^{(a)}_{(b)},
\end{equation}
and after some straightforward calculations,
\begin{equation}
-F_{ac}^{(a)}F^{c}_{b(a)}=
\xi_{ab}\left(3-\frac{\mu}{\rho^2}\right)+\left(\frac{\mu^2}{\rho^2}-2\rho^2\right)A_a^{(a)}A_b^{(b)}K_{(a)(b)}.
\end{equation}
The Einstein equations are then
$$
\begin{array}{r}
-\frac34\rho^2\xi_{ab}+\frac{\rho^4}{2\mu^2}A_a^{(a)}A_b^{(b)}K_{(a)(b)}=
\frac{\kappa^2\mu^2}{g^2}
\xi_{ab}\left(3-\frac{\mu}{\rho^2}\right)+\frac{\kappa^2}{g^2}\left(\frac{\mu^2}{\rho^2}-2\rho^2\right)A_a^{(a)}A_b^{(b)}K_{(a)(b)},
\end{array}$$
i.e.,
\begin{equation}
\left[\frac{\kappa^2}{g^2}\left(\frac{\mu^2}{\rho^2}-2\rho^2\right)-\frac{\rho^4}{2\mu^2}\right] A_a^{(a)}A_b^{(b)}K_{(a)(b)}+
\left[\frac{\kappa^2\mu^2}{g^2}
\left(3-\frac{\mu}{\rho^2}\right)+\frac34 \rho^2\right]\xi_{ab}=0.
\end{equation}

The Yang-Mills equations, after some tedious calculations, assume the form
\begin{equation}
\left(1-\frac{\mu}{\rho^2}\right)\left(\left[\rho^2+\mu\right]K^{(a)}_{(b)}A^{b(b)}-\frac32\rho^2 A^{b(a)}\right)=0.
\end{equation} 
The cosmological constant will be given by
\begin{equation}
\Lambda_m = -\frac1{4\kappa^2}R^a_a=\frac{3\rho^2d}{16\kappa^2}-\frac{\rho^4}{8\kappa^2\mu^2}\chi^{(a)(b)}K_{(a)(b)}.
\end{equation}
Presented this three sets of equations, one must now take a convenient choice of the parameters $(\mu,\rho,\Lambda_m)$ that obey them.
In order to such solution to be possible one must have \cite{luciani}
\begin{equation}\label{ui}
K^{(a)}_{(b)}A^{(b)}_b = cA^{(a)}_b,
\end{equation}
for real $c$.
The equations will then be,
\begin{eqnarray}
\left[\frac{\kappa^2}{g^2}\left(\frac{\mu^2}{\rho^2}-2\rho^2\right)-\frac{\rho^4}{2\mu^2}\right] c\frac{\mu^2}{\rho^2}+
\frac{\kappa^2\mu^2}{g^2}
\left(3-\frac{\mu}{\rho^2}\right)+\frac34 \rho^2
=0,\\
\left(1-\frac{\mu}{\rho^2}\right)\left[\left(1+\frac{\mu}{\rho^2}\right)c-\frac32\right]=0,\\
\Lambda_m=\frac{\rho^2d}{8\kappa^2}\left(\frac32-\frac{c\rho^2}{\mu^2}\right).
\end{eqnarray}

Two solutions are then easily found,
\begin{enumerate}
\item When $\mu=\rho^2$, we will have a solution with
\begin{eqnarray}
\mu = \frac{g^2}{2\kappa^2}\frac{c-\frac32}{2-c},\\
\Lambda_m = \frac{d}{8\kappa^2}\left(g^2\frac34 \frac{c-\frac32}{2-c}-c\right).
\end{eqnarray}
The dependence of $\Lambda_m$ with $\rho^2$ is given by
\begin{equation}
\Lambda_m(\rho^2)=\frac{3d}{16\kappa^2}\rho^2-\frac{dc(\rho^2)}{8\kappa^2},
\end{equation}
with
\begin{equation}
c(\rho^2)=\frac{2\rho^2+\frac{3g^2}{4\kappa^2}}{\rho^2+\frac{g^2}{\kappa^2}}.
\end{equation}

\item When $1+\frac{\mu}{\rho^2}=\frac3{2c}$, the Yang-Mills equation is satisfied and we shall have from the Einstein equation,
\begin{equation}
\mu = \frac{g^2}{\kappa^2} \frac{1-\frac12 c}{\left(\frac3{2c}-1\right)\left(c+\frac3{4c}-1\right)}.
\end{equation}
The cosmological constant will be given by
\begin{equation}
\Lambda_m=\frac{dc}{8\kappa^2}\left(\frac{g^2}{\kappa^2}\frac{1-\frac12 c}{
\left(1-\frac13 c\right)\left(\frac3{2c}-1\right)\left(c+\frac3{4c}-1\right)}-\frac1{c}\right).
\end{equation}
\end{enumerate}

It remains to verify if such solutions are consistent with the fact that the internal space is an homogeneous space, i.e.,
we must be able to solve (\ref{ui}) for a general homogeneous space. We can
simplify $K^{(a)}_{(b)}$ to
\begin{equation}
K^{(a)}_{(b)}=\delta^{(a)}_{(b)}-C^{(a)(c)(d)}C_{(b)(c)(d)},
\end{equation} 
and so the condition (\ref{ui}) can be restated as
\begin{equation}
C^{(a)(c)(d)}C_{(b)(c)(d)}=(1-c)\delta^{(a)}_{(b)}.
\end{equation}

There are two special cases for which this is verified: the one in which the
internal space is a Lie group, and that in which is a symmetric homogeneous
space.

In the first case, we shall have $\mathcal{H}=e$, having then $c=1$. We get,
\begin{equation}
  \mu=\frac{4g^2}{3\kappa^2},
\end{equation}
and
\begin{equation}
\Lambda_m=\frac1{8\kappa^2}\left(\frac{2g^2}{\kappa^2}-1\right).
\end{equation}

In the second case, we have (cf. \cite{luciani})
\begin{equation}
  C^{(a)(c)(d)}C_{(b)(c)(d)}=\frac12\delta^{(a)}_{(b)},
\end{equation}
and so $c=\frac12$,
\begin{equation}
  \mu=\frac{g^2}{3\kappa^2},
\end{equation}
and
\begin{equation}
\Lambda_m=-\frac{d}{16\kappa^2}\left(g^2+1\right). 
\end{equation}

A fibre bundle structure can then be dynamical achieved by the introduction of
Yang-Mills-type fields terms in an action of Hilbert-Einstein type defined
over a manifold without such type of structure.

\chapter{MODEL BUILDING}

\textit{The construction of realistic models within the Kaluza-Klein frame is analysed
and its principal problems discussed. In particular, the chiral and the
hierarchy problems are examined and some solutions presented. From the
construction of an effective field theory from the M-theory, the
Randall-Sundrum model can be obtained, solving the hierarchy problem.}

\section{Model Building}

\subsection{The Symmetry Breaking Scheme}

In chapter 3 we have presented the dimensional reduction process by which a
fibre bundle $\mathcal{F}(\mathcal{M},F,\mathcal{Y},p,\Psi)$ defined over a multidimensional universe
$\mathcal{M}$ that has also a fibre bundle structure,
$\mathcal{M}(V_n,M,\mathcal{G},\pi,\Phi)$, with a homogeneous space
$M=\mathcal{G}/\mathcal{H}$ as
internal space, can be reduced to a lower dimensional fibre bundle
$\mathcal{F}^r(\bar M,F,\mathcal{C},p^r,\Psi^r)$ defined over a manifold
$\bar\mathcal{M}\subset\mathcal{M}$, representative of our observed universe and that does not
possess a particular structure.

We have firstly reduced the fibre bundle of linear frames $\hat E(\mathcal{M})$
to that of adapted frames $\ddot E(\mathcal{M})$. This can then be reduced to
the fibre bundle of orthonormal frames $\tilde E(\bar\mathcal{M})$ over
$\bar\mathcal{M}$ times a principal fibre bundle
$\mathcal{W}(\bar\mathcal{M},\mathcal{N}(\mathcal{H})/\mathcal{H})$ with a
structure group $\mathcal{N}(\mathcal{H})/\mathcal{H}$. Physically this means
that an Hilbert-Einstein action defined over a multidimensional universe that
possesses a fibre bundle structure will conduce to an
Hilbert-Einstein-Yang-Mills action defined over the observed universe, being
the gauge group of the Yang-Mills field, $\mathcal{N}(\mathcal{H})/\mathcal{H}$.
This action will also contain several scalar fields that can, in special
conditions, break spontaneously this gauge group, conducing to the well known
effect of generation of masses for the gauge field.

We have secondly analysed the reduction process for a generic principal fibre
bundle $\mathcal{F}(\mathcal{M},\mathcal{Y})$ defined over a multidimensional
universe $\mathcal{M}(V_n,\mathcal{G}/\mathcal{H},\mathcal{G})$ when a
$\mathcal{H}$-invariant section $\sigma:V_n\rightarrow\mathcal{M}$ is given - and
so $\mathcal{N}/\mathcal{H}$ will be trivial. In
this case the reduced principal fibre bundle will be of the form
$\mathcal{F}^r(\bar\mathcal{M},\mathcal{C})$,
with $\bar\mathcal{M}=\sigma(V_n)$ and $\mathcal{C}$ the centraliser of the
$\mathcal{H}$-action in $\mathcal{Y}$ through a mapping $\bar\psi$.
If an infinitesimal connection form $\omega$ is defined over $\mathcal{F}$, we
shall have that over $\mathcal{F}^r$ there will also be defined an
infinitesimal connection form $\omega^r$ together with a scalar field
$\xi^r$. If one writes the action of $\omega$ over $\mathcal{F}$, this will be
reduced to an action over $\bar\mathcal{M}$ that will contain an Yang-Mills
term, being the gauge group $\mathcal{C}$, a scalar field kinetic term and a
scalar field potential $V(\phi)$. This scalar potential will lead, in
general, to a spontaneous symmetry breaking of the gauge group
$\mathcal{C}$ (cf. chapter 4). We have derived its explicit form for the case when the
internal space is a symmetric space. An extension of this analysis of symmetry breaking by dimensional reduction to the
supersymmetric case was given in \cite{manousselis1,manousselis2,manousselis3}.

Thirdly, we have reduced $\mathcal{G}$-invariant matter fields over
$\mathcal{M}(V_n,\mathcal{G}/\mathcal{H},\mathcal{G})$, and fourthly we have
performed a more general type of dimensional reduction by reducing matter
fields with no such symmetry requirement. For that purpose we have made a
Fourier decomposition of the matter fields into $\mathcal{G}$-invariant
components that could then be reduced by the methods presented before.

\textit{There will then occur, in general, two types of symmetry breaking in
non-Abelian Kaluza-Klein theories: a geometric symmetry breaking during the dimensional reduction
process and a spontaneous symmetry breaking resulting from that dimensional
reduction.}

These two types of symmetry breaking can be usefully used in the construction of
realistic models. One could consider, for instance, a large gauge group
$\mathcal{Y}\supset SU(3)\times SU(2)\times U(1)$ such that by geometric
symmetry breaking could be reduced to $\mathcal{C}=SU(3)\times SU(2)\times
U(1)$, and then to break it by the common spontaneous symmetry breaking
process to $\mathcal{R}=SU(3)\times U(1)$, being this induced by the scalar
field resulting from the previous dimensional reduction. This however can't be
done - at least for the conventional Kaluza-Klein theories - for fermion type
fields (this problem will be presented later on). Another problem is that we
are forcing all matter fields in the theory to be symmetric, and there is no
reason \textit{a priori} to do so.

If we do not retain the symmetry necessity of the matter fields, then we
should write all fields in the theory in terms of their Fourier symmetric
components. This however will inevitably conduce to a tower of symmetric
fields with increasing masses. Since such massive fields are not currently
observed and considering the resolution of the present detectors we have to
impose a very small volume for the internal space of the multidimensional
universe of the constructed model. The stability of such small space should
then be analysed. There are various types of stabilisation mechanisms but we
will only make a short remark of one induced at a quantum level.

\subsection{The Spontaneous Compactification}

A difficulty that could be risen is why should the multidimensional universe of our
theory have a fibre bundle structure? To bypass this difficulty one should
find such structure as a ground state of a more general model. One could then
consider small disturbances around this state and then to expand all fields
of the theory in its Fourier components, integrating next over the internal
space of such ground state. The result would be, as before a tower of states
with increasing masses.

With the spontaneous compactification process presented in the last chapter we
could achieve the desired fibre bundle structure with the wished internal
space for the multidimensional universe of the model. No need is then to
consider \textit{a priori} such structure since that can be obtained by such
dynamical mechanism.

\subsection{Quantum Kaluza-Klein Theories}

Through this work, only classical aspects (together with some generic quantum
aspects such as spontaneous symmetry breaking processes) were considered. There
is however a vasty literature available on the quantisation of Kaluza-Klein
theories (in special, for the abelian ones). Here we make only special
reference to the article \cite{appelquist} where a justification for the small
volume of the internal space of a Kaluza-Klein type model was presented by
extending the known Casimir effect to the gravitation field.
At a quantum level, due to the compact nature of the internal space there will
be a Casimir force that will decrease the volume of that space. In
\cite{appelquist} that analysis was only done for the abelian Kaluza-Klein
type models, but it is to suppose that a similar behaviour will be found for the
non-abelian ones.

\section{The Chiral Problem}

\subsection{The Non-Go Witten Theorem}

One of the main features of the SM and of SM type-like extensions is the existence of a left-right fermion asymmetry in the base space
$\bar V_4$, asymmetry that is derived from the V-A behaviour of the electroweak fermion currents. In this sense, at Weinberg-Salam energies 
($\sim 300$ GeV), left-handed fermions transform differently from right-handed ones under the gauge symmetry group $SU(3)\times SU(2)
\times U(1)$. This imediatly conduces to the severe constrain, due to gauge invariance, that fermions should be massless, acquiring mass
only through a spontaneously symmetry breaking of the electroweak symmetry gauge group $SU(2)\times U(1)$. This is usually 
done by introducing 
a scalar field in the theory (that is a doublet under $SU(2)$) that will present a non null vacuum expectation value (VEV), and that
its redefinition will, through the Higgs mechanism, break the $SU(2)\times U(1)$, then producing fermionic masses. Such type of 
symmetry breaking process can only produce masses that are close to the mass scale at which $SU(2)\times U(1)$ is broken 
($m_{EW}\sim 10^3$ GeV), and thus the relative light mass of fermions can by this way be explained.

Through the Kaluza-Klein scheme one can naturally identify a gauge field as a local infinitesimal connection form on a higher dimensional
universe that presents a principal fibre bundle structure (or a fibre bundle structure with an associated principal fibre bundle on
which the infinitesimal connection is defined). Gauge and internal symmetries are then considered as external symmetries 
(in a geometric sense). The construction of a realistic Kaluza-Klein theory \cite{witten} would then involve a differentiable 
fibre bundle $\mathcal{M}(V_4,\mathcal{G}/\mathcal{H},\mathcal{G},\pi,\Phi)$ whose internal space could be made an homogeneous
space - due to economical reason, cf. \cite{witten} - on which the action of an non-abelian group $\mathcal{G}$, such that
$$SU(3)\times SU(2)\times U(1)\subset\mathcal{G},$$
should be given.

Fermions should then be introduced \textit{a posteriori} as sections of an spinor bundle, $\underline{\tilde E}(\mathcal{M})$,
defined over $\mathcal{M}$. And at this point we are faced with a first problem: fermions should be massless in the reduced
universe, $\bar V_4$. The difficulty is obvious: if we start with a fermionic field $\psi$ (that is an scalar under general coordinate
transformations) in $\mathcal{M}$ that obeys the massless 
Dirac equation on $\mathcal{M}$:
\begin{equation}
        /\!\!\!\!\underline{\tilde D}\psi=\underline\gamma^{\hat\beta}\tilde{\underline\nabla}_{\hat\beta}\psi=0,
\end{equation}
and if we choose a local coordinate system at $z\in\mathcal{M}$ such that the first four coordinates are a local coordinate system
of $\bar V_4$ and the remaining a local coordinate system of the internal space at $z$, then the previous equation assumes the
form
\begin{equation}
        /\!\!\!\!\underline{\tilde D}\psi=\underline\gamma^{\beta}\tilde{\underline\nabla}_{\beta}\psi+
\underline\gamma^{b}\tilde{\underline\nabla}_{b}\psi=0,
\end{equation}
where, as in the previous chapters, the greek letters refer to the reduced space and the latin ones to the internal space.
We then see that the two terms appearing in this equation are, respectively, the Dirac operators of the fermionic field
in the reduced and the internal space, and so
\begin{equation}
  /\!\!\!\!\underline{\tilde D}^{(4)}\psi+ /\!\!\!\!\underline{\tilde D}^{(int)}\psi=0,
\end{equation}
with an obvious notation, and we imediatly see that the eigenvalue of the internal Dirac operator became the observed mass in
the reduced space.

A spectral analysis of the Dirac operator over the internal space shall then be done. Since in all Kaluza-Klein type theories
the internal space is compact, the spectrum of $/\!\!\!\!\underline{\tilde D}^{(int)}$ is discrete. Its non-null eigenvalues
will correspond to massive fermions on $\bar V_4$, and since as usually one takes the internal space volume to be of the order
of the Planck mass,
these eigenvalues will give extremely huge masses, and so such fermions, if to exist, are far from being observed by current detectors. 
The zero modes of 
$/\!\!\!\!\underline{\tilde D}^{(int)}$ will give origin to massless fermions on $\bar V_4$, those belonging to the SM.

Two pertinent questions can then be made: does the Dirac operator have zero modes for some specific internal space? And 
if it do so can the eigenvectors of such zero modes possess two different representations of the gauge group $SU(3)\times 
SU(2)\times U(1)$ - a left and a right-handed one?

An answer to the first question was first given for a special situation by Palla \cite{palla}  and a general treatment 
for positive curved internal spaces was first presented by Lichnerowicz \cite{lichnerowicz2}. 

Chapline, Manton, Slansky and Witten
\cite{chapline,chapline2,manton,witten} were the first to discuss the problem of obtaining a 
complex spectrum in Kaluza-Klein
theories and Witten \cite{witten2} gave its final answer: no such complex spectrum can be achieved in a 
multidimensional universe with a compact oriented differentiable
manifold without boundary as internal space. Several solutions were proposed \cite{witten,witten2,wetterich}
to solve the chiral problem, namely by abandoning the assumption that the internal space is compact without boundary 
\cite{witten2,wetterich}
or by modifying the riemannian geometry \cite{wetterich}. Another solution, proposed by Randjbar-Daemi, Salam and Strathdee
\cite{randjbar}, would be to consider additional gauge fields to the theory, as needed in non-Abelian Kaluza-Klein theories
to induce a spontaneous compactification, considering that the compactification involves a topologically non-trivial
configuration of such gauge fields. A similar exposition but with large extra
dimensions was given in \cite{dvali}. Another solution for the chiral problem
was provided by Hosotani in \cite{hosotani}, by observing that the Casimir
effect can induce, under certain circumstances, the spontaneous breakdown of
gauge symmetries with the desired chiral asymmetry.

\paragraph{The Atiyah-Hirzebruch-Witten Theorem}

By defining the character-value index of the Dirac operator as the number of
zero modes of that operator, we can present the two theorems deduced by
Witten in \cite{witten2}:

\textbf{Atiyah-Hirzebruch Theorem.} \textit{The character-valued index of the Dirac operator 
vanishes on any manifold with a continuous symmetry group.}

\textbf{Atiyah-Hirzebruch-Witten Theorem.} \textit{The character-valued index of the Rarita-Schwinger operator
vanish on any homogeneous space.}

These two non-go theorems represent a serious obstacle on the construction of
realistic Kaluza-Klein theories.

\subsection{The Wetterich Procedure}

The Wetterich evasion procedure \cite{wetterich} of the non-go Witten theorem consists in perform modifications of the 
conditions to which the theorem applies. 
The first procedure presented by Wetterich \cite{wetterich}, and Witten first suggested \cite{witten2}, is to abandon 
the assumption that the internal 
space is compact without boundary; the second, more radical procedure, is to modify the riemannian geometry of the total space. 
In this procedure one replaces the usual riemannian geometry constrains
\begin{eqnarray}
\gamma_{\hat\alpha\hat\beta}=e_{\hat\alpha}^{\hat b}e_{\hat\beta\hat b},\\
\nabla_{\hat\alpha}e_{\hat\beta\hat b}=0,\\
\hat S^{\hat\alpha}_{\hat\beta\hat\gamma}=0,
\end{eqnarray}
by the following generalised gravity conditions
\begin{eqnarray}
\gamma_{\hat\alpha\hat\beta}=e_{\hat\alpha}^{\hat b}e_{\hat\beta\hat b}-
k_{\hat\alpha\hat\beta},\\
\nabla_{\hat\alpha}e_{\hat\beta\hat b}=U_{\hat\alpha\hat\beta\hat b},\\
\hat S^{\hat\alpha}_{\hat\beta\hat\gamma}\neq 0.
\end{eqnarray}
These modifications must be so that in the base space $V_n$, one has a
ground state characterised by
\begin{eqnarray}
k_{\alpha\beta}=0,\\
U_{\alpha\beta b}=0,\\
S^\alpha_{\beta\gamma}=0.
\end{eqnarray}
In this way we have modified the geometry of the multidimensional universe
but one must get the usual riemannian geometry in the 
four-dimensional universe up to corrections of the order of the Planck mass.
Wetterich has shown that within such generalisation of the Riemann geometry it
is possible to construct realistic models - models that would include both
bosonic as fermionic massless particles. 



\section{The Hierarchy Problem}

\subsection{The Introduction of Extra-Dimensions}

The existence of two apparently fundamental energy scales in nature - the electroweak and the Planck scale - has been subject
to a detailed study in last past years. Explaining how can a occur at a so low level ($\frac{m_{EW}}{m_{Pl}}\approx 10^{-17}$)
a symmetry breaking, triggered by an elementary scalar, and be perturbatively
stable, has been one of the reasons to the construction of theories beyond the
SM. Such hierarchy stability could only, up to recent years, be explained by the introduction of techni-color or low-energy supersymmetry,
when a new hypothesis concerning the structure of the space-time itself was presented \cite{arkani} and that could solve the hierarchy
problem in a simpler way - by the introduction of a Kaluza-Klein type universe with a
small variant: only the gravity can propagate into the internal space.

By the introduction of extra, compact dimensions, the weak strength of gravity at the electroweak scale could be explained. If, for
simplicity we take such internal space to be a $d$-dimensional sphere of radius $a$, two test masses $m_1$ and $m_2$ would feel a
type like Gauss potential that would be given by
\begin{equation}
V(r)= \frac{1}{m_{Pl(4+d)}^{d+2}}\frac{m_1m_2}{r^{d+1}},
\end{equation}
where $m_{Pl(4+d)}$ dictates the Planck scale of this multidimensional universe. 

In order to explain why experimentally one obtains a gravitational potential proportional to $r^{-1}$, one must take $r>>R$,
\begin{equation}
V(r)= \frac{1}{m_{Pl(4+d)}^{d+2}R^d}\frac{m_1m_2}{r},
\end{equation}
where the Planck scale of our universe should be $m_{Pl}^2=m_{Pl(4+d)}^{d+2}R^d$. Since $R$ is, at this stage arbitrary,
 we can take $m_{Pl(4+d)}$
to be of the order of the electroweak scale, $m_{EW}$. In fact, we can take $m_{Pl(4+d)}=m_{EW}$, thus resolving the hierarchy
problem by taking the electroweak scale to be the only fundamental energy scale \cite{arkani,arkani2,antoniadis}.
We must then have for $R$ \cite{arkani},
\begin{equation}
R\approx 10^{30/n-17}\textrm{cm}\times\left(\frac{\textrm{1 TeV}}{m_{EW}}\right)^{1+2/n}.
\end{equation}
The case $n=1$ is excluded by the experience. All the other cases, $n\geq 2$ are possible. The case $n=2$, with 
$$R\approx 100 \mu\textrm{m}-1\textrm{mm},$$
is the more interesting one because in a very near future it can be subjected to experience, since deviations from the Newtonian gravity
will occur at distances of only one millimetre.

The apparent weakness of gravity is then explained by the propagation of the graviton in the bulk when the SM particles are
localised in our 4-dimensional world. In \cite{arkani} a model was proposed for the trapping of the SM fields by considering a
6-dimensional theory with a weak scale vortex, being the SM fields localised within its throat. Earlier models for trapping
SM particles on a domain wall in Kaluza-Klein theories were presented in \cite{rubakov1,rubakov2}. 

Another problem of this framework
is to explain the large size of the extra dimensions \cite{antoniadis,arkani2,arkani3}. A study of the effects of
possible flavour-violating in this model were already presented \cite{arkani4,berezhiani} and a study of the Kaluza-Klein
states originated from such large sized extra dimensions can be found in \cite{han}.

\subsection{The Membrane Solution: Randall-Sundrum Scheme}

The factorisation of the universe in a 4-dimensional manifold and a very large compact internal space has some unanswered questions,
as, for instance, the reason of such large size of the extra-dimensions. For this reason a simpler model was presented \cite{randall2},
in which only one small sized extra-dimension was considered. An extension of this model to a non-compact extra-dimension was
given in \cite{randall3}.

These type of models - called after the work of \cite{randall2} as
Randall-Sundrum models - will be analysed in the next chapter.

\section[Kaluza-Klein, Supergravity... and Randall-Sundrum]{Kaluza-Klein, Supergravity, Superstrings, M-theory and Randall-Sundrum}

With the advent of supersymmetry \cite{ferrara1,haag} and of
supergravity \cite{ferrara2,freedman1,freedman2,das,macdowell}, the unifying purpose of non-Abelian
Kaluza-Klein theories has assumed a secondary role, mainly due to all its
problems (cf. previous sections). Only after the introduction of supergravity
in multidimensional universes \cite{cremmer2,cremmer3,freund} the Kaluza-Klein
dimensional reduction techniques have been used directly in the construction
of grand unifying theories (GUT's). Spontaneous compactification assumes then
a decisive role in this type theories \cite{freund,duff}, and the
Kaluza-Klein type-like dimensional reduction process becomes the standard procedure in
the construction of effective field theories from multidimensional
supergravity theories.

Parallel to development of supergravity theories, the superstrings theories
have made enormous advances, arriving to a point in which multidimensional
supergravity could be derived from it, being in this way more general than
supergravity itself.

There are five types of consistent supersymmetric string theories, all defined in a 10-dimensional
space: type I ($O(32)$), heterotic ($O(32)$), heterotic ($E_8\times E_8$),
type IIA and type IIB. The type I superstring theory contains the open strings
and all the other types contain only close strings. Being different, these five
types of superstring theories become equivalent if we reduce the dimension of
the multidimensional universe. For instance, by compactifying one dimension on
a circle we can connect the two heterotic theories as well as the two type-II
theories.  The strong coupling limit of the six-dimensional toroidally
compactified heterotic string is given by the type IIA theory compactified on
K3 and vice-versa, and the strong coupling limit of the $O(32)$ heterotic
string theory is the type I theory and vice-versa, thus demonstrating the
equivalence between all the superstrings theories.

In recent years
\cite{witten4,witten5,witten6,horava1,horava2,witten3},
a more general 11-dimensional theory as been developed: the M-theory. It has
been shown \cite{horava1,horava2} that both heterotic $E_8\times E_8$
and type IIA superstring theories can be derived from this theory by
compactifying it over the orbifold $\mathcal{R}^{10}\times S^1/\mathcal{Z}_2$
and the bundle  $\mathcal{R}^{10}\times S^1$, respectively. M-theory is the most general physical
theory constructed so far and it is under an intense investigation.

One particular theme under investigation in M-theory is the determination of
alternative effective theories \cite{lukas}. These effective theories are
constructed by performing a generalised Kaluza-Klein dimensional reduction
process from eleven down to five dimensions. Only after this reduction to a
5-dimensional theory a final
dimensional reduction from five to four dimensions is performed - this being
rather different than that previously used, i.e., Kaluza-Klein type-like. The main reason for such
intermediate state of a 5-dimensional theory is that the scale of the fifth
dimension is larger than that of the Calabi-Yau manifold \cite{lukas}. The more interesting
aspect of these effective theories is that the 5-dimensional theory is a $N=1$
supergravity theory defined over a 5-dimensional orbifold with
four-dimensional boundaries. The Randall-Sundrum model has its roots in such dimensional reduction process
of the M-theory\footnote{in fact the Randall-Sundrum solution is a simply case of the solution obtained
in \cite{lukas}.}, having the conveniently property of solving the hierarchy
problem. It is not with a lack of humour that the Randall-Sundrum model is recovered from a
Kaluza-Klein-type dimensional reduction process from an eleven-dimensional
theory, after its compactification on a manifold (orbifold) with a fibre bundle
structure.


\chapter{RANDALL-SUNDRUM THEORIES}



\textit{The Randall-Sundrum model is presented and its principal properties
analysed. Some generalizations are given and the stability of the hierarchy discussed.}

\section{The Formalism}

\subsection{Basic Definitions}
The Randall-Sundrum geometric set is the following\footnote{Here we take a generalisation of the models proposed in 
\cite{sundrum1,sundrum2,randall2}}: let us consider a $m$-dimensional manifold $(\mathcal{M},\gamma)$ 
with no special structure \textit{a priori} (such as a fibre bundle structure, for instance) but that contains a $b$ number of riemannian manifolds
 $\{(V_{n_I}^I,g_I)\}_{I=1,\cdots,b}$ of lower dimension $n_I<m$ called branes. 
A point over $\mathcal{M}$ will be denoted by $z$ and a point over the brane $V_{n_I}^I$ will be denoted by $x^I$.
The bulk coordinates describing the position occupied by a point $x^I$ of $V_{n_I}^I$ will be dynamical fields and 
will be denoted by $z^I(x^I)$. In general, one identifies such submanifolds
with boundaries of the initial manifold.

\subsubsection{Induced Metrics on the Branes}
Since we describe the motion of a brane $V_{n_I}^I$ by the coordinates $z^I$ in $\mathcal{M}$, an induced metric on the
brane can be found from the metric of the total space $\mathcal{M}$. Let us consider two points $x^I$ and $x^I+dx^I$ on 
$V_{n_I}^I$. A distance can be introduced between these points using the bulk metric $\gamma$,
\begin{equation}
ds^2_I=\gamma_{\hat\alpha\hat\beta}(z^I(x^I))dz_I^{\hat\alpha}dz_I^{\hat\beta}=\gamma_{\hat\alpha\hat\beta}(z^I(x))
\partial_\alpha z_I^{\hat\alpha}(x^I)dx_I^\alpha\partial_\beta z_I^{\hat\beta}(x^I)dx_I^\beta,
\end{equation}
and from this an induced metric on the brane can be obtained,
\begin{equation}
g_{\alpha\beta}^I=\gamma_{\hat\alpha\hat\beta}(z^I)\partial_\alpha z_I^{\hat\alpha}\partial_\beta z_I^{\hat\beta}.
\end{equation}

%

\subsection{Matter Fields}

Within the Randall-Sundrum model matter fields are located in a particular
brane, being only gravity (for most the models) the one allowed to propagate in the bulk. This is the
main feature of the Randall-Sundrum model since it solves the hierarchy
problem (cf. chapter 5). We will then consider a matter field
configuration as a section of a fibre bundle
$\mathcal{F}_I(V_{n_I},F,\mathcal{Y})$ defined over a brane
$V_{n_I}\subset\mathcal{M}$.

\paragraph{Dynamical Localisation.} The idea of localisation of matter fields into submanifolds of a multidimensional
universe traces back to \cite{rubakov1,rubakov2}, where a
multidimensional universe of the form $\mathcal{M}=V_4\times\mathcal{R}$ was
considered. Matter fields of the scalar type were defined over $\mathcal{M}$ but a special
potential was used \cite{kubyshin}:
\begin{equation}
  V(\phi)=\frac\lambda{4}\left(\phi^2-\frac{m^2}{\lambda}\right)^2.
\end{equation}
For this potential there is a kink solution that depends only on the fifth
coordinate:
\begin{equation}
  \phi_0(y)=\frac{m}{\sqrt{\lambda}}\tanh\frac{m(y-y_0)}{\sqrt{2}},
\end{equation}
this solution being centred at $y=y_0$ and providing some type of localisation
around that coordinate point. This simple model provides an example of
dynamical localisation of fields on a submanifold of a multidimensional
universe. 


\section{Dimensional Reduction}

\subsection{Einstein Equations and Israel Junction Conditions}
The classical action of an Randall-Sundrum scheme will
be given by an Hilbert-Einstein term with a cosmological constant $\Lambda_m$ in $\mathcal{M}$, $b$ Gibbons-Hawking terms with
the respective brane cosmological constants
$\{\Lambda_{n_I}^I\}_{I=1,\cdots,b}$, and a matter field action of fields
located on the branes\footnote{we will consider $n_I=n=m-1$ in this section.} and on the bulk,
\begin{equation}\label{pop}
S[\gamma,g_1,\cdots,g_b,\phi]=S_{HE}[\gamma]+S_{GH}[g_1,\cdots,g_b]+S_{BM}[g_1,\cdots,g_b,\phi]+S_M[\gamma,\phi]
\end{equation}
with
\begin{eqnarray}
S_{HE}[\gamma]=\frac{1}{2\kappa^2}\int_\mathcal{M}d^mz\sqrt{-\gamma}\left(R-2\Lambda_m\right),\\
S_{GH}[g_1,\cdots,g_b]=\frac{1}{\kappa^2}
\sum_{I=1}^b\int_{V_{n_I}^I}d^{n_I}x_I\sqrt{-g_I}\left(K^I-\Lambda_{n_I}^I\right),\\
S_{BM}[g_1,\cdots,g_b,\phi]=\sum_{I=1}^b\int_{V^I_{n_I}}d^{n_I}x_I L_{BM}^I(\phi,\partial\phi,g_I),\\
S_M[\gamma,\phi]=\int_\mathcal{M}d^mz L_M(\phi,\partial\phi,\gamma),
\end{eqnarray}
and where $K^I$ is the trace of the extrinsic curvature of the $I$ brane, 
that can be written in terms of a normal vector $n^I$ to $V_{n_I}^I$ 
as \cite{kraus}
\begin{equation}
K^I_{\hat\alpha\hat\beta}=\tilde\nabla_{\hat\alpha}n_{\hat\beta}^I.
\end{equation}
If we take Gaussian normal coordinates to the brane $V_{n_I}$, we can simply
take
\begin{equation}\label{sdfd}
  K^I_{\hat\alpha\hat\beta}=\frac12\partial_{\eta}\gamma_{\hat\alpha\hat\beta},
\end{equation}
with $\delta_\eta$ the derivative along the normal coordinate to the brane.

By taking the variation of this action (\ref{pop}) with respect to $\{\gamma,g_I\}_{I=1,\cdots,b}$,
\begin{equation}
\begin{array}{r}
\delta S[\gamma,g_I]=\int_\mathcal{M}d^mz\sqrt{-\gamma}\left[-\frac{1}{2\kappa^2}\left(R_{\hat\alpha\hat\beta}-
\frac12\gamma_{\hat\alpha\hat\beta}R+\Lambda_m\gamma_{\hat\alpha\hat\beta}\right)+\frac12T_{\hat\alpha\hat\beta}
\right]\delta\gamma^{\hat\alpha\hat\beta}+\\
\sum_{I=1}^b\int_{V_{n_I}^I}d^{n_I}x_I\sqrt{-g_I}\left[-\frac1{2\kappa^2}\left(K^I_{\alpha\beta}-g^I_{\alpha\beta}K^I
+\Lambda_{n_I}^Ig^I_{\alpha\beta}\right)+
\frac12T^I_{\alpha\beta}\right]\delta g^{\alpha\beta},
\end{array}
\end{equation}
where
\begin{eqnarray}
T_{\hat\alpha\hat\beta}=\frac2{\sqrt{-\gamma}}\frac{\delta S_M[\gamma,\phi]}{\delta \gamma^{\alpha\beta}},\\
T^I_{\alpha\beta}=\frac2{\sqrt{-g^I}}\frac{\delta S_{BM}[g_1,\cdots,g_b,\phi]}{\delta g^{\alpha\beta}_I},
\end{eqnarray}
are the $m$-dimensional stress-energy tensor of the bulk matter fields and the $n_I$ dimensional stress-energy tensor 
of the matter fields on $V_{n_I}^I$, we obtain the Einstein equations,
\begin{equation}
\frac{\delta S[\gamma,g_I,\phi]}{\delta\gamma^{\hat\alpha\hat\beta}}=-\frac{1}{2\kappa^2}\sqrt{-\gamma}\left(R_{\hat\alpha\hat\beta}
-\frac12\gamma_{\hat\alpha\hat\beta}R+\Lambda_m\gamma_{\hat\alpha\hat\beta}\right)+\frac12 \sqrt{-\gamma}T_{\hat\alpha\hat\beta}=0,
\end{equation}
which must be subjected to the Israel junction conditions,
\begin{equation}
  \frac{\delta S[\gamma,g_I,\phi]}{\delta g^{\hat\alpha\hat\beta}_I}=-\frac1{2\kappa^2}\sqrt{-g_I}\left(
    \triangle[K^I_{\alpha\beta}-g^I_{\alpha\beta}
  K^I]+\Lambda^I_{n_I}g^I_{\alpha\beta}\right)+\frac12 \sqrt{-g_I}T^I_{\alpha\beta}=0,
\end{equation}
for $I=1,\cdots,b$.

In order to find a solution we have to solve the Einstein equation
\begin{equation}\label{opo1}
  R_{\hat\alpha\hat\beta}-\frac12\gamma_{\hat\alpha\hat\beta}R+\Lambda_m\gamma_{\hat\alpha\hat\beta}=\kappa^2T_{\hat\alpha\hat\beta},
\end{equation}
in the space between the branes - the bulk -, and then assure the jump over all the branes,
\begin{equation}\label{opo2}
\triangle[K^I_{\alpha\beta}-g^I_{\alpha\beta}K^I]+\Lambda^I_{n_I}g^I_{\alpha\beta}=\kappa^2 T^I_{\alpha\beta}.
\end{equation}

Another way of writing the Israel junction conditions is \cite{kraus},
\begin{equation}
\triangle K^I_{\alpha\beta}+\frac13\Lambda_{n_I}^Ig^I_{\alpha\beta}=\kappa^2\left(T^I_{\alpha\beta}-\frac13 T^{\alpha\;I}_\alpha 
g^I_{\alpha\beta}\right),
\end{equation}
where
\begin{equation}
\triangle K^I_{\alpha\beta}=K^{I\;+}_{\alpha\beta}-K^{I\;-}_{\alpha\beta},
\end{equation}
with
$$K^{I\;\pm}_{\alpha\beta}=\lim_{z\rightarrow (x^I,0^\pm)}K^I_{\alpha\beta},$$
being the limit taken along the normal $n^I$ to the brane.

\paragraph{The Randall-Sundrum solution.} Let us consider a 5-dimensional
universe of the form $\mathcal{M}=\bar\mathcal{M}\times S^1/\mathcal{Z}_2$,
where the internal space is a orbifold with a $\mathcal{Z}_2$ symmetry \cite{randall2}. We
will consider that there are two four-dimensional branes in this universe, and by choosing a
coordinate $y\in[0,d]$ for the internal space we will suppose that they lie
at the fixed points $y=0$ and $y=d$. In order to find the ground state for
this model, we consider the following action,
\begin{equation}
S[\gamma,g^\pm]=\frac1{2\kappa^2}\left(\int d^4x\int_{-d}^d
dy\sqrt{-\gamma}(R-2\Lambda_5)-2\Lambda^+_4\int d^4 x \sqrt{-g^+}-2\Lambda^-_4\int d^4 x \sqrt{-g^-}\right),
\end{equation}
where we have represented the brane at $y=0$ by the plus sign and that at
$y=d$ by the minus sign. We will then get the following system of equations,
\begin{equation}
R_{\hat\alpha\hat\beta}-\frac12\gamma_{\hat\alpha\hat\beta}R+\gamma_{\hat\alpha\hat\beta}\Lambda_5=0,
\end{equation}
and
\begin{equation}
\triangle K^{\pm}_{\alpha\beta}+\frac13\Lambda_4^\pm g^\pm_{\alpha\beta}=0,
\end{equation}
with
\begin{equation}
\triangle K^\pm_{\alpha\beta} = \frac12 \left(\partial_y
  g^\pm_{\alpha\beta}(y^\pm|_\mp)-\partial_y g^\pm_{\alpha\beta}(y^\pm|_\mp)\right),
\end{equation}
being $y^\pm|_\mp$ the limit taken from both sides of the position of the
brane.

Since we have chosen Gaussian normal coordinates for both branes that overlap,
we can simply take
\begin{equation}
  g^\pm = \gamma(y^\pm)|_\pm.
\end{equation}

The equations can then be written in terms of $\gamma$ only,
\begin{eqnarray}
  R_{\hat\alpha\hat\beta}-\frac12\gamma_{\hat\alpha\hat\beta}R+\gamma_{\hat\alpha\hat\beta}\Lambda_5=0, \\
\partial_y \gamma_{\alpha\beta}(y^\pm|_\mp)|_\pm-\partial_y
  \gamma_{\alpha\beta}(y^\pm|_\mp)|_\pm+\frac23\Lambda^\pm_4\gamma(y^\pm)|_\pm =0.
\end{eqnarray}

The Randall-Sundrum ansatz is
\begin{equation}\label{randalmetric}
\gamma=\left(\begin{array}{cc} e^{-2\phi(y)}\eta_{\alpha\beta} & 0 \\ 0 &
    1 \end{array}\right)
\end{equation}
By inserting it in the previous equations, we get for $\phi(y)$,
\begin{equation}
\phi(y) = |y|/l,
\end{equation}
with 
\begin{equation}
l= \sqrt{-\frac6{\Lambda_5}}
\end{equation}
and for $\Lambda_4^\pm$,
\begin{equation}
\Lambda_4^\pm=\pm \frac{\sqrt{-6\Lambda_5}}{\kappa^2}.
\end{equation}
We then conclude that the bulk of this model must be a slice of an $AdS_5$
geometry, being the observed universe the brane located at $y=y^-=d$.

\paragraph{Bulk's Black Hole Solutions.} We can solve the Einstein equations
together with the Israel conditions by choosing a known solution for the bulk and
then to introduce the wanted number of branes, that will be considered as
domain walls, into the bulk - their position unknown at the beginning, and not
fixed as in the Randall-Sundrum model - allowing them to move, separating
different regions of the bulk. We could, for instance, consider
Robertson-Walker type universes \cite{kraus,csaki},
\begin{equation}
\gamma=\left(\begin{array}{ccc} -h_k(y) & 0 & 0 \\ 0 & \frac{y^2}{l^2} g^k_{ij} &
    0 \\ 0 & 0 & h_k^{-1}(y)\end{array}\right),
\end{equation}
with $l$ given by the same expression as in the Randall-Sundrum, $g^k_{ab}$
being the metric corresponding to a unit three dimensional plane, an
hyperboloid or a sphere for $k=0,-1,+1$, respectively and
\begin{equation}
  h_k(y)=k+\frac{y^2}{l^2}-\frac\mu{y^2}.
\end{equation}
The geometry of the bulk assumes then a form of the geometry around a black
hole \cite{gomez}. The $\mu$ parameter can then be considered to be the mass of that black
hole. The singularity at $y=0$ will be hidden by an horizon at $y=y_h$, such
that $h(y_h)=0$. The case of a bulk's black hole in a six-dimensional case was
analysed in \cite{brower}.
More general metrics can be considered by supposing that the bulk black hole
possesses a charge $Q$ in relation to some Yang-Mills field defined on the
bulk \cite{csaki}. We would then have a AdS-Reissner-Nordstrom metric, with
\begin{equation}
  h_k(y)=k+\frac{y^2}{l^2}-\frac\mu{y^2}+\frac{Q^2}{y^4}.
\end{equation}
This situation however will be of no use to us.

By introducing only one brane - without matter on it, since we are only
determining
the ground state of the theory -, we would find out that its position along the
extra dimension, $y=R(t)$ would vary with time. In fact, we should get the
following motion equation for the brane\cite{kraus},
\begin{equation}
  \frac12 \frac{dR}{dt}+V(R)=-\frac{k}2,
\end{equation}
with
\begin{equation}
V(R)=\frac12\left(1-\left(\frac{\Lambda_4}{\Lambda_c}\right)\right)R^2-\frac{(\mu_++\mu_-)}{4R^2}-\frac1{32}\left(\frac{\Lambda_4}{\Lambda_c}\right)^{-2} \frac{l^2(\mu_+-\mu_-)^2}{R^6},
\end{equation}
with $\Lambda_4$ being the brane tension, $\mu_\pm$ the values of $\mu$ on
both sides of the brane and $$\Lambda_c=\frac{\sqrt{-6\Lambda_5}}{\kappa^2}.$$

By taking $k=\mu_\pm=0$ and $\Lambda_4=\Lambda_c$ we recover the
Randall-Sundrum model with a brane.

\subsection{Construction of the Effective Theory}

After obtaining a solution for the bulk metric $\gamma$, the metrics $g^I$
will be given by the restriction of $\gamma$ to the space occupied by the
branes. In general one can't simply take $g^I=\gamma|_{V_{n_I}}$ if one wishes
to use (\ref{sdfd}) since normal coordinates for one brane will not in general
be normal coordinates for another one. We use then different charts for the
different branes, being all normal to the brane that they are associated. In
general these different charts will not overlap, and so we can not simply make
$g^I=\gamma|_{V_{n_I}}$ for a given brane. By using gauge invariance we can
bypass this problem, adding to $\gamma$ $SO(1,m-1)$ gauge invariant terms
and then identifying its restriction to $V_{n_I}$ with $g^I$. This
gauge invariance of $\gamma$ will result into the introduction of various
scalar fields - the so called radion fields. 

\paragraph{The Linearised Theory of the Randall-Sundrum Model.}

Let us consider small perturbations $\tilde h_{\hat\alpha\hat\beta}$ around the Randall-Sundrum solution, with $\gamma_0$ given by (\ref{randalmetric}),
\begin{equation}
  \gamma_{\hat\alpha\hat\beta}=\gamma_{\hat\alpha\hat\beta}^0+\tilde h_{\hat\alpha\hat\beta}.
\end{equation}
For $\gamma_{\hat\alpha\hat\beta}$ to be also a solution of the field
equations, $\tilde h$ must obey 
\begin{equation}
\left(\frac{d^2}{dy^2}-\frac{4}{l^2}+e^{4\phi(y)}\partial_\mu\partial^\mu\right)\tilde h_{\alpha\beta}=0,
\end{equation}
where, for simplicity we have chosen the Randall-Sundrum gauge \cite{garriga,barvinsky}:
$$\tilde h_{44}=\tilde h_{\mu 4} = \partial_\nu\tilde h_{\mu}^\nu=\tilde h^\mu_\mu=0.$$

In order to apply the boundary conditions on the branes, we shall define two
charts such that for a given chart, one of the branes will be written in
Gaussian normal coordinates. Of course these two sets of charts will not
overlap completely and so we can't write the metrics in the branes directly in
terms of the bulk metric as was done before for the Randall-Sundrum model. 
We make then use of the coordinate transformation gauge invariance of $\gamma$
to write \cite{garriga,barvinsky} (this is the most general gauge transformation),
\begin{equation}
  \tilde h^\pm_{\alpha\beta}(x,y)=\tilde
  h_{\alpha\beta}(x,y)+l\partial_\alpha\partial_\beta\xi^\pm(x)+\frac2l\eta_{\alpha\beta}e^{-4\phi(y)}\xi^\pm(x)+\frac12e^{-4\phi(y)}\left(\partial_\alpha\xi_\beta^\pm+\partial_\beta\xi_\alpha^\pm\right),
\end{equation}
with $\xi^\pm(x)$ a two four-dimensional scalar field and $\xi^\pm_\alpha(x)$
two four-dimensional vector fields. In the Randall-Sundrum gauge the branes will not be located at $y=0$ and
$y=d$, instead, its location will be given by
$y^\pm=y^\pm_0-\xi^\pm(x^\alpha)$, where $y^\pm_0$ are the initials positions
of the branes. It is this
the reason for performing the last gauge transformation.

The Israel junction conditions take then the form
\begin{equation}
  \left(\frac{d}{dy}+\frac2l\right)\tilde h_{\alpha\beta}(x,y_\pm) = -2\partial_\alpha\partial_\beta\xi^\pm(x).
\end{equation}
By taking the trace of this equation, and considering that we are in the
Randall-Sundrum gauge, we get for the scalar fields \cite{garriga,barvinsky}- also called radion
fields -,
\begin{equation}
  \Box \xi^\pm(x)=0.
\end{equation}

In the presence of matter $T^\pm_{\alpha\beta}\neq 0$ on the branes, we should get
\cite{garriga,barvinsky}
\begin{equation}
\left(\frac{d}{dy}+\frac2l\right)\tilde h_{\alpha\beta}(x,y_\pm) = \mp\kappa\left(T_{\alpha\beta}^\pm-\frac13\eta_{\alpha\beta}T^\pm\right)-2\partial_\alpha\partial_\beta\xi^\pm(x).
\end{equation}
and
\begin{equation}
  \Box\xi^\pm(x)=\pm\frac16\kappa T^\pm(x),
\end{equation}

In general, we redefine the radion fields by taking then to be dimensionless,
\begin{equation}
\psi^+(x)=\frac{\xi^+(x)}{l},\;\; \psi^-(x)=e^{-2d/l}\frac{\xi^-(x)}{l},
\end{equation}
with $d$ the distance between the branes.


\subsection{The Hierarchy}

\paragraph{The Randal-Sundrum Model.} Let us consider the Randal-Sundrum model
with a complex scalar field $\xi$ defined over the visible brane, i.e., the $V^-_4$ brane. Its
action will be given by
\begin{equation}
  S_s[\xi,\xi^\dagger]=\int_{V^-_4} d^4x\sqrt{-g^-}\left(g^{-\alpha\beta}
  D_\alpha\xi^\dagger D_\beta\xi-V(\xi)\right).
\end{equation}
By writing $\gamma_{\alpha\beta}(d)=g^-_{\alpha\beta}$ and taking
$\gamma_{\alpha\beta}(d)=e^{-2d/l}\eta_{\alpha\beta}$, we shall have,
\begin{equation}
  S_s[\xi,\xi^\dagger]=\int_{V^-_4} d^4x
  e^{-4\phi/l}\left(e^{2d/l}\eta^{\alpha\beta}D_\alpha\xi^\dagger
  D_\beta\xi-V(\xi)\right),
\end{equation}
and by rescaling the scalar field $\xi\rightarrow e^{d/l}\xi$, we shall get
\begin{equation}
  S_s[\xi,\xi^\dagger]=\int_{V^-_4} d^4x
  \left(\eta^{\alpha\beta}D_\alpha\xi^\dagger
  D_\beta\xi-V(e^{d/l}\xi)\right).
\end{equation}
If we take $V$ to lead to spontaneous symmetry breaking,
\begin{equation}
V(\xi)=\frac14c\left(\parallel\xi\parallel^2-\upsilon\right)^2+\beta,
\end{equation}
we shall have for the potential of the rescaled field,
\begin{equation}
  V(\xi)=\frac14c\left(\parallel\xi\parallel^2-e^{-2d/l}\upsilon\right)^2+\beta.
\end{equation}
We have then made a rescaling of the non-null expectation vacuum value of the
scalar field: $\upsilon\rightarrow e^{-2d/l}\upsilon$, thus changing the
scale at which the spontaneous symmetry breaking will occur. By conveniently
choosing the parameter $l$, we can take the Planck scale to be coincident with
the weak scale, obtaining only one fundamental scale, solving the
hierarchy problem.

\subsection{Quantum Effects}

\paragraph{Stabilization of the Radion Fields.} One of the questions that
could be risen after the construction of a type-like Randall-Sundrum model is
whether such a model would be stable. Taking, the Randall-Sundrum two-brane
solution, one could ask if the distance between the two branes could vary in
such a way that the full system could become unstable. By other words, the
radion fields, $\psi^\pm$, could be unstable. In order to verify if that is
so, let us construct their action:
\begin{equation}
S[\psi^\pm]=\int_{V_4^\pm} d^4x\sqrt{-g^\pm}\left(\frac12
  g^{\pm\alpha\beta}\partial_\alpha\psi^\pm\partial_\beta\psi^\pm\right).
\end{equation}
In general, both fields will be subjected to a non-null scalar field effective
potential, and so their effective action will be given by
\begin{equation}
\Gamma[\psi^\pm]=\int_{V_4^\pm} d^4x\sqrt{-g^\pm}\left(\frac12
  g^{\pm\alpha\beta}\partial_\alpha\psi^\pm\partial_\beta\psi^\pm+V^\pm_{eff}(\psi^\pm)\right).
\end{equation}
At a first stage, let us consider, as suggested in \cite{goldberger2}, the
existence of massless bulk fields and let us find their contribuition for
$V^\pm_{eff}$. Let us consider, for instance, a massless conformally coupled bulk's scalar field
$\Phi$:
\begin{equation}\label{opioi}
  \left(-\Box+\frac3{16}R\right)\Phi=0,
\end{equation}
where the $\Box$ is written using the Randall-Sundrum metric solution
(\ref{randalmetric}).
By taking the following redefinition of the scalar field,
\begin{equation}
\Phi\rightarrow\tilde\Phi=e^{-3\phi(y)}\Phi,
\end{equation}
we can rewrite (\ref{opioi}) as the following equation for $\tilde\Phi$,
\begin{equation}
\partial_{\hat\alpha}\partial^{\hat\alpha}\tilde\Phi=0,
\end{equation}
having for solution $\tilde\Phi=e^{ik\cdot z}$. By forcing $\tilde\Phi$ to the
boundary conditions
\begin{equation}
  \partial_4\tilde\Phi=0,
\end{equation}
for $y=y^\pm$. The eigeinvalues of
$\partial_{\hat\alpha}\partial^{\hat\alpha}$ will then be
\begin{equation}
  \lambda^2_{n,k_{(4)}} = k_{(4)}^2+\left(\frac{n\pi}{d}\right)^2,
\end{equation}
where $k_{(4)}$ denotes the four-momentum along the branes direction.

We can now determine the contribution of the one-loop quantum correction
(\ref{zeta}) for the scalar potential of $\tilde\Phi$. We have for the
zeta-function,
\begin{equation}
\zeta\left(-\frac{\partial_{\hat\alpha}\partial^{\hat\alpha}}{\mu^2}|s\right)=\sum_{n=0}^{+\infty}\int
\frac{d^4k}{(2\pi)^4}\left(\lambda^2_{n,k}\right)^{-s}.
\end{equation}
After performing the integration in $k_{(4)}$ we get
\begin{equation}
  \zeta\left(-\frac{\partial_{\hat\alpha}\partial^{\hat\alpha}}{\mu^2}|s\right)=\frac{\mu^{2s}}{16\pi^{2s-2}}\frac{\Gamma(s-2)}{\Gamma(s)}d^{2s-4}\zeta_R(2s-4),
\end{equation}
with
\begin{equation}
\zeta_R(s)=\sum_{n=0}^{+\infty}n^{-s},
\end{equation}
the Riemann zeta-function.

We shall then get for the one-loop term,
\begin{equation}
   \tilde\Gamma^{(1)}(\tilde\Phi)=\tilde\Gamma^{(1)}=-\frac{\pi^2}{2^5d^4}\left(\left[\ln\left(\frac{\mu
   d}{\pi}\right)\right]\zeta_R(-4)+\zeta_R'(-4)\right),
\end{equation}
where $d$ is the distance between the branes.
We can now make the rescaling for the $\Phi$ potential, getting
\begin{equation}
V(d)=\tilde V(d)e^{-3\phi(y)},
\end{equation}
with $V(d)$ the potential obtained from $\Gamma^{(1)}$.

The effective potential will be given by
\begin{equation}
V_{eff}(d)=\int_{-d}^d dy \sqrt{+e^{-4|y|/l}}e^{-3|y|/l}\tilde
V(d)=\frac{2l}{5}\left(e^{-5d/l}-1\right)\tilde V(d),
\end{equation}
i.e.,
\begin{equation}
V_{eff}(d)=\frac{l\pi^2}{2^4\cdot 5
  d^4}\left(1-e^{-5d/l}\right)\zeta'_R(-4),
\end{equation}
and where we have used $\zeta_R(-4)=0$.

If we take, for instance, the brane ``+'' to be fixed in the position
$y=0$, we shall have for the radion field
\begin{equation}
\xi_-(x)=d_0-d,
\end{equation}
where $d_0$ is the initial position of the brane ``-''. The radion field
$\xi_-$ will then be subjected to the effective potential,
\begin{equation}
V_{eff}(\xi_-)=\frac{l\pi^2}{2^4\cdot 5
  (d_0-\xi_-)^4}\left(1-e^{5\xi_-/l}e^{-5d_0/l}\right)\zeta'_R(-4).
\end{equation}

The radion field mass contribution due to the Casimir effect will be given by
\begin{equation}
\delta m_\xi^2=\frac{d^2
  V_{eff}}{d\xi^2}(0)=-\frac{\pi^2\zeta_R'(-4)}{16d_0^6l}e^{-5d_0/l}\left(5d_0^2+8d_0l+4l^2\left[1-
  e^{5d_0/l}\right]\right),
\end{equation}
and will have a negative value, thus leading to instability. If the calculation was performed for 
a Dirac spinor, the result would be the symmetric from this, conducing to stability. Depending on the 
bulks content, the radion, and so the hierarchy, can be stabilised - we must have a greater number of
fermionic fields than of bosons. In \cite{garriga2} it is shown that gravity conduces, by Casimir effect,
to a negative square mass for the radion, and thus can not, by itself, lead to stability. 
In \cite{goldberger2} a classical potential for the bulk's fields was introduced in order to force the 
Randall-Sundrum construction to be stable.

\chapter{CONCLUSIONS}

The wish of a theory that could unified all interactions by a common principle
has been a constant over Physic's history. The non-Abelian Kaluza-Klein
theories have been introduced as a premature response to that need, being
today more appreciated for their elegance than for their original purpose. They
find their beauty on their accurate mathematical description, reducing all
physical principles to a geometric construction where the universe is the most
likeble of all the places, containing no fundamental forces. This idilic image of
the world as been, however, overtrone by the recent unifying theories such as
Superstrings and Supergravity. However these become inaccessible 
 without the dimensional reduction process caracteristic to the Kaluza-Klein theories, and
constantly used in the construction of their effective field theories.

In this work we have analysed such process for both symmetric and general
fields, performing the dimensional reduction of gravity, matter and gauge
fields over a multidimensional universe with a fibre bundle structure. It has
been found that such process, when applied to a gauge field will inevitably conduce, first to a geometric
symmetry breaking of the initial gauge group, and secondly to spontaneous symmetry
breaking of the gauge group thus obtained, led by a scalar field that has its
origin in the dimensional reduction process itself. The potential responsable
for such symmetry breaking was explicitly calculated for the case in which the
internal space is symmetric.
It was also presented the spontaneous compactification mechanism by which a manifold without a special
structure can dynamicly achieve a fibre bundle structure, thus being able of
being reduced.

The construction of models within the Kaluza-Klein cenario was studied and its
dificulties presented. The chiral and the hierarchy problems were discussed
and some solutions given. By using the Kaluza-Klein dimensional reduction
process, the Randall-Sundrum cenario can be obtained, taking for the starting
point, the M-theory.

Some of the principal charateristics of Randall-Sundrum type-like models were
analysed and its main feature studied: the solution of the hierarchy problem. Spontaneous
symmetry breaking assumes through the Randall-Sundrum theory a fundamental
role in defining the fundamental scale of nature. The stability of the
hierarchy through the Randall-Sundrum model was partially analised: the Casimir
effect was presented as a stabilization mechanism. 
Randall-Sundrum appeals also by its rich phenomenology since it can be tested in the next generation
of particle acelerators (such as the LHC).


\bibliographystyle{plain}






\end{document}